\documentclass[a4paper]{book}

\usepackage[utf8]{inputenc}
\usepackage[T1]{fontenc}
\usepackage[nottoc,notlot]{tocbibind}
\usepackage{amsmath, amssymb, amsthm}
\usepackage{booktabs}
\usepackage[strict]{changepage}
\usepackage{multirow}
\usepackage{enumerate}
\usepackage{xspace}
\usepackage{hyperref}
\usepackage[all]{hypcap}
\usepackage{graphicx}
\usepackage{subfig}
\usepackage{listings}
\usepackage[table]{xcolor}
\usepackage{tikz}
\usepackage{pgfplots}
\usepackage[hmargin=3cm]{geometry}

\colorlet{graybg}{gray!10}
\colorlet{plot1}{red}
\colorlet{plot2}{green!75!black}
\colorlet{plot3}{blue}
\colorlet{plot4}{yellow!75!gray}
\colorlet{plot5}{magenta}
\colorlet{plot6}{cyan}
\colorlet{plot7}{orange}
\colorlet{plot8}{brown}

\colorlet{plot1bg}{plot1!20}
\colorlet{plot2bg}{plot2!20}
\colorlet{plot3bg}{plot3!20}
\colorlet{plot4bg}{plot4!20}

\usetikzlibrary{
    external,
    scopes,
    patterns,
    decorations.pathreplacing,
    calc,
    positioning,
    plotmarks,
    fadings,
    3d
}
\pgfplotsset{
    every axis/.append style={
        axis background/.style={
            fill=graybg
        },
        xmin=0,
        ymin=0
    },
    every axis legend/.append style={
        cells={
            anchor=west
        }
    },
    twocolplot/.style={
        width=.45\textwidth
    }
}
\pgfkeys{
    /tikz/external/mode=list and make
}
\tikzsetexternalprefix{figures/externalized/}
\tikzset{external/export=false}

\hypersetup{
    pdftitle={Hierarchical Performance Modelling for Ranking Dense Linear Algebra Algorithms},    
    pdfauthor={Elmar Peise},     
    pdfsubject={},   
    pdfkeywords={blocked algorithms} {performance modelling} {ranking}, 
    pdfnewwindow=true,      
    colorlinks=true,       
    linkcolor=blue!50!black,          
    citecolor=green!50!black,        
    filecolor={.5 0 0},      
    urlcolor=red!50!black           
}

\newcommand{\appendixref}[1]{\hyperref[#1]{Appendix~\ref*{#1}}}

\numberwithin{figure}{chapter}

\newcommand{\overparameq}[2]{%
\tikz[baseline=(n.base)] {
    \node[inner sep=0pt, label=90:] (n) {\texttt{\vphantom{T}#2}};
    \node[inner sep=0pt, anchor=south, above=1mm of n.north]
        {\footnotesize \textcolor{gray}{\texttt{\vphantom{g}#1}}};
}%
}
\newcommand{\tikzball}[1]{\tikz \path plot[mark=ball, ball color=#1] coordinates {(0, 0)};}
\newcommand{\tikzsquare}[1]{\tikz \filldraw[fill=#1] (0, 0) rectangle (.25, .25);}
\newcommand{\tikzline}[1]{\tikz[baseline=-.75ex] \draw[#1] (0, 0) -- (.5, 0);}

\usepackage[OMLmathsfit, sfdefault=iwona, scaled=1.0]{isomath}
\newcommand{\metric}[1]{\ensuremath{\text{\small\itshape\sffamily #1}}}
\newcommand{\smallmetric}[1]{\ensuremath{\text{\tiny\itshape\sffamily #1}}}
\renewcommand{\metric}[1]{\ensuremath{\mathsfit{#1}}}
\renewcommand{\smallmetric}[1]{\metric{#1}}
\newcommand{\gotoblas}{GotoBLAS2\xspace}

\newcommand*{\bigtimes}{\mathop{\raisebox{-.5ex}{\hbox{\huge{$\times$}}}}}

\theoremstyle{definition}
\newtheorem*{example}{Example}

\title{Hierarchical Performance Modeling for\\Ranking Dense Linear Algebra Algorithms}
\author{Elmar Peise}

\begin{document}
    \pagenumbering{roman}

\newgeometry{margin=3cm}
\begin{titlepage}
    \thispagestyle{empty}
    \centering
    \large


    \includegraphics[height=1.5cm]{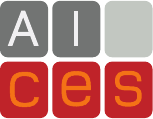}
    \hfill
    \includegraphics[height=1.5cm]{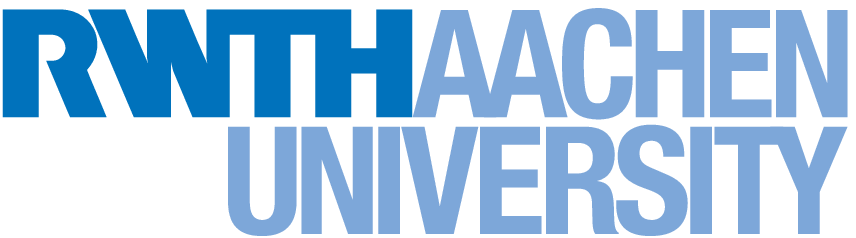}

    \rule{\linewidth}{.5pt}

    \vspace{1cm}

    {\sc Rheinisch-Westf\"alische Technische Hochschule Aachen}

    \vspace{.5cm}

    {\sc Aachen Institute for Advanced Study in\\Computational Engineering Science}

    \vfill

    {\sc \Large Master's Thesis}

    \vspace{.2cm}
    
    {\LARGE \bfseries Hierarchical Performance Modeling for\\Ranking Dense Linear Algebra Algorithms}

    \vspace{1.5cm}

    {\large Elmar Peise}

    \vfill

    {May 4\textsuperscript{th}, 2012}

    \vspace{1.5cm}

    \begin{minipage}{.3\textwidth}
        \centering
        {\sc Supervisor}\\
        Paolo Bientinesi
    \end{minipage}
    \hfill
    \begin{minipage}{.3\textwidth}
        \centering
        {\sc Co-Examiner}\\
        Martin B\"ucker
    \end{minipage}
\end{titlepage}
\restoregeometry

\newpage
{
    \thispagestyle{empty}
    \mbox{}
}

\newpage
{
    \thispagestyle{empty}

    \mbox{}

    \vfill

    I hereby declare that this thesis is entirely the result of my own work except where otherwise indicated.
    I have only used the resources given in the list of references.

    \vspace{2cm}

    Aachen, May 4\textsuperscript{th}, 2012 \hfill Elmar Peise \hspace{1cm}
}
    
\newpage
{
    \thispagestyle{empty}
    \mbox{}
}

    \tableofcontents

    \chapter{Introduction}
    \pagenumbering{arabic}
    \setcounter{page}{1}
    \label{sec:intro}
    A large class of dense linear algebra operations operations, such as LU decomposition or inversion of a triangular matrix, are usually performed by blocked algorithms.
For one such operation, typically, not only one but many algorithmic variants exist; depending on architecture, libraries, and problem size, each variant attains a different performances.
Our goal is to rank the algorithmic variants according to their performance for a given scenario \emph{without} executing them.

For this purpose, we analyze the routines upon which the algorithms are built and introduce a tool that, based on measurements, models their performance.
The generated performance models are then used to predict the performance of the considered algorithmic variants.
For a given scenario, these predictions allow us not only to rank the variants but to determine the optimal algorithmic block-size.

The strength of our approach mainly originates from the performance models.
Generated once for a given system, they can be used to analyze any number of blocked algorithms.
Besides reliably predicting and ranking algorithm performance, they yield further insight into how the performance is influenced by certain factors such as the block-size.

        \section{Motivating Example}
        \label{sec:intro.motivation}
        We consider an exemplary dense linear algebra operation: the inversion of a lower triangular matrix, $L \leftarrow L^{-1}$.
For this operation, there exist four different blocked algorithms (see \autoref{sec:intro.blockedalgs}).
For now, all we need to understand is that all of them take a lower triangular matrix $L \in \mathbb R^{n \times n}$ as an input and compute its inverse in place; they accept only one additional algorithm parameter: the algorithmic block-size $b$.

We use the following setup to analyze the four algorithms:
\begin{itemize}
    \item Their implementation (see \appendixref{app:blockedalgs.trinv} for the source code) is compiled with Intel's C Compiler (\texttt{icc}) version 12.0 \cite{iccpage}.
    \item They are executed on one core of an Intel Harpertown E5450 processor \cite{e5450} running at $2.99 \mathrm{GHz}$.
        This processor can issue $2$ double precision floating point operations\footnote{
            We consider a fused multiply add operation (FMA) $d \leftarrow a \cdot b + c$ to be \emph{one} floating point operation.
            It is the core of any dense matrix operation.
        } per clock cycle.
        Therefore, it can perform up to $\metric{peak\_flops/s} = 2 \times 2.99 \cdot 10^9$ floating point operations per second.
    \item The Intel's Math Kernel Library (MKL) \cite{mklpage} is used for the underlying Basic Linear Algebra Subroutines (BLAS) \cite{blas, blas3}.
\end{itemize}

\begin{figure}[t]
    \tikzset{external/export=true}
    \centering
    \subfloat[Execution time with $b = 96$ fixed]{
        \label{fig:intro.motivation.trinv:tn}
        \begin{tikzpicture}
            \begin{axis}[
                twocolplot,
                xlabel={matrix size $n$},
                ylabel={Time [s]},
                legend to name=fig:intro.motivation.trinv.legend,
                legend columns=-1,
                xtick={0,512,...,2048}
            ]
                \addlegendimage{plot1, only marks}
                \label{fig:intro.motivation.trinv:v1}
                \addlegendentry{variant 1}
                \addlegendimage{plot2, only marks}
                \label{fig:intro.motivation.trinv:v2}
                \addlegendentry{variant 2}
                \addlegendimage{plot3, only marks}
                \label{fig:intro.motivation.trinv:v3}
                \addlegendentry{variant 3}
                \addlegendimage{plot4, only marks}
                \label{fig:intro.motivation.trinv:v4}
                \addlegendentry{variant 4}

                \addplot[plot1, mark size=.2pt, only marks] file {figures/data/intro.motivation.trinv/trinv1.n.time.dat};
                \addplot[plot2, mark size=.2pt, only marks] file {figures/data/intro.motivation.trinv/trinv2.n.time.dat};
                \addplot[plot3, mark size=.2pt, only marks] file {figures/data/intro.motivation.trinv/trinv3.n.time.dat};
                \addplot[plot4, mark size=.2pt, only marks] file {figures/data/intro.motivation.trinv/trinv4.n.time.dat};
            \end{axis}
        \end{tikzpicture}
    }
    \hfill
    \subfloat[Efficiency with $b = 96$ fixed]{
        \label{fig:intro.motivation.trinv:en}
        \begin{tikzpicture}
            \begin{axis}[
                twocolplot,
                xlabel={matrix size $n$},
                ylabel={Efficiency [\%]},
                ymax=100,
                xtick={0,512,...,2048}
            ]
                \addplot[plot1, mark size=.2pt, only marks] file {figures/data/intro.motivation.trinv/trinv1.n.eff.dat};
                \addplot[plot2, mark size=.2pt, only marks] file {figures/data/intro.motivation.trinv/trinv2.n.eff.dat};
                \addplot[plot3, mark size=.2pt, only marks] file {figures/data/intro.motivation.trinv/trinv3.n.eff.dat};
                \addplot[plot4, mark size=.2pt, only marks] file {figures/data/intro.motivation.trinv/trinv4.n.eff.dat};
            \end{axis}
        \end{tikzpicture}
    }

    \subfloat[Execution time with $n = 1000$ fixed]{
        \label{fig:intro.motivation.trinv:tb}
        \begin{tikzpicture}
            \begin{axis}[
                twocolplot,
                xlabel={block-size},
                ylabel={Time [s]},
                xtick={0,128,...,512}
            ]
                \addplot[plot1, mark size=.2pt, only marks] file {figures/data/intro.motivation.trinv/trinv1.b.time.dat};
                \addplot[plot2, mark size=.2pt, only marks] file {figures/data/intro.motivation.trinv/trinv2.b.time.dat};
                \addplot[plot3, mark size=.2pt, only marks] file {figures/data/intro.motivation.trinv/trinv3.b.time.dat};
                \addplot[plot4, mark size=.2pt, only marks] file {figures/data/intro.motivation.trinv/trinv4.b.time.dat};
            \end{axis}
        \end{tikzpicture}
    }
    \hfill
    \subfloat[Efficiency with $n = 1000$ fixed]{
        \label{fig:intro.motivation.trinv:eb}
        \begin{tikzpicture}
            \begin{axis}[
                twocolplot,
                xlabel={block-size},
                ylabel={Efficiency [\%]},
                ymax=100,
                xtick={0,128,...,512},
            ]
                \addplot[plot1, mark size=.2pt, only marks] file {figures/data/intro.motivation.trinv/trinv1.b.eff.dat};
                \addplot[plot2, mark size=.2pt, only marks] file {figures/data/intro.motivation.trinv/trinv2.b.eff.dat};
                \addplot[plot3, mark size=.2pt, only marks] file {figures/data/intro.motivation.trinv/trinv3.b.eff.dat};
                \addplot[plot4, mark size=.2pt, only marks] file {figures/data/intro.motivation.trinv/trinv4.b.eff.dat};
            \end{axis}
        \end{tikzpicture}
    }

    \vspace{.5cm} 

    \tikzset{external/export=false}
    \ref*{fig:intro.motivation.trinv.legend}
    \caption{Inversion of a triangular matrix: execution time and efficiency.}
    \label{fig:intro.motivation.trinv}
\end{figure}
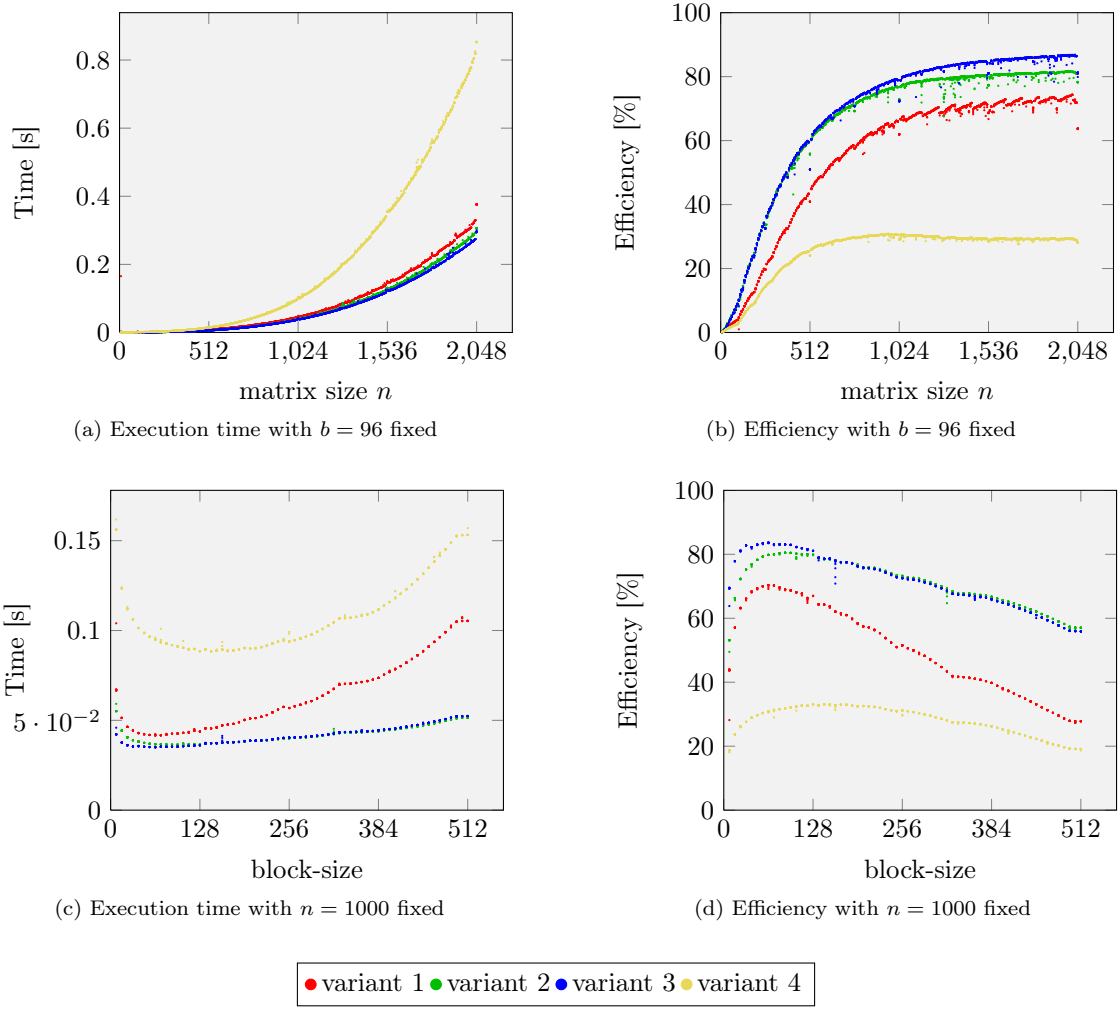

\autoref{fig:intro.motivation.trinv} shows the execution time and the efficiency\footnote{
    \autoref{sec:sampling.goal.metrics} contains details on the computation of efficiency.
} of the four algorithm variants.
In Figures \ref{fig:intro.motivation.trinv:tn} and \ref{fig:intro.motivation.trinv:en}, the algorithmic block-size is fixed to $96$ and the matrix size $n$ is varied.
The results show significant differences in performance between algorithms:
Variant~4~(\ref{fig:intro.motivation.trinv:v4}) takes more than twice as long compared to the other three variants; it reaches a maximum  efficiency of $29\%$.
Variant~1~(\ref{fig:intro.motivation.trinv:v1}) is also slower than variants 2~(\ref{fig:intro.motivation.trinv:v2}) and 3~(\ref{fig:intro.motivation.trinv:v3}) and attains an efficiency of $73\%$.
Variants 2~(\ref{fig:intro.motivation.trinv:v2}) and 3~(\ref{fig:intro.motivation.trinv:v3}) seem equally good for matrices up to size $n = 512$.
For larger matrices, variant~2~(\ref{fig:intro.motivation.trinv:v2}) becomes the fastest; ultimately, variants 2~(\ref{fig:intro.motivation.trinv:v2}) and 3~(\ref{fig:intro.motivation.trinv:v3}) reach efficiencies of $81\%$ and $86\%$, respectively.

In Figures \ref{fig:intro.motivation.trinv:tb} and \ref{fig:intro.motivation.trinv:eb}, the matrix size is fixed to $n = 1000$ and only the block-size varies.
For all variants, the efficiency decreases for very small and large block-sizes.
Variants 1~(\ref{fig:intro.motivation.trinv:v1}), 2~(\ref{fig:intro.motivation.trinv:v2}), and 3~(\ref{fig:intro.motivation.trinv:v3}) reach their peak efficiency for block-sizes close to $100$.

This example shows that in order to reach high efficiency, it is crucial to both choose the algorithmic variant as well as optimizing the block-size.
Due to the complexity of the architecture and the memory access patterns, it is not possible to determine the optimal configuration by only analyzing the algorithms mathematically.
On the contrary, the best choice depends on the matrix size, the underlying computational kernels, such as BLAS\footnote{
    An introduction to BLAS is given in \autoref{app:blas}.
}, and the processor architecture; changing these may lead to entirely different performance behavior.

In this thesis, we introduce tools and methods to analyze and model the performance of dense linear algebra kernels.
These tools allow us to perform the challenging task of ranking algorithmic variants according to their performance and determining the optimal block size.
Results for the example at hand are given in \autoref{sec:ranking.trinv}.

        \section{Related Work}
        \label{sec:intro.relwork}
        Several different approaches of using performance modeling in dense linear algebra already exist.

Iakymchuk et al. \cite{roman} model the performance of BLAS \cite{blas, blas3} analytically based on memory access patterns.
While their models represent the program execution very accurately, constructing them requires a high level of expertise on both the routines and the architecture.

Cuenca et al. \cite{solar} develop a self-optimizing linear algebra routines (SOLAR).
In their system, every routine is associated with performance information, which is hierarchically propagated to higher level routines (e.g., BLAS $\rightarrow$ LAPACK $\rightarrow$ ScaLAPACK); on each level, the information is used to tune the routines and associate according performance information.

Dongarra et al. \cite{hplmodeling} propose a modeling approach targeted at programs such as High Performance LINPACK (HPL) \cite{hplpage} and ScaLAPACK \cite{scalapack, scalapackpage}.
They employ sampling and polynomial fitting to construct their models and use them to extrapolate the performance of routines for larger problems and higher parallelism.

        \section{Outline of this Thesis}
        \label{sec:intro.outline}
        This thesis is structured as follows:

\begin{itemize}
    \item In \autoref{sec:intro.blockedalgs}, we give an introduction to blocked algorithms, the target of our predictions.

    \item In \autoref{sec:sampling}, we discuss the \emph{Sampler}, a tool that measures different performance metrics during the execution of dense linear algebra routines.

    \item In \autoref{sec:modeling}, we introduce the \emph{Modeler}, a framework that, based on the Sampler, generates analytical performance models for dense linear algebra routines.

    \item In \autoref{sec:ranking}, we use the performance models generated by the Modeler to predict the performance of blocked algorithms and rank them accordingly.

    \item In \autoref{sec:conclusion}, we conclude with a summary of our achievements and an outlook on possible future research directions based on this thesis.
\end{itemize}

        \section{Blocked Algorithms}
        \label{sec:intro.blockedalgs}
        Since our goal is to rank blocked algorithms, in this section, we introduce their structure.
We begin by studying blocked algorithms for a specific operation, the inversion of a triangular matrix (\ref{sec:intro.blockedalgs.trinv}), and then generalize the concepts to arbitrary operations (\autoref{sec:intro.blockedalgs.general}).

            \subsection{Triangular Inverse \texorpdfstring{$L \leftarrow L^{-1}$}{L <- inv(L)}}
            \label{sec:intro.blockedalgs.trinv}
            In \autoref{sec:intro.motivation}, we discussed the performance of four blocked algorithms for the inversion of a lower triangular matrix.
The algorithms take a lower triangular matrix $L \in \mathbb R^{n \times n}$ as an input and compute its inverse in place.

Within a blocked algorithm, the matrix $L$ is seen in a \emph{partitioned} form
$$
    L = \left(\begin{array}{cc}
        L_{TL} &0 \\
        L_{BL} &L_{BR}
    \end{array}\right),
$$
where $L_{TL} \in \mathbb R^{p \times p}$, $L_{BL} \in \mathbb R^{q \times p}$, and $L_{BL} \in \mathbb R^{q \times q}$ with $p + q = n$.
Initially, $p$ is $0$, or equivalently, $L_{TL}$ and $L_{BL}$ are "empty".
$L$ is then traversed from the top left corner along its diagonal 
\tikz[baseline=(A.base)] {
    \filldraw[fill=graybg] (0, 0) rectangle (1, 1);
    \draw[->, dotted] (.1, .9) -- (.9, .1);
    \node at (.5, .5) (A) {$L$};
} in steps of the algorithmic \emph{block-size}, increasing $p$ up to $n$.
During the traversal, the inverse of $L$ is computed in $L_{TL}$; this part of the matrix grows in size until $p = n$.
At this point, $L_{TL}$ is of the size $n \times n$ and contains the inverse of the original matrix $L$; the algorithm terminates.

Traversing $L$ not element-wise but in blocks of size $b$ has the advantage that BLAS Level-3 operations, such as \texttt{dgemm}, can be used to attain high performance.

\begin{figure}[t]
    \centering
    \begin{tikzpicture}
        \coordinate (pos) at (0, 0);
        \filldraw[fill=graybg] (pos) rectangle +(4, -4);
        \filldraw[fill=plot4bg] (pos) +(0, 0)  -- +(1, -1) -- +(0, -1) -- cycle;
        \filldraw[fill=plot3bg] (pos) +(0, -1) rectangle +(1, -4);
        \filldraw[fill=plot3bg] (pos) +(1, -1) -- +(4, -4) -- +(1, -4) -- cycle;
        \draw (pos) ++(0, -2) -- ++(2, 0) -- ++(0, -2);
        \draw[dashed] (pos) ++(1, 0) -- ++(0, -1);
        \draw[dashed] (pos) ++(2, 0) -- ++(0, -2);
        \foreach \x/\xl/\y/\yl in {
            .5/0/.5/0,
            .5/0/1.5/1, 1.5/1/1.5/1,
            .5/0/3/2,   1.5/1/3/2,   3/2/3/2
        }
            \path (pos) ++(\x, -\y) node {$L_{\yl\xl}$};
        \foreach \xa/\xb/\label in {0/1/p, 1/2/b, 2/4/r} {
            \draw[decorate, decoration=brace] (pos) ++(0,  .1) +(\xa,  0) -- +(\xb, 0) node[above, align=center, midway] {$\label$};
            \draw[decorate, decoration=brace] (pos) ++(-.1, 0) +(0, -\xb) -- +(0, -\xa) node[left, align=center, midway] {$\label$};
        }
        \draw[decorate, decoration=brace] (pos) ++(0, .6) +(0, 0) -- +(4, 0) node[above, align=center, midway] {$n$};

        \path (pos) ++(4.5, -2) node[rotate=90] {$\downarrow$ update $\downarrow$};

        \coordinate (pos) at (5, 0);
        \filldraw[fill=graybg] (pos) rectangle +(4, -4);
        \filldraw[fill=plot4bg] (pos) +(0, 0)  -- +(2, -2) -- +(0, -2) -- cycle;
        \filldraw[fill=plot3bg] (pos) +(0, -2) rectangle +(2, -4);
        \filldraw[fill=plot3bg] (pos) +(2, -2) -- +(4, -4) -- +(2, -4) -- cycle;
        \draw (pos) ++(0, -1) -- ++(1, 0) -- ++(0, -3);
        \foreach \x/\xl/\y/\yl in {
            .5/0/.5/0,
            .5/0/1.5/1, 1.5/1/1.5/1,
            .5/0/3/2,   1.5/1/3/2,   3/2/3/2
        }
            \path (pos) ++(\x, -\y) node {$L_{\yl\xl}$};

        \path (pos) ++(4.5, -2) node[rotate=90] {$\downarrow$ repartition $\downarrow$};

        \coordinate (pos) at (10, 0);
        \filldraw[fill=graybg] (pos) rectangle +(4, -4);
        \filldraw[fill=plot4bg] (pos) +(0, 0)  -- +(2, -2) -- +(0, -2) -- cycle;
        \filldraw[fill=plot3bg] (pos) +(0, -2) rectangle +(2, -4);
        \filldraw[fill=plot3bg] (pos) +(2, -2) -- +(4, -4) -- +(2, -4) -- cycle;
        \draw (pos) ++(0, -3) -- ++(3, 0) -- ++(0, -1);
        \draw[dashed] (pos) ++(2, 0) -- ++(0, -2);
        \draw[dashed] (pos) ++(3, 0) -- ++(0, -3);
        \foreach \x/\xl/\y/\yl in {
            1/0/1/0,
            1/0/2.5/1, 2.5/1/2.5/1,
            1/0/3.5/2, 2.5/1/3.5/2, 3.5/2/3.5/2
        }
            \path (pos) ++(\x, -\y) node {$L_{\yl\xl}$};
        \foreach \xa/\xb/\label in {0/2/p \leftarrow p + b, 2/3/b, 3/4/r}
            \draw[decorate, decoration=brace] (pos) ++(0, .1) +(\xa, 0) -- +(\xb, 0) node[above, align=center, midway] {$\label$};
    \end{tikzpicture}
    \caption{Triangular Inverse --- traversal of $L$.}
    \label{fig:intro.blockedalgs.trinv.traversal}
    \tikzset{external/export=false}
\end{figure}
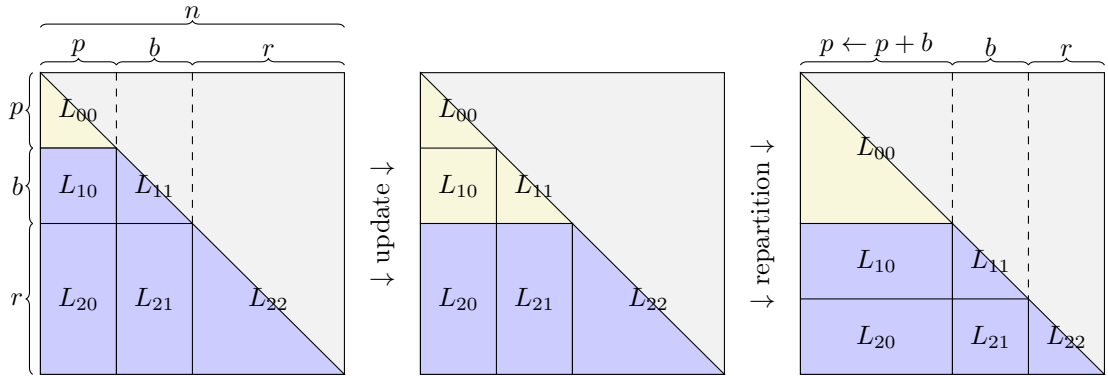

At each step of the matrix traversal (depicted in \autoref{fig:intro.blockedalgs.trinv.traversal}), $L$ is repartitioned as
$$
    \left(\begin{array}{c|c}
        L_{TL} &0      \\
        \hline
        L_{BL} &L_{BR}
    \end{array}\right)
    \rightarrow
    \left(\begin{array}{c|cc}
        L_{00} &0      &0      \\
        \hline
        L_{10} &L_{11} &0      \\
        L_{20} &L_{21} &L_{22}
    \end{array}\right),
$$
with $L_{11} \in \mathbb R^{b \times b}$, $L_{22} \in \mathbb R^{r \times r}$ (with $r = n - p - b$), and conforming sizes for the other submatrices.
(When $n$ is not divisible by $b$, $b$ is adjusted to $b \leftarrow n - p$ in the last step.)

At this point, the four algorithms perform different updates on the $3 \times 3$ partitioned form of the matrix:
\begin{center}
    \begin{tabular}{|>{\cellcolor{graybg}}p{.25\textwidth}|}
        \hline
        \multicolumn{1}{|>{\cellcolor{graybg}}c|}{Variant 1} \\
        \hline
        $L_{10} \leftarrow L_{10} L_{00}$ \\
        $L_{10} \leftarrow -L_{11}^{-1} L_{10}$ \\
        $L_{11} \leftarrow L_{11}^{-1}$ \\
        \hline
    \end{tabular}
    \hspace{.5cm}
    \begin{tabular}{|>{\cellcolor{graybg}}p{.25\textwidth}|}
        \hline
        \multicolumn{1}{|>{\cellcolor{graybg}}c|}{Variant 2} \\
        \hline
        $L_{21} \leftarrow L_{22}^{-1} L_{21}$ \\
        $L_{21} \leftarrow -L_{21} L_{11}^{-1}$ \\
        $L_{11} \leftarrow L_{11}^{-1}$ \\
        \hline
    \end{tabular}

    \vspace{.5cm}

    \begin{tabular}{|>{\cellcolor{graybg}}p{.25\textwidth}|}
        \hline
        \multicolumn{1}{|>{\cellcolor{graybg}}c|}{Variant 3} \\
        \hline
        $L_{21} \leftarrow -L_{21} L_{11}^{-1}$\\
        $L_{20} \leftarrow L_{21} L_{10} + L_{20}$\\
        $L_{10} \leftarrow L_{11}^{-1} L_{10}$\\
        $L_{11} \leftarrow L_{11}^{-1}$ \\
        \hline
    \end{tabular}
    \hspace{.5cm}
    \begin{tabular}{|>{\cellcolor{graybg}}p{.25\textwidth}|}
        \hline
        \multicolumn{1}{|>{\cellcolor{graybg}}c|}{Variant 4} \\
        \hline
        $L_{21} \leftarrow -L_{22}^{-1} L_{21}$ \\
        $L_{20} \leftarrow -L_{21} L_{10} + L_{20}$ \\
        $L_{10} \leftarrow L_{10} L_{00}$ \\
        $L_{11} \leftarrow L_{11}^{-1}$ \\
        \hline
    \end{tabular}
\end{center}
All but the last update statement in each algorithm map directly to one of the BLAS routines \texttt{dtrsm}, \texttt{dtrmm}, and \texttt{dgemm} (see \autoref{app:blas} for an introduction to BLAS).
The last update on the other hand is the inversion of $L_{11} \in \mathbb R^{b \times b}$ --- a recursive call to the inversion of a triangular matrix with a smaller input matrix of size $b \times b$.
This recursive invocation, which is typical for blocked algorithms, is performed by an \emph{unblocked} version of the algorithm (that is block-size $b = 1$).
In this version, the last update is the scalar operation $L_{11} = \frac1{L_{11}} \in \mathbb R$.

Once the updates have been performed, the $3 \times 3$ partitioning of the matrix $L$ is merged back into four quadrants
$$
    \left(\begin{array}{cc|c}
        L_{00} &0      &0      \\
        L_{10} &L_{11} &0      \\
        \hline
        L_{20} &L_{21} &L_{22}
    \end{array}\right)
    \rightarrow
    \left(\begin{array}{c|c}
        L_{TL} &0      \\
        \hline
        L_{BL} &L_{BR}
    \end{array}\right)
$$
and $p$ is incremented by $b$.
Unless $p = n$, the matrix is now again partitioned into
$$
    \left(\begin{array}{c|c}
        L_{TL} &0      \\
        \hline
        L_{BL} &L_{BR}
    \end{array}\right)
    \rightarrow
    \left(\begin{array}{c|cc}
        L_{00} &0      &0      \\
        \hline
        L_{10} &L_{11} &0      \\
        L_{20} &L_{21} &L_{22}
    \end{array}\right)
$$
and the matrix traversal proceeds.
When $p = n$, the algorithm terminates and $L^{-1}$ has been computed.
%

            \subsection{Generalization}
            \label{sec:intro.blockedalgs.general}
            In the previous section, we presented four blocked algorithms for the inversion of a lower triangular matrix.
We now generalize the principle of blocked algorithms step by step.

\begin{figure}[t]
    \centering
    \begin{tikzpicture}
        \coordinate (pos) at (0, 0);
        \fill[plot4bg] (pos) +(0, 0)  rectangle +(1, -1);
        \fill[plot3bg] (pos) +(0, -1) rectangle +(1, -4);
        \fill[plot3bg] (pos) +(1, 0)  rectangle +(4, -1);
        \fill[plot3bg] (pos) +(1, -1) rectangle +(4, -4);
        \foreach \x in {0, 1, 2, 4}
            \draw (pos) +(\x, 0) -- +(\x, -4) +(0, -\x) -- +(4, -\x);
        \foreach \x/\xl in {.5/0, 1.5/1, 3/2}
            \foreach \y/\yl in {.5/0, 1.5/1, 3/2}
                \path (pos) ++(\x, -\y) node {$A_{\yl\xl}$};
        \foreach \xa/\xb/\label in {0/1/p, 1/2/b, 2/4/r} {
            \draw[decorate, decoration=brace] (pos) ++(0,  .1) +(\xa,  0) -- +(\xb, 0) node[above, align=center, midway] {$\label$};
            \draw[decorate, decoration=brace] (pos) ++(-.1, 0) +(0, -\xb) -- +(0, -\xa) node[left, align=center, midway] {$\label$};
        }
        \draw[decorate, decoration=brace] (pos) ++(0, .6) +(0, 0) -- +(4, 0) node[above, align=center, midway] {$n$};

        \path (pos) ++(4.5, -2) node[rotate=90] {$\downarrow$ update $\downarrow$};

        \coordinate (pos) at (5, 0);
        \fill[plot4bg] (pos) +(0, 0)  rectangle +(2, -2);
        \fill[plot3bg] (pos) +(0, -2) rectangle +(2, -4);
        \fill[plot3bg] (pos) +(2, 0)  rectangle +(4, -2);
        \fill[plot3bg] (pos) +(2, -2) rectangle +(4, -4);
        \foreach \x in {0, 1, 2, 4}
            \draw (pos) +(\x, 0) -- +(\x, -4) +(0, -\x) -- +(4, -\x);
        \foreach \x/\xl in {.5/0, 1.5/1, 3/2}
            \foreach \y/\yl in {.5/0, 1.5/1, 3/2}
                \path (pos) ++(\x, -\y) node {$A_{\yl\xl}$};

        \path (pos) ++(4.5, -2) node[rotate=90] {$\downarrow$ repartition $\downarrow$};

        \coordinate (pos) at (10, 0);
        \fill[plot4bg] (pos) +(0, 0) rectangle +(2, -2);
        \fill[plot3bg] (pos) +(0, -2) rectangle +(2, -4);
        \fill[plot3bg] (pos) +(2, 0) rectangle +(4, -2);
        \fill[plot3bg] (pos) +(2, -2) rectangle +(4, -4);
        \foreach \x in {0, 2, 3, 4}
            \draw (pos) +(\x, 0) -- +(\x, -4) +(0, -\x) -- +(4, -\x);
        \foreach \x/\xl in {1/0, 2.5/1, 3.5/2}
            \foreach \y/\yl in {1/0, 2.5/1, 3.5/2}
                \path (pos) ++(\x, -\y) node {$A_{\yl\xl}$};
        \foreach \xa/\xb/\label in {0/2/p \leftarrow p + b, 2/3/b, 3/4/r}
            \draw[decorate, decoration=brace] (pos) ++(0, .1) +(\xa, 0) -- +(\xb, 0) node[above, align=center, midway] {$\label$};
    \end{tikzpicture}
    \caption{Blocked algorithms --- update and repartitioning for a square matrix.}
    \label{fig:intro.blockedalgs.general.square}
    \tikzset{external/export=false}
\end{figure}
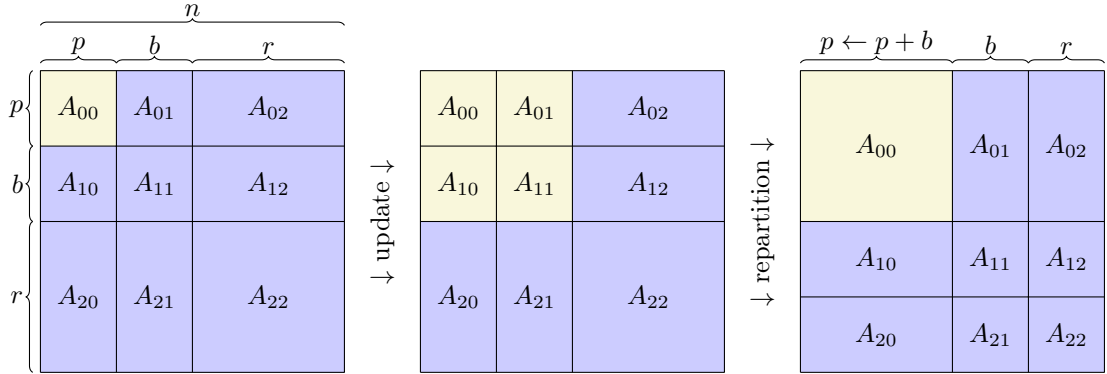

\paragraph{Full matrix.}
Extending blocked algorithms to non-triangular matrices is straightforward.
Assuming the same direction of traversal
\tikz[baseline=(A.base)] {
    \filldraw[fill=graybg] (0, 0) rectangle (1, 1);
    \draw[->, dotted] (.1, .9) -- (.9, .1);
    \node at (.5, .5) (A) {$A$};
},
an input matrix $A$ is partitioned as follows:
$$
    A =
    \left(\begin{array}{c|c}
        A_{TL} &A_{TR} \\
        \hline
        A_{BL} &A_{BR}
    \end{array}\right)
    \rightarrow
    \left(\begin{array}{c|cc}
        A_{00} &A_{01} &A_{02} \\
        \hline
        A_{10} &A_{11} &A_{12} \\
        A_{20} &A_{21} &A_{22}
    \end{array}\right).
$$
The corresponding update and repartitioning are shown in \autoref{fig:intro.blockedalgs.general.square}.

\paragraph{Traversal directions.}
Until now, we considered the case where the input matrix $A$ is traversed from the top left corner along its diagonal
\tikz[baseline=(A.base)] {
    \filldraw[fill=graybg] (0, 0) rectangle (1, 1);
    \draw[->, dotted] (.1, .9) -- (.9, .1);
    \node at (.5, .5) (A) {$A$};
}.
In principle, $A$ can be traversed in any direction:
\tikz[baseline=(A.base)] {
    \filldraw[fill=graybg] (0, 0) rectangle (1, 1);
    \draw[->, dotted] (.1, .5) -- (.9, .5);
    \node at (.5, .5) (A) {$A$};
},
\tikz[baseline=(A.base)] {
    \filldraw[fill=graybg] (0, 0) rectangle (1, 1);
    \draw[->, dotted] (.1, .9) -- (.9, .1);
    \node at (.5, .5) (A) {$A$};
},
\tikz[baseline=(A.base)] {
    \filldraw[fill=graybg] (0, 0) rectangle (1, 1);
    \draw[->, dotted] (.5, .9) -- (.5, .1);
    \node at (.5, .5) (A) {$A$};
},
\tikz[baseline=(A.base)] {
    \filldraw[fill=graybg] (0, 0) rectangle (1, 1);
    \draw[->, dotted] (.9, .9) -- (.1, .1);
    \node at (.5, .5) (A) {$A$};
},
\tikz[baseline=(A.base)] {
    \filldraw[fill=graybg] (0, 0) rectangle (1, 1);
    \draw[->, dotted] (.9, .5) -- (.1, .5);
    \node at (.5, .5) (A) {$A$};
},
\tikz[baseline=(A.base)] {
    \filldraw[fill=graybg] (0, 0) rectangle (1, 1);
    \draw[->, dotted] (.9, .1) -- (.1, .9);
    \node at (.5, .5) (A) {$A$};
},
\tikz[baseline=(A.base)] {
    \filldraw[fill=graybg] (0, 0) rectangle (1, 1);
    \draw[->, dotted] (.5, .1) -- (.5, .9);
    \node at (.5, .5) (A) {$A$};
}, or
\tikz[baseline=(A.base)] {
    \filldraw[fill=graybg] (0, 0) rectangle (1, 1);
    \draw[->, dotted] (.1, .1) -- (.9, .9);
    \node at (.5, .5) (A) {$A$};
}.
In all cases that traverse the matrix diagonally, it is partitioned into $2 \times 2$ and $3 \times 3$ as shown above.
In the other cases, the matrix is partitioned into two and then three matrices
$$
    A \rightarrow
    \left(\begin{array}{c}
        A_T \\
        A_B
    \end{array}\right)
    \rightarrow
    \left(\begin{array}{c}
        A_0 \\
        A_1 \\
        A_2
    \end{array}\right)
    \text{ or }
    A \rightarrow
    \left(\begin{array}{cc}
        A_L &A_R
    \end{array}\right)
    \rightarrow
    \left(\begin{array}{ccc}
        A_0 & A_1 & A_2
    \end{array}\right)
$$
and traversed in blocks of size $b$.

\paragraph{Multiple matrices.}
When a blocked algorithm operates on multiple matrices, each of them can potentially be traversed along different directions.
The block-size $b$, however, is constant across all matrices.

\paragraph{Non-square matrices.}
When non-square matrices are traversed horizontally or vertically, nothing changes compared to the square case.
When, on the other hand, a non-square matrix is traversed diagonally, there are two alternatives:
\begin{itemize}
    \item Using two different block-sizes, leading to rectangular blocks on the diagonal, or
    \item Traversing diagonally as far as possible and continuing along the remaining vertical or horizontal direction.
\end{itemize}
We consider the latter.

Without loss of generality, we assume that a matrix $A \in \mathbb R^{m \times n}$ with $m > n$ is traversed from the top left corner
\tikz[baseline=(A.base)] {
    \filldraw[fill=graybg] (0, 0) rectangle (.666, 1);
    \draw[->, dotted] (.1, .9) -- (.566, .433);
    \node at (.333, .5) (A) {$A$};
}.
The traversal of $A$ is identical to the square case until the rightmost column is reached.
At this point, the matrix is partitioned as follows:
$$
    A =
    \left(\begin{matrix}
        A_{TL} &\textcolor{black!50}{A_{TR}}\\
        A_{BL} &\textcolor{black!50}{A_{BR}}
    \end{matrix}\right),
$$
where $A_{TL} \in \mathbb R^{m \times m}$ and $A_{BL} \in \mathbb R^{(n-m) \times m}$; both $A_{TR}$ and $A_{BR}$ have a width of $0$ columns.

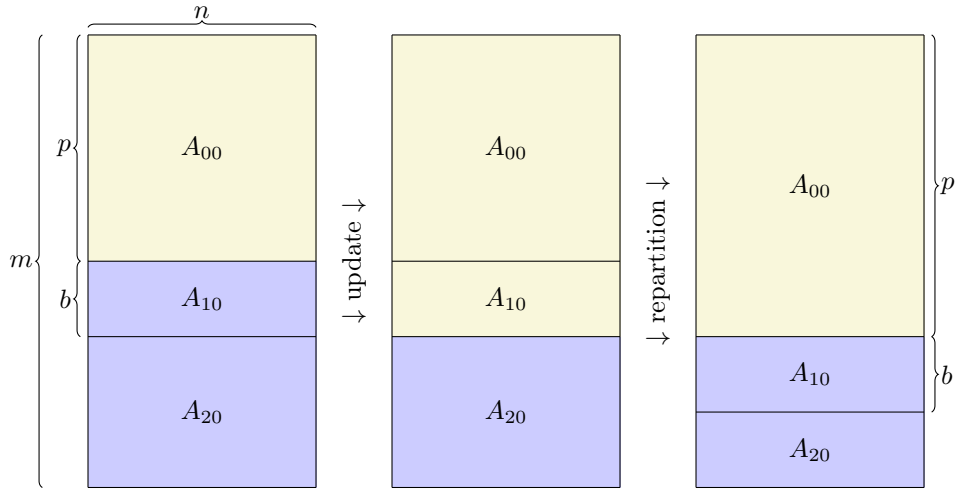
\begin{figure}[t]
    \centering
    \begin{tikzpicture}
        \coordinate (pos) at (0, 0);
        \fill[plot4bg] (pos) +(0, 0)  rectangle +(3, -3);
        \fill[plot3bg] (pos) +(0, -3) rectangle +(3, -6);
        \draw (pos) ++(0,0) -- +(0, -6) ++(3, 0) -- +(0, -6);
        \foreach \x in {0, 3, 4, 6}
            \draw (pos) ++(0, -\x) -- +(3, 0);
        \foreach \y/\yl in {1.5/0, 3.5/1, 5/2}
            \path (pos) ++(1.5, -\y) node {$A_{\yl0}$};
        \draw[decorate, decoration=brace] (pos) ++(0, .1) +(0, 0) -- +(3, 0) node[above, align=center, midway] {$n$};
        \draw[decorate, decoration=brace] (pos) ++(-.6, 0) +(0, -6) -- +(0, 0) node[left, align=center, midway] {$m$};
        \foreach \xa/\xb/\label in {0/3/p, 3/4/b}
            \draw[decorate, decoration=brace] (pos) ++(-.1, 0) +(0, -\xb) -- +(0, -\xa) node[left, align=center, midway] {$\label$};

        \path (pos) ++(3.5, -3) node[rotate=90] {$\downarrow$ update $\downarrow$};

        \coordinate (pos) at (4, 0);
        \fill[plot4bg] (pos) +(0, 0)  rectangle +(3, -4);
        \fill[plot3bg] (pos) +(0, -4) rectangle +(3, -6);
        \draw (pos) ++(0,0) -- +(0, -6) ++(3, 0) -- +(0, -6);
        \foreach \x in {0, 3, 4, 6}
            \draw (pos) ++(0, -\x) -- +(3, 0);
        \foreach \y/\yl in {1.5/0, 3.5/1, 5/2}
            \path (pos) ++(1.5, -\y) node {$A_{\yl0}$};

        \path (pos) ++(3.5, -3) node[rotate=90] {$\downarrow$ repartition $\downarrow$};

        \coordinate (pos) at (8, 0);
        \fill[plot4bg] (pos) +(0, 0)  rectangle +(3, -4);
        \fill[plot3bg] (pos) +(0, -4) rectangle +(3, -6);
        \draw (pos) ++(0,0) -- +(0, -6) ++(3, 0) -- +(0, -6);
        \foreach \x in {0, 4, 5, 6}
            \draw (pos) ++(0, -\x) -- +(3, 0);
        \foreach \y/\yl in {2/0, 4.5/1, 5.5/2}
            \path (pos) ++(1.5, -\y) node {$A_{\yl0}$};

        \foreach \xa/\xb/\label in {0/4/p, 4/5/b}
            \draw[decorate, decoration=brace] (pos) ++(.1, 0) +(3, -\xa) -- +(3, -\xb) node[right, align=center, midway] {$\label$};

    \end{tikzpicture}
    \caption{Blocked algorithms --- update and repartitioning for a non-square matrix.}
    \label{fig:intro.blockedalgs.general.nonsquare}
    \tikzset{external/export=false}
\end{figure}

From this point on, the matrix is repartitioned, where all submatrices that originate from $A_{TR}$ and $A_{TL}$ have $0$ columns:
$$
    A = 
    \left(\begin{array}{c}
        A_{TL}\\
        \hline
        A_{BL}
    \end{array}\right)
    =
    \left(\begin{array}{c}
        A_{00}\\
        \hline
        A_{10}\\
        A_{20}
    \end{array}\right).
$$
The new submatrices are of size $A_{00} \in \mathbb R^{p \times n}$, $A_{10} \in \mathbb R^{b \times n}$, and $A_{20} \in \mathbb R^{(m - p - b) \times n}$ (see \autoref{fig:intro.blockedalgs.general.nonsquare}).
The updates are applied to these submatrices as usual.
Those that involve empty matrices do not have any effect.

At the end of the iteration, the matrix is partitioned as
$$
    \left(\begin{array}{c}
        A_{00}\\
        A_{10}\\
        \hline
        A_{20}
    \end{array}\right)
    =
    \left(\begin{array}{c}
        A_{TL}\\
        \hline
        A_{BL}
    \end{array}\right)
    =
    A
$$
and we update $p \leftarrow p + b$.
The algorithm terminates, once $p = m$ (and inherently $p \geq n$).

    \chapter{Sampling}
    \label{sec:sampling}
    In this section, we describe the construction of a tool for performance measurements of dense linear algebra (DLA) routine executions: The \emph{Sampler}.

We start by specifying our goal and the desired functionality of the Sampler (\autoref{sec:sampling.goal}).
Subsequently, we conduct a set of experiments that will aid us in the design and implementation of the Sampler (\autoref{sec:sampling.experiments}).
In \autoref{sec:sampling.tool}, we introduce the Sampler and discuss its design and interface.

         \section{The Goal}
         \label{sec:sampling.goal}
         Our goal is to design a performance measurement tool that can be used as a basis for performance modeling.
In the following, we will specify the Sampler's desired functionality, discussing
\begin{itemize}
    \item the performance metrics that are available and of interest to us (\autoref{sec:sampling.goal.metrics}),
    \item the interface used to specify measurement requests (\autoref{sec:sampling.goal.input}), and
    \item the configuration of the environment for the measurements (\autoref{sec:sampling.goal.env}).
\end{itemize}

             \subsection{Performance Metrics}
             \label{sec:sampling.goal.metrics}
             Before designing our performance measurement tool, we first need to clarify and structure our understanding of performance.

We consider the performance of a routine execution to be a set of \emph{performance metrics}.
These describe several aspects of the execution and events occurring during its runtime.
We divide performance metrics into two types:
\begin{itemize}
    \item Intrinsic performance metrics --- henceforth called \emph{performance counters} --- are available through special CPU registers;
    \item \emph{Derived performance metrics} are computed from performance counters and further information on the execution environment.
\end{itemize}

The time stamp counter, the most fundamental performance counter, is a register that is incremented once per CPU cycle.
This register's value can be accessed directly through the x86 instruction \texttt{RDTSC} (read time stamp counter).
From now on, we use this highly accurate measure for execution time; the corresponding performance counter is referred to as \metric{ticks}.

In order to measure further hardware events, each CPU offers between 2 and 5 hardware counters that can be configured to count certain types of events.
We use the Performance Application Programming Interface version 4.2.1.0 (PAPI) \cite{papi} to access these event counter.

PAPI provides functions to configure, start, and read the counters.
It supports up to 107 different events; usually only a subset of these are available on a particular CPU.
These events include, but are not limited to:
\begin{description}
    \item[Cache access events.]
        For each cache level, PAPI can measure the number of cache reads, writes,  hits, misses, and total accesses.
        Data and instruction caches can be treated separately ore combined.
        On some CPUs, PAPI can distinguish between cache misses occurring during load and store instructions.
        On multicore CPUs, the occurrences of cache coherency related events such as accesses to shared cache lines or cache line invalidations can be counted through PAPI.

        We denote the number of Level-1 and Level-2 cache misses by the performance counters \metric{L1misses} and \metric{L2misses}, respectively.

    \item[Translation Lookaside Buffer (TLB) events.]
        PAPI provides counters for the number of TLB misses; data and instruction TLBs can again be treated separately or combined.
        We call the performance counter representing the total number of TLB misses \metric{TLBmisses}.

    \item[Instruction events.]
        PAPI can count the number of instructions of specific types, including: load, store, branch, and floating point instructions (with further subdivisions).
        The performance counter for the number of floating point operations is subsequently called \metric{flops}\footnote{
            \metric{flops} is the \emph{number} of floating point operations.
            The number of floating point operations \emph{per second} is consistently denoted by \metric{flops/s}.
            The property of a certain CPU that determines the maximum number of floating point operations \emph{available} per second is referred to as \metric{peak\_flops/s}.
        }.
        Further, the number of CPU cycles that specific instruction units are idle or stalling due to memory accesses can be measured separately.
\end{description}
The complete list of event counters and those which are available on a specific CPU can be accessed through the shell command \texttt{papi\_avail}.

Derived performance metrics are computed from the performance counter and (possibly) additional information, for instance the CPU frequency or cache sizes.
A list of common derived performance metrics is given below:
\begin{itemize}
    \item Floating point operations per second can be computed from \metric{flops}, \metric{ticks}, the CPU's frequency \metric{hz} and the CPU's available floating point instructions per cycle \metric{fpipc}:
        $$\metric{flops/s} = \frac{\metric{flops} \times \metric{hz}}{\metric{ticks} \times \metric{fpipc}}.$$
    \item The \emph{efficiency} of a routine execution is computed from \metric{ticks}, \metric{fpipc}, and the number of mathematical operations performed \metric{mops}\footnote{Routines often perform more \metric{flops} than an operation requires from a mathematical point of view.}:
        $$\metric{efficiency} = \frac{\metric{mops}}{\metric{ticks} \times \metric{fpipc}}.$$
    \item Level-1 cache miss ratio is obtained from \metric{L1misses} and the number of Level-1 cache accesses \metric{L1accesses}:
        $$\metric{L1missratio} = \frac{\metric{L1accesses}}{\metric{L1misses}}.$$
\end{itemize}

For the sampling tool, we only consider the directly measurable performance counters.

             \subsection{Routines and Arguments}
             \label{sec:sampling.goal.input}
             We are interested in sampling dense linear algebra routines, such as BLAS or unblocked algorithms.

When sampling a routine, we need to specify both its symbol and its arguments.
Let us consider the following BLAS call:
\begin{center}
\texttt{%
    dtrsm(%
    \overparameq{side}{R},
    \overparameq{uplo}{L},
    \overparameq{transA}{N},
    \overparameq{diag}{U},
    \overparameq{m}{512},
    \overparameq{n}{128},
    \overparameq{alpha}{0.37},
    \overparameq{A}{\textit A},
    \overparameq{ldA}{256},
    \overparameq{B}{\textit B},
    \overparameq{ldB}{512})%
}.
\end{center}
This invocation performs the operation $B \leftarrow 0.37 B A^{-1}$, where $A \in \mathbb R^{128 \times 128}$ is lower triangular and has leading dimension $256$ and $B \in \mathbb R^{512 \times 128}$ with leading dimension $512$.
We can execute the above call with different BLAS implementations.
Thus, the routine can only be identified by both its name and its implementation.

Since all BLAS implementations use the same interface, we cannot use several implementations in the same program.
The simplest possible solution is to generate separate sampling programs, each linked with a different BLAS implementations.
In each of these programs, a routine is uniquely identified by its name.

All parameters apart form \texttt A and \texttt B are basic data types such as characters, integers, and floating point numbers; \texttt A and \texttt B on the other hand are matrices.
At this point, we can exploit, that most dense linear algebra operations are mostly independent of their matrices' values; the floating point operations that are performed are solely determined by the other arguments.
This means, all we need to sample a routine are sufficiently large memory chunks assigned to matrices and vectors such as \texttt A and \texttt B.
Therefore, a request to sample the above call to \texttt{dtrsm} can be represented as the following tuple:
$$
    (\mathtt{dtrsm}, \mathtt R, \mathtt L, \mathtt N, \mathtt U, 512, 128, 0.37, 256 \times 128, 256, 512 \times 128, 512).
$$
Here, the data arguments have been replaced by their sizes in memory:
\texttt A has leading dimension $256$ and $128$ columns.
Thus, \texttt{dtrsm} accesses a range of $256 \times 128 = 32,768$ double precision floating point numbers for \texttt A.

             \subsection{Systems and Environment}
             \label{sec:sampling.goal.env}
             The remaining concern is to control the sampling environment.
This consists of the execution environment of the Sampler and the preconditions for each sample.
The relevant aspects are the system architecture and the configuration of the BLAS library, such as support for multithreading.

From within the sampler, we can influence the data locality of routine arguments, which may greatly influence performance.
We need a mechanism to control where the routine's arguments reside in memory.
Experiments in \autoref{sec:sampling.experiments} will give us further understanding regarding the influence of memory locality on performance.
This in turn will aid us in the design of our tool.

The last important requirements for the Sampler are minimal overhead and performance distortion.
Both can be achieved by a clean design and structure.

        \section{Experiments}
        \label{sec:sampling.experiments}
        We now consider at a series of preliminary experiments that guide our design of the Sampler.
In particular, we execute \texttt{dtrsm} to analyze the effects of various setups.

The system configuration remains unchanged, that is:
\begin{itemize}
    \item One core of an Intel Harpertown E5450 processor running at $2.99 \mathrm{GHz}$;
    \item Intel's C Compiler (\texttt{icc}).
\end{itemize}
We use the high performance BLAS implementations \gotoblas, MKL, and ATLAS\footnote{See \autoref{app:blas.implementations} for further detail on these implementations.}.

The first experiment (\autoref{sec:sampling.experiments.ref}) is discussed in greater detail and serves as a reference for modifications of the setup (\autoref{sec:sampling.experiments.mods}).

            \subsection{Reference Experiment}
            \label{sec:sampling.experiments.ref}
            In the first experiment, we repeatedly execute
\begin{center}
\texttt{%
    dtrsm(%
    \overparameq{side}{R},
    \overparameq{uplo}{L},
    \overparameq{transA}{N},
    \overparameq{diag}{U},
    \overparameq{m}{128},
    \overparameq{n}{96},
    \overparameq{alpha}{0.37},
    \overparameq{A}{\textit A},
    \overparameq{ldA}{128},
    \overparameq{B}{\textit B},
    \overparameq{ldB}{128})%
},
\end{center}
where \texttt A and \texttt B are assigned fixed memory locations across all repetitions.

We measure the performance counters \metric{ticks}, \metric{flops}, and \metric{L1misses}, and compute the derived performance metric \metric{efficiency}.
The measurements are performed as follows:
\begin{enumerate}
    \item The performance counters are initialized using PAPI;
    \item The routine is executed;
    \item The performance counters are read through PAPI;
    \item The results are written to a file;
\end{enumerate}
These steps are repeated 1000 times, collecting all results.

\begin{table}[t]
    \begin{center}
        \begin{tabular}{lrrrrrr}
            \toprule
                &\multicolumn{2}{c}{\gotoblas~(\ref{fig:sampling.experiments.ref.plot:goto})} &\multicolumn{2}{c}{MKL~(\ref{fig:sampling.experiments.ref.plot:mkl})}   &\multicolumn{2}{c}{ATLAS~(\ref{fig:sampling.experiments.ref.plot:atlas})} \\
                &1\textsuperscript{st}  &median                                         &1\textsuperscript{st}  &median                                     &1\textsuperscript{st}  &median \\
            \midrule
            \metric{ticks}         &$5,737,806$  &$497,790$ &$12,408,516$ &$451,026$    &$224,973,414$  &$1,081,404$\\
            \metric{flops}         &$611,992$    &$611,785$ &$646,350$    &$609,784$    &$1,217,632$    &$1,216,491$\\
            \metric{L1misses}      &$38,762$     &$38,428$  &$98,996$     &$28,690$     &$15,248$       &$13,925$\\
            \metric{efficiency}    &$5.30\%$     &$61.1\%$  &$2.45\%$     &$67.4\%$     &$0.135\%$      &$28.1\%$\\
            \bottomrule
        \end{tabular}
    \end{center}
    \caption{Initial measurement outliers.}
    \label{tab:sampling.experiments.ref.outliers}
\end{table}

The first execution of each series of experiments reproducibly yields a significantly higher and thus less efficient result.
\autoref{tab:sampling.experiments.ref.outliers} shows these series' first measurements and their medians.
Especially MKL's first result is very large.
These outliers are due to the BLAS implementations, which, when called for the first time, perform a number of initializations, such as memory allocations, system architecture identification, and algorithm selection.
From the observation of the initial outliers, we learn to discard the first experiment's result in each program invocation.
We do not integrate this idea into the Sampler, since we might later use it to analyze exactly this behavior.
However, we have to keep the outliers in mind when we use the Sampler to measure performance, for example, for the construction of performance models in \autoref{sec:modeling.models.create}.

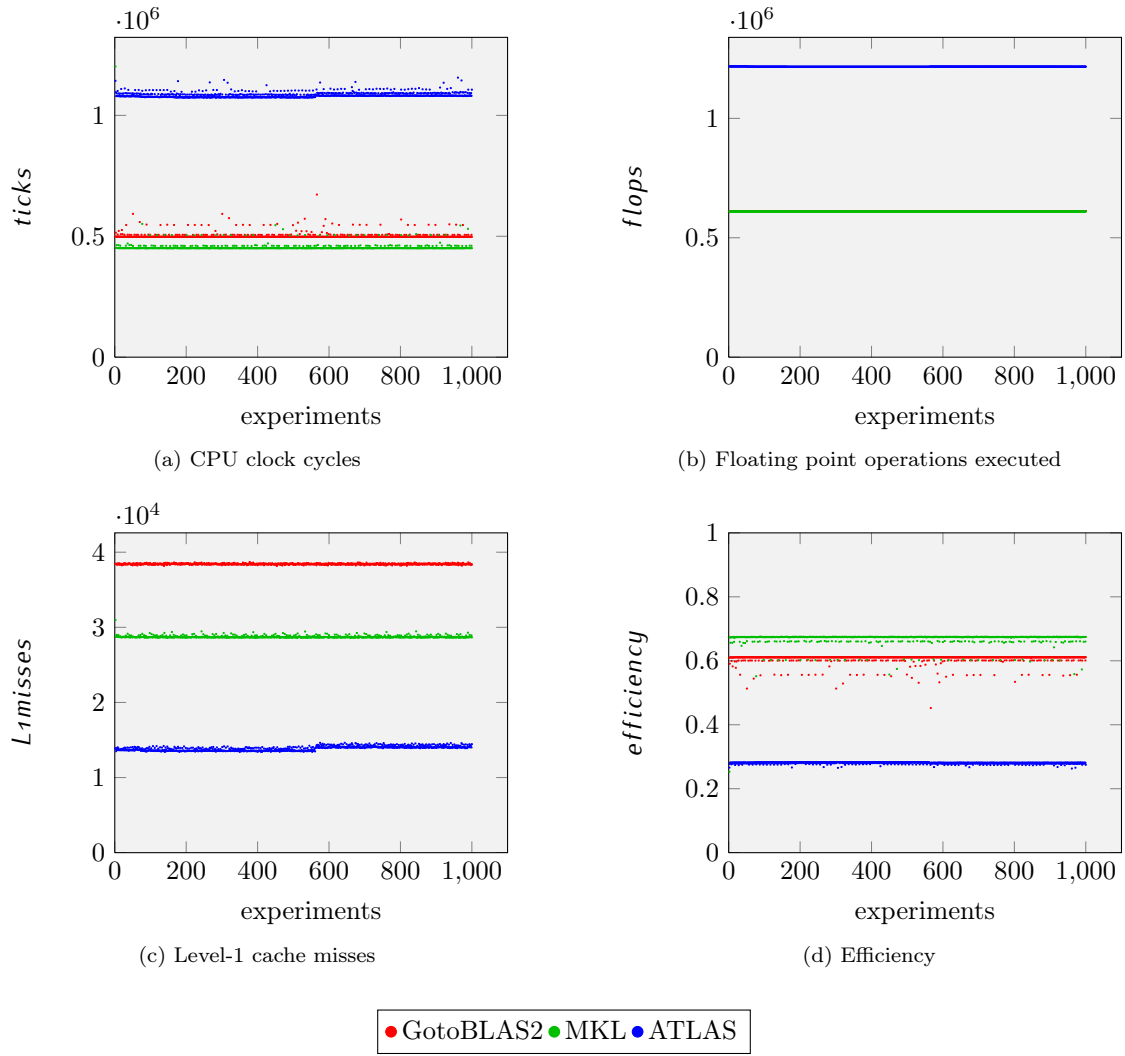
\begin{figure}[t]
    \tikzset{external/export=true}
    \centering
    \subfloat[CPU clock cycles]{
        \label{fig:sampling.experiments.ref:ticks}
        \begin{tikzpicture}
            \begin{axis}[
                twocolplot,
                xlabel={experiments},
                ylabel={\metric{ticks}},
                legend to name=fig:sampling.experiments.ref:legend,
                legend columns=-1,
            ]
                \addlegendimage{plot1, only marks}
                \label{fig:sampling.experiments.ref.plot:goto}
                \addlegendentry{\gotoblas}
                \addlegendimage{plot2, only marks}
                \label{fig:sampling.experiments.ref.plot:mkl}
                \addlegendentry{MKL}
                \addlegendimage{plot3, only marks}
                \label{fig:sampling.experiments.ref.plot:atlas}
                \addlegendentry{ATLAS}

                \addplot[color=plot1, mark size=.2pt, only marks] file {figures/data/sampling.experiments.ref/goto.ticks.dat};
                \addplot[color=plot2, mark size=.2pt, only marks] file {figures/data/sampling.experiments.ref/mkl.ticks.dat};
                \addplot[color=plot3, mark size=.2pt, only marks] file {figures/data/sampling.experiments.ref/atlas.ticks.dat};
            \end{axis}
        \end{tikzpicture}
    }
    \hfill
    \subfloat[Floating point operations executed]{
        \label{fig:sampling.experiments.ref:flops}
        \begin{tikzpicture}
            \begin{axis}[
                twocolplot,
                xlabel={experiments},
                ylabel={\metric{flops}},
            ]
                \addplot[color=plot1, mark size=.2pt, only marks] file {figures/data/sampling.experiments.ref/goto.flops.dat};
                \addplot[color=plot2, mark size=.2pt, only marks] file {figures/data/sampling.experiments.ref/mkl.flops.dat};
                \addplot[color=plot3, mark size=.2pt, only marks] file {figures/data/sampling.experiments.ref/atlas.flops.dat};
            \end{axis}
        \end{tikzpicture}
    }

    \subfloat[Level-1 cache misses]{
        \label{fig:sampling.experiments.ref:l1}
        \begin{tikzpicture}
            \begin{axis}[
                twocolplot,
                xlabel={experiments},
                ylabel={\metric{L1misses}},
            ]
                \addplot[color=plot1, mark size=.2pt, only marks] file {figures/data/sampling.experiments.ref/goto.l1.dat};
                \addplot[color=plot2, mark size=.2pt, only marks] file {figures/data/sampling.experiments.ref/mkl.l1.dat};
                \addplot[color=plot3, mark size=.2pt, only marks] file {figures/data/sampling.experiments.ref/atlas.l1.dat};
            \end{axis}
        \end{tikzpicture}
    }
    \hfill
    \subfloat[Efficiency]{
        \label{fig:sampling.experiments.ref:eff}
        \begin{tikzpicture}
            \begin{axis}[
                twocolplot,
                xlabel={experiments},
                ylabel={\metric{efficiency}},
                xmin=0,
                ymin=0,
                ymax=1
            ]
                \addplot[color=plot1, mark size=.2pt, only marks] file {figures/data/sampling.experiments.ref/goto.eff.dat};
                \addplot[color=plot2, mark size=.2pt, only marks] file {figures/data/sampling.experiments.ref/mkl.eff.dat};
                \addplot[color=plot3, mark size=.2pt, only marks] file {figures/data/sampling.experiments.ref/atlas.eff.dat};
            \end{axis}
        \end{tikzpicture}
    }

    \vspace{.5cm} 

    \tikzset{external/export=false}
    \ref*{fig:sampling.experiments.ref:legend}
    \caption{Repeated execution of \texttt{dtrsm}.}
    \label{fig:sampling.experiments.ref}
\end{figure}

All but the first outlier measurement are shown in \autoref{fig:sampling.experiments.ref}.
We observe the following:
For the faster implementations \gotoblas~(\ref{fig:sampling.experiments.ref.plot:goto}) and MKL~(\ref{fig:sampling.experiments.ref.plot:mkl}) the measured \metric{ticks} and, thus, the computed \metric{efficiency} fluctuate in a range of $8\%$.
\metric{flops} and \metric{L1misses} on the other hand are almost constant.
\gotoblas~(\ref{fig:sampling.experiments.ref.plot:goto}) and MKL~(\ref{fig:sampling.experiments.ref.plot:mkl}) attain efficiencies of $61\%$ and $67\%$, respectively; this relatively low performance is due to the small problem size of $96 \times 128$.

            \subsection{Modified Setups}
            \label{sec:sampling.experiments.mods}
            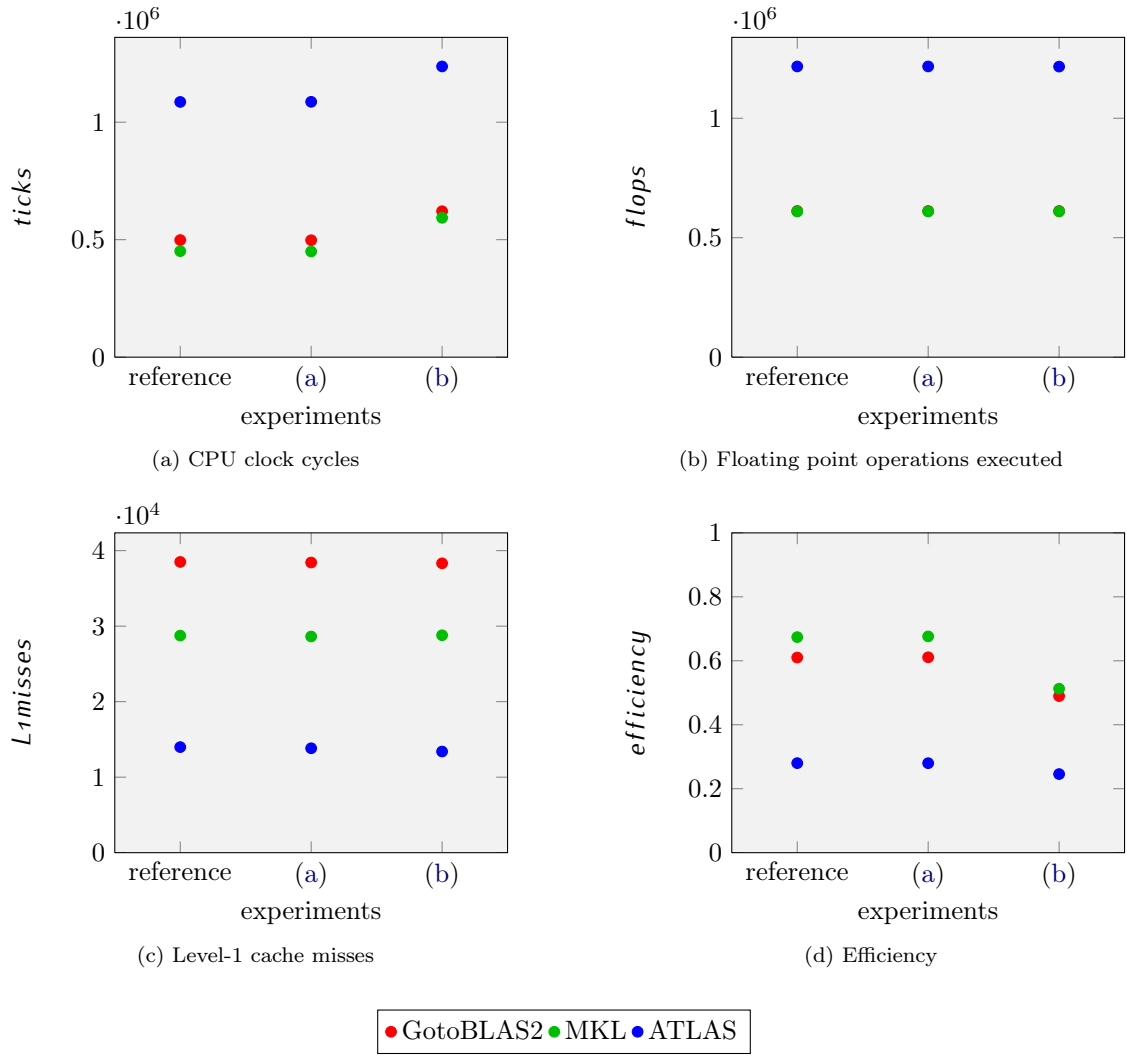
\begin{figure}[t]
    \tikzset{external/export=true}
    \centering
    \subfloat[CPU clock cycles]{
        \label{fig:sampling.experiments.mods.plot:ticks}
        \begin{tikzpicture}
            \begin{axis}[
                twocolplot,
                xlabel={experiments},
                ylabel={\metric{ticks}},
                legend to name=fig:sampling.experiments.mods.plot:legend,
                legend columns=-1,
                xmin=0.5,
                xmax=3.5,
                xtick={1,2,3},
                xticklabels={
                    reference,
                    (\ref{sampling.experiments.mods.sep}),
                    (\ref{sampling.experiments.mods.mem})}
            ]
                \addlegendimage{plot1, only marks}
                \label{fig:sampling.experiments.mods.plot:goto}
                \addlegendentry{\gotoblas}
                \addlegendimage{plot2, only marks}
                \label{fig:sampling.experiments.mods.plot:mkl}
                \addlegendentry{MKL}
                \addlegendimage{plot3, only marks}
                \label{fig:sampling.experiments.mods.plot:atlas}
                \addlegendentry{ATLAS}

                \addplot[color=plot1, only marks] file {figures/data/sampling.experiments.mods/goto.ticks.dat};
                \addplot[color=plot2, only marks] file {figures/data/sampling.experiments.mods/mkl.ticks.dat};
                \addplot[color=plot3, only marks] file {figures/data/sampling.experiments.mods/atlas.ticks.dat};
            \end{axis}
        \end{tikzpicture}
    }
    \hfill
    \subfloat[Floating point operations executed]{
        \label{fig:sampling.experiments.mods.plot:flops}
        \begin{tikzpicture}
            \begin{axis}[
                twocolplot,
                xlabel={experiments},
                ylabel={\metric{flops}},
                xmin=0.5,
                xmax=3.5,
                xtick={1,2,3},
                xticklabels={
                    reference,
                    (\ref{sampling.experiments.mods.sep}),
                    (\ref{sampling.experiments.mods.mem})},
            ]
                \addplot[color=plot1, only marks] file {figures/data/sampling.experiments.mods/goto.flops.dat};
                \addplot[color=plot2, only marks] file {figures/data/sampling.experiments.mods/mkl.flops.dat};
                \addplot[color=plot3, only marks] file {figures/data/sampling.experiments.mods/atlas.flops.dat};
            \end{axis}
        \end{tikzpicture}
    }

    \subfloat[Level-1 cache misses]{
        \label{fig:sampling.experiments.mods.plot:l1}
        \begin{tikzpicture}
            \begin{axis}[
                twocolplot,
                xlabel={experiments},
                ylabel={\metric{L1misses}},
                xmin=0.5,
                xmax=3.5,
                xtick={1,2,3},
                xticklabels={
                    reference,
                    (\ref{sampling.experiments.mods.sep}),
                    (\ref{sampling.experiments.mods.mem})},
            ]
                \addplot[color=plot1, only marks] file {figures/data/sampling.experiments.mods/goto.l1.dat};
                \addplot[color=plot2, only marks] file {figures/data/sampling.experiments.mods/mkl.l1.dat};
                \addplot[color=plot3, only marks] file {figures/data/sampling.experiments.mods/atlas.l1.dat};
            \end{axis}
        \end{tikzpicture}
    }
    \hfill
    \subfloat[Efficiency]{
        \label{fig:sampling.experiments.mods.plot:eff}
        \begin{tikzpicture}
            \begin{axis}[
                twocolplot,
                xlabel={experiments},
                ylabel={\metric{efficiency}},
                xmin=0.5,
                xmax=3.5,
                xtick={1,2,3},
                xticklabels={
                    reference,
                    (\ref{sampling.experiments.mods.sep}),
                    (\ref{sampling.experiments.mods.mem})},
                ymax=1
            ]
                \addplot[color=plot1, only marks] file {figures/data/sampling.experiments.mods/goto.eff.dat};
                \addplot[color=plot2, only marks] file {figures/data/sampling.experiments.mods/mkl.eff.dat};
                \addplot[color=plot3, only marks] file {figures/data/sampling.experiments.mods/atlas.eff.dat};
            \end{axis}
        \end{tikzpicture}
    }

    \vspace{.5cm} 

    \tikzset{external/export=false}
    \ref*{fig:sampling.experiments.mods.plot:legend}
    \caption{Modifications of the experiment setup for \texttt{dtrsm}.}
    \label{fig:sampling.experiments.mods.plot}
\end{figure}

We now modify the experiment setup used in \autoref{sec:sampling.experiments.ref} and investigate its influence on the measurements.
The results of the modifications for the same performance metrics as in the reference experiment are shown in \autoref{fig:sampling.experiments.mods.plot}.
Each experiment was executed 1000 times and the corresponding plots show the series' median.

\begin{enumerate}[(a)]
    \item \emph{Separation of sampling and IO}:
        \label{sampling.experiments.mods.sep}
        In this modification the 1000 repeated routine invocations are executed consecutively without writing the results to the output.
        Instead, the results are stored in memory and only written to the standard output stream after all 1000 samples were taken.
        As we can see in \autoref{fig:sampling.experiments.mods.plot}, this modification did not affect the performance measurements.
        However, we observe a slight decrease in the total runtime of the experiment series of $0.7$ seconds, or $0.7$ milliseconds ($2,137,850$ ticks) per sample.
    \item \emph{Random matrix memory locations}:
        \label{sampling.experiments.mods.mem}
        In this modification, the routine's matrix arguments are randomly taken from a $2 \mathrm{GB}$ large memory chunk.
        Using this configuration, the results differ significantly:
        While \metric{flops} and \metric{L1misses} remain unchanged, we experience an increase in execution time and decrease in efficiency.
        This is expected, since the number of floating point instructions is unchanged, while the memory accesses are to main memory instead of the CPU's caches.
        Both reusing the same memory locations to gain increased memory locality, as well as accessing memory areas which do not reside in the caches --- called \emph{cache trashing} --- are interesting scenarios.
\end{enumerate}

        \section{The Sampler}
        \label{sec:sampling.tool}
        With gained insights into the relevant factors for measuring performance, we turn to the Sampler.
This tool is designed to fulfill the requirements specified in \autoref{sec:sampling.goal}.
\autoref{sec:sampling.tool.design} gives an overview of the design and mechanisms used in the Sampler's implementation.
We then turn to its usage and interface in \autoref{sec:sampling.tool.usage}.
The tool is employed in a series of experiments in \autoref{sec:modeling.experiments}.

            \subsection{Design}
            \label{sec:sampling.tool.design}
            The Sampler is designed as a flexible lightweight performance measurement tool.
It is written in C and can, hence, directly interfaces to libraries such as BLAS \cite{blas, blas3} or LAPACK \cite{lapack, lapackpage}.

The used libraries and the list of routines that can be sampled are incorporated into the Sampler during building process.
Given a list of header files for the routines, specific object files are generated.
These contain information on the routines signatures necessary to read and execute corresponding sampling requests.
Linking these object files with the implementations of the routines yields the Sampler.
The same object files can be linked with different routine implementations (e.g., \gotoblas or MKL) to create separate Samplers for each of them.
In order to sample multithreaded routines, the implementations need to be configured either before linking or through the system environment at runtime, according to the libraries configuration system.

We discuss the structure of the Sampler by looking at its program flow.
The outline of the execution of the sampler is as follows:
\begin{enumerate}
    \item Initialization.
    \item While the end of input stream is not reached:
        \begin{enumerate}
            \item Read a set of sampling requests and prepare their execution;
            \item Execute and sample the routines;
            \item Write the results to the output.
        \end{enumerate}
    \item Finalization.
\end{enumerate}

We will now consider each of these stages in more detail.

\paragraph{Initialization.}
The initialization of the Sampler starts by reading a specified configuration file (a full list of possible configurations is given in \autoref{sec:sampling.tool.usage.config}).
In case PAPI is requested, the PAPI library is initialized.

Then, the Sampler sets up the input and output streams.
These are used to read sampling requests and write sampling results.
These streams, which by default are the programs standard IO streams, can be redirected to files.

Next, the Sampler allocates a large contiguous chunk of memory of configurable size.
This memory is used for the vector and matrix arguments of the sampled routines.
It is initialized with random double precision numbers between $0$ and $1$.

\paragraph{Reading sampling requests.}
Once the main loop is reached, the Sampler begins to read sampling requests from the input stream (their syntax is given in \autoref{sec:sampling.tool.usage.input}).
For each request, the Sampler's assigns memory locations within the preallocated memory chunk to sampled routine's vector and matrix arguments.
Different memory access patterns (such as cache trashing or in-cache) can be configured (see \autoref{sec:sampling.tool.usage.config}).

There are three conditions upon which the Sampler stops reading requests:
\begin{itemize}
    \item The (configurable) maximum number of routine executions that are sampled in one block is reached;
    \item A special command "\texttt{go}" was read from the input stream;
    \item The end of the input stream is reached.
\end{itemize}

\paragraph{Executing sampling requests.}
In this step, the Sampler executes all sampling requests read in the previous step.
The resulting measurements are stored in memory for each execution.
IO and sampling are separated in order to reduce possible interference and overhead.

\paragraph{Writing the results to the output.}
Now, the measurement results are written to the output stream (the output format is given in \autoref{sec:sampling.tool.usage}).
The structures used to handle the sampling requests are cleaned such that they can be reused in the next iteration of the main loop.

\paragraph{Finalization.}
After the main loop terminates, the Sampler frees the used memory.

            \subsection{Interface and Usage}
            \label{sec:sampling.tool.usage}
            The Sampler's interface is designed to be as clean and flexible as possible; the Sampler can be used either as an interactive tool in a shell or by other programs and scripts.
The interface consists of three major parts:
\begin{itemize}
    \item The configuration file (\autoref{sec:sampling.tool.usage.config}),
    \item The input stream format (\autoref{sec:sampling.tool.usage.input}), and
    \item The output stream format (\autoref{sec:sampling.tool.usage.output}).
\end{itemize}

                \subsubsection{Configuration File}
                \label{sec:sampling.tool.usage.config}
                The Sampler is invoked with only one argument --- the path to a configuration file:
\begin{center}
    \texttt{\textit{sampler} \textit{<configuration file>}}
\end{center}
The configuration file consists of several options, each in an individual line.
Lines beginning with a \texttt{\#} are ignored.
The following options are available:
\begin{itemize}
    \item \texttt{input = \textit{file}}: The input is to be read from \texttt{\textit{file}}.
        By default the input stream is the program's standard input stream.
    \item \texttt{output = \textit{file}}: The sampling results are to be redirected to \texttt{\textit{file}}.
        The default is the standard output.
    \item \texttt{usepapi = 1|0}: Whether PAPI is to be used for performance counter sampling (\texttt 1) or not (\texttt 0).
        Without PAPI, only \metric{ticks} is measured.
    \item \texttt{ncounters = \textit{n}}: The number of PAPI performance counters to be used.
        This number is inherently limited by the systems CPU architecture.
    \item \texttt{counters[\textit{i}] = \textit{counter\_id}}: The \texttt{\textit{i}}-th PAPI performance counter is set to \texttt{\textit{counter\_id}} --- any valid PAPI event name.
        On systems with PAPI, a full list of available event names is available through the shell command \texttt{papi\_avail}.
    \item \texttt{maxcalls = \textit{n}}: \texttt{\textit{n}} sampling requests are to be executed in a block.
    \item \texttt{mem\_size = \textit{n}}: The Sampler assigns memory chunks to routine arguments from an allocation of \texttt{\textit{n}} bytes.
    \item \texttt{mem\_policy = \textit{n}}: The memory policy is identified by a number between \texttt 0 and \texttt 3:
        \begin{description}
            \item[\texttt 0: Static.] Each routine is assigned disjoint memory chunks from the start of the Sampler's memory.
                Since this configuration will assign the same memory locations over and over again, this configuration can be used to simulate cache locality.
            \item[\texttt 1: Forward.] Each assigned memory chunk begins where the last ended, such that there is no overlap between chunks.
                Once the end of the Sampler's memory is used, the assignment starts over from its beginning.
                This configuration ensures that there are no overlaps between memory chunks in consecutive routine executions.
                It can therefore be used to achieve cache trashing.
            \item[\texttt 2: Backward.] The memory chunks are assigned in backward order, starting from the end of the Sampler's memory.
                The concept of this policy is the same as for the Forward mode (\texttt 1).
            \item[\texttt 3: Random.] The memory chunks are taken randomly from within the Sampler's memory.
                If the Sampler's memory is large enough, the chances of overlapping memory regions is minimal, leading to cache trashing.
                If on the other hand, the memory allocation is small, overlapping memory assignments are more likely, even for chunks assigned to the same routine execution.
        \end{description}
        Varying the memory policy will allow to represent certain scenarios in the estimation of performance in the 
    \item \texttt{mem\_align = \textit{n}}: Every assigned memory chunk will be aligned in memory by blocks of \texttt{\textit{n}} bytes.
        This configuration may be used to achieve cache alignment, ensuring that each matrix and vector starts in a new cache line.
    \item \texttt{state\_queries = 0|1}: When set to \texttt 1, the Sampler prints queries to the standard output, stating which routine type of arguments are expected (e.g., \texttt{int*} or \texttt{double*}).
    \item \texttt{show\_progress = 0|1}: When set to \texttt 1, the Sampler prints its progress to standard output.
    \item \texttt{matlab\_output = 0|1}: Activating this option leads to output in Matlab matrix notation.
        This is useful when Matlab is used to process the sampling results.
\end{itemize}

All options that take values \texttt 0 or \texttt 1 are \texttt 0 by default and may be omitted.
The options for PAPI counters are only valid with a previous \texttt{usepapi = 1}.
In this case, the number of specified performance counters has to match the number given in \texttt{ncounters = \textit{n}}.

                \subsubsection{Input Stream Format}
                \label{sec:sampling.tool.usage.input}
                The format of the input stream agrees with the considerations in \autoref{sec:sampling.goal.input}.
A request consists of a routine name followed by that routine's argument specifications.
The only exception is the command \texttt{go}, which may be issued between sampling requests.
Upon encountering this command, the Sampler immediately starts the sampling process, prints out the results, and only then continues to read its input stream.
This command is needed when the sampler is used interactively or by another program in order to initiate the sampling process.

A request to sample
\begin{center}
\texttt{%
    dtrsm(%
    \overparameq{side}{R},
    \overparameq{uplo}{L},
    \overparameq{transA}{N},
    \overparameq{diag}{U},
    \overparameq{m}{128},
    \overparameq{n}{96},
    \overparameq{alpha}{0.37},
    \overparameq{A}{\textit A},
    \overparameq{ldA}{128},
    \overparameq{B}{\textit B},
    \overparameq{ldB}{128})%
},
\end{center}
for instance is submitted by the following line on the input stream:
\begin{center}
    \texttt{dtrsm R L N U 128 96 v.37 16384 128 16384 128}.
\end{center}
From the routine's signature, the Sampler know how to interpret the arguments.
\begin{itemize}
    \item The capital characters are allowed for \texttt{char*} arguments.
    \item \texttt{int*} arguments take an integer value.
    \item For \texttt{float*} or \texttt{double*} arguments there are two possibilities:
        \begin{itemize}
            \item Given an integer, a memory chunk of this size will be assigned to this argument.
            \item Given the character \texttt v (for "value") followed by a floating point number, the number is passed to the routine as a single floating point number.
                This is useful to cover special cases such as $0$, $1$ or $-1$, which might trigger separate routine branches.
        \end{itemize}
\end{itemize}

                \subsubsection{Output Stream Format}
                \label{sec:sampling.tool.usage.output}
                The output format is very similar to the input format:
For each sampling request, the name of the routine is printed, followed by its \texttt{char*} and \texttt{int*} arguments\footnote{
    \texttt{float*} and \texttt{double*} arguments are omitted.
}.
After the arguments, the sampling results are written to the output stream, followed by a new-line character.

The first sampling result is always the performance counter \metric{ticks}.
It is followed by PAPI's performance counter results as specified in the configuration file.

For instance, the output for the request
\begin{center}
    \texttt{dtrsm R L N U 128 96 v0.37 16384 128 16384 128}
\end{center}
with PAPI counters \metric{flops} and \metric{L1misses} might be:
\begin{center}
    \begin{tikzpicture}
        \node (pos) {};
        \def\oldlabel{pos}
        \foreach \label/\text in {
            head/{dtrsm},
            c/\phantom{0},
            char/{R L N U},
            c/\phantom{0},
            int/{128 96 128 128},
            c/\phantom{0},
            ticks/{1079973},
            c/\phantom{0},
            flops/{1217382},
            c/\phantom{0},
            L1misses/{13509}
        } {
            \node[inner sep=0, anchor=base west, base right=0 of \oldlabel.base east]
                (\label) {\texttt{\vphantom{T}\text}};
            \xdef\oldlabel{\label}
        }

        \node[gray, anchor=south] at (head.north) {\footnotesize \vphantom{$f$}name};
        \draw[gray, decorate, decoration=brace] (char.north west) -- (char.north east)
            node[midway, anchor=south] {\footnotesize \vphantom{$f$}\texttt{char*}};
        \draw[gray, decorate, decoration=brace] (int.north west) -- (int.north east)
            node[midway, anchor=south] {\footnotesize \vphantom{$f$}\texttt{int*}};
        \foreach \counter in {ticks, flops, L1misses}
            \node[gray, anchor=south] at (\counter.north) {\footnotesize \vphantom{$f$}\metric{\counter}};
    \end{tikzpicture}
\end{center}

    \chapter{Modeling}
    \label{sec:modeling}
    In the previous Chapter we discussed the Sampler --- a tool for measuring the performance of DLA routines.
From measurements obtained by the Sampler, we now want to construct analytical performance models.
For this purpose, we introduce the \emph{Modeler}, that based on the Sampler, generates a certain type of performance models automatically.
These models form the base for our performance prediction and algorithm ranking (\autoref{sec:ranking}).

The Modeler is developed in the following design process:
\begin{itemize}
    \item We begin by studying the dependence of performance on various arguments of DLA routines (\autoref{sec:modeling.experiments}).
        The focus is BLAS, the most fundamental routines.
    \item Using the gained understanding on the dependencies, we develop the structure of the performance model that the Modeler shall generate (\autoref{sec:modeling.models}).
    \item With the desired model structure in mind, we develop the Modeler, which generates these models automatically (\autoref{sec:modeling.tool}).
    \item We conclude by evaluating how well the Modeler fulfills its objective --- the reliable generation of accurate performance models (\autoref{sec:modeling.res}).
        In the process, we illustrate how the Modeler can be configured to obtain models with certain properties.
\end{itemize}

        \section{Preliminary Experiments}
        \label{sec:modeling.experiments}
        In the experiments in \autoref{sec:sampling.experiments}, we have already determined a few interesting aspects of performance measurements:
\begin{enumerate}
    \item The first measurement is always an outlier;
    \item When a routine is executed repeatedly, most performance metrics produce fluctuating values;
    \item The performance changes significantly depending on whether the routine's arguments are in cache or main memory.
\end{enumerate}
All of these aspects are taken into account in the designs of our models in \autoref{sec:modeling.models}.
To avoid their influence in the experiments in this section, we take the following measures:
\begin{enumerate}
    \item The first measurement is always discarded;
    \item Each experiment is repeated 100 times and the median of this series is presented as a result;
    \item Successive routine invocations use the same memory locations for their arguments (static memory policy).
\end{enumerate}

We focus on the dependence of performance on the types of arguments encountered in BLAS.
(An introduction to BLAS is given in \autoref{app:blas}.)
We consider each argument type and its influence on performance individually:
\begin{itemize}
    \item Discrete\footnote{
        Meaning with a finite value range.
        Arguments that take integer values are referred to as continuous.
        The usage of "discrete" and "continuous" thus differs from their purely mathematical meaning.
    } arguments, such as \texttt{uplo} or \texttt{trans} (\autoref{sec:modeling.experiments.disc}),
    \item Size arguments \texttt m, \texttt n, and \texttt k (\autoref{sec:modeling.experiments.size}),
    \item Scalar arguments, for instance \texttt{alpha} or \texttt{beta} (\autoref{sec:modeling.experiments.scal}),
    \item Vector and matrix arguments, such as \texttt A, \texttt B, or \texttt x (\autoref{sec:modeling.experiments.mat}), and
    \item Leading dimension or increment arguments like \texttt{ldA} and \texttt{incx} (\autoref{sec:modeling.experiments.lda}).
\end{itemize}

Throughout the experiments in this section, we take measurements on one core of a Quad-Core AMD Opteron Processor 8356 \cite{amd} running at $2.30 \mathrm{GHz}$.
This processor can issue $2$ double precision floating point instructions per cycle; therefore, its theoretical peak performance is $\metric{peak\_flops/s} = 2 \times 2.30 \cdot 10^9$ operations per second.

While we consider \metric{ticks} as a representative performance counter in most experiments, the gained insights are also applicable for other performance counters, such as \metric{flops} or \metric{L1misses}.

            \subsection{Discrete Arguments}
            \label{sec:modeling.experiments.disc}
            In order to understand the influence of discrete arguments on performance, we consider \texttt{dtrsm} ($B \leftarrow A^{-1} B$, $A$ triangular).
This routine is an ideal candidate, since it covers all types of discrete arguments that appear in BLAS: \texttt{side}, \texttt{uplo}, \texttt{transA}, and \texttt{diag}.
For our experiments, we vary these arguments and fix the remaining ones:
\begin{center}
    \texttt{%
        dtrsm(%
        \overparameq{side}{\textit{side}},
        \overparameq{uplo}{\textit{uplo}},
        \overparameq{transA}{\textit{transA}},
        \overparameq{diag}{\textit{diag}},
        \overparameq{m}{256},
        \overparameq{n}{256},
        \overparameq{alpha}{0.5},
        \overparameq{A}{\textit A},
        \overparameq{ldA}{256},
        \overparameq{B}{\textit B},
        \overparameq{ldB}{256})%
    }.
\end{center}

\begin{figure}[t]
    \tikzset{external/export=true}
    \centering
    \begin{tikzpicture}
        \begin{axis}[
            width=\textwidth,
            height=.45\textwidth,
            ylabel={\metric{ticks}},
            legend to name=fig:modeling.experiments.disc.plot.plot:legend,
            legend columns=-1,
            xmin=0.5,
            xmax=16.5,
            xtick={1,2,3,4,5,6,7,8,9,10,11,12,13,14,15,16},
            xticklabels={
                \texttt{L L N N},
                \texttt{L L N U},
                \texttt{L L T N},
                \texttt{L L T U},
                \texttt{L U N N},
                \texttt{L U N U},
                \texttt{L U T N},
                \texttt{L U T U},
                \texttt{R L N N},
                \texttt{R L N U},
                \texttt{R L T N},
                \texttt{R L T U},
                \texttt{R U N N},
                \texttt{R U N U},
                \texttt{R U T N},
                \texttt{R U T U}},
            every x tick label/.append style={text width=.3cm, align=center},
            ymin=0
        ]
            \addplot[color=plot1, only marks] file {figures/data/modeling.experiments.disc.plot/goto.ticks.dat};
            \label{fig:modeling.experiments.disc.plot:goto}
            \addlegendentry{\gotoblas}

            \addplot[color=plot2, only marks] file {figures/data/modeling.experiments.disc.plot/mkl.ticks.dat};
            \label{fig:modeling.experiments.disc.plot:mkl}
            \addlegendentry{MKL}

            \addplot[color=plot3, only marks] file {figures/data/modeling.experiments.disc.plot/atlas.ticks.dat};
            \label{fig:modeling.experiments.disc.plot:atlas}
            \addlegendentry{ATLAS}
        \end{axis}
        \draw (0, 0) node[anchor=north east, align=left] {
            \texttt{side}\\
            \texttt{uplo}\\
            \texttt{transA}\\
            \texttt{diag}
        };
    \end{tikzpicture}

    \vspace{.5cm} 

    \tikzset{external/export=false}
    \ref*{fig:modeling.experiments.disc.plot.plot:legend}
    \caption{Dependence of \metric{ticks} on discrete arguments in \texttt{dtrsm}.}
    \label{fig:modeling.experiments.disc.plot}
\end{figure}
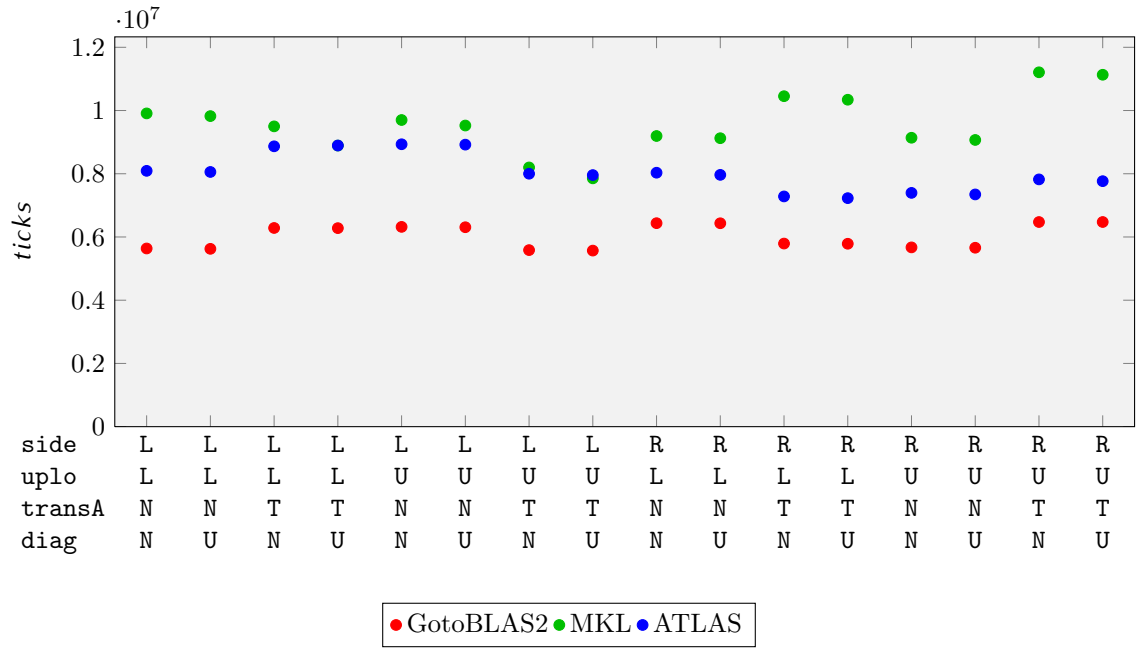

Measurements of the performance counter \metric{ticks} for all possible combinations of discrete argument values are shown in \autoref{fig:modeling.experiments.disc.plot}.
At a first glance, we can already see that every BLAS implementation shows a different behavior; we cannot find a consistent pattern across all of them.
However, we can find several similarities common to at least some of the implementations.

First of all, \texttt{diag} has only very little influence on performance in any of the implementations.
Only for MKL\footnote{
    Since MKL is designed for Intel architectures, it attains lower performance on this AMD system.
}~(\ref{fig:modeling.experiments.disc.plot:mkl}) and ATLAS~(\ref{fig:modeling.experiments.disc.plot:atlas}), we observe that $\mathtt{diag} = \mathtt U$ results in a slightly lower runtime.
(This is not unexpected, since \texttt U indicates that $A$ is unit-triangular, therefore requiring 256 floating point operations less.)

When we consider the discrete arguments \texttt{side}, \texttt{uplo}, and \texttt{transA}, we observe a coupled dependence.
In \autoref{fig:modeling.experiments.disc.plot}, we see that the pattern on the left ($\mathtt{side} = \mathtt L$) is inverted (mirrored along a horizontal line) with respect to the same pattern on the right ($\mathtt{side} = \mathtt R$).
(This is understandable, since \texttt{side} determines the order of operations --- $A^{-1} B$ for \texttt L and $B A^{-1}$ for \texttt R.)
Additionally, ATLAS~(\ref{fig:modeling.experiments.disc.plot:atlas}) is on average faster for $\mathtt{side} = \texttt R$, while MKL~(\ref{fig:modeling.experiments.disc.plot:mkl}) is slower for this case.

Considering $\mathtt{side} = \mathtt L$ (the left half of \autoref{fig:modeling.experiments.disc.plot}) and looking at \gotoblas~(\ref{fig:modeling.experiments.disc.plot:goto}) and ATLAS~(\ref{fig:modeling.experiments.disc.plot:atlas}), we find that when $\mathtt{uplo} = \mathtt L$, $\texttt{transA} = \mathtt N$ is faster than $\mathtt{transA} = \mathtt T$, while the opposite is the case for $\mathtt{uplo} = \mathtt R$;
for MKL~(\ref{fig:modeling.experiments.disc.plot:mkl}), $\mathtt{transA} = \mathtt T$ is faster than $\mathtt{transA} = \mathtt N$ independent of $\texttt{uplo}$.

All these observations have shown that \texttt{side}, \texttt{uplo}, and \texttt{transA} have a significant impact on performance and that the dependency cannot be described in a consistent manner across all implementations.
When we aim at modeling this dependency, our only option is to generate a separate model for each of combination of these parameters.

\texttt{diag}, on the other hand, was shown to only have a minor impact on performance, therefore, we are inclined to ignore this argument in order to reduce our model complexity.

            \subsection{Size Arguments}
            \label{sec:modeling.experiments.size}
            \begin{figure}[t]
    \tikzset{external/export=true}
    \centering
    \subfloat[CPU clock cycles]{
        \label{fig:modeling.experiments.size.mnk:ticks}
        \begin{tikzpicture}
            \begin{axis}[
                twocolplot,
                xlabel={\texttt m/\texttt n/\texttt k},
                ylabel={\metric{ticks}},
                legend to name=fig:modeling.experiments.size.mnk:legend,
                legend columns=4,
                xtick={0,128,...,512}
            ]
                \addplot[color=plot1] file {figures/data/modeling.experiments.size.mnk/goto.m.ticks.dat};
                \label{fig:modeling.experiments.size.mnk:goto}
                \addlegendentry{\texttt{\vphantom{pk}m}}

                \addplot[color=plot1, dotted] file {figures/data/modeling.experiments.size.mnk/goto.n.ticks.dat};
                \addlegendentry{\texttt{\vphantom{pk}n}}
                \addplot[color=plot1, dashed] file {figures/data/modeling.experiments.size.mnk/goto.k.ticks.dat};
                \addlegendentry{\texttt{\vphantom{pk}k}}
                \addlegendimage{empty legend}
                \addlegendentry{\gotoblas\vphantom{\texttt{pk}}}

                \addplot[color=plot2] file {figures/data/modeling.experiments.size.mnk/mkl.m.ticks.dat};
                \label{fig:modeling.experiments.size.mnk:mkl}
                \addlegendentry{\texttt{\vphantom{pk}m}}

                \addplot[color=plot2, dotted] file {figures/data/modeling.experiments.size.mnk/mkl.n.ticks.dat};
                \addlegendentry{\texttt{\vphantom{pk}n}}
                \addplot[color=plot2, dashed] file {figures/data/modeling.experiments.size.mnk/mkl.k.ticks.dat};
                \addlegendentry{\texttt{\vphantom{pk}k}}
                \addlegendimage{empty legend}
                \addlegendentry{MKL\vphantom{\texttt{pk}}}

                \addplot[color=plot3] file {figures/data/modeling.experiments.size.mnk/atlas.m.ticks.dat};
                \label{fig:modeling.experiments.size.mnk:atlas}
                \addlegendentry{\texttt{\vphantom{pk}m}}

                \addplot[color=plot3, dotted] file {figures/data/modeling.experiments.size.mnk/atlas.n.ticks.dat};
                \addlegendentry{\texttt{\vphantom{pk}n}}
                \addplot[color=plot3, dashed] file {figures/data/modeling.experiments.size.mnk/atlas.k.ticks.dat};
                \addlegendentry{\texttt{\vphantom{pk}k}}
                \addlegendimage{empty legend}
                \addlegendentry{ATLAS\vphantom{\texttt{pk}}}
            \end{axis}
        \end{tikzpicture}
    }
    \hfill
    \subfloat[Efficiency]{
        \label{fig:modeling.experiments.size.mnk:eff}
        \begin{tikzpicture}
            \begin{axis}[
                twocolplot,
                xlabel={\texttt m/\texttt n/\texttt k},
                ylabel={\metric{efficiency}},
                ymax=1,
                xtick={0,128,...,512}
            ]
                \addplot[color=plot1] file {figures/data/modeling.experiments.size.mnk/goto.m.eff.dat};
                \addplot[color=plot1, dotted] file {figures/data/modeling.experiments.size.mnk/goto.n.eff.dat};
                \addplot[color=plot1, dashed] file {figures/data/modeling.experiments.size.mnk/goto.k.eff.dat};

                \addplot[color=plot2] file {figures/data/modeling.experiments.size.mnk/mkl.m.eff.dat};
                \addplot[color=plot2, dotted] file {figures/data/modeling.experiments.size.mnk/mkl.n.eff.dat};
                \addplot[color=plot2, dashed] file {figures/data/modeling.experiments.size.mnk/mkl.k.eff.dat};

                \addplot[color=plot3] file {figures/data/modeling.experiments.size.mnk/atlas.m.eff.dat};
                \addplot[color=plot3, dotted] file {figures/data/modeling.experiments.size.mnk/atlas.n.eff.dat};
                \addplot[color=plot3, dashed] file {figures/data/modeling.experiments.size.mnk/atlas.k.eff.dat};
            \end{axis}
        \end{tikzpicture}
    }

    \vspace{.5cm} 

    \tikzset{external/export=false}
    \ref*{fig:modeling.experiments.size.mnk:legend}
    \caption{Dependence of \metric{ticks} and \metric{efficiency} on size arguments in \texttt{dgemm}.}
    \label{fig:modeling.experiments.size.mnk}
\end{figure}
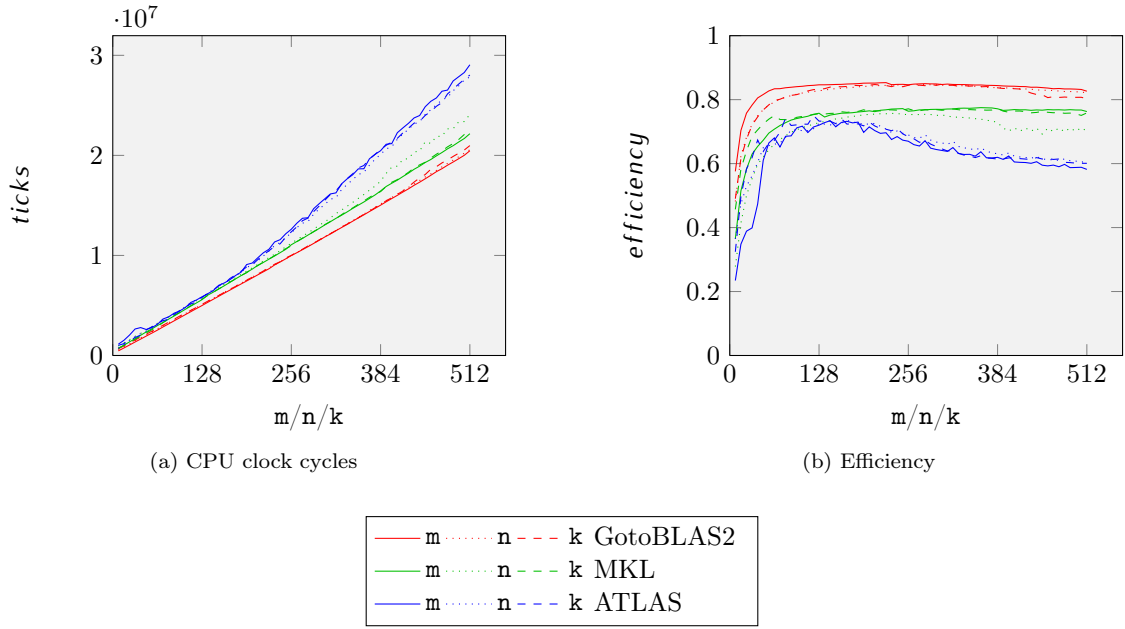

With the exception of \texttt{dgemm}~($C \leftarrow \alpha A B + \beta C$), all BLAS Level-3 routines have 2 size arguments; \texttt{dgemm} has 3: \texttt m, \texttt n, and \texttt k.
For this reason and since \texttt{dgemm} is usually the most optimized BLAS routine, we use it in the following experiments in order to understand the dependence of performance on the size arguments.
We consider the routine invocation
\begin{center}
    \texttt{%
        dgemm(%
        \overparameq{transA}{N},
        \overparameq{transB}{N},
        \overparameq{m}{\textit m},
        \overparameq{n}{\textit n},
        \overparameq{k}{\textit k},
        \overparameq{alpha}{0.5},
        \overparameq{A}{\textit A},
        \overparameq{ldA}{\textit m},
        \overparameq{B}{\textit B},
        \overparameq{ldB}{\textit k},
        \overparameq{beta}{0.5},
        \overparameq{C}{\textit C},
        \overparameq{ldC}{\textit m})%
    }
\end{center}
and perform three experiments; in each, we vary one of the size arguments \texttt m, \texttt n, and \texttt k (and the corresponding leading dimension) between $8$ and $512$ while the other two are fixed to $256$.
\autoref{fig:modeling.experiments.size.mnk} shows the performance metrics \metric{ticks} and \metric{efficiency} for these experiments.
From these plots, we can learn two important things:
\begin{itemize}
    \item Each of the three size arguments
        \texttt m~(\tikzline{}),
        \texttt n~(\tikzline{dotted}), and
        \texttt k~(\tikzline{dashed}) has a different influence on the performance of the routines.

        Since the number of mathematical operations for \texttt{dgemm} is $m n k + 2 m n$, one might expect similar dependencies on \texttt m and \texttt n and a different one for \texttt k.
        However, the results show that this is not the case:
        For \gotoblas~(\ref{fig:modeling.experiments.size.mnk:goto}), \texttt m is the argument which, compared to \texttt n and \texttt k,  produces significantly different behavior;
        the same is true for MKL~(\ref{fig:modeling.experiments.size.mnk:mkl}) and \texttt n.

        While in this experiment, we only varied one of the size arguments while the others remained fixed, in general, all size arguments may vary independently, leading to nonlinear influences on performance.
        The only possibility to capture this behavior is to represent the performance as a function defined on a three dimensional argument space -- one dimension for each size argument.

    \item The performance dependency on size parameters is generally nonlinear.
        Especially \autoref{fig:modeling.experiments.size.mnk:eff} shows the high degree of non-linearity --- in this \metric{efficiency} graph, a linear function would result in a clean hyperbola.

        However, it is possible to identify value ranges which are almost linear.
        For instance, we could approximate ATLAS's \metric{ticks}~(\ref{fig:modeling.experiments.size.mnk:atlas}) roughly by two linear functions: one ranging from $0$ to about $200$ and a second one for the remaining value range.
        In a similar fashion, we can approximate the performance dependency on single size arguments for \gotoblas~(\ref{fig:modeling.experiments.size.mnk:goto}) and MKL~(\ref{fig:modeling.experiments.size.mnk:mkl}) by several linear functions.

        If we extend this approach to accommodate for all sizes arguments simultaneously, we obtain multivariate piecewise polynomials.
\end{itemize}

\begin{figure}[t]
    \tikzset{external/export=true}
    \centering
    \subfloat[CPU clock cycles]{
        \label{fig:modeling.experiments.size.fine:ticks}
        \begin{tikzpicture}
            \begin{axis}[
                twocolplot,
                xlabel={\texttt m = \texttt n = \texttt k},
                ylabel={\metric{ticks}},
                legend to name=fig:modeling.experiments.size.fine:legend,
                legend columns=4,
                xtick={64,80,...,128},
                xmin={}
            ]
                \addplot[color=plot1] file {figures/data/modeling.experiments.size.fine/goto.ticks.dat};
                \label{fig:modeling.experiments.size.fine:goto}
                \addlegendentry{\gotoblas}

                \addplot[color=plot2] file {figures/data/modeling.experiments.size.fine/mkl.ticks.dat};
                \label{fig:modeling.experiments.size.fine:mkl}
                \addlegendentry{MKL}

                \addplot[color=plot3] file {figures/data/modeling.experiments.size.fine/atlas.ticks.dat};
                \label{fig:modeling.experiments.size.fine:atlas}
                \addlegendentry{ATLAS}

                \addplot[color=plot1, dotted] file {figures/data/modeling.experiments.size.fine/goto.ticks.8.dat};
                \label{fig:modeling.experiments.size.fine:goto8}
                \addplot[color=plot2, dotted] file {figures/data/modeling.experiments.size.fine/mkl.ticks.8.dat};
                \label{fig:modeling.experiments.size.fine:mkl8}
                \addplot[color=plot3, dotted] file {figures/data/modeling.experiments.size.fine/atlas.ticks.8.dat};
                \label{fig:modeling.experiments.size.fine:atlas8}

            \end{axis}
        \end{tikzpicture}
    }
    \hfill
    \subfloat[Efficiency]{
        \label{fig:modeling.experiments.size.fine:eff}
        \begin{tikzpicture}
            \begin{axis}[
                twocolplot,
                xlabel={$\mathtt m = \mathtt n = \mathtt k$},
                ylabel={\metric{efficiency}},
                ymax=1,
                xtick={64,80,...,128},
                xmin={}
            ]
                \addplot[color=plot1] file {figures/data/modeling.experiments.size.fine/goto.eff.dat};
                \addplot[color=plot2] file {figures/data/modeling.experiments.size.fine/mkl.eff.dat};
                \addplot[color=plot3] file {figures/data/modeling.experiments.size.fine/atlas.eff.dat};
                \addplot[color=plot1, dotted] file {figures/data/modeling.experiments.size.fine/goto.eff.8.dat};
                \addplot[color=plot2, dotted] file {figures/data/modeling.experiments.size.fine/mkl.eff.8.dat};
                \addplot[color=plot3, dotted] file {figures/data/modeling.experiments.size.fine/atlas.eff.8.dat};
            \end{axis}
        \end{tikzpicture}
    }

    \vspace{.5cm} 

    \tikzset{external/export=false}
    \ref*{fig:modeling.experiments.size.fine:legend}
    \caption{Dependence of \metric{ticks} and \metric{efficiency} on size arguments in \texttt{dgemm} at small scale.}
    \label{fig:modeling.experiments.size.fine}
\end{figure}
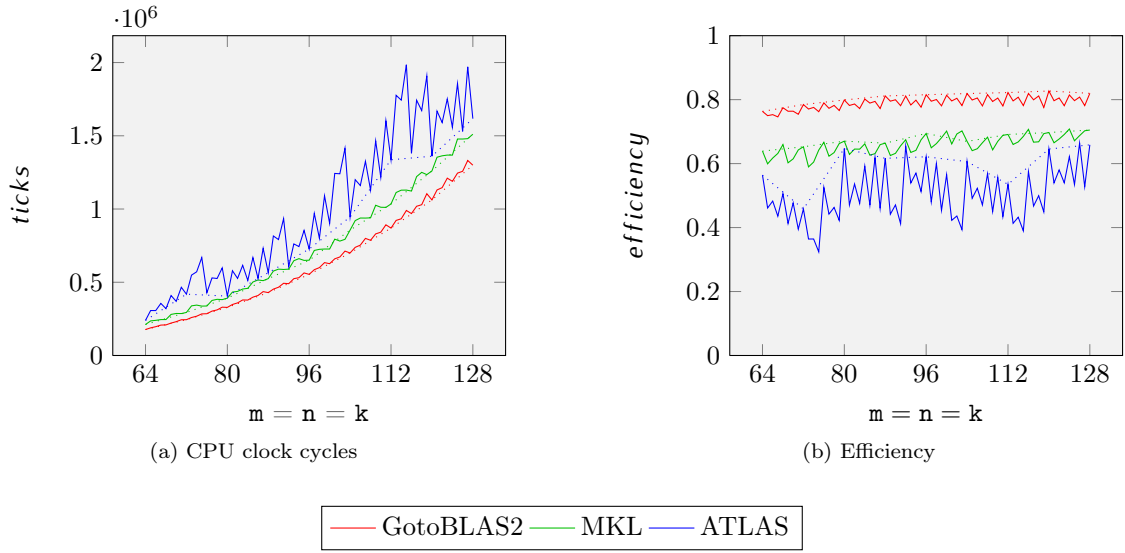

These experiments have analyzed the performance dependence on a fairly large scale.
Now, we take a closer look at a very fine scale: We once more consider \texttt{dgemm}; this time for $\mathtt m = \mathtt n = \mathtt k \in \{32, \ldots, 64\}$.
\metric{ticks} and \metric{efficiency} for these experiments are given in \autoref{fig:modeling.experiments.size.fine}.

These plots, especially for ATLAS~(\ref{fig:modeling.experiments.size.fine:atlas}), show fine oscillations.
While we do not want to represent these oscillations, we have to be aware of them during the generation of models.
For this purpose, we only consider samples in small intervals in order to obtain smooth performance dependencies (e.g., intervals of $8$ result in \tikzline{plot1, dotted}, \tikzline{plot2, dotted}, and \tikzline{plot3, dotted} in \autoref{fig:modeling.experiments.size.fine}).

            \subsection{Scalar Arguments}
            \label{sec:modeling.experiments.scal}
            In BLAS, scalar arguments multiply matrices and vectors, for instance $\alpha$ and $\beta$ in \texttt{dgemm} ($C \leftarrow \alpha A B + \beta C$).
This generally leads to one multiplication (floating point operation) per entry of the scaled matrix or vector.
There are a few exceptions:
Multiplications by $-1$, $0$, and $1$ do not require multiplications.
Especially $-1$ and $1$ are very common in application codes.

\begin{figure}[t]
    \tikzset{external/export=true}
    \centering
    \begin{tikzpicture}
        \begin{axis}[
            width=\textwidth,
            height=.45\textwidth,
            ylabel={\metric{ticks}},
            legend to name=fig:modeling.experiments.scal.plot:legend,
            legend columns=-1,
            xmin=0.5,
            xmax=16.5,
            xtick={1,2,3,4,5,6,7,8,9,10,11,12,13,14,15,16},
            xticklabels={
                $C \leftarrow 0$,
                $C \leftarrow A B$,
                $C \leftarrow -A B$,
                $C \leftarrow 0.5 A B$,
                $C \leftarrow C$,
                $C \leftarrow A B + C$,
                $C \leftarrow -A B + C$,
                $C \leftarrow 0.5 A B + C$,
                $C \leftarrow -C$,
                $C \leftarrow A B - C$,
                $C \leftarrow -A B - C$,
                $C \leftarrow 0.5 A B - C$,
                $C \leftarrow 0.5 C$,
                $C \leftarrow A B + 0.5 C$,
                $C \leftarrow -A B + 0.5 C$,
                $C \leftarrow 0.5 A B + 0.5 C$
            },
            every x tick label/.append style={rotate=90, text width=3cm},
        ]
            \addplot[color=plot1, only marks] file {figures/data/modeling.experiments.scal.plot/goto.ticks.dat};
            \label{fig:modeling.experiments.scal.plot:goto}
            \addlegendentry{\gotoblas}

            \addplot[color=plot2, only marks] file {figures/data/modeling.experiments.scal.plot/mkl.ticks.dat};
            \label{fig:modeling.experiments.scal.plot:mkl}
            \addlegendentry{MKL}

            \addplot[color=plot3, only marks] file {figures/data/modeling.experiments.scal.plot/atlas.ticks.dat};
            \label{fig:modeling.experiments.scal.plot:atlas}
            \addlegendentry{ATLAS}
        \end{axis}
    \end{tikzpicture}

    \vspace{.5cm} 

    \tikzset{external/export=false}
    \ref*{fig:modeling.experiments.scal.plot:legend}
    \caption{Dependence of \metric{ticks} on scalar arguments in \texttt{dgemm}.}
    \label{fig:modeling.experiments.scal.plot}
\end{figure}
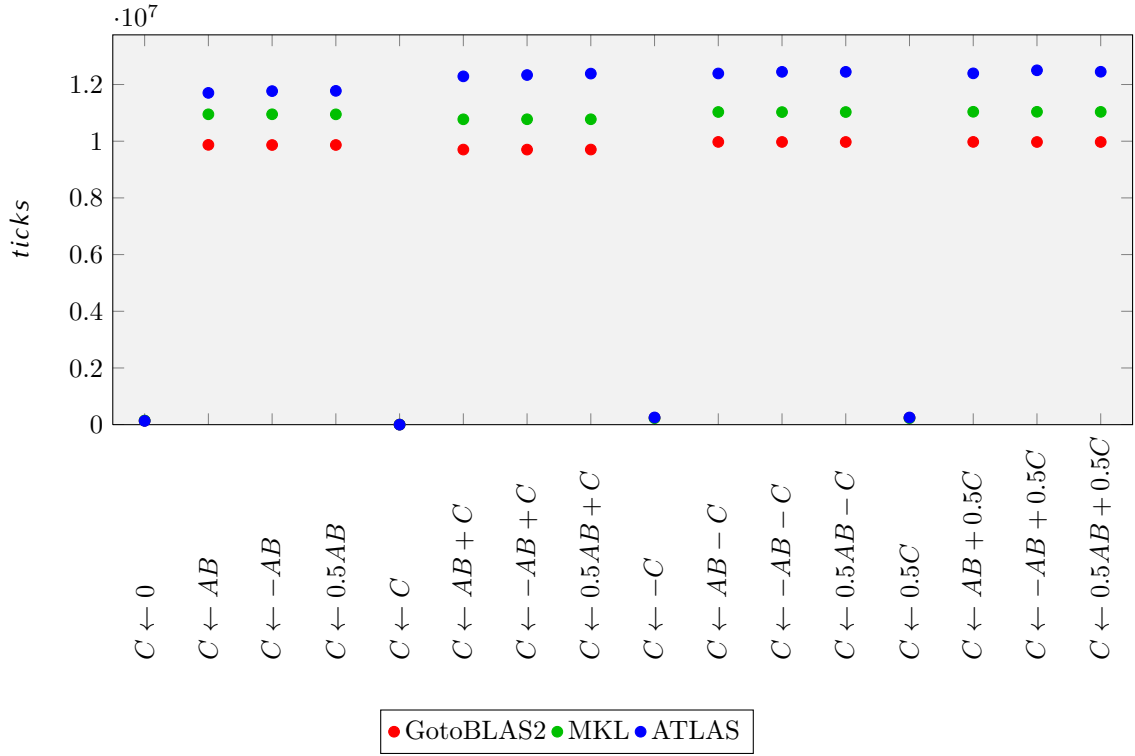

In the following experiment, we once more consider \texttt{dgemm}; this time with the following arguments:
\begin{center}
    \texttt{%
        dgemm(%
        \overparameq{transA}{N},
        \overparameq{transB}{N},
        \overparameq{m}{256},
        \overparameq{n}{256},
        \overparameq{k}{256},
        \overparameq{alpha}{\textit{alpha}},
        \overparameq{A}{\textit A},
        \overparameq{ldA}{256},
        \overparameq{B}{\textit B},
        \overparameq{ldB}{256},
        \overparameq{beta}{\textit{beta}},
        \overparameq{C}{\textit C},
        \overparameq{ldC}{256})%
    }.
\end{center}
For both \texttt{alpha} and \texttt{beta} we consider the values $0$, $1$, $-1$, and $0.5$ (as a representative for the general case).
The resulting \metric{ticks} are shown in \autoref{fig:modeling.experiments.scal.plot}.

Evidently, all implementations take advantage of the special cases where $\mathtt{alpha} = 0$: \texttt{dgemm} reduces to $C \leftarrow \beta C$ --- scaling a matrix.
Compared to matrix-matrix multiplication with complexity $O(n^3)$, this only requires $O(n^2)$ floating point operations.
As a result the execution time drops significantly across all implementations.
If now, $\mathtt{beta} = 1$, no work has to be done at all and the execution time is effectively $0$.

The other values for \texttt{alpha} ($1$, $-1$, and $0.5$) do not result in different execution times.

Considering \texttt{beta}, we see that \gotoblas~(\ref{fig:modeling.experiments.scal.plot:goto}) and MKL~(\ref{fig:modeling.experiments.scal.plot:mkl}) show a behavior different from ATLAS~(\ref{fig:modeling.experiments.scal.plot:atlas}):
\gotoblas~(\ref{fig:modeling.experiments.scal.plot:goto}) and MKL~(\ref{fig:modeling.experiments.scal.plot:mkl}) show a decreased execution time only for $\mathtt{alpha} = 1$, while ATLAS~(\ref{fig:modeling.experiments.scal.plot:atlas}) apparently only take advantage of $\mathtt{alpha} = 0$.

We can conclude that some special values of scalar arguments trigger different branches in BLAS routines, similar to discrete arguments.
However, Since the exceptional case $\texttt{alpha} = 0$ does not appear in any well written application, the overall influence of scalar arguments on performance is rather small compared to the impact of other arguments.
Therefore, we decide not to represent this dependency.

            \subsection{Vector and Matrix Arguments}
            \label{sec:modeling.experiments.mat}
            As discussed earlier, the entries of matrix and vector arguments have no effect on the computation.
Therefore, these arguments do not influence any performance metrics.

            \subsection{Leading Dimension and Increment Arguments}
            \label{sec:modeling.experiments.lda}
            Leading dimension and increment arguments specify the distance of adjacent entries in a row of a matrix or vector.
They can potentially influence the access patterns of both the CPU's cache and main memory, and thus affect performance.

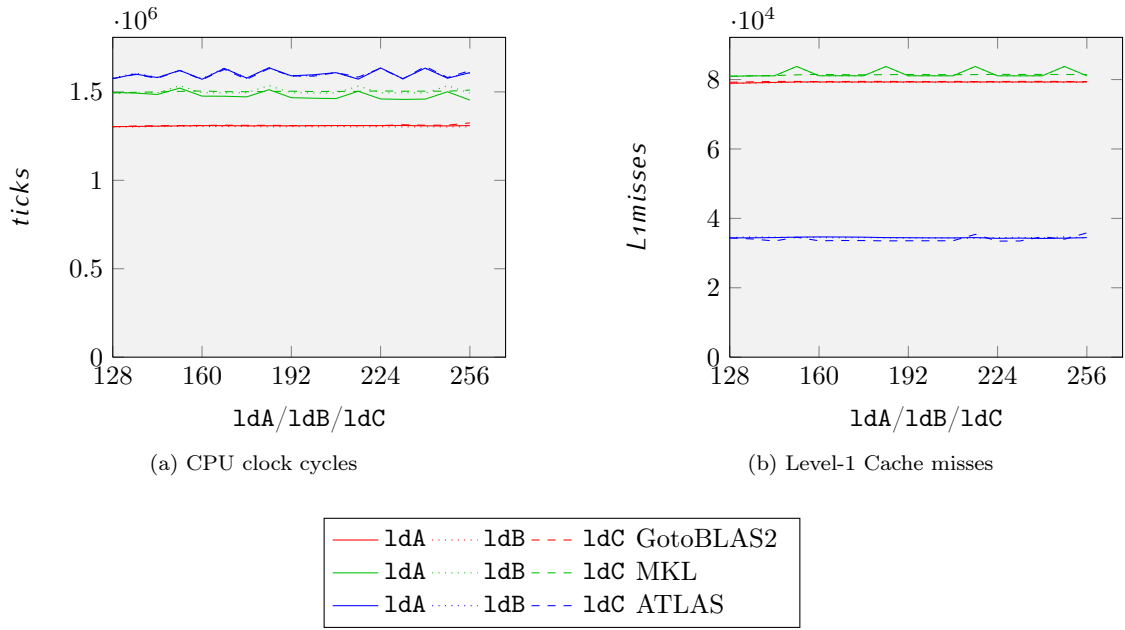
\begin{figure}[t]
    \tikzset{external/export=true}
    \centering
    \subfloat[CPU clock cycles]{
        \label{fig:modeling.experiments.lda.plot:ticks}
        \begin{tikzpicture}
            \begin{axis}[
                twocolplot,
                xlabel={\texttt{ldA}/\texttt{ldB}/\texttt{ldC}},
                ylabel={\metric{ticks}},
                legend to name=fig:modeling.experiments.lda.plot:legend,
                legend columns=4,
                xmin=128,
                xtick={128,160,...,256}
            ]
                \addplot[color=plot1] file {figures/data/modeling.experiments.lda.plot/goto.ldA.ticks.dat};
                \label{fig:modeling.experiments.lda.plot:goto}
                \addlegendentry{\texttt{\vphantom{pk}ldA}}

                \addplot[color=plot1, dotted] file {figures/data/modeling.experiments.lda.plot/goto.ldB.ticks.dat};
                \addlegendentry{\texttt{\vphantom{pk}ldB}}
                \addplot[color=plot1, dashed] file {figures/data/modeling.experiments.lda.plot/goto.ldC.ticks.dat};
                \addlegendentry{\texttt{\vphantom{pk}ldC}}
                \addlegendimage{empty legend}
                \addlegendentry{\gotoblas\vphantom{\texttt{pk}}}

                \addplot[color=plot2] file {figures/data/modeling.experiments.lda.plot/mkl.ldA.ticks.dat};
                \label{fig:modeling.experiments.lda.plot:mkl}
                \addlegendentry{\texttt{\vphantom{pk}ldA}}

                \addplot[color=plot2, dotted] file {figures/data/modeling.experiments.lda.plot/mkl.ldB.ticks.dat};
                \addlegendentry{\texttt{\vphantom{pk}ldB}}
                \addplot[color=plot2, dashed] file {figures/data/modeling.experiments.lda.plot/mkl.ldC.ticks.dat};
                \addlegendentry{\texttt{\vphantom{pk}ldC}}
                \addlegendimage{empty legend}
                \addlegendentry{MKL\vphantom{\texttt{pk}}}

                \addplot[color=plot3] file {figures/data/modeling.experiments.lda.plot/atlas.ldA.ticks.dat};
                \label{fig:modeling.experiments.lda.plot:atlas}
                \addlegendentry{\texttt{\vphantom{pk}ldA}}

                \addplot[color=plot3, dotted] file {figures/data/modeling.experiments.lda.plot/atlas.ldB.ticks.dat};
                \addlegendentry{\texttt{\vphantom{pk}ldB}}
                \addplot[color=plot3, dashed] file {figures/data/modeling.experiments.lda.plot/atlas.ldC.ticks.dat};
                \addlegendentry{\texttt{\vphantom{pk}ldC}}
                \addlegendimage{empty legend}
                \addlegendentry{ATLAS\vphantom{\texttt{pk}}}
            \end{axis}
        \end{tikzpicture}
    }
    \hfill
    \subfloat[Level-1 Cache misses]{
        \label{fig:modeling.experiments.lda.plot:l1}
        \begin{tikzpicture}
            \begin{axis}[
                twocolplot,
                xlabel={\texttt{ldA}/\texttt{ldB}/\texttt{ldC}},
                ylabel={\metric{L1misses}},
                xmin=128,
                xtick={128,160,...,256}
            ]
                \addplot[color=plot1] file {figures/data/modeling.experiments.lda.plot/goto.ldA.l1.dat};
                \addplot[color=plot1, dotted] file {figures/data/modeling.experiments.lda.plot/goto.ldB.l1.dat};
                \addplot[color=plot1, dashed] file {figures/data/modeling.experiments.lda.plot/goto.ldC.l1.dat};

                \addplot[color=plot2] file {figures/data/modeling.experiments.lda.plot/mkl.ldA.l1.dat};
                \addplot[color=plot2, dotted] file {figures/data/modeling.experiments.lda.plot/mkl.ldB.l1.dat};
                \addplot[color=plot2, dashed] file {figures/data/modeling.experiments.lda.plot/mkl.ldC.l1.dat};

                \addplot[color=plot3] file {figures/data/modeling.experiments.lda.plot/atlas.ldA.l1.dat};
                \addplot[color=plot3, dotted] file {figures/data/modeling.experiments.lda.plot/atlas.ldB.l1.dat};
                \addplot[color=plot3, dashed] file {figures/data/modeling.experiments.lda.plot/atlas.ldC.l1.dat};
            \end{axis}
        \end{tikzpicture}
    }

    \vspace{.5cm} 

    \tikzset{external/export=false}
    \ref*{fig:modeling.experiments.lda.plot:legend}
    \caption{Dependence of \metric{ticks} and \metric{L1misses} on leading dimension arguments in \texttt{dgemm}.}
    \label{fig:modeling.experiments.lda.plot}
\end{figure}

To study the actual influence of these arguments, we once more consider \texttt{dgemm} ($C \leftarrow \alpha A B + \beta C$) and vary the three leading dimension arguments \texttt{ldA}, \texttt{ldB}, and \texttt{ldC} separately between 128 and 256.
The other arguments are fixed:
\begin{center}
    \texttt{%
        dgemm(%
        \overparameq{transA}{N},
        \overparameq{transB}{N},
        \overparameq{m}{128},
        \overparameq{n}{128},
        \overparameq{k}{128},
        \overparameq{alpha}{0.5},
        \overparameq{A}{\textit A},
        \overparameq{ldA}{\textit{ldA}},
        \overparameq{B}{\textit B},
        \overparameq{ldB}{\textit{ldB}},
        \overparameq{beta}{0.5},
        C,
        ldC)%
    }.
\end{center}
Measurements of the resulting \metric{ticks} and \metric{L1misses} are given in \autoref{fig:modeling.experiments.lda.plot}.

Again, we can observe different behaviors across our BLAS implementations:
\gotoblas's performance (\ref{fig:modeling.experiments.lda.plot:goto}) is independent of the leading dimension arguments, while MKL~(\ref{fig:modeling.experiments.lda.plot:mkl}) and ATLAS~(\ref{fig:modeling.experiments.lda.plot:atlas}) show small oscillations for increasing leading dimensions.
For MKL~(\ref{fig:modeling.experiments.lda.plot:mkl}), for instance, we find that \metric{ticks} decrease slightly for increasing \texttt{ldA} and spikes again at constant intervals of $32$; \metric{L1misses} increase correspondingly.

Overall, however, we conclude that the leading dimension arguments only have a minor influence on performance compared to other arguments.
Modeling this influence would increase the complexity of our models tremendously; we usually decide to not represent it.

        \section{The Targeted Models}
        \label{sec:modeling.models}
        Based on the observations made in the previous sections, we now introduce the type of the performance models we aim to generate.
We begin by defining the model structure for the general case in \autoref{sec:modeling.models.create} and then show how a concrete model of this type may be evaluated in \autoref{sec:modeling.models.eval}.

            \subsection{Creation and Reasoning}
            \label{sec:modeling.models.create}
            In order to devise the structure for our performance models, we first recall all significant observations that we have to account for:
\begin{enumerate}[(a)]
    \item \label{itm:modeling.models.create:impl} Performance varies significantly between different implementations of BLAS (\autoref{sec:sampling.experiments} and \appendixref{app:blas.implementations}).
    \item \label{itm:modeling.models.create:fluct} Repeated executions of a single routine with fixed arguments results in fluctuating performance (\autoref{sec:sampling.experiments}).
    \item \label{itm:modeling.models.create:memloc} Performance is influenced by the memory locality (e.g., cache or main memory) of the matrix (and vector) arguments (\autoref{sec:sampling.experiments}).
    \item \label{itm:modeling.models.create:disc} Many discrete arguments and special values for scalar arguments trigger separate branches in BLAS routines.
        These branches can potentially lead to a completely different performance behavior (Sections \ref{sec:modeling.experiments.disc} and \ref{sec:modeling.experiments.scal}).
    \item \label{itm:modeling.models.create:size} Performance generally depends non-linearly on size arguments are generally non-linear; however, it can often be represented by multivariate piecewise polynomials (\autoref{sec:modeling.experiments.size}).
    \item \label{itm:modeling.models.create:lda} Leading dimension arguments only have a minor influence on performance (\autoref{sec:modeling.experiments.lda}).
\end{enumerate}

Before we turn to the construction of models, we need to clarify what we understand under the term "performance model":
\begin{center}
    \begin{minipage}{.8\textwidth}
        \centering
        \emph{A \emph{performance model} for a DLA routine is a function that, given values for the routine's arguments, provides estimates for a set of performance metrics for its execution in a certain environment.}
    \end{minipage}
\end{center}

This definition separates the observations enumerated above into two categories:
\begin{itemize}
    \item Implementation and memory locality ((\ref{itm:modeling.models.create:impl}) and (\ref{itm:modeling.models.create:memloc})) depend on the execution environment.
        Therefore, we will create different models for different BLAS implementations and memory locality situations.
    \item Fluctuations and argument values ((\ref{itm:modeling.models.create:fluct}) and (\ref{itm:modeling.models.create:disc}) through (\ref{itm:modeling.models.create:lda})) are covered by a single performance model.
\end{itemize}

To build the structure our models, we begin by considering the routine arguments.
Since some of these only have very little influence on performance, we only account for a subset of arguments in our models, called the model \emph{parameters}.

The aspects that can most significantly influence the performance behavior are discrete arguments and special values for scalar arguments (\ref{itm:modeling.models.create:disc}): they can trigger the execution of different code branches.
To reduce the model complexity and the effort spent on generating it, we select a subset of these --- the \emph{discrete parameters}. The argument type determines if it is modeled; \texttt{diag} and scalar arguments are usually omitted.
In our models, each combination of discrete parameter values --- referred to as a \emph{(discrete) case} --- is treated separately.
Example: If we consider three discrete parameters with two possible values each, we have $2^3 = 8$ cases; each of them is modeled independently.

For every case, we introduce a separate (sub-)model for each performance metric.
The remaining aspects (\ref{itm:modeling.models.create:size}), (\ref{itm:modeling.models.create:lda}), and (\ref{itm:modeling.models.create:size}) are treated for each case and each performance metric separately.

We now turn to the dependence of performance on size and leading dimension arguments ((\ref{itm:modeling.models.create:size}) and (\ref{itm:modeling.models.create:lda})), which can both take non-negative integer values.
In our models we offer to represent both argument types.
However, we have observed that leading dimension arguments ((\ref{itm:modeling.models.create:lda})) only have a minor influence on performance.
Therefore, we once more select a subset of size and leading dimension parameters (usually only the former) called \emph{continuous parameters}\footnote{
    Continuous not in the mathematical sense but in contrast to discrete parameters, which can only take two values.
}.

In principle, continuous parameters can take arbitrarily large values.
We, however, limit our models to a certain range --- for example, values between 8 and 1024.
Additionally, in order to avoid the small scale performance oscillations (\autoref{sec:modeling.experiments.size}), we restrict the allowed parameter values to multiples of \metric{mingap}: a small step size, such as 4, 8 or 16.
The collection of the sets of allowed values for all continuous parameters spans a product space, referred to as \emph{(continuous) parameter space}.
Example: Two continuous parameters within $\{8, 16, \ldots, 1024\}$ span the parameter space $\{8, 16, \ldots, 1024\}^2$. 

As suggested in \autoref{sec:modeling.experiments.size}, our choice to represent the dependencies on continuous parameters are multivariate piecewise polynomials.
In 1D, these are several intervals of the parameter axis each associated with a polynomial.
In higher dimension, however, piecewise polynomials could in general consist of arbitrary regions of the parameter space associated with multivariate polynomials.
We restrict the type of regions to (hyper-)cuboids with faces parallel to the parameter axes (intervals in 1D, rectangles in 2D, and cuboids in 3D).
These can be represented by two points, determining lower and upper limit of the region along each coordinate direction.

It remains to incorporate aspect (\ref{itm:modeling.models.create:fluct}): The fluctuations of performance counters for fixed argument values.
These fluctuations can only be represented in a statistical or probabilistic fashion.
In principle, we would like to create a probability distribution for each performance counter at every point.
Since this is not possible in practice, we limit our models to a vector of statistical quantities: minimum, average, median, standard deviation, maximum, and possibly others.
Therefore, we associate each polynomial region with a vector-valued polynomial, providing these quantities.
As a result, the model for one discrete case is a vector-valued multivariate piecewise polynomial.

            \subsection{Evaluation of a Model at a Point}
            \label{sec:modeling.models.eval}
            We now show how a model of the previously introduced type is evaluated given values for the modeled routine's arguments.
We take \texttt{dtrsm} ($B \leftarrow A^{-1} B$, $A$ triangular) as an example:
\begin{center}
    \texttt{dtrsm(side, uplo, transA, diag, m, n, alpha, A, ldA, B, ldB)}.
\end{center}
We consider a performance model with the following properties:
\begin{itemize}
    \item \texttt{side}, \texttt{uplo}, and \texttt{transA} are its discrete parameters, resulting in 8 cases;
    \item \texttt m and \texttt n are the continuous parameters, both taken from the range $\{8, \ldots, 1024\}$;
    \item \texttt{diag}, \texttt{alpha}, \texttt{ldA}, and \texttt{ldB} are neglected;
    \item It provides estimates for \metric{ticks}, \metric{flops}, and \metric{L1misses};
    \item It represents the performance through the statistical quantities minimum, average, standard deviation, and maximum.
\end{itemize}

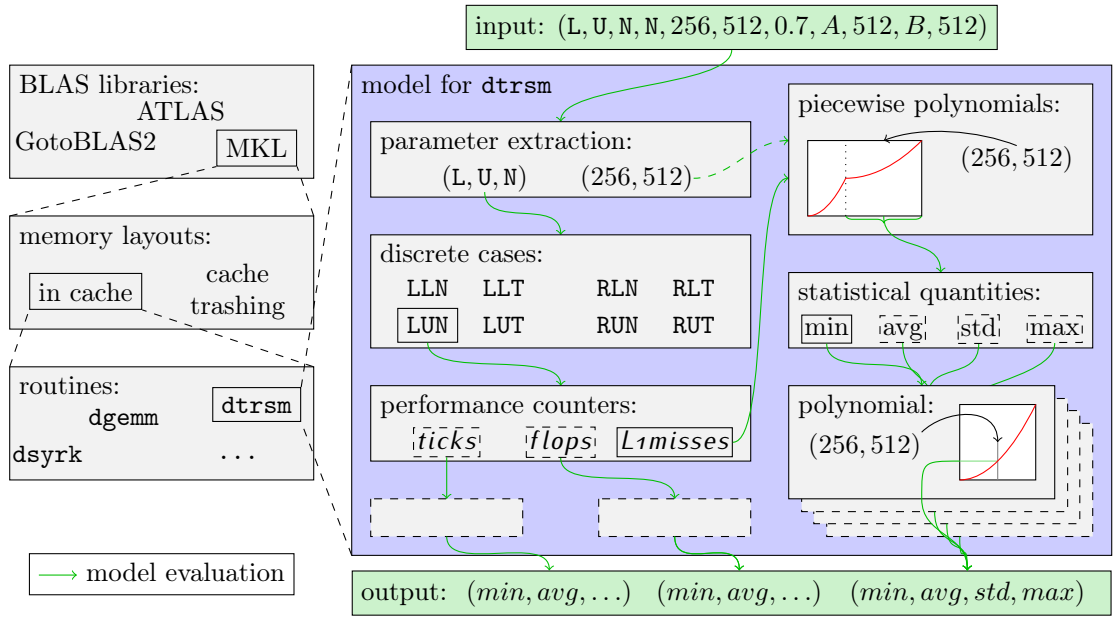
\begin{figure}[t]
    \tikzset{external/export=true}
    \centering
    \begin{tikzpicture}
        \coordinate (pos) at (0, 0);

        \filldraw[fill=graybg] (pos) rectangle ++(4, -1.5);
        \path (pos) node[anchor=north west] {BLAS libraries:};
        \path (pos) ++(3.25, -1.1) node[draw] (mkl) {MKL};
        \path (pos) ++(1, -1) node {\gotoblas};
        \path (pos) ++(2.25, -.6) node {ATLAS};

        \draw[dashed] (pos) ++(0, -2) -- (mkl.south west);
        \draw[dashed] (pos) ++(4, -2) -- (mkl.south east);

        \path (pos) ++(0, -2) coordinate (pos);
        \filldraw[fill=graybg] (pos) rectangle ++(4, -1.5);
        \path (pos) node[anchor=north west] {memory layouts:};
        \path (pos) ++(1, -1) node[draw] (incache) {in cache};
        \path (pos) ++(3, -1) node[text width=1.5cm, align=center] {cache trashing};

        \draw[dashed] (pos) ++(0, -2) -- (incache.south west);
        \draw[dashed] (pos) ++(4, -2) -- (incache.south east);

        \path (pos) ++(0, -2) coordinate (pos);
        \filldraw[fill=graybg] (pos) rectangle ++(4, -1.5);
        \path (pos) node[anchor=north west] {routines:};
        \path (pos) ++(1.5, -.7) node {\texttt{dgemm}};
        \path (pos) ++(.5, -1.2) node {\texttt{dsyrk}};
        \path (pos) ++(3.25, -.5) node[draw] (dtrsm) {\texttt{dtrsm}};
        \path (pos) ++(3, -1.2) node {\texttt{...}};

        \draw[dashed] (pos) ++(4.5, 4) -- (dtrsm.north east);
        \draw[dashed] (pos) ++(4.5, -2.5) -- (dtrsm.south east);
        
        \path (pos) ++(4.5, 4) coordinate (pos) coordinate (model);
        \filldraw[fill=plot3bg] (pos) rectangle ++(10, -6.5);
        \path (pos) node[anchor=north west] {model for \texttt{dtrsm}};

        \path (pos) ++(5, .5) node[draw, fill=plot2bg] (input) {input: $(\mathtt L, \mathtt U, \mathtt N, \mathtt N, 256, 512, 0.7, A, 512, B, 512)$};

        \path (pos) ++(.25, -.75) coordinate (pos) coordinate (args);
        \draw[<-, plot2] (args) ++(2.5, 0) .. controls ++(0, .5) and ++(0, -.5) .. (input.south);
        \filldraw[fill=graybg] (pos) rectangle ++(5, -1);
        \path (pos) node[anchor=north west] {parameter extraction:};
        \path (pos) ++(1.5, -.75) node[inner sep=0pt] (discarg) {$(\mathtt L, \mathtt U, \mathtt N)$};
        \path (pos) ++(3.5, -.75) node[inner sep=0] (contarg) {$(256, 512)$};

        \path (pos) ++(0, -1.5) coordinate (pos) coordinate (cases);
        \draw[<-, plot2] (cases) ++(2.5, 0) .. controls ++(0, .5) and ++(0, -.5) .. (discarg.south);
        \filldraw[fill=graybg] (pos) rectangle ++(5, -1.5);
        \path (pos) node[anchor=north west] {discrete cases:};
        \foreach \side/\xa in {L/0, R/2.5}
            \foreach \uplo/\yb in {L/0, U/-.5}
                \foreach \trans/\xc in {N/0, T/1}
                    \draw (pos) ++(.75,-.7) ++(\xc, 0) ++ (\xa, \yb) node (\side\uplo\trans) {\texttt{\side\uplo\trans}};
        \path (LUN) node[fill=graybg, draw] (lun) {\texttt{LUN}};

        \path (pos) ++(0, -2) coordinate (pos) coordinate (counters);
        \draw[<-, plot2] (counters) ++(2.5, 0) .. controls ++(0, .5) and ++(0, -.5) .. (lun.south);
        \filldraw[fill=graybg] (pos) rectangle ++(5, -1);
        \path (pos) node[anchor=north west] {performance counters:};
        \path (pos) ++(1, -.75) node[inner sep=1pt, draw, dashed] (ticks) {\metric{\vphantom{p}ticks}};
        \path (pos) ++(2.5, -.75) node[inner sep=1pt, draw, dashed] (flops) {\metric{flops}};
        \path (pos) ++(4, -.75) node[inner sep=1pt, draw] (l1) {\metric{\vphantom{p}L1misses}};

        \path (pos) ++(0, -1.5) coordinate (pos2);
        \filldraw[fill=graybg,dashed] (pos2) rectangle ++(2, -.5);
        \filldraw[fill=graybg,dashed] (pos2) ++(3, 0) rectangle ++(2, -.5);
        \draw[<-, plot2] (pos2) ++(1, 0) .. controls ++(0, .5) and ++(0, -.5) .. (ticks.south);
        \draw[<-, plot2] (pos2) ++(4, 0) .. controls ++(0, .5) and ++(0, -.5) .. (flops.south);

        \path (pos) ++(5.5, 4) coordinate (pos) coordinate (ppol);
        \draw[<-, plot2] (ppol) ++(0, -1.25) .. controls ++(-.5, 0) and ++(.5, 0) .. (l1.east);
        \draw[<-, plot2, dashed] (ppol) ++(0, -.75) .. controls ++(-.5, 0) and ++(.5, 0) .. (contarg.east);
        \filldraw[fill=graybg] (pos) rectangle ++(4, -2);
        \path (pos) node[anchor=north west] {piecewise polynomials:};
        \path (pos) ++(.25, -1.75) coordinate (plotbase) ++(1, 0) coordinate (pb1a);
        \filldraw[fill=white] (plotbase) rectangle ++(1.5, 1);
        \draw[plot1] (plotbase) parabola ++(.5, .5) parabola ++(1, .5);
        \draw[dotted] (plotbase) ++(.5, 0) -- ++(0, 1);
        \path (pos) ++(3, -1) node[inner sep=0] (coords) {$(256, 512)$};
        \path (plotbase) ++(1, 1) coordinate (target);
        \draw[->] (coords.north) .. controls ++(-.5, .2) and ++(.5, .2) .. (target);

        \path (pos) ++(0, -2.5) coordinate (pos) coordinate (stat);
        \draw[decorate, decoration=brace, plot2] (pb1a) ++(.5, 0) -- ++(-1, 0);
        \draw[<-, plot2] (stat) ++(2, 0) .. controls ++(0, .5) and ++(0, -.5) .. (pb1a);
        \filldraw[fill=graybg] (pos) rectangle ++(4, -1);
        \path (pos) node[anchor=north west] {statistical quantities:};
        \path (pos) ++(.5, -.75) node[inner sep=1pt, draw] (min) {\vphantom{gt}min};
        \path (pos) ++(1.5, -.75) node[inner sep=1pt, draw, dashed] (std) {\vphantom{gt}avg};
        \path (pos) ++(2.5, -.75) node[inner sep=1pt, draw, dashed] (avg) {\vphantom{gt}std};
        \path (pos) ++(3.5, -.75) node[inner sep=1pt, draw, dashed] (max) {\vphantom{gt}max};

        \path (pos) ++(0, -1.5) coordinate (pos) ++(1.75, -1.5) coordinate (polss);
        \draw[<-, plot2] (pos) ++(1.75, 0) .. controls ++(0, .5) and ++(0, -.5) .. (min.south);
        \draw[<-, plot2] (pos) ++(1.83, -.17) .. controls ++(0, .5) and ++(0, -.5) .. (avg.south);
        \draw[<-, plot2] (pos) ++(1.92, -.33) .. controls ++(0, .5) and ++(0, -.5) .. (std.south);
        \draw[<-, plot2] (pos) ++(2, -.5) .. controls ++(0, .5) and ++(0, -.5) .. (max.south);
        \filldraw[fill=graybg, dashed] (pos) ++(.5, -.5) rectangle ++(3.5, -1.5);
        \filldraw[fill=graybg, dashed] (pos) ++(.33, -.33) rectangle ++(3.5, -1.5);
        \filldraw[fill=graybg, dashed] (pos) ++(.17, -.17) rectangle ++(3.5, -1.5);
        \filldraw[fill=graybg] (pos) rectangle ++(3.5, -1.5);
        \path (pos) node[anchor=north west] {polynomial:};
        \path (pos) ++(1, -.8) node[inner sep=0] (coords) {$(256, 512)$};
        \path (pos) ++(2.25, -1.25) coordinate (plotbase);
        \filldraw[fill=white] (plotbase) rectangle ++(1, 1);
        \draw[gray] (plotbase) ++(.5, 0) -- ++(0, .5); 
        \draw[plot2!50] (plotbase) ++(.5, .25) -- ++(-.5, 0); 
        \draw[<-] (plotbase) ++(.5, .5) .. controls ++(0, .5) and ++(.3, .3) .. (coords.north east);
        \draw[plot1] (plotbase) parabola ++(1, 1);
        \draw[plot2] (plotbase) ++(0, .25) .. controls ++(-.5, 0) and ++(0, .5) .. (polss.center);

        \path (model) ++(0, -7) coordinate (output);
        \path (output) node[anchor=west, align=left, text width=9.75cm, draw, fill=plot2bg] {\vphantom{$()$}output:};
        \path (output) ++(2.6, 0) node (outticks) {$(min, avg, \ldots)$};
        \path (output) ++(5.1, 0) node (outflops) {$(min, avg, \ldots)$};
        \path (output) ++(8.1, 0) node (outl1) {$(min, avg, std, max)$};
        \draw[->, plot2] (pos2) ++(1, -.5) .. controls ++(0, -.5) and ++(0, .5) .. (outticks.north);
        \draw[->, plot2] (pos2) ++(4, -.5) .. controls ++(0, -.5) and ++(0, .5) .. (outflops.north);
        \draw[->, plot2] (pos2) ++(4, -.5) .. controls ++(0, -.5) and ++(0, .5) .. (outflops.north);
        \draw[->, plot2] (polss) ++(0, 0) .. controls ++(0, -.5) and ++(0, .5) .. (outl1.north);
        \draw[->, plot2] (polss) ++(.17, -.17) .. controls ++(0, -.5) and ++(0, .5) .. (outl1.north);
        \draw[->, plot2] (polss) ++(.33, -.33) .. controls ++(0, -.5) and ++(0, .5) .. (outl1.north);
        \draw[->, plot2] (polss) ++(.5, -.5) .. controls ++(0, -.5) and ++(0, .5) .. (outl1.north);

        \node[draw] at (2, -6.75) {
            \tikz[baseline=-.5ex] \draw[->, plot2] (0, 0) -- (.5, 0); model evaluation\vphantom{g}
        };
    \end{tikzpicture}
    \caption{Structure and categorization of performance models and their evaluation.}
    \label{fig:modeling.models.eval.flow}
    \tikzset{external/export=false}
\end{figure}

\autoref{fig:modeling.models.eval.flow} visualizes the evaluation of this model for the exemplary argument values
\begin{center}
    \texttt{%
        dtrsm(%
        \overparameq{side}{L},
        \overparameq{uplo}{U},
        \overparameq{transA}{N},
        \overparameq{diag}{N},
        \overparameq{m}{256},
        \overparameq{n}{512},
        \overparameq{alpha}{0.7},
        A,
        \overparameq{ldA}{512},
        B,
        \overparameq{ldB}{512})%
    }.
\end{center}

Given this tuple of arguments $(\mathtt L, \mathtt U, \mathtt N, \mathtt N, 256, 512, A, 512, B, 512)$, the first step in the model evaluation is to extract the model parameters.
In our case, these are:
$$
    (\mathtt{side}, \mathtt{uplo}, \mathtt{transA}, \mathtt m, \mathtt n) = (\mathtt L, \mathtt U, \mathtt N, 256, 512).
$$
These are separates into discrete and continuous parameters:
\begin{itemize}
    \item The discrete parameters $(\mathtt{size}, \mathtt{uplo}, \mathtt{transA}) = (\mathtt L, \mathtt U, \mathtt N)$ determine the case;
    \item The continuous parameters $(\mathtt m, \mathtt n) = (256, 512)$ identify a point in the two dimensional continuous parameter space.
\end{itemize}

For the case $(\mathtt L, \mathtt U, \mathtt N)$, the model has one vector valued multivariate piecewise polynomial $P: \{8, \ldots, 1024\} \rightarrow \mathbb N^4$ for each performance counter; the result vector contains the statistical quantities.
$P$ is represented by a set of rectangular regions that cover the two dimensional parameter space, each associated with one vector valued piecewise polynomial.
The point $(256, 512)$ lies within at least one of these regions.
For the case of overlapping regions, the model accuracy of each of them is stored along with its polynomial; the region with the most accurate model is selected.

For the identified region, the associated vector valued polynomial is evaluated at the point $(256, 512)$.
The resulting vector contains the quantities minimum, average, standard deviation, and maximum for one performance counter.

The process --- identifying the region and evaluating its polynomial --- is applied to all performance counters: \metric{ticks}, \metric{flops}, and \metric{L1misses}.
The result of the model evaluation is a vector with the statistic quantities for each of them.

        \section{The Modeler}
        \label{sec:modeling.tool}
        In the previous section, we have described the structure of our targeted performance models.
We now introduce the tool that automatically generates these models: the \emph{Modeler}.

It is implemented in Python mainly for the following reasons:
\begin{itemize}
    \item Python has extensive high-level built-in functionality for lists and dictionaries, including functional programming constructs;
    \item It is completely object oriented;
    \item It can easily interface with other programs (in our case the Sampler);
    \item Its scientific library SciPy provides mathematical tools, such as least-squares solvers.
\end{itemize}

The Modeler uses an iterative approach, starting with an initial set of samples and a very rough or small  models.
Depending on the accuracy, more samples are taken to expand or refine the models.
The main program flow is as follows.
\begin{enumerate}
    \item Initialization:
        \begin{itemize}
            \item Read a configuration file.
            \item Initialize one instance of the Sampler.
            \item For every routine that needs to be modeled, create a separate \emph{Routine Modeler} (RModeler).
        \end{itemize}
    \item While not all RModelers have completed modeling their associated routines' performance do the following:
        \begin{enumerate}
            \item Retrieve a list of sampling requests from each of the RModelers;
            \item Execute the sampling requests by the Sampler;
            \item Let the RModelers refine their models using the sampling results.
        \end{enumerate}
    \item Finalization: Retrieve the generated models form the RModelers and write them to a file.
\end{enumerate}

The Modeler is limited to one Sampler; models for different BLAS implementations and memory locality situations can be obtained through the Modeler configuration file.

The Modeler consists of several, separately discussed components:
\begin{itemize}
    \item The Sampler Interface (\autoref{sec:modeling.tool.sampler});
    \item The RModelers (\autoref{sec:modeling.tool.rmodeler});
    \item The \emph{Piecewise Polynomial Modelers} (PModelers) as part of the RModelers (\autoref{sec:modeling.tool.pmodeler}).
\end{itemize}

            \subsection{Sampler Interface}
            \label{sec:modeling.tool.sampler}
            The Sampler Interface (SI) serves as a Python wrapper for the Sampler, handling and passing on sampling requests and returning the resulting measurements.
It keeps a list of all measurements in a \emph{memory file}; when the Modeler is executed repeatedly with the same Sampler configuration, measurements from this file are used to reduce the time spent on sampling.

Upon initialization, the SI loads the memory file corresponding to its Sampler's configuration;  if no such file exists, a new memory file is created.
Then, the Sampler process is started and OS pipes are used to exchange data.

The RModelers pass sampling requests to the SI in the form of tuples, for instance
$$(\mathtt{dtrsm}, \mathtt L, \mathtt U, \mathtt N, \mathtt N, {256}, {512}, \mathtt v.5, 131072, 512, 262144, 512).$$
All incoming requests are collected and processed together, once the Modeler instructs the SI to do so.

When this happens, the SI first scans its memory file for results matching the collected requests; the matches are added to an initially empty result list.
Each entry in the memory file is served at most once per execution of the Modeler; Identical sampling requests receive different sampling results.

Those remaining requests that cannot be served from the memory file are passed to the Sampler.
Every resulting measurement is both stored in the memory file and added to the result list.
Once all sampling requests are covered by this list, each RModeler --- identified by its routine's name --- is passed exactly those results that correspond to its routine.

Upon termination, the Sampler Interface stores its memory file.

            \subsection{Routine Modeler}
            \label{sec:modeling.tool.rmodeler}
            A Routine Modeler (RModeler) creates a performance model for a single routine.
Its main task is to provide a layer of abstraction between two components: the Sampler Interface and the Piecewise Polynomial Modelers (PModelers), whose job is to create models of the form
$$
    f: \mathbb N^\text{\#continuous parameters} \rightarrow \mathbb N^\text{\#statistical quantities}.
$$
The abstraction, as seen from the Modeler and the Sampler Interface, consists of the following stages:
\begin{description}
    \item[Stage 1] Selection of parameters from the routine's argument list;
    \item[Stage 2] Separation of discrete and continuous parameters;
    \item[Stage 3] Treatment of the discrete cases;
    \item[Stage 4] Handling of PModelers for each case and performance counter.
\end{description}

The RModeler provides three major functions, each of which covers the four stages of abstraction:
\begin{itemize}
    \item Generating of sampling requests for the Sampler Interface (\autoref{sec:modeling.tool.rmodeler.requests});
    \item Passing the obtained sampling result to the PModelers in order to improve their models (\autoref{sec:modeling.tool.rmodeler.results});
    \item Constructing an exporting the final model (\autoref{sec:modeling.tool.rmodeler.assembly}).
\end{itemize}

                \subsubsection{Generation of Sampling Requests}
                \label{sec:modeling.tool.rmodeler.requests}
                The PModelers generate sampling requests in order to improve their models.
These requests are processed by the RModeler according to the four stages of abstraction and then passed to the Sampler Interface.

Each PModeler constructs a piecewise polynomial model $f$ for one performance counter $p_i$ for one of the discrete cases $c_j$ of the form 
$$
    f: \mathbb N^\text{\#continuous parameters} \rightarrow \mathbb N^\text{\#statistical quantities}.
$$
In every iteration of the Modeler's main loop, it generates a list of requests $l_i^j$; the requests are points within the continuous parameter space, for instance $(256, 512)$.
The RModeler needs to transform these tuples into a complete sampling requests, such as
$$
    (\mathtt{dtrsm}, \mathtt L, \mathtt U, \mathtt N, \mathtt N, \mathbf{256}, \mathbf{512}, \mathtt v.5, 131072, 512, 262144, 512).
$$
This is done as follows:
\begin{description}
    \item[Stage 4] The RModeler begins by merging the request lists $l_j^i$ of all PModelers for the same discrete case $c_j$ into one list $l_j$.
        It is not necessary to keep track of which PModeler issued which requests, since all performance counters are measured for each sampling request.
        Every PModeler is afterwards given as many results to construct its model from as possible --- even those which it did not request.
        
        When $l_j$ is created, all duplicates across \emph{different} PModelers are removed:
        When two or more PModelers request samples at the same point, the maximum number of samples at this point appears in $l_j$ --- not the sum of all requests.

        Next, $l_j$ is compared with the list of sampling results for the case $c_j$ that the RModeler has already received for previous sampling requests; it is possible that a PModeler requests samples at points that were already sampled for other PModelers.
        Every point in $l_j$ that is found in the sampling results is removed from the list as many times as measurements are available for it already.
        This ensures that each PModeler receives at least as many samples at each point as specified.
        (It also means that when a PModeler wants to increase the number of samples at a point it previously requested, it needs to specify the total number of samples it requires --- not the difference.)

        The output of this stage of abstraction is a list of sampling points $l_j$ for each discrete case $c_j$.
    \item[Stage 3] Next, the RModeler creates a list that contains all lists $l_j$, each associated with its case $c_j$: $((l_1, c_1), (l_2, c_2), \ldots)$.
    \item[Stage 2] Then, the discrete parameters for each case are incorporated into the sampling requests, resulting in a single list of requests $l$.
        Requests in this list are of the form $(c_i, p)$, for instance $(\mathtt L, \mathtt U, \mathtt N, 256, 512)$ when three discrete parameters constitute each case.
    \item[Stage 1]  In the last step, the RModeler needs to add the routine name and values for the missing routine arguments to each request in $l$, for example,
        $$(\mathtt{dtrsm}, \texttt{\bfseries L}, \texttt{\bfseries U}, \texttt{\bfseries N}, \mathtt N, \mathbf{256}, \mathbf{512}, \mathtt v.5, 131072, 512, 262144, 512).$$

        An argument that does not correspond to a parameter is assigned a value according to its type:
        \begin{description}
            \item[Discrete] arguments are given a default value.
            \item[Size] arguments are also given a default value.
                However, usually all size arguments are model parameters.
            \item[Scalar] arguments receive a default value, such as $\mathtt v.5$. (It is also possible to configure these as discrete parameters, for example, with values $-1$, $0$, $1$, and $.5$.)
            \item[Leading dimension] arguments are handled in one of two ways:
                \begin{itemize}
                    \item They are given a large default value, such as 2500.
                        This represents the situation where the corresponding matrices are submatrices of a larger matrix.
                    \item They are set exactly to the height of the corresponding matrix, which is given by one of the size arguments.
                        In order to determine the correct size argument the dependency between size and leading dimension arguments (which may in turn depend on the discrete arguments, such as \texttt{transA} or \texttt{side}) is encoded in the Python representation of the routine signature.
                \end{itemize}
            \item[Increment] arguments are set to either $1$ or a larger default value to represent the access of a row of a matrix.
            \item[Matrix (or vector)] arguments are set to the number of element that their matrices (or vectors) occupy in memory.
                This number is the product of the leading dimension (vector length) argument and the width of the matrix (1).
                Which size argument is involved may again depend on the discrete arguments and is determined by the Python representation of the routines signature.
        \end{description}
\end{description}
The result of these four steps is a list of sampling requests that is passed to the Sampler Interface.

                \subsubsection{Processing Sampling Results}
                \label{sec:modeling.tool.rmodeler.results}
                When an RModeler receives results from the Sampler Interface, each of them consist of the sampling request and one measurement for each performance counter.

Before the RModeler can present the results to the PModelers, they need to be processed according to the four stages of abstraction:
\begin{description}
    \item[Stage 1] The RModeler begins by extracting the parameters from the sampling requests, which consists of values for all routine arguments.
    \item[Stage 2] These are then split into discrete and continuous parameters.
    \item[Stage 3] The results are added to the list of all results for the corresponding case, which is determined by the discrete parameters.
    \item[Stage 4] Each PModeler is then given a list of results corresponding to its case, containing only the measurements for its performance counter.
        The RModelers immediately use the results to improve or advance their models.
\end{description}

                \subsubsection{Assembly of the Model}
                \label{sec:modeling.tool.rmodeler.assembly}
                When all models are complete, each RModeler retrieves the piecewise polynomials from its PModelers and assembles a model object.
This independent object is then written to a file and later used independently of the Modeler.

The structure of this model was introduced in \autoref{sec:modeling.models} and resembles the four stages of abstraction.
This becomes apparent when we consider its evaluation for given routine arguments:
\begin{description}
    \item[Stage 1] The model parameters are extracted
    \item[Stage 2] They are separated into discrete and continuous parameters;
    \item[Stage 3] The discrete parameters determine a case;
    \item[Stage 4] For this case, all piecewise polynomials are evaluated at the continuous parameters.
\end{description}

            \subsection{Piecewise Polynomial Modelers}
            \label{sec:modeling.tool.pmodeler}
            In \autoref{sec:modeling.experiments.size}, we saw that performance depends on the size arguments of BLAS routines in the form of piecewise polynomials.
Constructing such polynomials by hand is rather simple given a large number of samples.
By contrast, the Piecewise Polynomial Modelers (PModelers) build models that closely fit the performance automatically and with as few samples as possible.

The process for modeling performance involves three steps.
First, a set of sampling points is chosen for an initial region within the parameter space.
The measurements resulting from these sampling points is modeled through least squares fitting as a polynomial.
According to the polynomial's accuracy, the region is reshaped, more regions are generated or the model is accepted.
These steps are repeated until the whole parameter space is covered with accurate models.

We offer two PModeler implementations, which differ in both their choice of sampling points and the strategy they use to cover the parameter space with regions.
Before we present the implementations in Sections \ref{sec:modeling.tool.exp} and \ref{sec:modeling.tool.ref}, we discuss their common basis.

                \subsubsection{Polynomial Fitting through Least Squares}
                \label{sec:modeling.tool.pmodeler.lsfit}
                The approximation of a set of sampling results by polynomials through least squares fitting is a task common to all Piecewise Polynomial Modelers.
Every performance counter is represented by a set of statistical quantities, such as minimum or average; each quantity is fitted by a separate polynomial.
In this section, we introduce polynomial fitting for a single quantity.

The fitting procedure has the following inputs:
\begin{itemize}
    \item A list of coordinates $\{\mathbf x_1, \ldots, \mathbf x_n\} \in \mathbb N^d$ from the $d$ dimensional parameter space;
    \item A value for each coordinate $\mathbf v = (v_1, \ldots, v_n)^T$;
    \item A set of monomials\footnote{
            A polynomial with only one non-zero term, such as $1$, $x_1$, or $x_1 x_2^2$.
        } $\{m_1, \ldots, m_b\} \subset \{f:\mathbb N^d \rightarrow \mathbb N\}$ as a base for the polynomials.
\end{itemize}
Its goal is to find a polynomial $P(\mathbf x) = \sum\limits_{i = 1}^b a_i m_i(\mathbf x)$ with coefficient vector $\mathbf a = (a_1, \ldots, a_b)^T \in \mathbb R^b$ such that $\sum\limits_{i = 1}^n \bigl(P(\mathbf x_i) - v_i\bigr)^2$ is minimal. 
This is know as a \emph{least squares problem} and can be written as $\underset{\mathbf a}{\operatorname{arg\,min}} ||X \mathbf a - \mathbf v||^2$, where
$$
    X = 
    \left(\begin{matrix}
        m_1(\mathbf x_1)    &m_2(\mathbf x_1)   &\cdots &m_b(\mathbf x_1)\\
        m_1(\mathbf x_2)    &m_2(\mathbf x_2)   &\cdots &m_b(\mathbf x_2)\\
        \vdots              &\vdots             &\ddots &\vdots\\           
        m_1(\mathbf x_n)    &m_2(\mathbf x_n)   &\cdots &m_b(\mathbf x_n)\\
    \end{matrix}\right).
$$

Its solution can be obtained as $\mathbf a = (X^T X)^{-1} X^T \mathbf v$ or by the means of singular value decomposition (SVD).
We use the function \texttt{linalg.lstsq()} provided by Python's SciPy package, which is based on SVD.

For small values of $\mathbf x_i$ and $v_i$, this least squares solver yields very accurate results.
If, however, the $\mathbf x_i$ and $v_i$ are concentrated in a small area far away from the origin, the solution accuracy decreases significantly.
This is not a problem of the solution method but rather due to the conditioning of the least squares problem.

\begin{figure}[t]
    \centering
    \begin{tikzpicture}
        \draw[->] (-2, 0) -- (7, 0) node [anchor=west] {$\mathbf x$};
        \draw[->] (0, -1) -- (0, 5) node [anchor=south] {$v$};
        \node[anchor=north east] at (0,0) {$0$};

        \path plot[mark=ball, ball color=plot1] coordinates {
            (4, 3.5)
            (5, 3)
            (6, 4)
            (7, 5)
        };

        \path plot[mark=ball, ball color=plot2] coordinates {
            ($(4, 3.5) - 1/4*(22, 15.5)$)
            ($(5, 3)   - 1/4*(22, 15.5)$)
            ($(6, 4)   - 1/4*(22, 15.5)$)
            ($(7, 5)   - 1/4*(22, 15.5)$)
        };

        \foreach \x/\y in {4/3.5, 5/3, 6/4, 7/5}
            \draw[->, dotted] (\x, \y) -- +($-1/4*(22, 15.5)$);
        \path (4, 3.5) -- +($-1/4*(22, 15.5)$) node[midway, above, sloped] {coordinate translation};

        \draw[plot3, domain=1.5:5.5] plot (\x - 14/4, 0.375 * \x * \x - 2.075 * \x + 6.075 - 15.5/4);
        \draw[plot4, domain=1.5:5.5] plot (\x + 2, 0.375 * \x * \x - 2.075 * \x + 6.075);

        \node[draw, align=left] at (5.75, 1.25) {
            \tikz[baseline=-.5ex] {\path plot[mark=ball, ball color=plot1] coordinates {(0,0)}; \path (-.25, 0) (.25, 0);}
            $(\mathbf x_i, v_i)$\\
            \tikz[baseline=-.5ex] {\path plot[mark=ball, ball color=plot2] coordinates {(0,0)}; \path (-.25, 0) (.25, 0);}
            $(\widetilde{\mathbf x}_i, \widetilde v_i)$\\
            \tikz[baseline=-.5ex] \draw[plot3] plot coordinates {(-.25,0) (.25,0)};
            $\widetilde P$\\
            \tikz[baseline=-.5ex] \draw[plot4] plot coordinates {(-.25,0) (.25,0)};
            $P$
        };

    \end{tikzpicture}
    \caption{Translation of the least squares problem.}
    \label{fig:modeling.tool.pmodeler.lstrans}
    \tikzset{external/export=false}
\end{figure}
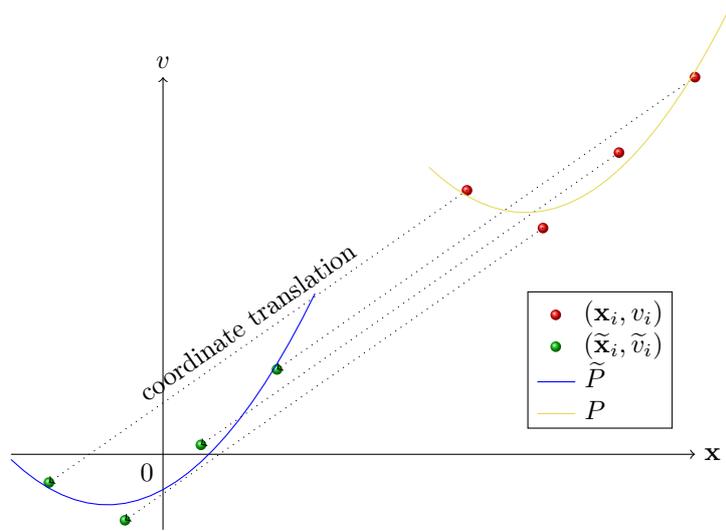

In order to improve the solution accuracy, we need to improve the conditioning  of $X$.
For this purpose, we translate the problem to the origin, solve the translated problem, and translate the solution back to the original problem (see \autoref{fig:modeling.tool.pmodeler.lstrans}).
We begin by translating the coordinates $\mathbf x_i$ and values $v_i$~(\tikzball{plot1}) to $\widetilde{\mathbf x}_i$ and $\widetilde v_i$~(\tikzball{plot2}) such that they are evenly distributed around the origin:
$$
    \widetilde{\mathbf x}_i = \mathbf x_i - \frac1n \sum\limits_{j = 1}^n \mathbf x_j
    \text{ and }
    \widetilde v_i = v_i - \frac1n \sum\limits_{j = 1}^n v_j.
$$
These quantities form a new least squares problem $\underset{\widetilde{\mathbf a}}{\operatorname{arg\,min}} \bigl\|\widetilde X \widetilde{\mathbf a} - \widetilde{\mathbf v}\bigr\|^2$; its solution defines a polynomial $\widetilde P(\mathbf x) = \sum\limits_{j = 1}^b \widetilde a_i m_i(\mathbf x)$~(\tikzline{plot3}).
The solution $P$~(\tikzline{plot4}) to the original problem is now obtained through a second translation:
$$
    P(\mathbf x) = \widetilde P\Bigl(\mathbf x - \frac1n \sum\limits_{j = 1}^n \mathbf x_j\Bigr) + \frac1n \sum\limits_{j = 1}^n v_j.
$$

In addition to this coordinate translation, our least squares fitting mechanism rounds coefficient to close rational numbers or discards relatively small coefficient according to a configurable threshold.
This is especially useful for performance counters such as \metric{flops}, where all coefficients are usually rational numbers with small denominators.

                \subsubsection{Approximation Accuracy}
                \label{sec:modeling.tool.pmodeler.acc}
                The accuracy of a polynomial approximation $P$ for coordinates $\mathbf x_i$ and values $v_i$ is determined by the local errors $e_i = P(\mathbf x_i) - v_i$.
The used least squares approach minimizes $\sum\limits_{i = 1}^n e_i^2$; this quantity could be used to measure the approximation accuracy.
We, however, use the \emph{maximum relative error} across all $\mathbf x_i$:
$$
    e_\text{relmax} = \max\limits_{1 \leq i \leq n} \frac{|e_i|}{v_i}.
$$

For a vector valued polynomial representing the statistical quantities, we select the approximation error in one of the statistical quantities to represent the approximation's accuracy; usually, we choose the median.

            \subsection{PModeler: Model Expansion}
            \label{sec:modeling.tool.exp}
            In this section, we introduce the first of two Piecewise Polynomial Modelers (PModelers) presented in this thesis.
The \emph{Model Expansion} strategy creates piecewise polynomials as follows:
\begin{itemize}
    \item It begins by modeling a small region in a corner of the parameter space through a single polynomial (\autoref{sec:modeling.tool.exp.initial});
    \item This region is expanded as far as possible (\autoref{sec:modeling.tool.exp.exp}), as long as
        \begin{itemize}
            \item its polynomial's approximation accuracy is above a given threshold, and
            \item it stays within the boundaries of the parameter space;
        \end{itemize}
    \item Once a region cannot be extended further, new adjacent regions are generated (\autoref{sec:modeling.tool.exp.gen}).
\end{itemize}
The process of modeling, expansion, and region generation is repeated for all regions until the whole parameter space is covered.

Models are expanded either from the origin of the parameter space towards its maximum or in the opposite direction.
The initial domain is accordingly chosen at the minimum (bottom left) or maximum (top right) corner of the parameter space.
Without loss of generality, we introduce Model Expansion by considering the expansion away from the origin.

                \subsubsection{Initial Model}
                \label{sec:modeling.tool.exp.initial}
                The size of the initial region is chosen with both the resulting model and the expense of sampling in mind: Smaller regions lead to a higher resolution of the model, while larger --- thus, fewer --- regions require less sampling.
We found a length of $64$ or $128$ along each parameter direction to be a good compromise.

Within the initial region, the sampling points are chosen on a regular grid.
For a targeted polynomial order $o$, this grid has least $o + 1$ points along each parameter direction; for a $d$-dimensional parameters-space, the grid consists of $(o + 1)^d$ points.
At the expense of more sampling requests, the grid density can be increased to generate smoother models.

Once a first model is created, its approximation error is computed.
If this error is below the configurable threshold, the region is expanded (\autoref{sec:modeling.tool.exp.exp}); otherwise, new adjacent regions are created immediately (\autoref{sec:modeling.tool.exp.gen}).

                \subsubsection{Region Expansion}
                \label{sec:modeling.tool.exp.exp}
                While the approximation error of a region is below the target threshold, the region is expanded as far as possible, as long as
\begin{itemize}
    \item The parameter space boundary is not reached, and
    \item The approximation error stays above the threshold.
\end{itemize}
Expanding away from the origin, initially, only the lower limit $\mathbf b = (b_1, \ldots, b_d)$ (its base) of the region's extent is fixed.
An upper limit, denoted by $\mathbf c = (c_1, \ldots, c_d)$, has yet to be determined.
However, we can already give both a lower and an upper bound for $\mathbf c$:
The lower bound $\mathbf l = (l_1, \ldots, l_d)$ is the upper limit of the initial region's extent;
the upper bound $\mathbf u = (u_1, \ldots, u_d)$ is the maximum corner of the parameter space.
During the process of model expansion, the bounds $\mathbf l$ and $\mathbf u$ are increased and decreased, respectively, until they converge to one point $\mathbf c$.

We denote the region of the parameter space spanned by two points $\mathbf a = (a_1, \ldots, a_d)$ and $\mathbf b = (b_1, \ldots, b_d)$ by
$$
    R_\mathbf a^\mathbf b = \{(x_1, \ldots, x_d) \in \mathbb N^d| \forall i \in \{1, \ldots, d\} . a_i \leq x_i \leq b_i\}.
$$

At this point either the end of the parameter is reached or the approximation error is just below the threshold.

In each step of the model expansion, a set of sampling points $P \subset R_\mathbf b^\mathbf u$ is chosen.
For this purpose, a parameter value $p_i$ is selected along each parameter direction $i$ by means of the first of the following rules that is applicable:
\begin{enumerate}[(a)]
    \item \label{itm:modeling.tool.exp.rule1} When $\frac{u_i - l_i}2 \geq \metric{maxgap}$, select $p_i = u_i + \metric{maxgap}$;
    \item \label{itm:modeling.tool.exp.rule2} In the first step, when $u_i - l_i \geq \metric{maxgap}$, select $p_i = u_i$ (this reduces the number of steps in case that $c_i = u_i$);
    \item \label{itm:modeling.tool.exp.rule3} When $l_i + \metric{mingap} \geq u_i$, select $p_i = u_i$;
    \item \label{itm:modeling.tool.exp.rule4} Otherwise, select $p_i = \left\lfloor\frac{l_i + u_i}2\right\rfloor_\smallmetric{mingap}$, resulting in a binary search pattern\footnote{
        $\lfloor x \rfloor_y := \max\limits \{i \cdot y | i \in \mathbb N, i \cdot y \leq x \}$ denotes the largest multiple of $y$ that is smaller or equal to $x$.
    }.
\end{enumerate}
This results in one sampling value $p_i$ along each direction $i$.

Next, a set of sampling points $P$ is generated from the sampling values $p_i$:
$$
    P = \biggl(\bigtimes\limits_{i = 1}^d \{b_i, l_i, p_i\}\biggr) \setminus \biggl(\bigtimes\limits_{i = 1}^d \{b_i, l_i\}\biggr).
$$

\begin{figure}[t]
    \centering
    \subfloat[first step]{
        \label{fig:modeling.tool.exp.samples:first}
        \begin{tikzpicture}[scale=.75]
            \filldraw[fill=plot2] (0, 0) rectangle (2, 2);

            \foreach \x/\l in {0/b, 2/l, 7/u, 4/p}
                \draw[dotted] (\x, -.25) node[anchor=north] (\l1) {$\l_1$} -- ++(0, 5.5);
            \foreach \y/\l in {0/b, 2/l, 5/u, 3.5/p}
                \draw[dotted] (-.25, \y) node[anchor=east] (\l2) {$\l_2$} -- ++(7.5, 0);

            \draw[decorate, decoration=brace] (2, 5.5) -- ++(2, 0) node[midway, above] {\metric{maxgap}};

            \path plot[mark size=2.66pt, mark=ball, ball color=plot1] coordinates {(0, 3.5) (2, 3.5) (4, 3.5) (4, 2) (4, 0)};
        \end{tikzpicture}
    }
    \hfill
    \subfloat[second step]{
        \label{fig:modeling.tool.exp.samples:second}
        \begin{tikzpicture}[scale=.75]
            \filldraw[fill=plot2!80!plot1] (0, 0) rectangle (4, 2);

            \foreach \x/\l in {0/b, 4/l, 7/u, 5.5/p}
                \draw[dotted] (\x, -.25) node[anchor=north] (\l1) {$\l_1$} -- ++(0, 4);
            \foreach \y/\l in {0/b, 2/l, 3.3/u, 2.65/p}
                \draw[dotted] (-.25, \y) node[anchor=east] (\l2) {$\l_2$} -- ++(7.5, 0);

            \path plot[mark=ball, mark size=2.66, ball color=plot1] coordinates {(0, 2.65) (4, 2.65) (5.5, 2.65) (5.5, 2) (5.5, 0)};
        \end{tikzpicture}
    }

    \vspace{.5cm}

    \tikzset{external/export=false}
    \fbox{
        \tikz[baseline=(x.base), scale=.75] {
            \draw[<->] (-1, 0) -- (1, 0);
            \node[anchor=north, inner sep=0] (x) at (0, -.1) {\metric{\vphantom{i}maxgap}};
        }
        \hspace{.2cm}
        \tikz[baseline=(x.base), scale=.75] {
            \draw[<-] (-.1, 0) -- ++(-.25, 0);
            \draw[<-] (.1, 0) -- ++(.25, 0);
            \node[anchor=north, inner sep=0] (x) at (0, -.1) {\metric{mingap}};
        }
        \hspace{.2cm}
        \tikz {
            \fill[fill=plot2] (0, 0) rectangle (.25, .25);
            \fill[fill=plot2!80!plot1] (0, 0) -- (.25, .25) -- (.25, 0) -- cycle;
            \draw (0, 0) rectangle (.25, .25);
        }
        model
        \hspace{.2cm}
        \tikzball{plot1}
        $P$
    }
    \caption{Choice of sampling points for Model Expansion.}
    \label{fig:modeling.tool.exp.samples}
\end{figure}
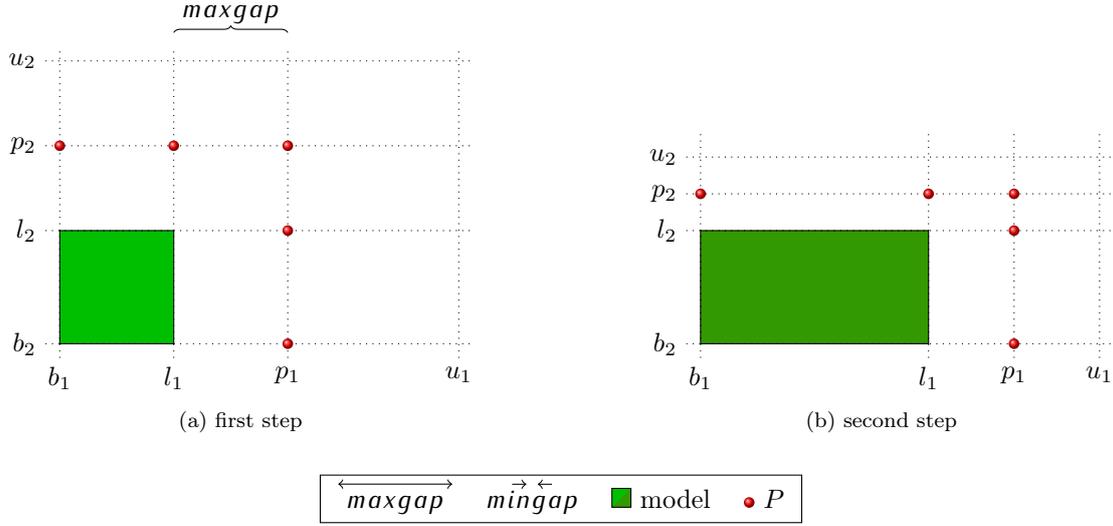

\begin{example}
Let us study an example illustrating the construction of $P$ (\autoref{fig:modeling.tool.exp.samples:first}).
We consider a two-dimensional parameter space ($d = 2$) with an initial region~(\tikzsquare{plot2}) in the lower left corner.
This region determines the base $\mathbf b = (b_1, b_2)$ and the initial lower approximation bound $\mathbf l = (l_1, l_2)$ for $\mathbf c = (c_1, c_2)$; the upper bound $\mathbf u = (u_1, u_2)$ is given by the upper limit of the parameter space.

The parameter values $p_1$ and $p_2$ are determined as follows (see \autoref{fig:modeling.tool.exp.samples:first}):
\begin{itemize}
    \item For $p_1$, rule (\ref{itm:modeling.tool.exp.rule1}) applies, since $\frac{u_1 - l_1}2 \geq \metric{maxgap}$,  resulting in  $p_1 = l_1 + \metric{maxgap}$.
    \item For $p_2$, none of the first three rules apply, since (\ref{itm:modeling.tool.exp.rule2}) $\frac{u_2 - l_2}2 < \metric{maxgap}$, (\ref{itm:modeling.tool.exp.rule3}) $u_2 - l_2 > \metric{maxgap}$, and (\ref{itm:modeling.tool.exp.rule3}) $u_2 - l_2 > \metric{mingap}$.
        Therefore,  $p_2 = \left\lfloor\frac{l_2 + u_2}2\right\rfloor_\smallmetric{mingap}$ is chosen according to rule (\ref{itm:modeling.tool.exp.rule4}).
\end{itemize}
Given $p_1$ and $p_2$, $P$~(\tikzball{plot1}) is constructed:
\begin{eqnarray*}
    P   &=  &\bigl(\{b_1, l_1, p_1\} \times \{b_2, l_2, p_2\}\bigr) \setminus \bigl(\{b_1, l_1\} \times \{b_2, l_2\}\bigr) \nonumber\\
        &=  &\bigl\{(b_1, b_2), (b_1, l_2), (b_1, p_2), (p_1, b_2), (p_1, l_2), (p_1, p_2), (l_1, b_2), (l_1, l_2), (l_1, p_2)\bigr\} \nonumber\\
        &   &\setminus \bigl\{(b_1, b_2), (b_1, l_2), (l_1, b_2), (l_1, l_2)\bigr\} \nonumber\\
        &=  &\bigl\{(b_1, p_2), (p_1, b_2), (p_1, l_2), (p_1, p_2), (l_1, p_2)\bigr\}. \nonumber
\end{eqnarray*}
The set $P$ forms a hull around the initial region (\tikzsquare{plot2}) with points on all intersections of $b_i$, $l_i$, and $p_i$.
\hfill \qed
\end{example}

Once the sampling results for the points $P$ are available, the expansion of the model along each direction $i$ is considered separately.
For direction $i$, the region of the model is tentatively expanded only along this direction up to and including $p_i$, resulting in a region $R_\mathbf{b}^{(l_1, \ldots, l_{i-1}, p_i, l_{i+1}, \ldots, l_d)}$.
Using all available sampling results within this region --- in particular, including those at $p_i$ --- a polynomial model is created through least squares fitting.
Depending on how accurately the obtained polynomial approximates the sampling results, the lower or the upper bound $l_i$ or $u_i$ is updated:
\begin{itemize}
    \item If the approximation error is below the threshold, region is expanded up to and including $p_i$ by setting $l_i \leftarrow p_i.$.
    \item Otherwise, we know that $c_i$ must be below $p_i$; therefore,  $u_i$ is set \metric{mingap} away from  $p_i$, since this is the largest value that $c_i$ can still attain: $u_i \leftarrow p_i - \metric{mingap}$.
\end{itemize}

When the updated bounds $l_i$ and $u_i$ coincide, the maximum extent of the region along direction $i$ is reached: $c_i = l_i = u_i$.
The process of sampling and modifying $\mathbf l$ and $\mathbf u$ is repeated until all values of $\mathbf c$ are determined in the same way.
Once this is the case, new regions are generated based on the current region's extent (\autoref{sec:modeling.tool.exp.gen}).

\begin{example}
Returning to the previous in \autoref{fig:modeling.tool.exp.samples:first}, we now consider the expansion of the region given the new sampling results.
First, the region is tentatively expanded along parameter direction 1.
For this purpose, the region $R_\mathbf{b}^{(p_1, l_2)}$ is considered; it reaches from $b_1$ to $p_1$ and $b_2$ to $l_2$ in parameter directions 1 and 2, respectively.
For this region, both the samples used for the construction of the initial model and the new samples at $(p_1, b1)$ and $(p_1, l_2)$ are available.
From all of these samples, a new polynomial is generated through least squares fitting.
Let's say that this model's approximation error is below the threshold.
Consequently, $l_1$ is updated to: $l_1 \leftarrow p_1$.

The updated value for $l_1$ is now used for the tentative expansion along the second parameter direction up to $p_2$.
This region $R_\mathbf b^{(l_1, p_2)}$ contains both the initial model's sampling points and as all new sampling points (\tikzball{plot1}).
Let's say that the polynomial model created from all these points has an error that exceeds the threshold.
As a consequence, the region's expansion along parameter direction 2 cannot include $p_2$.
Hence, $u_2$ is set to the largest possible value that the region's extent can attain: $u_2 \leftarrow p_2 - \metric{mingap}$.
The resulting situation is shown in \autoref{fig:modeling.tool.exp.samples:second}.

Since both $l_1 < u_1$ and $l_2 < u_2$, another step of the model expansion process follows; this time the sampling points shown in \autoref{fig:modeling.tool.exp.samples:second} are chosen.
In this next step, the range of possible values for $\mathbf c$ (spanned by $\mathbf l$ and $\mathbf u$)is significantly smaller than before.

After only a few more steps, $\mathbf l$ and $\mathbf u$ will converge, yielding $\mathbf c$.
\hfill \qed
\end{example}

                \subsubsection{Region Generation}
                \label{sec:modeling.tool.exp.gen}
                \begin{figure}[t]
    \centering
    \begin{tikzpicture}
        \filldraw[fill=plot2!60!plot1] (0, 0) rectangle ++(3, 2);
        \node[align=center] at (1.5, 1) {completed\\region};
        
        \filldraw[fill=plot2, opacity=.25] (3.1, 0) rectangle ++(3, 3);
        \filldraw[fill=plot2, opacity=.25] (0, 2.1) rectangle ++(5, 2);

        \filldraw[fill=plot2] (3.1, 0) rectangle ++(1, 1);
        \draw[->, dotted] (3.55, 1) -- ++(0, 1);
        \draw[->, dotted] (4.1, .5) -- ++(1, 0);

        \filldraw[fill=plot2] (0, 2.1) rectangle ++(1, 1);
        \draw[->, dotted] (.5, 3.1) -- ++(0, .75);
        \draw[->, dotted] (1, 2.55) -- ++(1, 0);

        \path plot[mark=ball, ball color=red] coordinates {(3.1, 0) (0, 2.1)};

        \node[align=center] (gen) at (-1, 1) {new\\regions};
        \draw[->] (gen) .. controls ++(0, 1) and ++(-1, 0) .. (-.1, 2.55);
        \draw[->] (gen) .. controls ++(0, -2) and ++(0, -1) .. (3.55, -.1);

    \end{tikzpicture}
    \caption{Basic concept of region generation in Model Expansion.}
    \label{fig:modeling.tool.exp.gen.idea}
    \tikzset{external/export=false}
\end{figure}
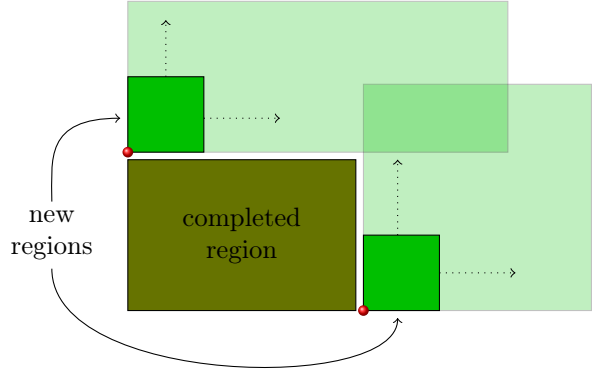

Once the extent of a region has been maximized, new neighboring regions are generated.
The principle behind this generation is shown in \autoref{fig:modeling.tool.exp.gen.idea}: Along each expansion direction, one new region is generated right next to the maximized region.
Unfortunately, it is not that simple.
If we consider the two shaded regions in \autoref{fig:modeling.tool.exp.gen.idea} to be the result of the expansion process, we find that they overlap.
In this case, it is undesirable to generate new regions right next to these, because they would lead to more overlapping regions and thus redundant modeling.
A suitable solution for this particular case would be to generate a new region in the top right corner, bordering both regions.
In this section, we introduce a procedure that generalizes this approach.

Once more considering the expansion away from the origin, the inputs of this procedure are the following:
\begin{itemize}
    \item A region $R_{\mathbf b^\ast}^{\mathbf c^\ast}$ that was extended as far as possible,
    \item The set of all other regions $\mathcal R$, and
    \item The limits of the parameter space in the expansion directions.
\end{itemize}
From these inputs, the procedure generates a set of new regions.
Each of these regions is identified by its base point $\mathbf b$ --- the lower limit of the new region's initial extent.
We now describe how the set $S$ containing these base points is created.

First, an initial set of points $S$ is chosen; this set is then iteratively refined.
The initial set consists of one point $\mathbf p = (b^\ast_1, \ldots, b^\ast_{i-1}, c^\ast_i + \metric{mingap}, b^\ast_{i+1}, \ldots, b^\ast_d)$ for each parameter direction $i$.
(\tikzball{plot1} in the situation in \autoref{fig:modeling.tool.exp.gen.idea}).
Then the procedure iteratively updates $S$ by considering the regions $R \in \mathcal R$ until $S$ stays the same for all regions.

A given region $R$ can be of two types:
\begin{itemize}
    \item $R$ is in the process of expansion, or
    \item $R$ has been expanded as far as possible.
\end{itemize}

In the first case $R$'s upper limit $\mathbf c$ is not fixed yet: only a lower bound $\mathbf l$ and an upper bound $\mathbf u$ are known, such that $\mathbf c \in R_{\mathbf l}^{\mathbf u}$ (see \autoref{sec:modeling.tool.exp.exp}).
With $R$'s base $\mathbf b$ and its upper bound $\mathbf u$, its maximum possible extent is $R_{\mathbf b}^{\mathbf u}$.
Every point $\mathbf p \in S$ that lies within this maximum extent is remove from $S$:
$$
    S \leftarrow S \setminus R_{\mathbf b}^{\mathbf u}.
$$
Since the extent of $R$ is not fixed yet, $\mathbf p$ cannot be selected right next to it; such a base point is created, once the expansion of $R$ is complete.

In the case of a region $R = R_{\mathbf b}^{\mathbf{c}} \in \mathcal R$ that has been expanded as far as possible, its upper limit $\mathbf c$ is known.
The points $S \cap R_{\mathbf b}^{\mathbf c}$, which lie within this region are replaced by a new set of points $S'$.
$S$ is then updated by
$$
    S \leftarrow (S \cap R_{\mathbf b}^{\mathbf c}) \cup S'.
$$

For this purpose, each point $\mathbf p \in S$ is associated with a direction $p_i$; for the points in the initial set $S$ these are the directions $i$, along which the base of $R_{\mathbf b^\ast}^{\mathbf c^\ast}$ was shifted to generate them.
$S'$ is created from the points $\mathbf p \in S$ which lie within $R_{\mathbf b}^{\mathbf c}$.
For each such point $\mathbf p$, one new point is created for each coordinate direction $j \neq i_{\mathbf p}$ by shifting $\mathbf p$ up to $c_j + \metric{maxgap}$ along this direction.
$S'$ is, hence, given by
$$
    S' = \bigcup\limits_{\mathbf p \in S \cap R_{\mathbf b}^{\mathbf c}} \bigl\{(p_1, \ldots, p_{j-1}, c_j + \metric{mingap}, p_{j+1}, \ldots, p_d) \bigm| j \in \{1, \ldots, d\} \setminus \{i_{\mathbf p}\} \bigr\}.
$$
Each point $S'$ is associated with the directions $i_p$ of the point $\mathbf p$ from which it was generated.

The update of $S$ through $S'$ is the reason why all regions $\mathcal R$ repeatedly until $S$ remains unchanged:
The new points within $S'$ can lie within other regions, which were already processed.
Through the repeated processing, it is guaranteed that none of the suggested new base points $S$ lies within any other region.
The termination of the process is assured since we have a finite number of regions $\mathcal R$  and the new points $S'$ are always chosen further along the expansion direction as the original point $\mathbf p$.

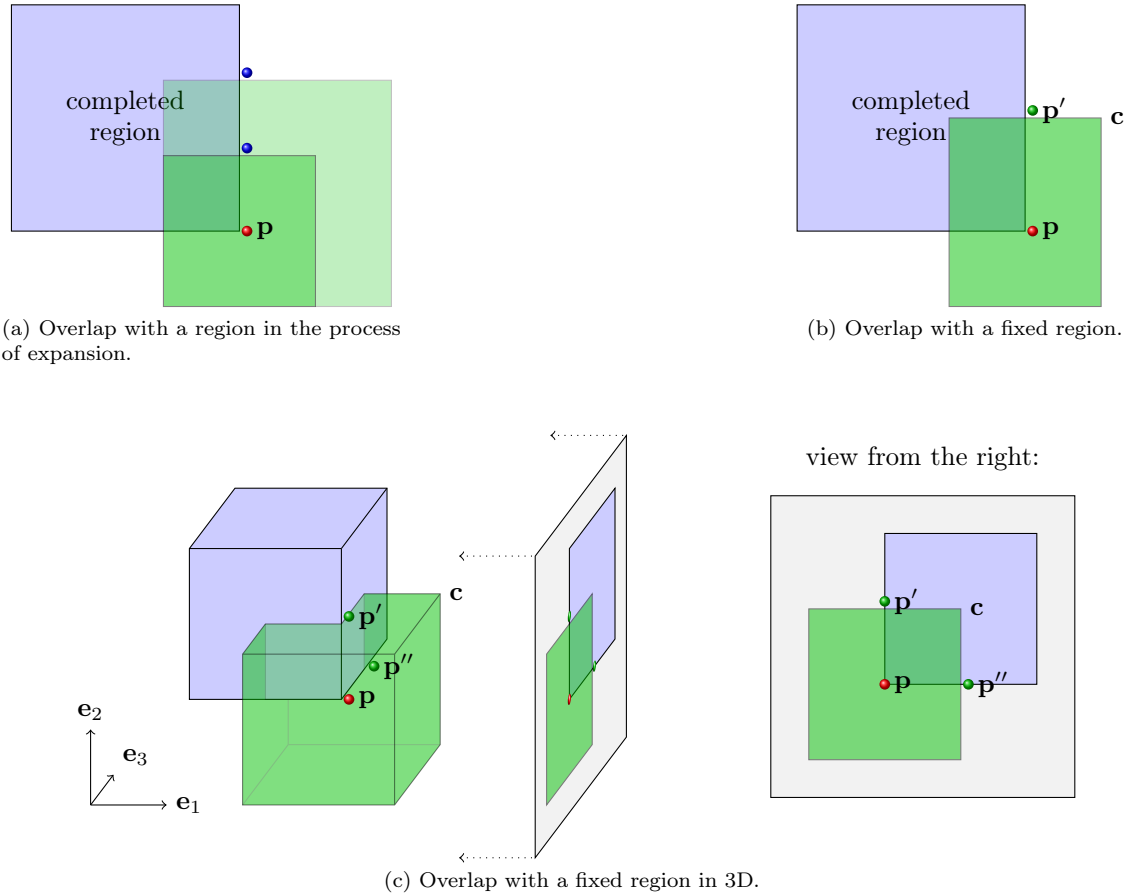
\begin{figure}[t]
    \centering
    \subfloat[Overlap with a region in the process of expansion.]{
        \label{fig:modeling.tool.exp.gen.ex:1}
        \begin{tikzpicture}
            \filldraw[fill=plot3bg] (0, 0) rectangle ++(3, 3);
            \node[align=center] at (1.5, 1.5) {completed\\region};
            
            \begin{scope}[transparency group, opacity=.25]
                \fill[fill=plot2] (2, 1) -- (2, 2) -- (5, 2) -- (5, -1) -- (4, -1) -- (4, 1) -- cycle;
                \draw (2, 1) -- (2, 2) -- (5, 2) -- (5, -1) -- (4, -1);
            \end{scope}
            \filldraw[fill=plot2, opacity=.5] (2, -1) rectangle ++(2, 2);

            \path plot[mark=ball, ball color=plot1] coordinates {(3.1, 0)};
            \node[anchor=west] at (3.1, 0) {$\mathbf p$};

            \path plot[mark=ball, ball color=plot3] coordinates {(3.1, 1.1) (3.1, 2.1)};
        \end{tikzpicture}
    }
    \hfill
    \subfloat[Overlap with a fixed region.]{
        \label{fig:modeling.tool.exp.gen.ex:2}
        \begin{tikzpicture}
            \filldraw[fill=plot3bg] (0, 0) rectangle ++(3, 3);
            \node[align=center] at (1.5, 1.5) {completed\\region};
            
            \filldraw[fill=plot2, opacity=.5] (2, -1) rectangle ++(2, 2.5);

            \path plot[mark=ball, ball color=plot1] coordinates {(3.1, 0)};
            \node[anchor=west] at (3.1, 0) {$\mathbf p$};

            \path plot[mark=ball, ball color=plot2] coordinates {(3.1, 1.6)};
            \node[anchor=west] at (3.1, 1.6) {$\mathbf p'$};

            \node[anchor=west] at (4, 1.5) {$\mathbf c$};
        \end{tikzpicture}
    }

    \vspace{.5cm}

    \subfloat[Overlap with a fixed region in 3D.]{
        \label{fig:modeling.tool.exp.gen.ex:3}
        \begin{tikzpicture}[z={(-.3, -.4)}]

            \draw[->] (-1, -1, 3) -- ++(1, 0, 0)  node[anchor=west]  {$\mathbf e_1$};
            \draw[->] (-1, -1, 3) -- ++(0, 1, 0)  node[anchor=south] {$\mathbf e_2$};
            \draw[->] (-1, -1, 3) -- ++(0, 0, -1) node[anchor=south west]  {$\mathbf e_3$};


            \filldraw[fill=plot3bg] (0, 2, 0) -- (0, 2, 2) -- (0, 0, 2) -- (2, 0, 2) -- (2, 0, 0) -- (2, 2, 0) -- cycle;
            \draw (2, 2, 2) -- (0, 2, 2);
            \draw (2, 2, 2) -- (2, 0, 2);
            \draw (2, 2, 2) -- (2, 2, 0);

            \begin{scope}[transparency group, opacity=.25]
                \fill[fill=plot2] (3, -1, 1) -- (3, 1, 1) -- (2, 1, 1) -- (2, 0, 1) -- (2, 0, 2) -- (1, 0, 2) -- (1, 1, 2) -- (1, 1, 3) -- (1, -1, 3) -- (3, -1, 3) -- cycle;
                \draw (1, -1, 1) -- (3, -1, 1);
                \draw (1, -1, 1) -- (1, 1, 1 |- 2, 0, 2);
                \draw (1, -1, 1) -- (1, -1, 3);

                \draw (1, 1, 2) -- (1, 0, 2) -- (2, 0, 2) -- (2, 0, 1) -- (2, 1, 1,);
            \end{scope}

            \begin{scope}[transparency group, opacity=.333]
                \fill[fill=plot2] (3, -1, 1) -- (3, 1, 1) -- (2, 1, 1) -- (2, 1, 2) -- (1, 1, 2) -- (1, 1, 3) -- (1, -1, 3) -- (3, -1, 3) -- cycle;
            \end{scope}

            \begin{scope}[transparency group, opacity=.5]
                \draw (3, -1, 1) -- (3, 1, 1) -- (2, 1, 1) -- (2, 1, 2) -- (1, 1, 2) -- (1, 1, 3) -- (1, -1, 3) -- (3, -1, 3) -- cycle;
                \draw (3, 1, 3) -- (1, 1, 3);
                \draw (3, 1, 3) -- (3, -1, 3);
                \draw (3, 1, 3) -- (3, 1, 1);
            \end{scope}

            \path plot[mark=ball, ball color=plot1] coordinates {(2.1, 0, 2)};
            \node[anchor=west] at (2.1, 0, 2) {$\mathbf p$};

            \path plot[mark=ball, ball color=plot2] coordinates {(2.1, 1.1, 2) (2.1, 0, .9)};
            \node[anchor=west] at (2.1, 1.1, 2) {$\mathbf p'$};
            \node[anchor=west] at (2.1, 0, .9) {$\mathbf p''$};

            \node[anchor=west] at (3, 1, 1) {$\mathbf c$};

            \draw[<-, dotted] (4,  2.5, -.5) -- ++(1, 0, 0);
            \draw[<-, dotted] (4,  2.5, 3.5) -- ++(1, 0, 0);
            \draw[<-, dotted] (4, -1.5, 3.5) -- ++(1, 0, 0);

            \begin{scope}[canvas is zy plane at x=5, xshift=2cm, xscale=-1]
                \filldraw[fill=graybg] (-1.5, -1.5) rectangle (2.5, 2.5);
                \filldraw[fill=plot3bg] (0, 0) rectangle (2, 2);
                \filldraw[fill=plot2, opacity=.5] (-1, -1) rectangle (1, 1);

                \path plot[mark=ball, ball color=plot1] coordinates {(0, 0)};
                \path plot[mark=ball, ball color=plot2] coordinates {(1.1, 0) (0, 1.1)};
            \end{scope}

            \begin{scope}[canvas is xy plane at z=1.5, xshift=9cm]
                \node at (.5, 3) {view from the right:};
                \filldraw[fill=graybg] (-1.5, -1.5) rectangle (2.5, 2.5);
                \filldraw[fill=plot3bg] (0, 0) rectangle (2, 2);
                \filldraw[fill=plot2, opacity=.5] (-1, -1) rectangle (1, 1);

                \path plot[mark=ball, ball color=plot1] coordinates {(0, 0)};
                \node[anchor=west] at (0, 0) {$\mathbf p$};

                \path plot[mark=ball, ball color=plot2] coordinates {(1.1, 0) (0, 1.1)};
                \node[anchor=west] at (0, 1.1) {$\mathbf p'$};
                \node[anchor=west] at (1.1, 0) {$\mathbf p''$};

                \node[anchor=west] at (1, 1) {$\mathbf c$};
            \end{scope}
        \end{tikzpicture}
    }
    \caption{Generation of new regions' base points $S$.}
    \label{fig:modeling.tool.exp.gen.ex}
    \tikzset{external/export=false}
\end{figure}

\begin{example}
Let us consider examples of the treatment of points in $S$ in different situations regarding the overlap of other regions, as shown in \autoref{fig:modeling.tool.exp.gen.ex}.
In each example, we have a completed regions $R^\ast$~(\tikzsquare{plot3bg}), an overlapping region $R$~(\tikzsquare{plot2!50}) one a point $p$~(\tikzball{plot1}), which was generated by shifting the base of the completed region $R^\ast$~(\tikzsquare{plot3bg}) along direction $i_{\mathbf p} = 1$ to the right.
\begin{itemize}
    \item \autoref{fig:modeling.tool.exp.gen.ex:1}: In this scenario, the region $R$~(\tikzsquare{plot2!50}) is still within the process of expansion.
        Creating a new region at $\mathbf p$ would lead to a potentially large overlap with $R$.
        However, since the final extent of $R$ is unknown, $\mathbf p$ cannot be shifted to another position (e.g., \tikzball{blue}) at the side of $R$.
        Therefore, $\mathbf p$ is discarded without a replacement.
        Once $R$ is expanded as far as possible, a new region will be generated somewhere between the indicated points (\tikzball{blue}).
    \item In \autoref{fig:modeling.tool.exp.gen.ex:2} $\mathbf p$~(\tikzball{plot1}) lies within a region $R$~(\tikzsquare{plot2!50}) with a fixed limit $\mathbf c$.
        Therefore, $\mathbf p$ is shifted along the side of $R^\ast$~(\tikzsquare{plot3bg}) to a point $\mathbf p'$~(\tikzball{plot2}), which is \metric{mingap} away from $R$.
        With $\mathbf p = (p_1, p_2)$ and $\mathbf c = (c_1, c_2)$, we obtain $\mathbf p' = (p_1, c_2 + \metric{mingap})$.
    \item \autoref{fig:modeling.tool.exp.gen.ex:3} shows a 3D version of the scenario from the previous example.
        From $\mathbf p = (p_1, p_2, p_3)$~(\tikzball{plot1}) and $\mathbf c = (c_1, c_2, c_3)$, two new points $\mathbf p' = (p_1, c_2 + \metric{mingap}, p_3)$ and $\mathbf p'' = (p_1, p_2, c_3 + \metric{mingap})$ are generated.
\hfill \qed
\end{itemize}
\end{example}

Once $S$ is not changed anymore by any of the regions $R \in \mathcal R$, all points that exceed the size of the parameter space are removed from it.
The resulting set is gives the base points for the new regions.

            \subsection{PModeler: Adaptive Refinement}
            \label{sec:modeling.tool.ref}
            Performance depends on continuous parameters in a spatially nonuniform fashion:
In some regions of the parameters space --- usually for large parameter values --- this dependency is rather regular and smooth, while in others it can be very irregular, containing jumps, kinks, and curvature changes.
In order to represent such structures to different degrees of detail, we now introduce an approach based on adaptive refinement.
Similar approaches are commonly used in a variety of disciplines such as mesh generation and optimization.
In our context, the idea is to begin with a simple and regular model constructed from a coarse grid of samples; then, the model's quality and accuracy are evaluated and it is refined in the insufficiently approximated regions by locally increasing the grid resolution.
These steps are applied recursively to the refined regions, until either the accuracy suffices across the whole domain or a given resolution limit is reached.

In the beginning, a single region is created spanning the whole parameter space.
This region is then sampled at points on a regular grid; for polynomial approximations of order $o$, this grid consists of at least $o + 1$ points along each direction.
From the resulting samples, a first polynomial approximation is computed through least squares fitting.
If the approximation error of this polynomial is below the error bound, the model is already completed.
However, it is very unlikely that one polynomial is accurate enough to represent the performance of the entire parameter space.

When the accuracy of a region is insufficient, it is divided into four roughly equally sized\footnote{
    The splitting point's coordinates are a multiple of \metric{mingap} in order to avoid the small scale oscillations of performance counters.
} quadrants in the 2D case ($2^d$ hyper-cuboids in $d$ dimensions).
Those quadrants, whose size along any coordinate direction is smaller than a configurable minimum size are discarded; the others form new regions.
The new regions are then sampled on a regular grid with the same number of points as for the initial region.
New polynomials are fitted to the regions samples and their approximation error is computed.
When this error is above the threshold, the regions are further refined recursively; otherwise, the model for the considered region is complete.

\begin{figure}[t]
    \centering
    \begin{tikzpicture}
        \draw[help lines] (0, 0) rectangle (3, 3);
        \foreach \x in {0, 1, 2, 3}
            \foreach \y in {0, 1, 2, 3}
                \path plot[mark=ball, ball color=plot1] coordinates {(\x, \y)};
        \begin{scope}[xshift=7cm]
            \draw[help lines] (0, 0) rectangle (3, 3);
            \foreach \x in {0, 1, 2, 3}
                \foreach \y in {0, 1, 2, 3}
                    \path plot[mark=ball, ball color=plot1] coordinates {(\x, \y)};
            \begin{scope}[scale=.5]
                \draw[help lines] (0, 0) rectangle (3, 3);
                \foreach \x in {0, 1, 2, 3}
                    \foreach \y in {0, 1, 2, 3}
                        \path plot[mark=ball, ball color=plot2] coordinates {(\x, \y)};
            \end{scope}
        \end{scope}
        \draw[->] (4, 1.5) -- ++(2,0) node[midway, above] {refinement};
    \end{tikzpicture}
    \caption{Sampling points distributed on a rectangular grid.}
    \label{fig:modeling.tool.ref.grid}
    \tikzset{external/export=false}
\end{figure}
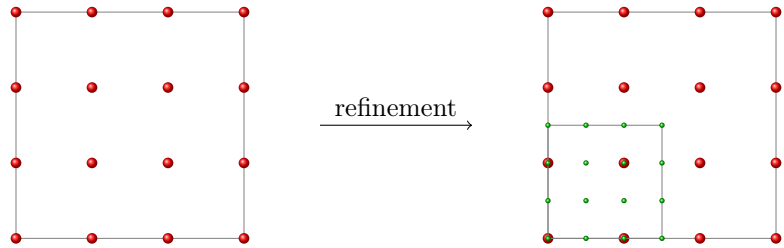

In each step of the refinement, all sampling points of the coarse model are incorporated in the model of the refined regions.
In fact, the sampling grid on the refined regions covers all previous sampling points within the same region (see \autoref{fig:modeling.tool.ref.grid} for the 2D case); sampling for these points is not repeated.

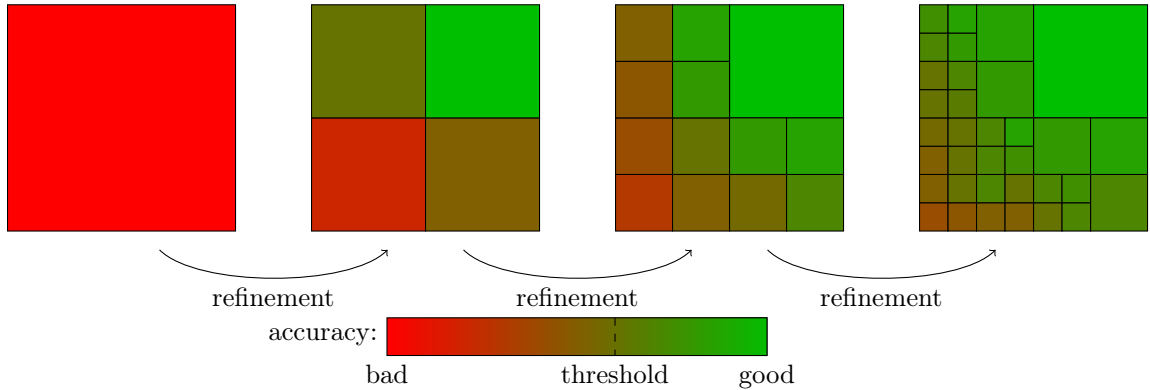
\begin{figure}[t]
    \centering
    \begin{tikzpicture}
        \coordinate (pos) at (0, 0);
        \filldraw[fill=plot1!100!plot2] (pos) rectangle ++(3, 3);

        \path (pos) ++(4, 0) coordinate (pos);

        \foreach \x/\y/\c in {
            0/0/80,     .5/0/50,
            0/.5/40,    .5/.5/0}
            \filldraw[fill=plot1!\c!plot2] (pos) ++($3*(\x, \y)$) rectangle ++(1.5, 1.5);

        \path (pos) ++(4, 0) coordinate (pos);

        \foreach \x/\y/\c in {
                        .5/.5/0}
            \filldraw[fill=plot1!\c!plot2] (pos) ++($3*(\x, \y)$) rectangle ++(1.5, 1.5);
        \foreach \x/\y/\c in {
            0/0/70,     .25/0/50,   .5/0/45,    .75/0/30,
            0/.25/60,   .25/.25/40, .5/.25/20,  .75/.25/15,
            0/.5/55,    .25/.5/20,
            0/.75/50,   .25/.75/15}
            \filldraw[fill=plot1!\c!plot2] (pos) ++($3*(\x, \y)$) rectangle ++(.75, .75);

        \path (pos) ++(4, 0) coordinate (pos);

        \foreach \x/\y/\c in {
                        .5/.5/0}
            \filldraw[fill=plot1!\c!plot2] (pos) ++($3*(\x, \y)$) rectangle ++(1.5, 1.5);
        \foreach \x/\y/\c in {
                                                .75/0/30,
                                    .5/.25/20,  .75/.25/15,
                        .25/.5/20,
                        .25/.75/15}
            \filldraw[fill=plot1!\c!plot2] (pos) ++($3*(\x, \y)$) rectangle ++(.75, .75);
        \foreach \x/\y/\c in {
            0/0/60,     .125/0/55,      .25/0/50,       .375/0/50,      .5/0/40,    .625/0/30,
            0/.125/50,  .125/.125/40,   .25/.125/30,    .375/.125/40,   .5/.125/30, .625/.125/25,
            0/.25/50,   .125/.25/40,    .25/.25/30,     .375/.25/25,
            0/.375/45,  .125/.375/40,   .25/.375/30,    .375/.375/15,
            0/.5/40,    .125/.5/35,
            0/.625/40,  .125/.625/30,
            0/.75/30,   .125/.75/20,
            0/.875/25,  .125/.875/15}
            \filldraw[fill=plot1!\c!plot2] (pos) ++($3*(\x, \y)$) rectangle ++(.375, .375);

         \coordinate (pos) at (2, -.25);
         \draw[->] (pos) .. controls ++(.5, -.5) and ++(-.5, -.5) .. ++(3, 0) node[midway, below] {refinement};
         \path (pos) ++(4, 0) coordinate (pos);
         \draw[->] (pos) .. controls ++(.5, -.5) and ++(-.5, -.5) .. ++(3, 0) node[midway, below] {refinement};
         \path (pos) ++(4, 0) coordinate (pos);
         \draw[->] (pos) .. controls ++(.5, -.5) and ++(-.5, -.5) .. ++(3, 0) node[midway, below] {refinement};
    \end{tikzpicture}

    \tikzset{external/export=false}
    \begin{tikzpicture}
         \coordinate (pos) at (0, 0);
         \filldraw[fill=plot1] (pos) +(-2.5, 0) rectangle +(2.5, -.5);
         \filldraw[path fading=west, fill=plot2] (pos) +(-2.5, 0) rectangle +(2.5, -.5);
         \path (pos) ++(-2.5, -.25) node[anchor=east] {accuracy:};
         \path (pos) ++(-2.5, -.5) node[anchor=north] {bad};
         \path (pos) ++(2.5, -.5) node[anchor=north] {good};
         \draw[dashed] (pos) ++(-2.5, 0) ++($.6*(5, 0)$) -- ++(0, -.5) node[anchor=north] {threshold};

         \path (pos) ++(2.5, -.25) node[anchor=west] {\phantom{accuracy:}};
    \end{tikzpicture}
    \caption{Adaptive Refinement example.}
    \label{fig:modeling.tool.ref.refine}
\end{figure}

\begin{example}
An example of Adaptive Refinement for the 2D case is shown in \autoref{fig:modeling.tool.ref.refine}.
The polynomial approximation for the initial region spanning the entire parameter space is very inaccurate (\tikzsquare{plot1}, 1\textsuperscript{st} square).
Therefore, it is refined, generating four new regions; new samples are generated for these, leading to new polynomial approximations (2\textsuperscript{nd} square).
Here, the error in the top right quadrant (\tikzsquare{plot2}) is already below the threshold (\tikzsquare{plot1!40!plot2});
The other quadrants are not accurate enough and are further refined (3\textsuperscript{rd} square).
Now, several regions are below the error threshold; the others are refined once more (4\textsuperscript{th} square).
Although some of the resulting regions are still of insufficient accuracy, they are not further refined since we do not wish to generate any smaller models.
\hfill\qed
\end{example}

        \section{Results}
        \label{sec:modeling.res}
        In the previous sections, we introduced the type of performance models we are interested in and the Modeler --- a tool that automatically generates these models.
We now study how the Modeler is used and configured to generate the models.

We consider \texttt{dtrsm} ($B \leftarrow A^{-1} B$, $A$ triangular):
\begin{center}
    \texttt{dtrsm(side, uplo, transA, diag, side, m, n, alpha, A, ldA, B, ldB)}.
\end{center}
This BLAS Level-3 routine has four discrete arguments (\texttt{side} through \texttt{diag}), two size arguments (\texttt m and \texttt n), one scalar argument (\texttt{alpha}) and operates on two matrices (\texttt A and \texttt B with corresponding leading dimensions \texttt{ldA} and \texttt{ldB}).
From these arguments, we select
\begin{itemize}
    \item the discrete parameters \texttt{side}, \texttt{uplo}, and \texttt{transA}, and
    \item the continuous parameters \texttt m and \texttt n.
\end{itemize}
The size parameters are modeled for values between $8$ and $1024$ with $\metric{mingap} = 8$, that is, $\mathtt m, \mathtt n \in \{8, 16, 24, \ldots, 1024\}$.
The remaining arguments \texttt{diag}, \texttt{alpha}, \texttt{ldA}, and \texttt{ldB} are not modeled.
They receive the following values:
\begin{itemize}
    \item $\mathtt{diag} = \mathtt N$;
    \item $\mathtt{alpha} = 0.5$;
    \item $\mathtt{ldA} = \mathtt{ldB} = 2500$ (representing the access of submatrices \texttt A and \texttt B).
\end{itemize}
We model the performance counters \texttt{flops} and \texttt{ticks}.

All samples for the modeling process are taken on one core of a Quad-Core AMD Opteron Processor 8356 \cite{amd} running at $2.30 \mathrm{GHz}$.
We use \gotoblas \cite{gotopage, gotoblas} and the static memory policy for high memory locality (in-cache).

Although both performance counters are modeled simultaneously, we discuss the resulting models separately in the following sections.

            \subsection{\texorpdfstring{\metric{flops}}{flops}}
            \label{sec:modeling.res.flops}
            For the performance counter \metric{flops} (number of floating point operations performed by a routine) we can expect a constant value across routine executions with the same arguments, since the \texttt{dtrsm}'s instructions are data independent.
Therefore, each sampling point is sampled only once; instead of a range of statistical quantities, we only need to consider one value at each point.

                \subsubsection{Model Expansion.}
                \label{sec:modeling.res.flops.me}
                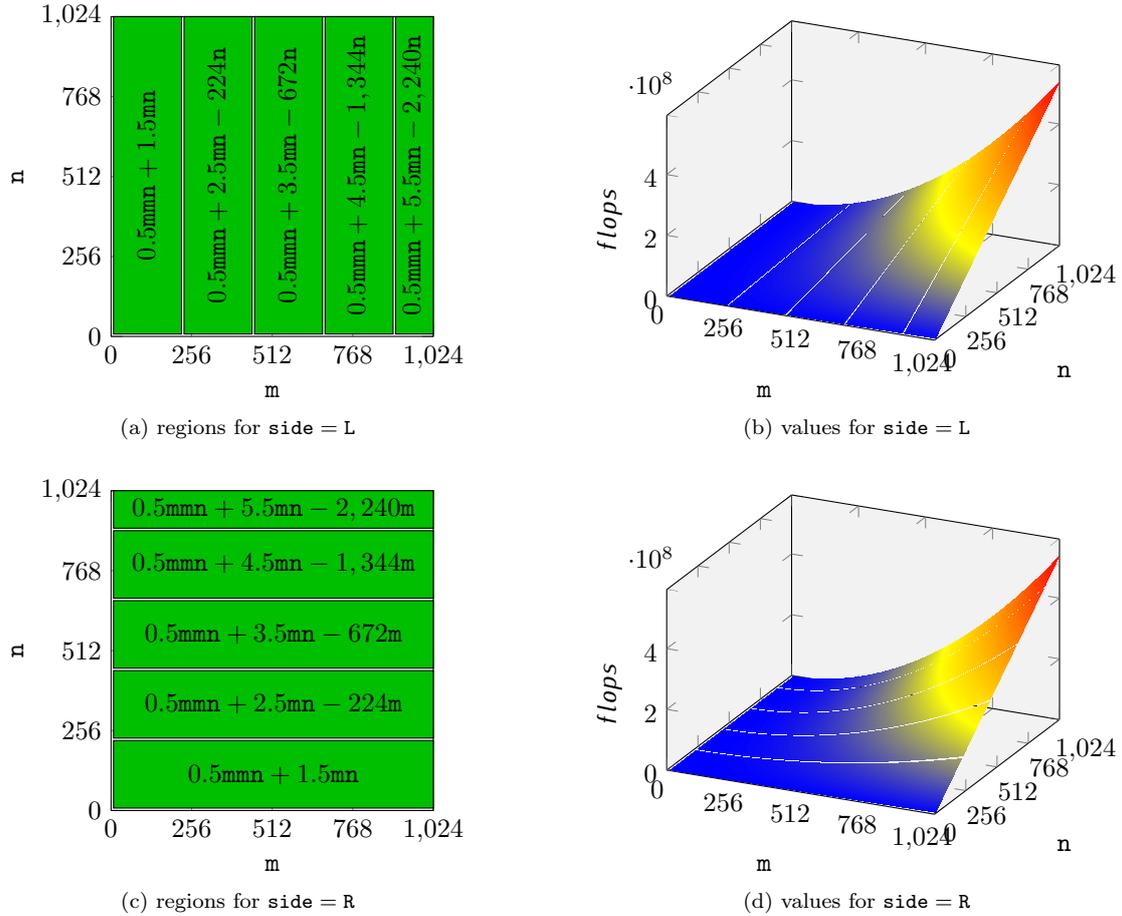
\begin{figure}[t]
    \centering
    \subfloat[regions for $\mathtt{side} = \mathtt L$]{
        \label{fig:modeling.res.flops.me:Lr}
        \begin{tikzpicture}
            \begin{axis}[
                twocolplot,
                axis equal image=true,
                xlabel={\texttt m},
                ylabel={\texttt n},
                xtick={0,256,...,1024},
                ytick={0,256,...,1024},
                xmax=1024,
                ymax=1024
            ]
                \coordinate (oo) at (axis description cs: 0, 0);
                \coordinate (oi) at (axis description cs: 0, 1);
                \coordinate (io) at (axis description cs: 1, 0);
                \coordinate (ii) at (axis description cs: 1, 1);
            \end{axis}

            \foreach \ax/\ay/\bx/\by/\t in {
                0.0078125/0.0078125/0.21875/1.0/{$0.5 \mathtt m \mathtt m \mathtt n + 1.5 \mathtt m \mathtt n$},
                0.2265625/0.0078125/0.4375/1.0/ {$0.5 \mathtt m \mathtt m \mathtt n + 2.5 \mathtt m \mathtt n - 224 \mathtt n$},
                0.4453125/0.0078125/0.65625/1.0/{$0.5 \mathtt m \mathtt m \mathtt n + 3.5 \mathtt m \mathtt n - 672 \mathtt n$},
                0.6640625/0.0078125/0.875/1.0/  {$0.5 \mathtt m \mathtt m \mathtt n + 4.5 \mathtt m \mathtt n - 1,344 \mathtt n$},
                0.8828125/0.0078125/1.0/1.0/    {$0.5 \mathtt m \mathtt m \mathtt n + 5.5 \mathtt m \mathtt n - 2,240 \mathtt n$}
            } {
                \filldraw[fill=plot2] ($\ax*(io) + \ay*(oi)$) rectangle ($\bx*(io) + \by*(oi)$) node[midway, rotate=90] {\t};
            }
        \end{tikzpicture}
    }
    \hfill
    \subfloat[values for $\mathtt{side} = \mathtt L$]{
        \label{fig:modeling.res.flops.me:Lv}
        \begin{tikzpicture}
            \begin{axis}[
                twocolplot,
                xlabel={\texttt m},
                ylabel={\texttt n},
                zlabel={\metric{flops}},
                xtick={0,256,...,1024},
                ytick={0,256,...,1024},
                xmax=1024,
                ymax=1024,
                zmin=0
            ]
                \addplot3[surf, shader=interp, domain=8:224,    y domain=8:1024] {0.5*x*x*y + 1.5*x*y};
                \addplot3[surf, shader=interp, domain=232:448,  y domain=8:1024] {0.5*x*x*y + 2.5*x*y - 244*y};
                \addplot3[surf, shader=interp, domain=456:672,  y domain=8:1024] {0.5*x*x*y + 3.5*x*y - 672*y};
                \addplot3[surf, shader=interp, domain=680:896,  y domain=8:1024] {0.5*x*x*y + 4.5*x*y - 1344*y};
                \addplot3[surf, shader=interp, domain=904:1024, y domain=8:1024] {0.5*x*x*y + 5.5*x*y - 2240*y};
            \end{axis}
        \end{tikzpicture}
    }

    \subfloat[regions for $\mathtt{side} = \mathtt R$]{
        \label{fig:modeling.res.flops.me:Rr}
        \begin{tikzpicture}
            \begin{axis}[
                twocolplot,
                axis equal image=true,
                xlabel={\texttt m},
                ylabel={\texttt n},
                xtick={0,256,...,1024},
                ytick={0,256,...,1024},
                xmax=1024,
                ymax=1024
            ]
                \coordinate (oo) at (axis description cs: 0, 0);
                \coordinate (oi) at (axis description cs: 0, 1);
                \coordinate (io) at (axis description cs: 1, 0);
                \coordinate (ii) at (axis description cs: 1, 1);
            \end{axis}

            \foreach \ax/\ay/\bx/\by/\t in {
                0.0078125/0.0078125/1.0/0.21875/{$0.5 \mathtt m \mathtt m \mathtt n + 1.5 \mathtt m \mathtt n$},
                0.0078125/0.2265625/1.0/0.4375/ {$0.5 \mathtt m \mathtt m \mathtt n + 2.5 \mathtt m \mathtt n - 224 \mathtt m$},
                0.0078125/0.4453125/1.0/0.65625/{$0.5 \mathtt m \mathtt m \mathtt n + 3.5 \mathtt m \mathtt n - 672 \mathtt m$},
                0.0078125/0.6640625/1.0/0.875/  {$0.5 \mathtt m \mathtt m \mathtt n + 4.5 \mathtt m \mathtt n - 1,344 \mathtt m$},
                0.0078125/0.8828125/1.0/1.0/    {$0.5 \mathtt m \mathtt m \mathtt n + 5.5 \mathtt m \mathtt n - 2,240 \mathtt m$}
            }
                \filldraw[fill=plot2] ($\ax*(io) + \ay*(oi)$) rectangle ($\bx*(io) + \by*(oi)$) node[midway] {\t};
        \end{tikzpicture}
    }
    \hfill
    \subfloat[values for $\mathtt{side} = \mathtt R$]{
        \label{fig:modeling.res.flops.me:Rv}
        \begin{tikzpicture}
            \begin{axis}[
                twocolplot,
                xlabel={\texttt m},
                ylabel={\texttt n},
                zlabel={\metric{flops}},
                xtick={0,256,...,1024},
                ytick={0,256,...,1024},
                xmax=1024,
                ymax=1024,
                zmin=0
            ]
                \addplot3[surf, shader=interp, y domain=8:224,    domain=8:1024] {0.5*x*x*y + 1.5*x*y};
                \addplot3[surf, shader=interp, y domain=232:448,  domain=8:1024] {0.5*x*x*y + 2.5*x*y - 244*x};
                \addplot3[surf, shader=interp, y domain=456:672,  domain=8:1024] {0.5*x*x*y + 3.5*x*y - 672*x};
                \addplot3[surf, shader=interp, y domain=680:896,  domain=8:1024] {0.5*x*x*y + 4.5*x*y - 1344*x};
                \addplot3[surf, shader=interp, y domain=904:1024, domain=8:1024] {0.5*x*x*y + 5.5*x*y - 2240*x};
            \end{axis}
        \end{tikzpicture}
    }

    \caption{\metric{flops} model for \texttt{dtrsm} with Model Expansion.}
    \label{fig:modeling.res.flops.me}
    \tikzset{external/export=false}
\end{figure}

The models generated by Model Expansion are shown in \autoref{fig:modeling.res.flops.me}.
The discrete parameters \texttt{uplo} and \texttt{transA} show no influence on \texttt{flops}; therefore we only consider the cases $\mathtt{side} = \mathtt L$  and $\mathtt{side} = \mathtt R$.
For both, we show the generated regions and their polynomials.
The approximation error is 0\% across the whole domain --- our model represents \metric{flops} exactly.

Considering the distribution of regions, in the case of $\mathtt{side} = \mathtt L$ (Figures \ref{fig:modeling.res.flops.me:Lr} and \ref{fig:modeling.res.flops.me:Lv}), we have regions in intervals of 224 along the dimension \texttt m.
Each of them is associated with a polynomial model of the form $0.5 \mathtt m \mathtt m \mathtt n + \alpha \mathtt m \mathtt n + \beta \mathtt n$.
Starting at $1.5$, $\alpha$ increases by $1$ in every interval, while $\beta$ can be expressed as $\sum\limits_{i=1}^n 224 i$ in the $n$-th region.
The observed regions are due to \gotoblas's blocked matrix access.

Considering $\mathtt{side} = \texttt R$ (Figures \ref{fig:modeling.res.flops.me:Rr} and \ref{fig:modeling.res.flops.me:Rv}), we find the same structure as in the previous case along the \texttt n-axis.
This is the case, since \texttt{side} change the order of the operands from $B \leftarrow 0.5 A^{-1} B$ to $B \leftarrow 0.5 B A^{-1}$, thus changing the blocked data access pattern.

                \subsubsection{Adaptive Refinement}
                \label{sec:modeling.res.flops.ar}
                \begin{figure}[t]
    \newcommand{\regionsplot}[1]{
        \begin{tikzpicture}
            \begin{axis}[
                twocolplot,
                axis equal image=true,
                xlabel={\texttt m},
                ylabel={\texttt n},
                xtick={0,256,...,1024},
                ytick={0,256,...,1024},
                xmax=1024,
                ymax=1024
            ]
                \coordinate (oi) at (axis description cs: 0, 1);
                \coordinate (io) at (axis description cs: 1, 0);
            \end{axis}

            \foreach \ax/\ay/\bx/\by/\e in {#1}
                \filldraw[fill=plot2] ($\ax/1024*(io) + \ay/1024*(oi)$) rectangle ($\bx/1024*(io) + \by/1024*(oi)$);
        \end{tikzpicture}
    }
    \centering
    \regionsplot{
        8/8/1024/1024/0.0948300567993,
        8/8/512/512/0.0132446864591,
        8/512/512/1024/0.012730998697,
        512/8/1024/512/0.00503054082508,
        512/512/1024/1024/0.00496583628213,
        8/8/256/256/0.0388762186117,
        8/256/256/512/0.0388762214014,
        256/8/512/256/0.0044352452151,
        256/256/512/512/0.0300355554152,
        8/512/256/768/0.0124687643923,
        8/768/256/1024/0.00868499446567,
        256/512/512/768/0.00428433358233,
        256/768/512/1024/0.00430587051829,
        512/8/768/256/0.00618689872132,
        512/256/768/512/0.0463617995478,
        768/8/1024/256/0.00591411822461,
        768/256/1024/512/0.00591411814979,
        512/512/768/768/0.0057713050436,
        512/768/768/1024/0.00547806479774,
        768/512/1024/768/0.00591411833164,
        768/768/1024/1024/0.00591411810225,
        8/8/128/128/0,
        8/128/128/256/0,
        128/8/256/128/0.00887892499837,
        128/128/256/256/0.00838313391602,
        8/256/128/384/0,
        8/384/128/512/0,
        128/256/256/384/0.00871883841466,
        128/384/256/512/0.00797898281577,
        256/8/384/128/0,
        256/128/384/256/0,
        384/8/512/128/0.0113026376677,
        384/128/512/256/0.0113026376677,
        256/256/384/384/0,
        256/384/384/512/0,
        384/256/512/384/0.0113026376677,
        384/384/512/512/0.0113026376677,
        8/512/128/640/0,
        8/640/128/768/0,
        128/512/256/640/0.00884214356625,
        128/640/256/768/0.00779199414591,
        8/768/128/896/0,
        8/896/128/1024/0,
        128/768/256/896/0.00910248588121,
        128/896/256/1024/0.00759714622989,
        256/512/384/640/0,
        256/640/384/768/0,
        384/512/512/640/0.0113026376677,
        384/640/512/768/0.0113026376677,
        256/768/384/896/0,
        256/896/384/1024/0,
        384/768/512/896/0.0113026376677,
        384/896/512/1024/0.0113026376677,
        512/8/640/128/0,
        512/128/640/256/0,
        640/8/768/128/0.00938203148245,
        640/128/768/256/0.00913840621637,
        512/256/640/384/0,
        512/384/640/512/0,
        640/256/768/384/0.00864833678068,
        640/384/768/512/0.00904472963467,
        768/8/896/128/0,
        768/128/896/256/3.97222055961e-10,
        896/8/1024/128/0,
        896/128/1024/256/5.38701749523e-10,
        768/256/896/384/0,
        768/384/896/512/1.45199591264e-10,
        896/256/1024/384/1.74231758855e-10,
        896/384/1024/512/0,
        512/512/640/640/0,
        512/640/640/768/0,
        640/512/768/640/0.00904375197722,
        640/640/768/768/0.00967171309426,
        512/768/640/896/0,
        512/896/640/1024/0,
        640/768/768/896/0.00904278290662,
        640/896/768/1024/0.00653267433525,
        768/512/896/640/8.16795223475e-11,
        768/640/896/768/0,
        896/512/1024/640/7.28527727006e-11,
        896/640/1024/768/4.29628357527e-10,
        768/768/896/896/6.463389454e-11,
        768/896/896/1024/0,
        896/768/1024/896/1.33069897386e-10,
        896/896/1024/1024/1.62901082159e-10,
        128/8/192/64/0,
        128/64/192/128/0,
        192/8/256/64/0,
        192/64/256/128/0,
        128/128/192/192/0,
        128/192/192/256/0,
        192/128/256/192/0,
        192/192/256/256/0,
        128/256/192/320/0,
        128/320/192/384/0,
        192/256/256/320/0,
        192/320/256/384/0,
        128/384/192/448/0,
        128/448/192/512/0,
        192/384/256/448/0,
        192/448/256/512/0,
        384/8/448/64/0,
        384/64/448/128/0,
        448/8/512/64/0,
        448/64/512/128/0,
        384/128/448/192/0,
        384/192/448/256/0,
        448/128/512/192/0,
        448/192/512/256/0,
        384/256/448/320/0,
        384/320/448/384/0,
        448/256/512/320/0,
        448/320/512/384/0,
        384/384/448/448/0,
        384/448/448/512/0,
        448/384/512/448/0,
        448/448/512/512/0,
        128/512/192/576/0,
        128/576/192/640/0,
        192/512/256/576/0,
        192/576/256/640/0,
        128/640/192/704/0,
        128/704/192/768/0,
        192/640/256/704/0,
        192/704/256/768/0,
        128/768/192/832/0,
        128/832/192/896/0,
        192/768/256/832/0,
        192/832/256/896/0,
        128/896/192/960/0,
        128/960/192/1024/0,
        192/896/256/960/0,
        192/960/256/1024/0,
        384/512/448/576/0,
        384/576/448/640/0,
        448/512/512/576/0,
        448/576/512/640/0,
        384/640/448/704/0,
        384/704/448/768/0,
        448/640/512/704/0,
        448/704/512/768/0,
        384/768/448/832/0,
        384/832/448/896/0,
        448/768/512/832/0,
        448/832/512/896/0,
        384/896/448/960/0,
        384/960/448/1024/0,
        448/896/512/960/0,
        448/960/512/1024/0,
        640/8/704/64/0.000621019912921,
        640/64/704/128/0,
        704/8/768/64/0,
        704/64/768/128/0,
        640/128/704/192/0,
        640/192/704/256/0,
        704/128/768/192/0,
        704/192/768/256/0,
        640/256/704/320/0,
        640/320/704/384/0.0434891959437,
        704/256/768/320/0,
        704/320/768/384/1.33908025974e-10,
        640/384/704/448/0,
        640/448/704/512/0,
        704/384/768/448/0,
        704/448/768/512/0,
        640/512/704/576/0,
        640/576/704/640/0,
        704/512/768/576/1.17504243826e-10,
        704/576/768/640/2.46082588618e-10,
        640/640/704/704/7.02490081699e-06,
        640/704/704/768/0,
        704/640/768/704/2.07193269301e-10,
        704/704/768/768/0,
        640/768/704/832/0,
        640/832/704/896/0,
        704/768/768/832/1.3174805864e-10,
        704/832/768/896/9.55812001693e-11,
        640/896/704/960/0,
        640/960/704/1024/0.0155693654608,
        704/896/768/960/5.82785457716e-11,
        704/960/768/1024/1.51514270254e-10
    }

    \caption{\metric{flops} model for \texttt{dtrsm} with Adaptive Refinement --- Regions for $\mathtt{side} = \mathtt L$.}
    \label{fig:modeling.res.flops.ar}
    \tikzset{external/export=false}
\end{figure}

For $\texttt{side} = \mathtt L$, Adaptive Refinement covers the parameter space with the regions shown in \autoref{fig:modeling.res.flops.ar}.
As for Model Expansion, these regions are arranged in a clear pattern, representing \metric{ticks} perfectly.
However, there are many rather small regions.
This is the due to the strict $2 \times 2$ subdivision pattern that is used; a comparison with Model Expansion \autoref{fig:modeling.res.flops.me:Lr} shows that the small regions are required to exactly fill up the regions caused by \gotoblas's blocked matrix access.

Since we do not gain any accuracy through the small regions, we conclude that Model Expansion is more suitable to model \metric{flops}.
The accuracy of moth modeling strategies is perfect, but Adaptive Refinement requires a lot more samples than Model Expansion and does not represent the structures found in \metric{ticks} in a clean way.

            \subsection{\texorpdfstring{\metric{ticks}}{ticks}}
            \label{sec:modeling.res.ticks}
            Turning to \metric{ticks}, we are faced with a completely different performance behavior.
\metric{ticks} shows both a not clearly polynomial dependence on the continuous parameters and fluctuations when one routine execution is sampled repeatedly.
To accommodate for the latter, we make use of the statistical facilities of the Modeler:
10 samples are taken at each sampling point used to compute the statistical minimum, average, standard deviation, and median.
We focus our analysis on the median, which we consider the most representative statistical quantity.

We focus on the discrete case $(\mathtt{side}, \mathtt{uplo}, \mathtt{transA}) = (\mathtt L, \mathtt L, \mathtt N)$.
While the performance dependencies of the other cases are slightly different, they show the same structures and general behavior.
Therefore, comparing different cases does not yield any insights regarding the modeling quality.

                \subsubsection{Model Expansion}
                \label{sec:modeling.res.ticks.me}
                Model Expansion has a few interesting configuration options for \metric{ticks}
\begin{itemize}
    \item The relative error bound,
    \item The direction of expansion (away from the origin or towards it), and
    \item The initial size of regions.
\end{itemize}

\begin{figure}[t]
    \newcommand{\regionsplot}[1]{
        \begin{tikzpicture}
            \begin{axis}[
                twocolplot,
                axis equal image=true,
                xlabel={\texttt m},
                ylabel={\texttt n},
                xtick={0,256,...,1024},
                ytick={0,256,...,1024},
                xmax=1024,
                ymax=1024
            ]
                \coordinate (oi) at (axis description cs: 0, 1);
                \coordinate (io) at (axis description cs: 1, 0);
            \end{axis}

            \foreach \ax/\ay/\bx/\by/\e in {#1} {
                \pgfmathparse{100 * max(0, min(1, 1 * (1 - 5*\e)))}
                \filldraw[fill=plot2!\pgfmathresult!plot1, opacity=.8] ($\ax/1024*(io) + \ay/1024*(oi)$) rectangle ($\bx/1024*(io) + \by/1024*(oi)$);
            }
        \end{tikzpicture}
    }
    \centering
    \subfloat[error bound = 10\%, expansion direction = $\nearrow$, initial width = 128]{
        \label{fig:modeling.res.ticks.me.regions:a10.sg32.dr}
        \regionsplot{
            8/8/104/104/1.00469552302,
            112/8/224/504/0.0925378117453,
            8/112/104/208/0.0899962466816,
            8/216/104/312/0.136738988956,
            8/320/104/416/0.205671004464,
            8/424/104/520/0.337808607959,
            232/8/424/248/0.0895894822351,
            112/512/216/944/0.0888837091473,
            8/528/104/624/0.11464252529,
            8/632/104/1008/0.0832475826006,
            432/8/528/104/0.341494934763,
            232/256/368/352/0.0985248840035,
            112/952/208/1024/0.151603768446,
            224/512/320/608/0.120270481454,
            8/1016/136/1024/0.088493281342,
            536/8/824/120/0.0990907381512,
            432/112/704/344/0.0981517499468,
            216/952/312/1024/0.124083089475,
            224/616/704/848/0.0969178070244,
            328/512/1000/1016/0.0967287902334,
            232/360/328/456/0.124388784778,
            376/256/1024/392/0.092724544515,
            336/360/1016/1008/0.0958524737678,
            232/464/328/560/0.129467784303,
            712/128/1024/1024/0.0869214186458,
            832/8/928/104/0.105363710864,
            832/112/1024/1024/0.0552884313051,
            936/8/1024/624/0.0994502179859,
            224/856/448/1008/0.0953467774158,
            320/1016/800/1024/0.0750460199508
        }
    }
    \hfill
    \subfloat[error bound = 10\%, expansion direction = $\swarrow$, initial width = 128]{
        \label{fig:modeling.res.ticks.me.regions:a10.sg32.dl}
        \regionsplot{
            144/408/1024/1024/0.064230745417,
            536/8/1024/400/0.0614626491845,
            40/928/136/1024/0.176948662506,
            32/192/136/920/0.0749014556158,
            8/928/32/1024/0.0462208519538,
            168/168/528/400/0.0884647795771,
            8/824/24/920/0.106224004915,
            8/720/24/816/0.108707652278,
            8/616/24/712/0.143674342052,
            160/48/528/160/0.0943284348788,
            64/64/160/400/0.0511439972596,
            8/512/24/608/0.137464637655,
            8/272/24/504/0.0945303973329,
            8/88/56/184/0.127791203696,
            56/8/152/56/0.0829470155189,
            8/8/56/80/0.0571283006961,
            184/8/528/40/0.0678138682488,
            8/168/24/264/0.200775124444,
            80/8/176/40/0.104301590893
        }
    }

    \subfloat[error bound = 5\%, expansion direction = $\swarrow$, initial width = 128]{
        \label{fig:modeling.res.ticks.me.regions:a5.sg32.dl}
        \regionsplot{
            512/192/1024/1024/0.0482120288048,
            152/552/504/1024/0.0439441743443,
            744/8/1024/184/0.045209148553,
            368/48/736/184/0.0457191034119,
            224/88/504/544/0.0498135669515,
            48/928/144/1024/0.146546362558,
            8/928/40/1024/0.0366798615078,
            48/824/144/920/0.17609281073,
            48/720/144/816/0.113799537273,
            8/720/40/816/0.0575027405058,
            24/304/144/712/0.0464559420081,
            640/8/736/40/0.0517294670242,
            8/824/40/920/0.164491358635,
            528/8/632/40/0.0298791729969,
            56/216/216/544/0.0427817539938,
            144/8/360/80/0.0471980240611,
            8/616/16/712/0.128335814423,
            424/8/520/40/0.850090470963,
            8/512/16/608/0.129485329486,
            40/8/136/80/0.126668878985,
            320/8/416/40/0.163671309602,
            8/408/16/504/0.0994956515629,
            8/304/16/400/0.108574361292,
            8/200/48/296/0.143784132967,
            56/112/216/208/0.0433732976348,
            8/96/48/192/0.194636783622,
            120/8/216/104/0.0870420981146,
            8/8/48/88/0.478092293419,
            16/8/112/104/0.470426091606
        }
    }
    \hfill
    \subfloat[error bound = 5\%, expansion direction = $\swarrow$, initial width = 64]{
        \label{fig:modeling.res.ticks.me.regions:a5.sg16.dl}
        \regionsplot{
            472/176/1024/1024/0.0469040916926,
            416/976/464/1024/0.0532795630622,
            720/8/1024/168/0.0470819522827,
            232/880/408/1024/0.0497572243788,
            288/656/464/968/0.0498514160808,
            536/48/712/168/0.0443373474728,
            176/976/224/1024/0.065508044443,
            120/976/168/1024/0.153568721594,
            176/888/224/968/0.0451571203487,
            232/824/280/872/0.0448652020619,
            256/592/464/648/0.0482581518478,
            64/976/112/1024/0.19073780162,
            8/976/56/1024/0.030969136151,
            664/8/712/40/0.0636646129475,
            480/120/528/168/0.0654420816894,
            120/920/168/968/0.135854507789,
            176/832/224/880/0.0676723878788,
            528/8/656/40/0.0346553619416,
            480/64/528/112/0.1197166747,
            424/120/472/168/0.0887150917635,
            176/776/224/824/0.0704173023976,
            120/824/168/872/0.182387707018,
            232/688/280/816/0.0474240593529,
            8/920/56/968/0.0841096128778,
            120/864/168/912/0.137462167925,
            64/920/112/968/0.185772047315,
            424/64/472/112/0.0965392941138,
            480/8/528/56/0.367512858912,
            368/120/416/168/0.101162750184,
            328/536/464/584/0.0498703150081,
            64/776/112/824/0.115998379257,
            64/824/112/872/0.143706965844,
            8/864/56/912/0.130987223145,
            64/864/112/912/0.154085349155,
            424/8/472/56/0.304757246881,
            8/776/56/824/0.0641376888399,
            64/720/112/768/0.0712288951234,
            8/808/56/856/0.230965541805,
            368/8/416/56/0.134330569443,
            8/720/56/768/0.10112949836,
            64/664/112/712/0.0538589130631,
            312/8/360/56/0.121931171322,
            8/664/56/712/0.156454833983,
            64/608/112/656/0.0705459585713,
            248/8/304/56/0.0400464335912,
            8/608/56/656/0.0643646899951,
            40/168/112/600/0.0410735636557,
            120/768/168/816/0.137948959692,
            176/720/224/768/0.0931684574737,
            232/632/280/680/0.0644471485231,
            416/480/464/528/0.0521212790588,
            272/536/320/584/0.0850841148138,
            120/712/168/760/0.163786022049,
            176/664/224/712/0.0864073377341,
            232/480/280/528/0.0661705198597,
            200/576/248/624/0.0458205996204,
            216/536/264/584/0.0840573987667,
            360/480/408/528/0.0478712933274,
            280/176/464/472/0.0428624697569,
            104/656/168/704/0.0128350463565,
            176/608/224/656/0.114285333573,
            192/8/240/56/0.0567630475192,
            8/552/32/600/0.14955288781,
            64/112/112/160/0.0513714476488,
            304/416/352/528/0.0451269835135,
            120/576/168/624/0.160086525026,
            144/552/192/600/0.139674626966,
            160/520/208/568/0.106307490237,
            120/600/168/648/0.149455854243,
            88/520/136/568/0.102469845805,
            8/496/32/544/0.128770882757,
            104/496/152/544/0.0838182418266,
            120/464/168/512/0.111771677753,
            8/112/56/160/0.0763583825661,
            56/8/112/104/0.037526133229,
            144/464/192/512/0.163603094003,
            160/464/208/512/0.162289374408,
            368/64/416/112/0.120067151782,
            136/8/184/56/0.0488337557606,
            312/120/360/168/0.133510920059,
            224/424/272/472/0.116687089085,
            8/440/32/488/0.127346382278,
            176/32/360/112/0.0428413754193,
            256/120/304/168/0.113527610467,
            224/368/272/416/0.108301386057,
            168/424/216/472/0.158876071371,
            224/312/272/360/0.125560501131,
            248/480/296/528/0.0588965805361,
            168/360/216/408/0.115684613096,
            200/8/248/168/0.047411023609,
            56/176/160/360/0.0417173775091,
            168/56/216/352/0.0455897036574,
            224/256/272/304/0.0984174766414,
            168/368/216/416/0.1242892291,
            176/480/224/528/0.110404288417,
            112/144/160/456/0.0458260637867,
            8/384/32/432/0.1010340302,
            224/200/272/248/0.0813359915597,
            224/144/272/192/0.111661921397,
            80/8/128/24/0.0389232293718,
            8/312/32/360/0.152206169895,
            8/56/48/104/0.0682498281356,
            80/8/128/48/0.0889719448527,
            8/8/48/48/0.194705559387,
            112/88/160/136/0.0554784620143,
            8/328/32/376/0.141131602765,
            112/32/160/80/0.0518663101012,
            8/256/32/304/0.0908893775485,
            8/200/32/248/0.123700153613,
            8/144/32/192/0.140849060573
        }
    }

    \vspace{.5cm}
    \tikzset{external/export=false}

    \begin{tikzpicture}
        \coordinate (pos) at (0, 0);
        \filldraw[fill=plot1] (pos) +(-2.5, 0) rectangle +(2.5, -.5);
        \filldraw[path fading=west, fill=plot2] (pos) +(-2.5, 0) rectangle +(2.5, -.5);
        \path (pos) ++(-2.5, -.25) node[anchor=east] {error:};
        \path (pos) ++(2.5, -.25) node[anchor=west] {\phantom{error:}};
        \path (pos) ++(-2.5, -.5) node[anchor=north] {0.2};
        \path (pos) ++(2.5, -.5) node[anchor=north] {0};
        \draw[dashed] (pos) ++(-2.5, 0) ++($.5*(5, 0)$) -- ++(0, -.5) node[anchor=north] {0.1};
        \draw[dashed] (pos) ++(-2.5, 0) ++($.75*(5, 0)$) -- ++(0, -.5) node[anchor=north] {0.05};
    \end{tikzpicture}

    \caption{\metric{ticks} model for \texttt{dtrsm} with Model Expansion.}
    \label{fig:modeling.res.ticks.me.regions}
\end{figure}
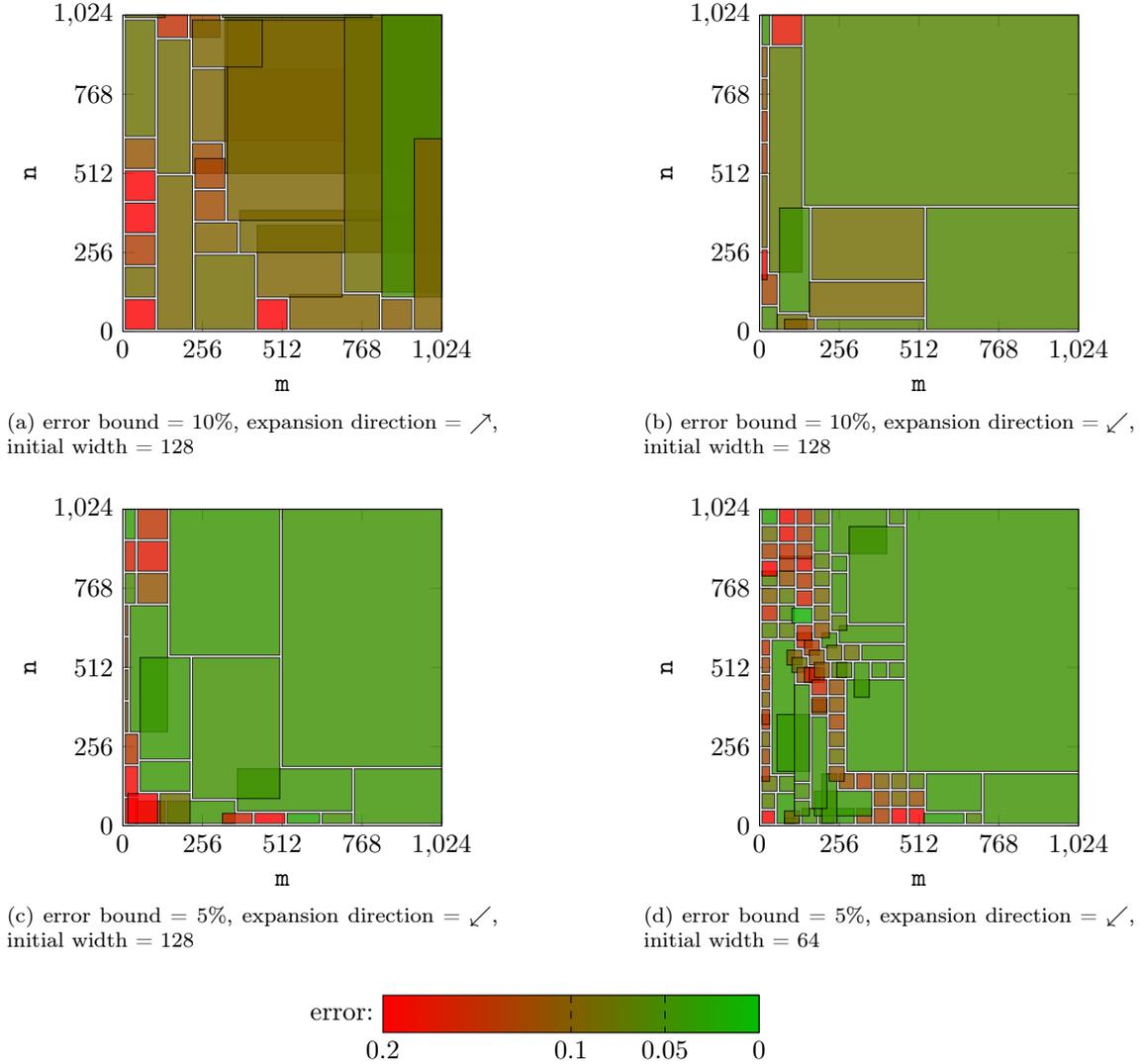

We address these options' influence by considering a series of models generated with different configurations.
The plots in \autoref{fig:modeling.res.ticks.me.regions} show how these models cover the parameter space with regions along with their relative errors.

In \autoref{fig:modeling.res.ticks.me.regions:a10.sg32.dr}, we begin with the following configuration:
\begin{itemize}
    \item The error bound is $10\%$;
    \item The direction of expansion is away from the origin ($\nearrow$);
    \item New regions are initially of size $128 \times 128$.
\end{itemize}
We observe that smaller and less accurately modeled regions are generated towards the left side of the parameter space.
Towards the top right corner, the regions become larger and the relative error decreases.
In this part of the parameter space, we also find several areas which are modeled by two or more overlapping regions\footnote{
    When the model is evaluated at a point covered by multiple regions, the most accurate model is selected.
}.

For the second Modeler configuration, we flip the direction of model expansion (\autoref{fig:modeling.res.ticks.me.regions:a10.sg32.dl}).
We observe the following changes:
\begin{itemize}
    \item Especially towards the top right, the generated regions are larger;
    \item These regions are of higher accuracy, although the error bound was not modified;
    \item Fewer regions overlap.
\end{itemize}
The average relative error improves from $10.4\%$ to $6.99\%$ while the number of required sampling points decreases from $65,220$ to $32,680$.
We conclude that it is preferable to expand the models towards the origin ($\swarrow$).

In \autoref{fig:modeling.res.ticks.me.regions:a5.sg32.dl} we now reduce the error bound to $5\%$.
As a result, the average model error improves from $6.98\%$ to $5.79\%$.
This comes at a cost of an increase in the number of samples from $32,580$ to $53,550$.
As in the previous cases, the accuracy of the models decreases as the parameter values become smaller; the least accurate models appear for small values of \texttt m.

Finally, we now decrease the size of initial models from $128 \times 128$ to $64 \times 64$ (\autoref{fig:modeling.res.ticks.me.regions:a5.sg16.dl}).
While this has little influence in the upper right part of the parameter space, lots of small less accurately modeled regions are generated for smaller parameter values.
This reveals a band of inaccurately modeled regions stretching from the top left corner of the parameter space ($(\mathtt m, \mathtt n) \approx (8, 1024)$) to the middle of its lower bound ($(\mathtt m, \mathtt n) \approx (512, 8)$); these artifacts were previously not visible --- the sampling was too coarse.
The finer sampling and the smaller models come at the cost of $84,710$ additional samples --- making a total of $138,290$ samples.
Due to the now revealed behavior of \metric{ticks}, the average error increases from $5.79\%$ to $5.96\%$.

                \subsubsection{Adaptive Refinement}
                \label{sec:modeling.res.ticks.ar}
                \begin{figure}[t]
    \newcommand{\regionsplot}[1]{
        \begin{tikzpicture}
            \begin{axis}[
                twocolplot,
                axis equal image=true,
                xlabel={\texttt m},
                ylabel={\texttt n},
                xtick={0,256,...,1024},
                ytick={0,256,...,1024},
                xmax=1024,
                ymax=1024
            ]
                \coordinate (oi) at (axis description cs: 0, 1);
                \coordinate (io) at (axis description cs: 1, 0);
            \end{axis}

            \foreach \ax/\ay/\bx/\by/\e in {#1} {
                \pgfmathparse{100 * max(0, min(1, 1 * (1 - 5*\e)))}
                \filldraw[fill=plot2!\pgfmathresult!plot1] ($\ax/1024*(io) + \ay/1024*(oi)$) rectangle ($\bx/1024*(io) + \by/1024*(oi)$);
            }
        \end{tikzpicture}
    }
    \centering
    \subfloat[error bound = 10\%, minimum width = 64]{
        \label{fig:modeling.res.ticks.ar.regions:a10.sg16}
        \regionsplot{
            8/8/1024/1024/298.25175291,
            8/8/512/512/5.31541192201,
            8/512/512/1024/5.36510601979,
            512/8/1024/512/0.254527699311,
            512/512/1024/1024/0.0179230028203,
            8/8/256/256/26.3145938504,
            8/256/256/512/6.13051537593,
            256/8/512/256/0.184679799091,
            256/256/512/512/0.0945001444503,
            8/512/256/768/1.67466544363,
            8/768/256/1024/0.738471349737,
            256/512/512/768/0.0262264805115,
            256/768/512/1024/0.00555637450589,
            512/8/768/256/0.153933691237,
            512/256/768/512/0.0150583005612,
            768/8/1024/256/0.0742773110493,
            768/256/1024/512/0.0117442584896,
            8/8/128/128/0.162557810175,
            8/128/128/256/0.138480799956,
            128/8/256/128/0.135481300661,
            128/128/256/256/0.15139880522,
            8/256/128/384/0.133690975314,
            8/384/128/512/0.581951712758,
            128/256/256/384/0.164681936985,
            128/384/256/512/0.102461212965,
            256/8/384/128/0.359645636395,
            256/128/384/256/0.107514770332,
            384/8/512/128/0.227385152439,
            384/128/512/256/0.10941593078,
            8/512/128/640/0.302395580312,
            8/640/128/768/1.43114894497,
            128/512/256/640/0.099394517086,
            128/640/256/768/0.15063017347,
            8/768/128/896/0.439737692737,
            8/896/128/1024/0.300650240482,
            128/768/256/896/0.133382895281,
            128/896/256/1024/0.0799613603409,
            512/8/640/128/0.157317756437,
            512/128/640/256/0.0553483736826,
            640/8/768/128/0.0788105426409,
            640/128/768/256/0.0232717611259,
            64/64/128/128/0.040754303298,
            64/128/128/192/0.0735726196423,
            64/192/128/256/0.0388847450564,
            128/64/192/128/0.0211557774325,
            192/64/256/128/0.0188786005001,
            128/128/192/192/0.0224043803283,
            128/192/192/256/0.026607883907,
            192/128/256/192/0.103221595743,
            192/192/256/256/0.0574209308441,
            64/256/128/320/0.0220965954127,
            64/320/128/384/0.0231499004738,
            64/384/128/448/0.0290166644936,
            64/448/128/512/0.0407691040216,
            128/256/192/320/0.0375663907743,
            128/320/192/384/0.0592638522889,
            192/256/256/320/0.120620705544,
            192/320/256/384/0.0529540011421,
            128/384/192/448/0.125456665596,
            128/448/192/512/0.136589393041,
            192/384/256/448/0.133965343081,
            192/448/256/512/0.0844543066018,
            256/64/320/128/0.0675774564232,
            320/64/384/128/0.061630307236,
            256/128/320/192/0.140107182032,
            256/192/320/256/0.117808639433,
            320/128/384/192/0.120446283677,
            320/192/384/256/0.0344093465829,
            384/64/448/128/0.117624008367,
            448/64/512/128/0.049856478824,
            384/128/448/192/0.122104648542,
            384/192/448/256/0.0773610622445,
            448/128/512/192/0.0710688485002,
            448/192/512/256/0.0396227067803,
            64/512/128/576/0.0445361006083,
            64/576/128/640/0.0346780796547,
            64/640/128/704/0.153193722831,
            64/704/128/768/0.0933242890008,
            128/640/192/704/0.160680452895,
            128/704/192/768/0.152079469914,
            192/640/256/704/0.0371690196719,
            192/704/256/768/0.0582696975023,
            64/768/128/832/0.123927393837,
            64/832/128/896/0.119489941716,
            64/896/128/960/0.163609649076,
            64/960/128/1024/0.161539355046,
            128/768/192/832/0.184120640679,
            128/832/192/896/0.0993804453485,
            192/768/256/832/0.0371189441238,
            192/832/256/896/0.0439350424172,
            512/64/576/128/0.0485380817039,
            576/64/640/128/0.0132057649732
        }
    }
    \hfill
    \subfloat[error bound = 5\%, minimum width = 64]{
        \label{fig:modeling.res.ticks.ar.regions:a5.sg16}
        \regionsplot{
            8/8/1024/1024/298.25175291,
            8/8/512/512/5.31541192201,
            8/512/512/1024/5.36510601979,
            512/8/1024/512/0.254527699311,
            512/512/1024/1024/0.0179230028203,
            8/8/256/256/26.3145938504,
            8/256/256/512/6.13051537593,
            256/8/512/256/0.184679799091,
            256/256/512/512/0.0945001444503,
            8/512/256/768/1.67466544363,
            8/768/256/1024/0.738471349737,
            256/512/512/768/0.0262264805115,
            256/768/512/1024/0.00555637450589,
            512/8/768/256/0.153933691237,
            512/256/768/512/0.0150583005612,
            768/8/1024/256/0.0742773110493,
            768/256/1024/512/0.0117442584896,
            8/8/128/128/0.162557810175,
            8/128/128/256/0.138480799956,
            128/8/256/128/0.135481300661,
            128/128/256/256/0.15139880522,
            8/256/128/384/0.133690975314,
            8/384/128/512/0.581951712758,
            128/256/256/384/0.164681936985,
            128/384/256/512/0.102461212965,
            256/8/384/128/0.359645636395,
            256/128/384/256/0.107514770332,
            384/8/512/128/0.227385152439,
            384/128/512/256/0.10941593078,
            256/256/384/384/0.113442616169,
            256/384/384/512/0.101742936041,
            384/256/512/384/0.0418828433823,
            384/384/512/512/0.0109782204121,
            8/512/128/640/0.302395580312,
            8/640/128/768/1.43114894497,
            128/512/256/640/0.099394517086,
            128/640/256/768/0.15063017347,
            8/768/128/896/0.439737692737,
            8/896/128/1024/0.300650240482,
            128/768/256/896/0.133382895281,
            128/896/256/1024/0.0799613603409,
            512/8/640/128/0.157317756437,
            512/128/640/256/0.0553483736826,
            640/8/768/128/0.0788105426409,
            640/128/768/256/0.0232717611259,
            768/8/896/128/0.0350013031103,
            768/128/896/256/0.024017495526,
            896/8/1024/128/0.00618543011681,
            896/128/1024/256/0.00657932345272,
            64/64/128/128/0.040754303298,
            64/128/128/192/0.0735726196423,
            64/192/128/256/0.0388847450564,
            128/64/192/128/0.0211557774325,
            192/64/256/128/0.0188786005001,
            128/128/192/192/0.0224043803283,
            128/192/192/256/0.026607883907,
            192/128/256/192/0.103221595743,
            192/192/256/256/0.0574209308441,
            64/256/128/320/0.0220965954127,
            64/320/128/384/0.0231499004738,
            64/384/128/448/0.0290166644936,
            64/448/128/512/0.0407691040216,
            128/256/192/320/0.0375663907743,
            128/320/192/384/0.0592638522889,
            192/256/256/320/0.120620705544,
            192/320/256/384/0.0529540011421,
            128/384/192/448/0.125456665596,
            128/448/192/512/0.136589393041,
            192/384/256/448/0.133965343081,
            192/448/256/512/0.0844543066018,
            256/64/320/128/0.0675774564232,
            320/64/384/128/0.061630307236,
            256/128/320/192/0.140107182032,
            256/192/320/256/0.117808639433,
            320/128/384/192/0.120446283677,
            320/192/384/256/0.0344093465829,
            384/64/448/128/0.117624008367,
            448/64/512/128/0.049856478824,
            384/128/448/192/0.122104648542,
            384/192/448/256/0.0773610622445,
            448/128/512/192/0.0710688485002,
            448/192/512/256/0.0396227067803,
            256/256/320/320/0.0994010139553,
            256/320/320/384/0.0537707594503,
            320/256/384/320/0.0291926282937,
            320/320/384/384/0.0115904950289,
            256/384/320/448/0.121956714078,
            256/448/320/512/0.0415113432217,
            320/384/384/448/0.00994769405619,
            320/448/384/512/0.00637406985477,
            64/512/128/576/0.0445361006083,
            64/576/128/640/0.0346780796547,
            64/640/128/704/0.153193722831,
            64/704/128/768/0.0933242890008,
            128/512/192/576/0.174347432251,
            128/576/192/640/0.123089031368,
            192/512/256/576/0.0598069646487,
            192/576/256/640/0.0154364808006,
            128/640/192/704/0.160680452895,
            128/704/192/768/0.152079469914,
            192/640/256/704/0.0371690196719,
            192/704/256/768/0.0582696975023,
            64/768/128/832/0.123927393837,
            64/832/128/896/0.119489941716,
            64/896/128/960/0.163609649076,
            64/960/128/1024/0.161539355046,
            128/768/192/832/0.184120640679,
            128/832/192/896/0.0993804453485,
            192/768/256/832/0.0371189441238,
            192/832/256/896/0.0439350424172,
            128/896/192/960/0.114318853219,
            128/960/192/1024/0.0740986673843,
            192/896/256/960/0.0677961991778,
            192/960/256/1024/0.0499630712646,
            512/64/576/128/0.0485380817039,
            576/64/640/128/0.0132057649732,
            512/128/576/192/0.0666165425769,
            512/192/576/256/0.0171133757638,
            576/128/640/192/0.00781440102144,
            576/192/640/256/0.00449335612066,
            640/64/704/128/0.0111739118211,
            704/64/768/128/0.0157050659385
        }
    }

    \subfloat[error bound = 10\%, minimum width = 32]{
        \label{fig:modeling.res.ticks.ar.regions:a10.sg8}
        \regionsplot{
            8/8/1024/1024/298.25175291,
            8/8/512/512/5.31541192201,
            8/512/512/1024/5.36510601979,
            512/8/1024/512/0.254527699311,
            512/512/1024/1024/0.0179230028203,
            8/8/256/256/26.3145938504,
            8/256/256/512/6.13051537593,
            256/8/512/256/0.184679799091,
            256/256/512/512/0.0945001444503,
            8/512/256/768/1.67466544363,
            8/768/256/1024/0.738471349737,
            256/512/512/768/0.0262264805115,
            256/768/512/1024/0.00555637450589,
            512/8/768/256/0.153933691237,
            512/256/768/512/0.0150583005612,
            768/8/1024/256/0.0742773110493,
            768/256/1024/512/0.0117442584896,
            8/8/128/128/0.162557810175,
            8/128/128/256/0.138480799956,
            128/8/256/128/0.135481300661,
            128/128/256/256/0.15139880522,
            8/256/128/384/0.133690975314,
            8/384/128/512/0.581951712758,
            128/256/256/384/0.164681936985,
            128/384/256/512/0.102461212965,
            256/8/384/128/0.359645636395,
            256/128/384/256/0.107514770332,
            384/8/512/128/0.227385152439,
            384/128/512/256/0.10941593078,
            8/512/128/640/0.302395580312,
            8/640/128/768/1.43114894497,
            128/512/256/640/0.099394517086,
            128/640/256/768/0.15063017347,
            8/768/128/896/0.439737692737,
            8/896/128/1024/0.300650240482,
            128/768/256/896/0.133382895281,
            128/896/256/1024/0.0799613603409,
            512/8/640/128/0.157317756437,
            512/128/640/256/0.0553483736826,
            640/8/768/128/0.0788105426409,
            640/128/768/256/0.0232717611259,
            8/8/64/64/0.0619046198102,
            8/64/64/128/0.239305548006,
            64/8/128/64/0.0722380052926,
            64/64/128/128/0.040754303298,
            8/128/64/192/0.154478739858,
            8/192/64/256/0.0873066720597,
            64/128/128/192/0.0735726196423,
            64/192/128/256/0.0388847450564,
            128/8/192/64/0.0300153191569,
            128/64/192/128/0.0211557774325,
            192/8/256/64/0.0546866959418,
            192/64/256/128/0.0188786005001,
            128/128/192/192/0.0224043803283,
            128/192/192/256/0.026607883907,
            192/128/256/192/0.103221595743,
            192/192/256/256/0.0574209308441,
            8/256/64/320/0.196357261143,
            8/320/64/384/0.127477635469,
            64/256/128/320/0.0220965954127,
            64/320/128/384/0.0231499004738,
            8/384/64/448/0.030985027995,
            8/448/64/512/0.123844756209,
            64/384/128/448/0.0290166644936,
            64/448/128/512/0.0407691040216,
            128/256/192/320/0.0375663907743,
            128/320/192/384/0.0592638522889,
            192/256/256/320/0.120620705544,
            192/320/256/384/0.0529540011421,
            128/384/192/448/0.125456665596,
            128/448/192/512/0.136589393041,
            192/384/256/448/0.133965343081,
            192/448/256/512/0.0844543066018,
            256/8/320/64/0.0170026209351,
            256/64/320/128/0.0675774564232,
            320/8/384/64/0.0832918646484,
            320/64/384/128/0.061630307236,
            256/128/320/192/0.140107182032,
            256/192/320/256/0.117808639433,
            320/128/384/192/0.120446283677,
            320/192/384/256/0.0344093465829,
            384/8/448/64/0.253959585343,
            384/64/448/128/0.117624008367,
            448/8/512/64/0.217233927739,
            448/64/512/128/0.049856478824,
            384/128/448/192/0.122104648542,
            384/192/448/256/0.0773610622445,
            448/128/512/192/0.0710688485002,
            448/192/512/256/0.0396227067803,
            8/512/64/576/0.150063988305,
            8/576/64/640/0.0953199266134,
            64/512/128/576/0.0445361006083,
            64/576/128/640/0.0346780796547,
            8/640/64/704/0.125616953956,
            8/704/64/768/0.0289027014333,
            64/640/128/704/0.153193722831,
            64/704/128/768/0.0933242890008,
            128/640/192/704/0.160680452895,
            128/704/192/768/0.152079469914,
            192/640/256/704/0.0371690196719,
            192/704/256/768/0.0582696975023,
            8/768/64/832/0.0706219677349,
            8/832/64/896/0.0940032724796,
            64/768/128/832/0.123927393837,
            64/832/128/896/0.119489941716,
            8/896/64/960/0.0941453218375,
            8/960/64/1024/0.0945905472816,
            64/896/128/960/0.163609649076,
            64/960/128/1024/0.161539355046,
            128/768/192/832/0.184120640679,
            128/832/192/896/0.0993804453485,
            192/768/256/832/0.0371189441238,
            192/832/256/896/0.0439350424172,
            512/8/576/64/0.171196717023,
            512/64/576/128/0.0485380817039,
            576/8/640/64/0.00887925751019,
            576/64/640/128/0.0132057649732,
            32/64/64/96/0.0750571862169,
            32/96/64/128/0.0692583554975,
            32/128/64/160/0.0768395093728,
            32/160/64/192/0.0420888404133,
            192/128/224/160/0.0124217313588,
            192/160/224/192/0.0392831028293,
            224/128/256/160/0.0210954161758,
            224/160/256/192/0.0920742796773,
            32/256/64/288/0.0640897013178,
            32/288/64/320/0.103399579235,
            32/320/64/352/0.0674666275691,
            32/352/64/384/0.0326013359802,
            32/448/64/480/0.0290598715311,
            32/480/64/512/0.0304204194019,
            192/256/224/288/0.0424192833617,
            192/288/224/320/0.0695992769116,
            224/256/256/288/0.0359789766592,
            224/288/256/320/0.085349165793,
            128/384/160/416/0.00557462909307,
            128/416/160/448/0.019870107866,
            160/384/192/416/0.0406581528893,
            160/416/192/448/0.0770525747057,
            128/448/160/480/0.0349085534583,
            128/480/160/512/0.0218132599726,
            160/448/192/480/0.0777166992387,
            160/480/192/512/0.104962409202,
            192/384/224/416/0.0790276251408,
            192/416/224/448/0.0833804018348,
            224/384/256/416/0.0347804048981,
            224/416/256/448/0.115902353289,
            256/128/288/160/0.0154316258447,
            256/160/288/192/0.0794563435213,
            288/128/320/160/0.044490295764,
            288/160/320/192/0.0913072019133,
            256/192/288/224/0.0604556829126,
            256/224/288/256/0.0302746495345,
            288/192/320/224/0.0780166280993,
            288/224/320/256/0.0844885170149,
            320/128/352/160/0.0688854119076,
            320/160/352/192/0.0739104870842,
            352/128/384/160/0.0321834350837,
            352/160/384/192/0.0795528401677,
            384/32/416/64/0.10011849673,
            416/32/448/64/0.174239442807,
            384/64/416/96/0.0877624395127,
            384/96/416/128/0.0335643333765,
            416/64/448/96/0.116223961799,
            416/96/448/128/0.0934155356675,
            448/32/480/64/0.149689495786,
            480/32/512/64/0.154425275736,
            384/128/416/160/0.04001777697,
            384/160/416/192/0.0603027068114,
            416/128/448/160/0.0625189426867,
            416/160/448/192/0.10003709818,
            32/512/64/544/0.111069188543,
            32/544/64/576/0.0753533245659,
            32/640/64/672/0.0311502757156,
            32/672/64/704/0.0273856278414,
            64/640/96/672/0.0232428773465,
            64/672/96/704/0.0379586829023,
            96/640/128/672/0.072361747574,
            96/672/128/704/0.0768970240205,
            128/640/160/672/0.0485741654494,
            128/672/160/704/0.0693467266045,
            160/640/192/672/0.0826660347724,
            160/672/192/704/0.126175252254,
            128/704/160/736/0.0959031471928,
            128/736/160/768/0.0424735333105,
            160/704/192/736/0.098872510684,
            160/736/192/768/0.123502366669,
            64/768/96/800/0.05920513391,
            64/800/96/832/0.0859849374678,
            96/768/128/800/0.0886655237753,
            96/800/128/832/0.0908317398233,
            64/832/96/864/0.118547185217,
            64/864/96/896/0.0698542585139,
            96/832/128/864/0.0835402787763,
            96/864/128/896/0.161320460398,
            64/896/96/928/0.07775397853,
            64/928/96/960/0.121471132977,
            96/896/128/928/0.0908587129755,
            96/928/128/960/0.106188079556,
            64/960/96/992/0.160615958984,
            64/992/96/1024/0.0913276727709,
            96/960/128/992/0.114254825633,
            96/992/128/1024/0.185370955846,
            128/768/160/800/0.109812829554,
            128/800/160/832/0.122860844546,
            160/768/192/800/0.113066022369,
            160/800/192/832/0.110211328397,
            512/32/544/64/0.180202689571,
            544/32/576/64/0.0252174492978
        }
    }
    \hfill
    \subfloat[error bound = 5\%, minimum width = 32]{
        \label{fig:modeling.res.ticks.ar.regions:a5.sg8}
        \regionsplot{
            8/8/1024/1024/298.25175291,
            8/8/512/512/5.31541192201,
            8/512/512/1024/5.36510601979,
            512/8/1024/512/0.254527699311,
            512/512/1024/1024/0.0179230028203,
            8/8/256/256/26.3145938504,
            8/256/256/512/6.13051537593,
            256/8/512/256/0.184679799091,
            256/256/512/512/0.0945001444503,
            8/512/256/768/1.67466544363,
            8/768/256/1024/0.738471349737,
            256/512/512/768/0.0262264805115,
            256/768/512/1024/0.00555637450589,
            512/8/768/256/0.153933691237,
            512/256/768/512/0.0150583005612,
            768/8/1024/256/0.0742773110493,
            768/256/1024/512/0.0117442584896,
            8/8/128/128/0.162557810175,
            8/128/128/256/0.138480799956,
            128/8/256/128/0.135481300661,
            128/128/256/256/0.15139880522,
            8/256/128/384/0.133690975314,
            8/384/128/512/0.581951712758,
            128/256/256/384/0.164681936985,
            128/384/256/512/0.102461212965,
            256/8/384/128/0.359645636395,
            256/128/384/256/0.107514770332,
            384/8/512/128/0.227385152439,
            384/128/512/256/0.10941593078,
            256/256/384/384/0.113442616169,
            256/384/384/512/0.101742936041,
            384/256/512/384/0.0418828433823,
            384/384/512/512/0.0109782204121,
            8/512/128/640/0.302395580312,
            8/640/128/768/1.43114894497,
            128/512/256/640/0.099394517086,
            128/640/256/768/0.15063017347,
            8/768/128/896/0.439737692737,
            8/896/128/1024/0.300650240482,
            128/768/256/896/0.133382895281,
            128/896/256/1024/0.0799613603409,
            512/8/640/128/0.157317756437,
            512/128/640/256/0.0553483736826,
            640/8/768/128/0.0788105426409,
            640/128/768/256/0.0232717611259,
            768/8/896/128/0.0350013031103,
            768/128/896/256/0.024017495526,
            896/8/1024/128/0.00618543011681,
            896/128/1024/256/0.00657932345272,
            8/8/64/64/0.0619046198102,
            8/64/64/128/0.239305548006,
            64/8/128/64/0.0722380052926,
            64/64/128/128/0.040754303298,
            8/128/64/192/0.154478739858,
            8/192/64/256/0.0873066720597,
            64/128/128/192/0.0735726196423,
            64/192/128/256/0.0388847450564,
            128/8/192/64/0.0300153191569,
            128/64/192/128/0.0211557774325,
            192/8/256/64/0.0546866959418,
            192/64/256/128/0.0188786005001,
            128/128/192/192/0.0224043803283,
            128/192/192/256/0.026607883907,
            192/128/256/192/0.103221595743,
            192/192/256/256/0.0574209308441,
            8/256/64/320/0.196357261143,
            8/320/64/384/0.127477635469,
            64/256/128/320/0.0220965954127,
            64/320/128/384/0.0231499004738,
            8/384/64/448/0.030985027995,
            8/448/64/512/0.123844756209,
            64/384/128/448/0.0290166644936,
            64/448/128/512/0.0407691040216,
            128/256/192/320/0.0375663907743,
            128/320/192/384/0.0592638522889,
            192/256/256/320/0.120620705544,
            192/320/256/384/0.0529540011421,
            128/384/192/448/0.125456665596,
            128/448/192/512/0.136589393041,
            192/384/256/448/0.133965343081,
            192/448/256/512/0.0844543066018,
            256/8/320/64/0.0170026209351,
            256/64/320/128/0.0675774564232,
            320/8/384/64/0.0832918646484,
            320/64/384/128/0.061630307236,
            256/128/320/192/0.140107182032,
            256/192/320/256/0.117808639433,
            320/128/384/192/0.120446283677,
            320/192/384/256/0.0344093465829,
            384/8/448/64/0.253959585343,
            384/64/448/128/0.117624008367,
            448/8/512/64/0.217233927739,
            448/64/512/128/0.049856478824,
            384/128/448/192/0.122104648542,
            384/192/448/256/0.0773610622445,
            448/128/512/192/0.0710688485002,
            448/192/512/256/0.0396227067803,
            256/256/320/320/0.0994010139553,
            256/320/320/384/0.0537707594503,
            320/256/384/320/0.0291926282937,
            320/320/384/384/0.0115904950289,
            256/384/320/448/0.121956714078,
            256/448/320/512/0.0415113432217,
            320/384/384/448/0.00994769405619,
            320/448/384/512/0.00637406985477,
            8/512/64/576/0.150063988305,
            8/576/64/640/0.0953199266134,
            64/512/128/576/0.0445361006083,
            64/576/128/640/0.0346780796547,
            8/640/64/704/0.125616953956,
            8/704/64/768/0.0289027014333,
            64/640/128/704/0.153193722831,
            64/704/128/768/0.0933242890008,
            128/512/192/576/0.174347432251,
            128/576/192/640/0.123089031368,
            192/512/256/576/0.0598069646487,
            192/576/256/640/0.0154364808006,
            128/640/192/704/0.160680452895,
            128/704/192/768/0.152079469914,
            192/640/256/704/0.0371690196719,
            192/704/256/768/0.0582696975023,
            8/768/64/832/0.0706219677349,
            8/832/64/896/0.0940032724796,
            64/768/128/832/0.123927393837,
            64/832/128/896/0.119489941716,
            8/896/64/960/0.0941453218375,
            8/960/64/1024/0.0945905472816,
            64/896/128/960/0.163609649076,
            64/960/128/1024/0.161539355046,
            128/768/192/832/0.184120640679,
            128/832/192/896/0.0993804453485,
            192/768/256/832/0.0371189441238,
            192/832/256/896/0.0439350424172,
            128/896/192/960/0.114318853219,
            128/960/192/1024/0.0740986673843,
            192/896/256/960/0.0677961991778,
            192/960/256/1024/0.0499630712646,
            512/8/576/64/0.171196717023,
            512/64/576/128/0.0485380817039,
            576/8/640/64/0.00887925751019,
            576/64/640/128/0.0132057649732,
            512/128/576/192/0.0666165425769,
            512/192/576/256/0.0171133757638,
            576/128/640/192/0.00781440102144,
            576/192/640/256/0.00449335612066,
            640/8/704/64/0.0257599041096,
            640/64/704/128/0.0111739118211,
            704/8/768/64/0.0124800861408,
            704/64/768/128/0.0157050659385,
            32/32/64/64/0.380022522361,
            32/64/64/96/0.0750571862169,
            32/96/64/128/0.0692583554975,
            64/32/96/64/0.0843851089688,
            96/32/128/64/0.0682786326089,
            32/128/64/160/0.0768395093728,
            32/160/64/192/0.0420888404133,
            32/192/64/224/0.202012705837,
            32/224/64/256/0.129051647789,
            64/128/96/160/0.0873286353723,
            64/160/96/192/0.0841735280631,
            96/128/128/160/0.0249055235109,
            96/160/128/192/0.0215936670384,
            192/32/224/64/0.0240242259305,
            224/32/256/64/0.0447838348853,
            192/128/224/160/0.0124217313588,
            192/160/224/192/0.0392831028293,
            224/128/256/160/0.0210954161758,
            224/160/256/192/0.0920742796773,
            192/192/224/224/0.0330637860859,
            192/224/224/256/0.0601019344243,
            224/192/256/224/0.0863724502723,
            224/224/256/256/0.0612683968244,
            32/256/64/288/0.0640897013178,
            32/288/64/320/0.103399579235,
            32/320/64/352/0.0674666275691,
            32/352/64/384/0.0326013359802,
            32/448/64/480/0.0290598715311,
            32/480/64/512/0.0304204194019,
            128/320/160/352/0.0259624614847,
            128/352/160/384/0.0123658973745,
            160/320/192/352/0.0409727561911,
            160/352/192/384/0.0457251605004,
            192/256/224/288/0.0424192833617,
            192/288/224/320/0.0695992769116,
            224/256/256/288/0.0359789766592,
            224/288/256/320/0.085349165793,
            192/320/224/352/0.0571864194555,
            192/352/224/384/0.106896825958,
            224/320/256/352/0.125422611226,
            224/352/256/384/0.0935247338886,
            128/384/160/416/0.00557462909307,
            128/416/160/448/0.019870107866,
            160/384/192/416/0.0406581528893,
            160/416/192/448/0.0770525747057,
            128/448/160/480/0.0349085534583,
            128/480/160/512/0.0218132599726,
            160/448/192/480/0.0777166992387,
            160/480/192/512/0.104962409202,
            192/384/224/416/0.0790276251408,
            192/416/224/448/0.0833804018348,
            224/384/256/416/0.0347804048981,
            224/416/256/448/0.115902353289,
            192/448/224/480/0.147228782683,
            192/480/224/512/0.120286385172,
            224/448/256/480/0.0877474222098,
            224/480/256/512/0.0729303897831,
            256/64/288/96/0.0226971521618,
            256/96/288/128/0.029034839936,
            288/64/320/96/0.0436886280205,
            288/96/320/128/0.0551453308189,
            320/32/352/64/0.0632157812816,
            352/32/384/64/0.111557677848,
            320/64/352/96/0.0559835976033,
            320/96/352/128/0.101740012057,
            352/64/384/96/0.118092362449,
            352/96/384/128/0.0742544200236,
            256/128/288/160/0.0154316258447,
            256/160/288/192/0.0794563435213,
            288/128/320/160/0.044490295764,
            288/160/320/192/0.0913072019133,
            256/192/288/224/0.0604556829126,
            256/224/288/256/0.0302746495345,
            288/192/320/224/0.0780166280993,
            288/224/320/256/0.0844885170149,
            320/128/352/160/0.0688854119076,
            320/160/352/192/0.0739104870842,
            352/128/384/160/0.0321834350837,
            352/160/384/192/0.0795528401677,
            384/32/416/64/0.10011849673,
            416/32/448/64/0.174239442807,
            384/64/416/96/0.0877624395127,
            384/96/416/128/0.0335643333765,
            416/64/448/96/0.116223961799,
            416/96/448/128/0.0934155356675,
            448/32/480/64/0.149689495786,
            480/32/512/64/0.154425275736,
            384/128/416/160/0.04001777697,
            384/160/416/192/0.0603027068114,
            416/128/448/160/0.0625189426867,
            416/160/448/192/0.10003709818,
            384/192/416/224/0.0680193576287,
            384/224/416/256/0.0216015485076,
            416/192/448/224/0.0599404709984,
            416/224/448/256/0.0653554938143,
            448/128/480/160/0.0780217975866,
            448/160/480/192/0.0598745669004,
            480/128/512/160/0.0258713136664,
            480/160/512/192/0.0615562991838,
            256/256/288/288/0.0438816829646,
            256/288/288/320/0.0852866901113,
            288/256/320/288/0.0804167292696,
            288/288/320/320/0.00467272694366,
            256/320/288/352/0.11991449009,
            256/352/288/384/0.0654515239879,
            288/320/320/352/0.00967595431971,
            288/352/320/384/0.00775814840027,
            256/384/288/416/0.087028129364,
            256/416/288/448/0.115431312022,
            288/384/320/416/0.00277308888571,
            288/416/320/448/0.00394416011847,
            32/512/64/544/0.111069188543,
            32/544/64/576/0.0753533245659,
            32/576/64/608/0.0454215027221,
            32/608/64/640/0.0322968116744,
            32/640/64/672/0.0311502757156,
            32/672/64/704/0.0273856278414,
            64/640/96/672/0.0232428773465,
            64/672/96/704/0.0379586829023,
            96/640/128/672/0.072361747574,
            96/672/128/704/0.0768970240205,
            64/704/96/736/0.0345618284905,
            64/736/96/768/0.0553336434889,
            96/704/128/736/0.125866681016,
            96/736/128/768/0.0823904834494,
            128/512/160/544/0.0472317989847,
            128/544/160/576/0.0634756066262,
            160/512/192/544/0.0960619459354,
            160/544/192/576/0.108854744647,
            128/576/160/608/0.0920567190574,
            128/608/160/640/0.081312573393,
            160/576/192/608/0.101228486793,
            160/608/192/640/0.100625872801,
            192/512/224/544/0.0913894898232,
            192/544/224/576/0.05927210597,
            224/512/256/544/0.0650761796754,
            224/544/256/576/0.0619922826756,
            128/640/160/672/0.0485741654494,
            128/672/160/704/0.0693467266045,
            160/640/192/672/0.0826660347724,
            160/672/192/704/0.126175252254,
            128/704/160/736/0.0959031471928,
            128/736/160/768/0.0424735333105,
            160/704/192/736/0.098872510684,
            160/736/192/768/0.123502366669,
            192/704/224/736/0.118167745674,
            192/736/224/768/0.0545477426099,
            224/704/256/736/0.0199812369939,
            224/736/256/768/0.0325271696742,
            32/768/64/800/0.0268724671824,
            32/800/64/832/0.0374417840306,
            32/832/64/864/0.0225006747624,
            32/864/64/896/0.0357646671666,
            64/768/96/800/0.05920513391,
            64/800/96/832/0.0859849374678,
            96/768/128/800/0.0886655237753,
            96/800/128/832/0.0908317398233,
            64/832/96/864/0.118547185217,
            64/864/96/896/0.0698542585139,
            96/832/128/864/0.0835402787763,
            96/864/128/896/0.161320460398,
            32/896/64/928/0.0410556400039,
            32/928/64/960/0.0262725745093,
            32/960/64/992/0.0493911245614,
            32/992/64/1024/0.0334477649821,
            64/896/96/928/0.07775397853,
            64/928/96/960/0.121471132977,
            96/896/128/928/0.0908587129755,
            96/928/128/960/0.106188079556,
            64/960/96/992/0.160615958984,
            64/992/96/1024/0.0913276727709,
            96/960/128/992/0.114254825633,
            96/992/128/1024/0.185370955846,
            128/768/160/800/0.109812829554,
            128/800/160/832/0.122860844546,
            160/768/192/800/0.113066022369,
            160/800/192/832/0.110211328397,
            128/832/160/864/0.144054713642,
            128/864/160/896/0.109725375371,
            160/832/192/864/0.0827917345754,
            160/864/192/896/0.0770422214038,
            128/896/160/928/0.124569814913,
            128/928/160/960/0.147359805601,
            160/896/192/928/0.0572367801822,
            160/928/192/960/0.0913847918846,
            128/960/160/992/0.139853482973,
            128/992/160/1024/0.112103554679,
            160/960/192/992/0.0547835271002,
            160/992/192/1024/0.0740677573859,
            192/896/224/928/0.0312136106369,
            192/928/224/960/0.0311544185264,
            224/896/256/928/0.0309444316331,
            224/928/256/960/0.0249998899328,
            512/32/544/64/0.180202689571,
            544/32/576/64/0.0252174492978,
            512/128/544/160/0.0602665400166,
            512/160/544/192/0.0655850334558,
            544/128/576/160/0.00554339607738,
            544/160/576/192/0.00413646085618
        }
    }

    \vspace{.5cm}
    \tikzset{external/export=false}

    \begin{tikzpicture}
        \coordinate (pos) at (0, 0);
        \filldraw[fill=plot1] (pos) +(-2.5, 0) rectangle +(2.5, -.5);
        \filldraw[path fading=west, fill=plot2] (pos) +(-2.5, 0) rectangle +(2.5, -.5);
        \path (pos) ++(-2.5, -.25) node[anchor=east] {error:};
        \path (pos) ++(2.5, -.25) node[anchor=west] {\phantom{error:}};
        \path (pos) ++(-2.5, -.5) node[anchor=north] {0.2};
        \path (pos) ++(2.5, -.5) node[anchor=north] {0};
        \draw[dashed] (pos) ++(-2.5, 0) ++($.5*(5, 0)$) -- ++(0, -.5) node[anchor=north] {0.1};
        \draw[dashed] (pos) ++(-2.5, 0) ++($.75*(5, 0)$) -- ++(0, -.5) node[anchor=north] {0.05};
    \end{tikzpicture}

    \caption{\metric{ticks} model for \texttt{dtrsm} with Adaptive Refinement.}
    \label{fig:modeling.res.ticks.ar.regions}
\end{figure}
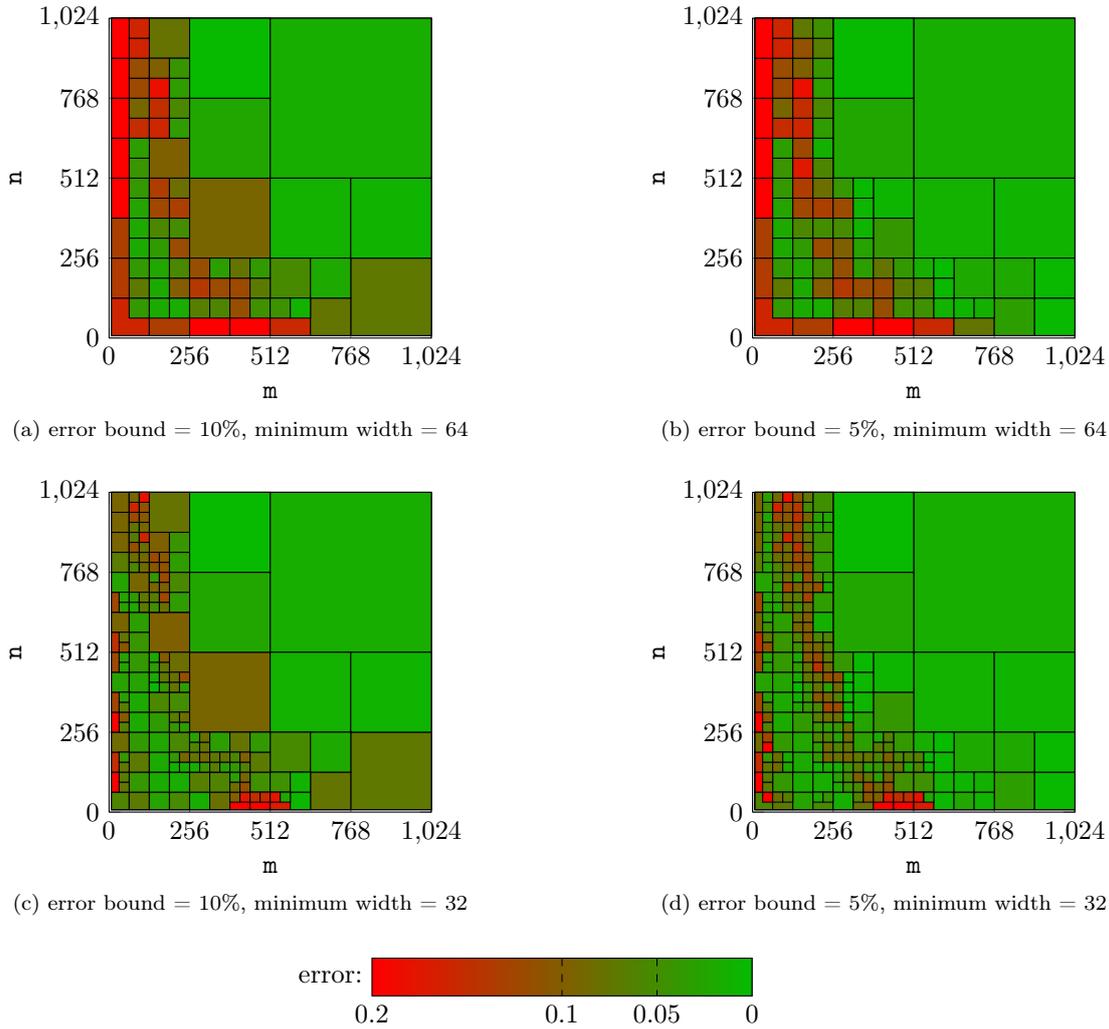

The generation and distribution of regions in Adaptive Refinement is governed by two configuration options:
\begin{itemize}
    \item The relative error bound that decides upon whether a certain region is further refined, and
    \item The minimum region size which determines the maximum depth of recursive refinements.
\end{itemize}
We now study the influence of these options individually.
The models' regions resulting from varying configurations are shown in \autoref{fig:modeling.res.ticks.ar.regions}.

For all configurations, the minimum region size does not allow to generate refined regions at the lower end of the parameter space.
This is because the minimum value for \texttt m and \texttt n is $8$ and not $0$.
The seemingly rectangular regions are parts of larger regions that were, for this reason, only partially refined.

The first model in \autoref{fig:modeling.res.ticks.ar.regions:a10.sg16} was generated with an error bound of $10\%$ and a minimum region size of $64 \times 64$.
The result shows an overall distribution of regions similar to Model Expansion: smaller and less accurately modeled regions are predominant for smaller parameter values --- especially for \texttt{m}.
Interestingly, we can already see a structured band of less accurate models, which was only visible in the most accurate model generated by Model Expansion (\autoref{fig:modeling.res.ticks.me.regions:a5.sg16.dl}).

For our next model, we decrease the error bound to $5\%$, resulting in the regions shown in \autoref{fig:modeling.res.ticks.ar.regions:a5.sg16}.
To attain the higher accuracy, several of the regions in the previous model are further refined, especially on the left side.
The increased number of regions is covered by $81,890$ samples --- $20,010$ more than previously.
The higher accuracy requirements lead to a decrease in the average error from $7.21\%$ to $6.32\%$.

For both values of the error bound ($10\%$ and $5\%$), we now decrease the minimum region size to $32 \times 32$ (Figures \ref{fig:modeling.res.ticks.ar.regions:a10.sg8} and \ref{fig:modeling.res.ticks.ar.regions:a5.sg8}).
This leads to the generation of very many tiny regions.
The error bound of $10\%$ ($5\%$) leads to an average error of $4.29\%$ ($3.17\%$) at the cost of $134,160$ ($227,820$) samples.

When we modify the accuracy and region size options, as shown in these results, Adaptive Refinement has an attractive property:
All samples that were taken for the less accurate models can be reused in the generation of the finer models (made possible by the memory file of the Modeler's Sampler Interface).
A user of the Modeler can take advantage of this property and dynamically improve the models if he considers their accuracy to be insufficient.

                \subsubsection{Comparison}
                \label{sec:modeling.res.ticks.comp}
                \begin{table}[t]
    \centering
    \subfloat[Model Expansion]{
        \label{tab:modeling.res.ticks.comp:me}
        \begin{tabular}{cccccc}
            \toprule
                &error  &expansion  &initial    &           &average\\
                &bound  &direction  &width      &\#samples  &error\\
            \midrule
            (a) &$10\%$ &$\nearrow$ &$128$      &$65,220$   &$10.4\%$\\
            (b) &$10\%$ &$\swarrow$ &$128$      &$32,680$   &$6.98\%$\\
            (c) &$5\%$  &$\swarrow$ &$128$      &$53,550$   &$5.79\%$\\
            (d) &$5\%$  &$\swarrow$ &$64$       &$138,290$  &$5.96\%$\\
            \bottomrule
        \end{tabular}
    }
    \vspace{.2cm}
    \subfloat[Adaptive Refinement]{
        \label{tab:modeling.res.ticks.comp:ar}
        \begin{tabular}{ccccc}
            \toprule
                &error   &minimum   &           &average\\
                &bound   &width     &\#samples  &error\\
            \midrule
            (a) &$10\%$  &$64$      &$61,880$   &$7.21\%$\\
            (b) &$5\%$   &$64$      &$81,890$   &$6.32\%$\\
            (c) &$10\%$  &$32$      &$134,160$  &$4.29\%$\\
            (d) &$5\%$   &$32$      &$227,820$  &$3.17\%$\\
            \bottomrule
        \end{tabular}
    }
    \caption{Comparison of Model Expansion and Adaptive Refinement.}
    \label{tab:modeling.res.ticks.comp}
\end{table}

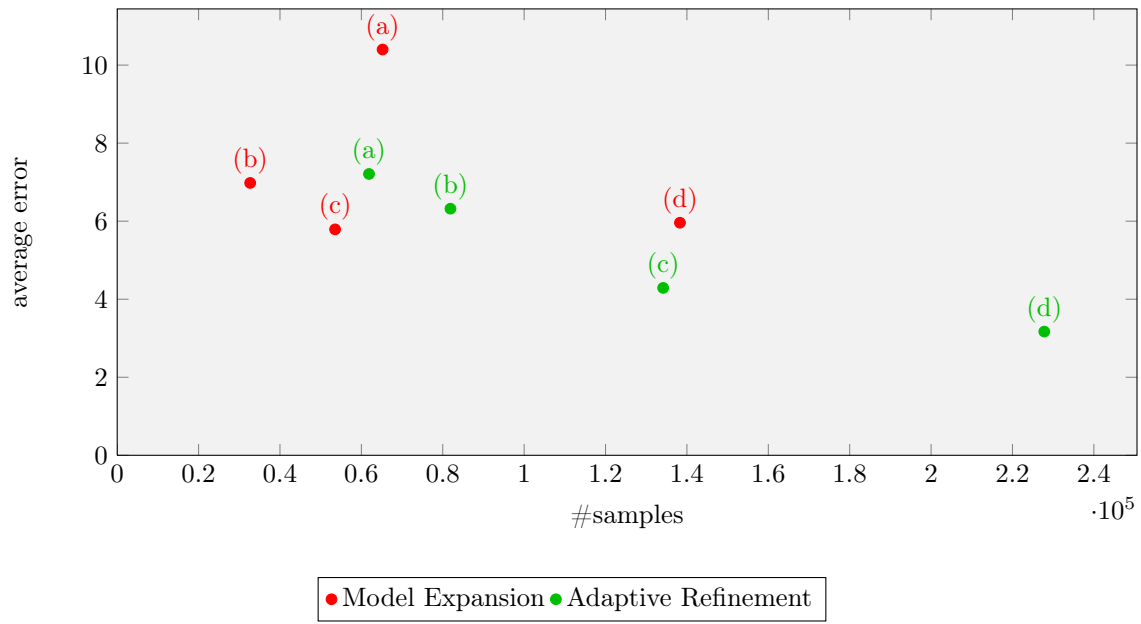
\begin{figure}[t]
    \centering
    \begin{tikzpicture}
        \begin{axis}[
            width=\textwidth,
            height=.5\textwidth,
            xlabel={\#samples},
            ylabel={average error},
            legend to name=fig:modeling.res.ticks.comp.smpacc:legend,
            legend columns=-1,
            nodes near coords
        ]
            \addplot[color=plot1, only marks, mark=*, point meta=explicit symbolic] coordinates {
              (65220, 10.4) [(a)]
              (32680, 6.98) [(b)]
              (53550, 5.79) [(c)]
              (138290, 5.96) [(d)]
            };
            \addlegendentry{Model Expansion}

            \addplot[color=plot2, only marks, mark=*, point meta=explicit symbolic] coordinates {
              (61880, 7.21) [(a)]
              (81890, 6.32) [(b)]
              (134160, 4.29) [(c)]
              (227820, 3.17) [(d)]
            };
            \addlegendentry{Adaptive Refinement}
        \end{axis}
    \end{tikzpicture}

    \vspace{.5cm}

    \ref*{fig:modeling.res.ticks.comp.smpacc:legend}
    \caption{Comparison of modeling methods for \metric{ticks}.}
    \label{fig:modeling.res.ticks.comp.smpacc}
    \tikzset{external/export=false}
\end{figure}

\autoref{tab:modeling.res.ticks.comp} reports the number of needed samples and the attained average error for the \metric{ticks} models presented in the previous sections.
\autoref{fig:modeling.res.ticks.comp.smpacc} visually compares the number of samples and accuracy of these models; we are interested in models that are as accurate as possible from a minimal number of samples.

For fewer samples, Model Expansion generates more accurate models (\textcolor{plot1}{(b)} and \textcolor{plot1}{(d)}).
However, we have to keep in mind that these models were not representing the fine scale behavior of \metric{ticks} very well.
When we are willing to use more samples, Adaptive Refinement generates more accurate models (\textcolor{plot2}{(c)}).
With huge amount of samples, this PModeler can generate very accurate models (\textcolor{plot2}{(d)}).

For the models used to predict the performance of blocked algorithms, we choose Adaptive Refinement with configuration \textcolor{plot2}{(c)}: $10\%$ error bound and $32 \times 32$ minimum regions size.
This configuration is a good compromise between the model accuracy and the number of samples.

    \chapter{Prediction and Ranking}
    \label{sec:ranking}
    We now have all necessary tools ready to tackle our main goal: ranking blocked algorithms based on performance predictions.
To predict the performance of a blocked algorithm, we analyze its sequence of subroutine invocations.
We use performance models generated by the Modeler to estimate the performance counters for these invocations.
These estimates are then accumulated, resulting in the prediction of the algorithm's performance.
The probabilistic nature of the performance model allows us to give detailed information on the expected ranges of the algorithm's performance.

Like the Modeler, our prediction tool is written in Python.
The automatically generated performance models can therefore be used directly.
The structure of the blocked algorithms, on the other hand, needs to be represented in a format that we can interpret within Python.

After we describe the blocked algorithm representation in \autoref{sec:ranking.algs}, we apply our performance prediction and ranking to three operations:
\begin{itemize}
    \item Inversion of a triangular matrix, $L \leftarrow L^{-1}$ (\autoref{sec:ranking.trinv});
    \item LU decomposition, $L U \leftarrow A$ (\autoref{sec:ranking.lu});
    \item Solution of the Sylvester Equation $L X + X U = C$ for $X$ (\autoref{sec:ranking.sylv}).
\end{itemize}

        \section{Representation of Blocked Algorithms}
        \label{sec:ranking.algs}
        To predict the performance of a certain blocked algorithm for a given set of arguments, we need a list of all subroutine invocations.
In this list, each invocation is represented by a tuple consisting of the invoked routine's name and its arguments.
To obtain performance estimates, these tuples can be passed directly to the performance models.
The list of invocations is generated by a Python function that mimics the execution of the blocked algorithm.

\begin{example}
We consider the blocked algorithm variant 1 for the inversion of a triangular matrix $L \leftarrow L^{-1}$ (\autoref{sec:intro.blockedalgs.trinv}).
The algorithm traverses $L$ from the top left corner
\tikz[baseline=(A.base)] {
    \filldraw[fill=graybg] (0, 0) rectangle (1, 1);
    \draw[->, dotted] (.1, .9) -- (.9, .1);
    \node at (.5, .5) (A) {$L$};
}
and performs the following update statements:
\begin{center}
    \begin{tabular}{|>{\cellcolor{graybg}}p{.25\textwidth}|}
        \hline
        \multicolumn{1}{|>{\cellcolor{graybg}}c|}{Variant 1} \\
        \hline
        $L_{10} \leftarrow L_{10} L_{00}$ \\
        $L_{10} \leftarrow -L_{11}^{-1} L_{10}$ \\
        $L_{11} \leftarrow L_{11}^{-1}$ \\
        \hline
    \end{tabular}
\end{center}

\begin{lstlisting}[
    caption={Inversion of a triangular matrix --- variant 1.},
    label=lst:ranking.algs.trinv1,
    float=t,
    captionpos=b
    ]
int trinv1_(char* diag, int* n, double* A, int* ldA, int* blocksize) {
	if (*n == 1) {
		if (diag[0] == 'N')
			*A = 1 / *A;
		return 0;
	}

	int ione = 1;
   	double one = 1;
	double minusone = -1;

    int p = 0;
    while (p < *n) {
 	    int b = *blocksize;
 	    if (p + b > *n)
		    b = *n - p;

#define A00 (A)
#define A10 (A + p)
#define A11 (A + *ldA * p + p)

	    dtrmm_("R", "L", "N", diag, &b, &p, &one, A00, ldA, A10, ldA);
	    dtrsm_("L", "L", "N", diag, &b, &p, &minusone, A11, ldA, A10, ldA);
		trinv1_(diag, &b, A11, ldA, &ione);

	    p += b;
    }
	return 0;
}
\end{lstlisting}

An implementation of this algorithm in C is given in \autoref{lst:ranking.algs.trinv1}.
The update $L_{11} \leftarrow L_{11}^{-1}$ is mapped to a recursive routine invocation with block-size 1.
This leads to the following recursion levels:
\begin{itemize}
    \item The main algorithm invocation performs BLAS Level-3 operations on submatrices.
    \item At the first level of recursion, the argument $\mathtt{blocksize} = \mathtt 1$ leads to matrix-vector operation.
        These are performed by the same BLAS Level-3 operations with one of the size parameters equal to 1.
    \item on the second recursion level, we have $n = 1$ --- the input matrix \texttt A consists of a single scalar.
        The inversion of this scalar is handled at the very top of the routine: When $\mathtt{diag} = \mathtt N$, its reciprocal is computed.
\end{itemize}

\begin{table}[t]
    \centering
    \begin{tabular}{lll}
        \toprule
        update &subroutine invocation\\
        \midrule
        $L_{10} \leftarrow L_{10} L_{00}$
        &$(\mathtt{dtrmm},  \mathtt R, \mathtt L, \mathtt N, \mathtt N, 100, 0,   \mathtt v1,  \cdot, 300, \cdot, 300)$ \\
        $L_{10} \leftarrow -L_{11}^{-1} L_{10}$
        &$(\mathtt{dtrsm},  \mathtt L, \mathtt L, \mathtt N, \mathtt N, 100, 0,   \mathtt v-1, \cdot, 300, \cdot, 300)$ \\
        $L_{11} \leftarrow L_{11}^{-1}$
        &$(\mathtt{trinv1}, \mathtt N, 100, \cdot, 300, 1)                                                            $ \\
        $L_{10} \leftarrow L_{10} L_{00}$
        &$(\mathtt{dtrmm},  \mathtt R, \mathtt L, \mathtt N, \mathtt N, 100, 100, \mathtt v1,  \cdot, 300, \cdot, 300)$ \\
        $L_{10} \leftarrow -L_{11}^{-1} L_{10}$
        &$(\mathtt{dtrsm},  \mathtt L, \mathtt L, \mathtt N, \mathtt N, 100, 100, \mathtt v-1, \cdot, 300, \cdot, 300)$ \\
        $L_{11} \leftarrow L_{11}^{-1}$
        &$(\mathtt{trinv1}, \mathtt N, 100, \cdot, 300, 1)                                                            $ \\
        $L_{10} \leftarrow L_{10} L_{00}$ 
        &$(\mathtt{dtrmm},  \mathtt R, \mathtt L, \mathtt N, \mathtt N, 100, 200, \mathtt v1,  \cdot, 300, \cdot, 300)$ \\
        $L_{10} \leftarrow -L_{11}^{-1} L_{10}$
        &$(\mathtt{dtrsm},  \mathtt L, \mathtt L, \mathtt N, \mathtt N, 100, 200, \mathtt v-1, \cdot, 300, \cdot, 300)$ \\
        $L_{11} \leftarrow L_{11}^{-1}$
        &$(\mathtt{trinv1}, \mathtt N, 100, \cdot, 300, 1)                                                            $ \\
        \bottomrule
    \end{tabular}
    \caption{Subroutine invocation of \texttt{dtrsm(N, 300, \textit A, 300, 100)}.}
    \label{tbl:ranking.algs.trinv}
\end{table}

When the algorithm is executed with the arguments
\begin{center}
    \texttt{%
        trinv1(%
        \overparameq{diag}{N},
        \overparameq{n}{300},
        \overparameq{A}{\textit A},
        \overparameq{ldA}{300},
        \overparameq{blocksize}{100})%
    },
\end{center}
that is, for a matrix of size $300 \times 300$ and a block-size of $100$, we obtain the routine invocations listed in \autoref{tbl:ranking.algs.trinv}.
The matrix arguments do not need to be specified: Their size can be computed from the size and leading dimension arguments and the matrix entries are irrelevant to performance.
Within Python, the invocation list is represented as follows:

\begin{minipage}{\textwidth}
    \begin{lstlisting}[language=Python,aboveskip=0cm,belowskip=0cm]
[['dtrmm', 'R', 'L', 'N', 'N', 100, 0, 'v1', None, 300, None, 300],
['dtrsm', 'L', 'L', 'N', 'N', 100, 0, 'v-1', None, 300, None, 300],
['trinv1', 'N', 100, None, 300, 1], 
['dtrmm', 'R', 'L', 'N', 'N', 100, 100, 'v1', None, 300, None, 300],
['dtrsm', 'L', 'L', 'N', 'N', 100, 100, 'v-1', None, 300, None, 300],
['trinv1', 'N', 100, None, 300, 1],
['dtrmm', 'R', 'L', 'N', 'N', 100, 200, 'v1', None, 300, None, 300],
['dtrsm', 'L', 'L', 'N', 'N', 100, 200, 'v-1', None, 300, None, 300],
['trinv1', 'N', 100, None, 300, 1]]
    \end{lstlisting}
\end{minipage}
\hfill \qed
\end{example}

        \section{Triangular Inverse \texorpdfstring{$L \leftarrow L^{-1}$}{L <- inv(L)}}
        \label{sec:ranking.trinv}
        The first operation we consider is the inversion of a triangular matrix, $L \leftarrow L^{-1}$.
In \autoref{sec:intro.blockedalgs.trinv}, we presented four blocked algorithms that perform this operation with the following update statements:
\begin{center}
    \begin{tabular}{|>{\cellcolor{graybg}}p{.25\textwidth}|}
        \hline
        \multicolumn{1}{|>{\cellcolor{graybg}}c|}{Variant 1} \\
        \hline
        $L_{10} \leftarrow L_{10} L_{00}$ \\
        $L_{10} \leftarrow -L_{11}^{-1} L_{10}$ \\
        $L_{11} \leftarrow L_{11}^{-1}$ \\
        \hline
    \end{tabular}
    \hspace{.5cm}
    \begin{tabular}{|>{\cellcolor{graybg}}p{.25\textwidth}|}
        \hline
        \multicolumn{1}{|>{\cellcolor{graybg}}c|}{Variant 2} \\
        \hline
        $L_{21} \leftarrow L_{22}^{-1} L_{21}$ \\
        $L_{21} \leftarrow -L_{21} L_{11}^{-1}$ \\
        $L_{11} \leftarrow L_{11}^{-1}$ \\
        \hline
    \end{tabular}

    \vspace{.5cm}

    \begin{tabular}{|>{\cellcolor{graybg}}p{.25\textwidth}|}
        \hline
        \multicolumn{1}{|>{\cellcolor{graybg}}c|}{Variant 3} \\
        \hline
        $L_{21} \leftarrow -L_{21} L_{11}^{-1}$\\
        $L_{20} \leftarrow L_{21} L_{10} + L_{20}$\\
        $L_{10} \leftarrow L_{11}^{-1} L_{10}$\\
        $L_{11} \leftarrow L_{11}^{-1}$ \\
        \hline
    \end{tabular}
    \hspace{.5cm}
    \begin{tabular}{|>{\cellcolor{graybg}}p{.25\textwidth}|}
        \hline
        \multicolumn{1}{|>{\cellcolor{graybg}}c|}{Variant 4} \\
        \hline
        $L_{21} \leftarrow -L_{22}^{-1} L_{21}$ \\
        $L_{20} \leftarrow -L_{21} L_{10} + L_{20}$ \\
        $L_{10} \leftarrow L_{10} L_{00}$ \\
        $L_{11} \leftarrow L_{11}^{-1}$ \\
        \hline
    \end{tabular}
\end{center}
These algorithms' C implementations used for the following performance prediction is given in \appendixref{app:blockedalgs.trinv}.

\begin{figure}[t]
    \tikzset{external/export=true}
    \centering
    \subfloat[\metric{ticks}]{
        \label{fig:ranking.trinv.measured:ticks}
        \begin{tikzpicture}
            \begin{axis}[
                twocolplot,
                xlabel={\texttt{n}},
                ylabel={\metric{ticks}},
                xtick={0,256,...,1024},
                legend to name=fig:ranking.trinv.measured.legend,
                legend columns=-1
            ]
                \addlegendimage{plot1, only marks}
                \addlegendentry{variant 1}
                \addlegendimage{plot2, only marks}
                \addlegendentry{variant 2}
                \addlegendimage{plot3, only marks}
                \addlegendentry{variant 3}
                \addlegendimage{plot4, only marks}
                \addlegendentry{variant 4}

                \addplot[plot1, mark size=.2pt, only marks, restrict x to domain=0:1024] file {figures/data/ranking.trinv.measured/trinv1.time.dat};
                \addplot[plot2, mark size=.2pt, only marks, restrict x to domain=0:1024] file {figures/data/ranking.trinv.measured/trinv2.time.dat};
                \addplot[plot3, mark size=.2pt, only marks, restrict x to domain=0:1024] file {figures/data/ranking.trinv.measured/trinv3.time.dat};
                \addplot[plot4, mark size=.2pt, only marks, restrict x to domain=0:1024] file {figures/data/ranking.trinv.measured/trinv4.time.dat};
            \end{axis}
        \end{tikzpicture}
    }
    \hfill
    \subfloat[\metric{efficiency}]{
        \label{fig:ranking.trinv.measured:eff}
        \begin{tikzpicture}
            \begin{axis}[
                twocolplot,
                xlabel={\texttt{n}},
                ylabel={\metric{efficiency}},
                ymax=1,
                xtick={0,256,...,1024}
            ]
                \addplot[plot1, mark size=.2pt, only marks, restrict x to domain=0:1024] file {figures/data/ranking.trinv.measured/trinv1.eff.dat};
                \addplot[plot2, mark size=.2pt, only marks, restrict x to domain=0:1024] file {figures/data/ranking.trinv.measured/trinv2.eff.dat};
                \addplot[plot3, mark size=.2pt, only marks, restrict x to domain=0:1024] file {figures/data/ranking.trinv.measured/trinv3.eff.dat};
                \addplot[plot4, mark size=.2pt, only marks, restrict x to domain=0:1024] file {figures/data/ranking.trinv.measured/trinv4.eff.dat};
            \end{axis}
        \end{tikzpicture}
    }

    \vspace{.5cm} 

    \tikzset{external/export=false}
    \ref*{fig:ranking.trinv.measured.legend}
    \caption{\metric{ticks} and \metric{efficiency} for \texttt{trinv\textit i(N, \textit n, \textit A, \textit n, 96)}.}
    \label{fig:ranking.trinv.measured}
\end{figure}

In \autoref{fig:ranking.trinv.measured}, we present performance measurements of these algorithms with the arguments
\begin{center}
    \texttt{%
        trinv\textit i(%
        \overparameq{diag}{N},
        \overparameq{n}{\textit n},
        \overparameq{A}{\textit A},
        \overparameq{ldA}{\textit n},
        \overparameq{blocksize}{96})%
    },
\end{center}
varying the matrix size $\mathtt n \in \{8, 16, \ldots, 1024\}$, taking 10 measurements at each point.
We show the metrics \metric{ticks} and \metric{efficiency}, where
$$
    \metric{efficiency} = \frac{\frac16 \mathtt n^3 + \frac12 \mathtt n^2 + \frac13 \mathtt n}{2 \cdot \metric{ticks}}.
$$

We now turn to the prediction of these measurements \metric{ticks}, for now, considering the median of our models.
For our first performance estimate, we use performance models for \texttt{dtrsm}, \texttt{dtrmm}, \texttt{dgemm}, and the unblocked versions of the blocked algorithms\footnote
    {The models for the unblocked versions are limited to values of $\texttt{size}$ up to $256$ --- large enough for the unblocked algorithm invocations.
}.
These are generated with a cache-trashing Sampler and the Modeler configuration selected in \autoref{sec:modeling.res.ticks.comp}
For each algorithm execution, we generate consider the list of subroutine invocations, containing calls to BLAS and the algorithms' unblocked versions.
We evaluate the performance models for these routine invocations and accumulate the obtained estimates, resulting in our performance prediction.

\begin{figure}[t]
    \tikzset{external/export=true}
    \centering
    \subfloat[cache-trashing]{
        \label{fig:ranking.trinv.estimates:trashing}
        \begin{tikzpicture}
            \begin{axis}[
                twocolplot,
                xlabel={\texttt{n}},
                ylabel={\metric{efficiency}},
                xtick={0,256,...,1024},
                ymax=1,
                legend to name=fig:ranking.trinv.estimates.legend,
                legend columns=4
            ]
                \addplot[plot1, restrict x to domain=0:1024] file {figures/data/ranking.trinv.estimates/trinv1.eff.dat};
                \label{fig:ranking.trinv.estimates:var1}
                \addlegendentry{variant 1\vphantom{gt/}}

                \addplot[plot2, restrict x to domain=0:1024] file {figures/data/ranking.trinv.estimates/trinv2.eff.dat};
                \label{fig:ranking.trinv.estimates:var2}
                \addlegendentry{variant 2\vphantom{gt/}}
                
                \addplot[plot3, restrict x to domain=0:1024] file {figures/data/ranking.trinv.estimates/trinv3.eff.dat};
                \label{fig:ranking.trinv.estimates:var3}
                \addlegendentry{variant 3\vphantom{gt/}}
                
                \addplot[plot4, restrict x to domain=0:1024] file {figures/data/ranking.trinv.estimates/trinv4.eff.dat};
                \label{fig:ranking.trinv.estimates:var4}
                \addlegendentry{variant 4\vphantom{gt/}}

                \addlegendimage{empty legend}
                \addlegendentry{prediction:\vphantom{gt/}}

                \addlegendimage{gray}
                \label{fig:ranking.trinv.estimates:median}
                \addlegendentry{median\vphantom{gt/}}

                \addlegendimage{draw=none, fill=gray, opacity=.25, area legend}
                \label{fig:ranking.trinv.estimates:minmax}
                \addlegendentry{min/max\vphantom{gt/}}

                \addlegendimage{gray, dashed}
                \label{fig:ranking.trinv.estimates:avg}
                \addlegendentry{average\vphantom{gt/}}

                \addplot[plot1, mark size=.2pt, only marks, restrict x to domain=0:1024] file {figures/data/ranking.trinv.measured/trinv1.eff.dat};
                \addplot[plot2, mark size=.2pt, only marks, restrict x to domain=0:1024] file {figures/data/ranking.trinv.measured/trinv2.eff.dat};
                \addplot[plot3, mark size=.2pt, only marks, restrict x to domain=0:1024] file {figures/data/ranking.trinv.measured/trinv3.eff.dat};
                \addplot[plot4, mark size=.2pt, only marks, restrict x to domain=0:1024] file {figures/data/ranking.trinv.measured/trinv4.eff.dat};
            \end{axis}
        \end{tikzpicture}
    }
    \hfill
    \subfloat[In-cache model]{
        \label{fig:ranking.trinv.estimates:ic}
        \begin{tikzpicture}
            \begin{axis}[
                twocolplot,
                xlabel={\texttt{n}},
                ylabel={\metric{efficiency}},
                ymax=1,
                xtick={0,256,...,1024},
            ]
                \addplot[plot1, restrict x to domain=0:1024] file {figures/data/ranking.trinv.estimates/trinv1.eff.ic.dat};
                \addplot[plot2, restrict x to domain=0:1024] file {figures/data/ranking.trinv.estimates/trinv2.eff.ic.dat};
                \addplot[plot3, restrict x to domain=0:1024] file {figures/data/ranking.trinv.estimates/trinv3.eff.ic.dat};
                \addplot[plot4, restrict x to domain=0:1024] file {figures/data/ranking.trinv.estimates/trinv4.eff.ic.dat};
                \addplot[plot1, mark size=.2pt, only marks, restrict x to domain=0:1024] file {figures/data/ranking.trinv.measured/trinv1.eff.dat};
                \addplot[plot2, mark size=.2pt, only marks, restrict x to domain=0:1024] file {figures/data/ranking.trinv.measured/trinv2.eff.dat};
                \addplot[plot3, mark size=.2pt, only marks, restrict x to domain=0:1024] file {figures/data/ranking.trinv.measured/trinv3.eff.dat};
                \addplot[plot4, mark size=.2pt, only marks, restrict x to domain=0:1024] file {figures/data/ranking.trinv.measured/trinv4.eff.dat};
            \end{axis}
        \end{tikzpicture}
    }

    \subfloat[In-cache model zoomed]{
        \label{fig:ranking.trinv.estimates:iczoom}
        \begin{tikzpicture}
            \begin{axis}[
                twocolplot,
                xlabel={\texttt{n}},
                ylabel={\metric{efficiency}},
                xmin=512,
                xtick={512,640,...,1024},
                ymin=.5,
                ymax=.8,
            ]
                \addplot[plot1, restrict x to domain=0:1024] file {figures/data/ranking.trinv.estimates/trinv1.eff.ic.dat};
                \addplot[plot2, restrict x to domain=0:1024] file {figures/data/ranking.trinv.estimates/trinv2.eff.ic.dat};
                \addplot[plot3, restrict x to domain=0:1024] file {figures/data/ranking.trinv.estimates/trinv3.eff.ic.dat};
                \addplot[plot4, restrict x to domain=0:1024] file {figures/data/ranking.trinv.estimates/trinv4.eff.ic.dat};
                \addplot[plot1, mark size=.2pt, only marks, restrict x to domain=0:1024] file {figures/data/ranking.trinv.measured/trinv1.eff.dat};
                \addplot[plot2, mark size=.2pt, only marks, restrict x to domain=0:1024] file {figures/data/ranking.trinv.measured/trinv2.eff.dat};
                \addplot[plot3, mark size=.2pt, only marks, restrict x to domain=0:1024] file {figures/data/ranking.trinv.measured/trinv3.eff.dat};
                \addplot[plot4, mark size=.2pt, only marks, restrict x to domain=0:1024] file {figures/data/ranking.trinv.measured/trinv4.eff.dat};
            \end{axis}
        \end{tikzpicture}
    }
    \hfill
    \subfloat[In-cache model statistical prediction]{
        \label{fig:ranking.trinv.estimates:ranges}
        \begin{tikzpicture}
            \begin{axis}[
                twocolplot,
                xlabel={\texttt{n}},
                ylabel={\metric{efficiency}},
                xmin=512,
                xtick={512,640,...,1024},
                ymin=.5,
                ymax=.8
            ]
                \addplot[opacity=.25, draw=none, fill=plot1, restrict x to domain=0:1024] file {figures/data/ranking.trinv.estimates/trinv1.eff.ranges.dat} \closedcycle;
                \addplot[opacity=.25, draw=none, fill=plot2, restrict x to domain=0:1024] file {figures/data/ranking.trinv.estimates/trinv2.eff.ranges.dat} \closedcycle;
                \addplot[opacity=.25, draw=none, fill=plot3, restrict x to domain=0:1024] file {figures/data/ranking.trinv.estimates/trinv3.eff.ranges.dat} \closedcycle;
                \addplot[opacity=.25, draw=none, fill=plot4, restrict x to domain=0:1024] file {figures/data/ranking.trinv.estimates/trinv4.eff.ranges.dat} \closedcycle;
                \addplot[plot1, restrict x to domain=0:1024] file {figures/data/ranking.trinv.estimates/trinv1.eff.ic.dat};
                \addplot[plot2, restrict x to domain=0:1024] file {figures/data/ranking.trinv.estimates/trinv2.eff.ic.dat};
                \addplot[plot3, restrict x to domain=0:1024] file {figures/data/ranking.trinv.estimates/trinv3.eff.ic.dat};
                \addplot[plot4, restrict x to domain=0:1024] file {figures/data/ranking.trinv.estimates/trinv4.eff.ic.dat};
                \addplot[plot1, dashed, restrict x to domain=0:1024] file {figures/data/ranking.trinv.estimates/trinv1.eff.avg.dat};
                \addplot[plot2, dashed, restrict x to domain=0:1024] file {figures/data/ranking.trinv.estimates/trinv2.eff.avg.dat};
                \addplot[plot3, dashed, restrict x to domain=0:1024] file {figures/data/ranking.trinv.estimates/trinv3.eff.avg.dat};
                \addplot[plot4, dashed, restrict x to domain=0:1024] file {figures/data/ranking.trinv.estimates/trinv4.eff.avg.dat};
                \addplot[plot1, mark size=.2pt, only marks, restrict x to domain=0:1024] file {figures/data/ranking.trinv.measured/trinv1.eff.dat};
                \addplot[plot2, mark size=.2pt, only marks, restrict x to domain=0:1024] file {figures/data/ranking.trinv.measured/trinv2.eff.dat};
                \addplot[plot3, mark size=.2pt, only marks, restrict x to domain=0:1024] file {figures/data/ranking.trinv.measured/trinv3.eff.dat};
                \addplot[plot4, mark size=.2pt, only marks, restrict x to domain=0:1024] file {figures/data/ranking.trinv.measured/trinv4.eff.dat};
            \end{axis}
        \end{tikzpicture}
    }

    \vspace{.5cm} 

    \tikzset{external/export=false}
    \ref*{fig:ranking.trinv.estimates.legend}
    \caption{\metric{efficiency} predictions \texttt{trinv\textit i(N, \textit n, \textit A, \textit n, 96)}.}
    \label{fig:ranking.trinv.estimates}
\end{figure}

Since the \metric{efficiency} graphs for the four algorithms (\autoref{fig:ranking.trinv.measured:eff}) give a better impression of their performance, we present the \metric{efficiency} computed from our obtained \metric{ticks} estimates in \autoref{fig:ranking.trinv.estimates:trashing}.
These predictions are already accurate enough to determine that variants 4~(\ref{fig:ranking.trinv.estimates:var4}) and 1~(\ref{fig:ranking.trinv.estimates:var1}) are less efficient than variants 2 (\ref{fig:ranking.trinv.estimates:var2}) and 3~(\ref{fig:ranking.trinv.estimates:var3}).
However, in our predictions, variant 2~(\ref{fig:ranking.trinv.estimates:var2}) erroneously becomes more efficient than variant 3~(\ref{fig:ranking.trinv.estimates:var3}) for increasing $n$; this is \emph{not} the case in the measurements.
Furthermore, all \metric{efficiency} predictions are too low; This results from overestimating \metric{ticks}.
This overestimation is due to the memory locality situations used during the generation of our performance models; the matrix arguments were placed in main memory.

For our second set of predictions, we use a performance model generated with an in-cache Sampler configuration.
The results are shown in \autoref{fig:ranking.trinv.estimates:ic}; a zoom-in on the upper right part of the plot ($\mathtt n \geq 512$ and $0.5 \leq \metric{efficiency} \leq 0.8$) is given in \autoref{fig:ranking.trinv.estimates:iczoom}.
Compared to our previous predictions, these results show a significantly closer to the measurements.
Furthermore, all algorithm variants are now ranked correctly.

Up to this point, we presented estimates for the median of the performance counter.
In \autoref{fig:ranking.trinv.estimates:ranges} we additionally take the quantities average, minimum, and maximum into account.
The range between minimum and maximum (\ref{fig:ranking.trinv.estimates:minmax}) covers almost all measurements and gives a good idea on the expected results; however, they are very broad --- they overlap and even include each other.
The average (\ref{fig:ranking.trinv.estimates:avg}) is closer to the measured algorithm performance than the previously used median.
Relying on the average predictions, however, is dangerous, since they are obtained for models generated with an error bound on the median and are influenced by outliers.

\paragraph{Block-size.}
So far we predicted performance fore varying matrix sizes with highly satisfactory results.
We now turn to our second aspect of interest: determining the most efficient block-size.

\begin{figure}[t]
    \tikzset{external/export=true}
    \centering
    \subfloat[\metric{ticks}]{
        \label{fig:ranking.trinv.esimates:time}
        \begin{tikzpicture}
            \begin{axis}[
                twocolplot,
                xlabel={\texttt{n}},
                ylabel={\metric{ticks}},
                ymax=.5e9,
                xtick={0,64,...,256},
                legend to name=fig:ranking.trinv.b.legend,
                legend columns=4
            ]
                \addlegendimage{plot1}
                \label{fig:ranking.trinv.b:var1}
                \addlegendentry{variant 1\vphantom{gt/}}

                \addlegendimage{plot2}
                \label{fig:ranking.trinv.b:var2}
                \addlegendentry{variant 2\vphantom{gt/}}

                \addlegendimage{plot3}
                \label{fig:ranking.trinv.b:var3}
                \addlegendentry{variant 3\vphantom{gt/}}

                \addlegendimage{plot4}
                \label{fig:ranking.trinv.b:var4}
                \addlegendentry{variant 4\vphantom{gt/}}

                \addlegendimage{empty legend}
                \addlegendentry{prediction:\vphantom{gt/}}

                \addlegendimage{gray}
                \label{fig:ranking.trinv.b:median}
                \addlegendentry{median\vphantom{gt/}}

                \addlegendimage{draw=none, fill=gray, opacity=.25, area legend}
                \label{fig:ranking.trinv.b:minmax}
                \addlegendentry{min/max\vphantom{gt/}}

                \addlegendimage{gray, dashed}
                \label{fig:ranking.trinv.b:avg}
                \addlegendentry{average\vphantom{gt/}}

                \addplot[opacity=.25, draw=none, fill=plot1, restrict x to domain=0:256] file {figures/data/ranking.trinv.b/trinv1.time.minmax.dat} \closedcycle;
                \addplot[opacity=.25, draw=none, fill=plot2, restrict x to domain=0:256] file {figures/data/ranking.trinv.b/trinv2.time.minmax.dat} \closedcycle;
                \addplot[opacity=.25, draw=none, fill=plot3, restrict x to domain=0:256] file {figures/data/ranking.trinv.b/trinv3.time.minmax.dat} \closedcycle;
                \addplot[opacity=.25, draw=none, fill=plot4, restrict x to domain=0:256] file {figures/data/ranking.trinv.b/trinv4.time.minmax.dat} \closedcycle;
                \addplot[plot1, restrict x to domain=0:256] file {figures/data/ranking.trinv.b/trinv1.time.med.dat};
                \addplot[plot2, restrict x to domain=0:256] file {figures/data/ranking.trinv.b/trinv2.time.med.dat};
                \addplot[plot3, restrict x to domain=0:256] file {figures/data/ranking.trinv.b/trinv3.time.med.dat};
                \addplot[plot4, restrict x to domain=0:256] file {figures/data/ranking.trinv.b/trinv4.time.med.dat};
                \addplot[plot1, dashed, restrict x to domain=0:256] file {figures/data/ranking.trinv.b/trinv1.time.avg.dat};
                \addplot[plot2, dashed, restrict x to domain=0:256] file {figures/data/ranking.trinv.b/trinv2.time.avg.dat};
                \addplot[plot3, dashed, restrict x to domain=0:256] file {figures/data/ranking.trinv.b/trinv3.time.avg.dat};
                \addplot[plot4, dashed, restrict x to domain=0:256] file {figures/data/ranking.trinv.b/trinv4.time.avg.dat};
                \addplot[plot1, mark size=.2pt, only marks, restrict x to domain=0:256] file {figures/data/ranking.trinv.b/trinv1.time.meas.dat};
                \addplot[plot2, mark size=.2pt, only marks, restrict x to domain=0:256] file {figures/data/ranking.trinv.b/trinv2.time.meas.dat};
                \addplot[plot3, mark size=.2pt, only marks, restrict x to domain=0:256] file {figures/data/ranking.trinv.b/trinv3.time.meas.dat};
                \addplot[plot4, mark size=.2pt, only marks, restrict x to domain=0:256] file {figures/data/ranking.trinv.b/trinv4.time.meas.dat};
            \end{axis}
        \end{tikzpicture}
    }
    \hfill
    \subfloat[\metric{efficiency}]{
        \label{fig:ranking.trinv.esimates:eff}
        \begin{tikzpicture}
            \begin{axis}[
                twocolplot,
                xlabel={\texttt{n}},
                ylabel={\metric{efficiency}},
                ymin=0,
                ymax=1,
                xtick={0,64,...,256},
            ]
                \addplot[opacity=.25, draw=none, fill=plot1, restrict x to domain=0:256] file {figures/data/ranking.trinv.b/trinv1.eff.minmax.dat} \closedcycle;
                \addplot[opacity=.25, draw=none, fill=plot2, restrict x to domain=0:256] file {figures/data/ranking.trinv.b/trinv2.eff.minmax.dat} \closedcycle;
                \addplot[opacity=.25, draw=none, fill=plot3, restrict x to domain=0:256] file {figures/data/ranking.trinv.b/trinv3.eff.minmax.dat} \closedcycle;
                \addplot[opacity=.25, draw=none, fill=plot4, restrict x to domain=0:256] file {figures/data/ranking.trinv.b/trinv4.eff.minmax.dat} \closedcycle;
                \addplot[plot1, restrict x to domain=0:256] file {figures/data/ranking.trinv.b/trinv1.eff.med.dat};
                \addplot[plot2, restrict x to domain=0:256] file {figures/data/ranking.trinv.b/trinv2.eff.med.dat};
                \addplot[plot3, restrict x to domain=0:256] file {figures/data/ranking.trinv.b/trinv3.eff.med.dat};
                \addplot[plot4, restrict x to domain=0:256] file {figures/data/ranking.trinv.b/trinv4.eff.med.dat};
                \addplot[plot1, dashed, restrict x to domain=0:256] file {figures/data/ranking.trinv.b/trinv1.eff.avg.dat};
                \addplot[plot2, dashed, restrict x to domain=0:256] file {figures/data/ranking.trinv.b/trinv2.eff.avg.dat};
                \addplot[plot3, dashed, restrict x to domain=0:256] file {figures/data/ranking.trinv.b/trinv3.eff.avg.dat};
                \addplot[plot4, dashed, restrict x to domain=0:256] file {figures/data/ranking.trinv.b/trinv4.eff.avg.dat};
                \addplot[plot1, mark size=.2pt, only marks, restrict x to domain=0:256] file {figures/data/ranking.trinv.b/trinv1.eff.meas.dat};
                \addplot[plot2, mark size=.2pt, only marks, restrict x to domain=0:256] file {figures/data/ranking.trinv.b/trinv2.eff.meas.dat};
                \addplot[plot3, mark size=.2pt, only marks, restrict x to domain=0:256] file {figures/data/ranking.trinv.b/trinv3.eff.meas.dat};
                \addplot[plot4, mark size=.2pt, only marks, restrict x to domain=0:256] file {figures/data/ranking.trinv.b/trinv4.eff.meas.dat};
            \end{axis}
        \end{tikzpicture}
    }

    \vspace{.5cm} 

    \tikzset{external/export=false}
    \ref*{fig:ranking.trinv.b.legend}
    \caption{\metric{efficiency} predictions \texttt{trinv\textit i(N, 1016, \textit A, 1016, \textit{blocksize})}.}
    \label{fig:ranking.trinv.b}
\end{figure}

For this purpose, we now fix the matrix size to\footnote{
    We chose $1016$, since we observed outliers for the algorithms' performance measurements at $1024$ which we are not able to explain.
} $\mathtt n = 1016$ and vary the block-size:
\begin{center}
    \texttt{%
        trinv\textit i(%
        \overparameq{diag}{N},
        \overparameq{n}{1016},
        \overparameq{A}{\textit A},
        \overparameq{ldA}{1016},
        \overparameq{blocksize}{\textit{blocksize}})%
    }.
\end{center}
The resulting predictions (\autoref{fig:ranking.trinv.b}) represent the measurements very well for the most efficient block-sizes between $48$ and $128$.
Variant~3~(\ref{fig:ranking.trinv.b:var3}) --- the fastest --- attains its highest performance for a block-size of 64 both in our prediction and the measurements.

The quality of our prediction decreases for very large and very small block-sizes.
\begin{description}
    \item[Small block-sizes.] The algorithms invoke subroutines very often with long and skinny matrices.
        For such matrices, BLAS routines usually do not attain high performance; this is carried forward to the blocked algorithms.
        Additionally, our models are less accurate for small argument values, further decreasing the prediction accuracy.
    \item[Large block-size.] The unblocked versions of the algorithm increasingly contribute to the overall performance.
        Since these unblocked versions are less efficient, the overall performance decreases; \metric{ticks} are overestimated only slightly worse than for block-sizes around $100$.
\end{description}
For our goal --- determining the fastest algorithm configuration --- the low accuracy in regions with low performance is not a major problem.

In the predictions of the maximum and average \metric{ticks} for variant~4~(\ref{fig:ranking.trinv.b:var4}), we further observe severe inaccuracies for block-sizes between $8$ and $160$.
These are due to outliers in the measurements that the corresponding models were constructed upon.
These outliers do not carry forward to the minimum and the median of \metric{ticks}.

        \section{LU Decomposition \texorpdfstring{$L U \leftarrow A$}{L U <- A}}
        \label{sec:ranking.lu}
        The second operation we study is the LU decomposition $L U = A$ of a square matrix $A$.
It is very commonly used to solve linear systems $A X = B$: With $L U = A$, we can write this as $L U X = B$ and apply \texttt{dtrsm} (solution of a triangular system) twice to obtain $X = U^{-1} L^{-1} B$.

There are five blocked algorithms for the LU decomposition.
They take $A$ as an input and compute $L$ and $U$ in place: The unit-diagonal lower triangular $L$ is stored in the strictly lower triangular part of $A$, whereas the non-unit-diagonal $U$ is stored in the upper triangular part of $A$.
Within the scheme of the blocked algorithms (\autoref{sec:intro.blockedalgs}), the five algorithmic variants perform the following updates:
\newcommand{\tril}{\tikz \draw (0, 0) -- (.25, 0) -- (0, .25) -- cycle;}
\newcommand{\triu}{\tikz \draw (.25, .25) -- (.25, 0) -- (0, .25) -- cycle;}
\begin{center}
    \begin{tabular}{|>{\cellcolor{graybg}}p{.25\textwidth}|}
        \hline
        \multicolumn{1}{|>{\cellcolor{graybg}}c|}{Variant 1} \\
        \hline
        $A_{01} \leftarrow (\tril A_{00})^{-1} A_{01}$ \\
        $A_{10} \leftarrow A_{10} (\triu A_{00})^{-1}$ \\
        $A_{11} \leftarrow A_{11} - A_{10} A_{01}$ \\
        $A_{11} \leftarrow LU(A_{11})$ \\
        \hline
    \end{tabular}
    \hspace{.5cm}                                      
    \begin{tabular}{|>{\cellcolor{graybg}}p{.25\textwidth}|}
        \hline
        \multicolumn{1}{|>{\cellcolor{graybg}}c|}{Variant 2} \\
        \hline
        $A_{10} \leftarrow A_{10} (\triu A_{00})^{-1}$ \\
        $A_{11} \leftarrow A_{11} - A_{10} A_{01}$ \\
        $A_{11} \leftarrow LU(A_{11})$ \\
        $A_{12} \leftarrow A_{12} - A_{10} A_{02}$ \\
        $A_{12} \leftarrow (\tril A_{11})^{-1} A_{12}$ \\
        \hline
    \end{tabular}
    \hspace{.5cm}                                      
    \begin{tabular}{|>{\cellcolor{graybg}}p{.25\textwidth}|}
        \hline
        \multicolumn{1}{|>{\cellcolor{graybg}}c|}{Variant 3} \\
        \hline
        $A_{01} \leftarrow (\tril A_{00})^{-1} A_{01}$ \\
        $A_{11} \leftarrow A_{11} - A_{10} A_{01}$ \\
        $A_{11} \leftarrow LU(A_{11})$ \\
        $A_{21} \leftarrow A_{21} - A_{20} A_{01}$ \\
        $A_{21} \leftarrow A_{21} (\triu A_{11})^{-1}$ \\
        \hline
    \end{tabular}
    
    \vspace{.5cm}

    \begin{tabular}{|>{\cellcolor{graybg}}p{.25\textwidth}|}
        \hline
        \multicolumn{1}{|>{\cellcolor{graybg}}c|}{Variant 4} \\
        \hline
        $A_{11} \leftarrow A_{11} - A_{10} A_{01}$ \\
        $A_{11} \leftarrow LU(A_{11})$ \\
        $A_{12} \leftarrow A_{12} - A_{10} A_{02}$ \\
        $A_{12} \leftarrow (\tril A_{11})^{-1} A_{12}$ \\
        $A_{21} \leftarrow A_{21} - A_{20} A_{01}$ \\
        $A_{21} \leftarrow A_{21} (\triu A_{11})^{-1}$ \\
        \hline
    \end{tabular}
    \hspace{.5cm}                                      
    \begin{tabular}{|>{\cellcolor{graybg}}p{.25\textwidth}|}
        \hline
        \multicolumn{1}{|>{\cellcolor{graybg}}c|}{Variant 5} \\
        \hline
        $A_{11} \leftarrow LU(A_{11})$ \\
        $A_{12} \leftarrow (\tril A_{11})^{-1} A_{12}$ \\
        $A_{21} \leftarrow A_{21} (\triu A_{11})^{-1}$ \\
        $A_{22} \leftarrow A_{22} - A_{21} A_{12}$ \\
        \hline
    \end{tabular}
\end{center}
Here, $\tril A$ and $\triu A$ denote the (strictly) lower and upper triangular part of $A$, respectively; $LU(A)$ denotes the LU decomposition itself --- an recursive application of the algorithm with block-size 1 to a submatrix.
Our C implementation of these algorithms is given in \appendixref{app:blockedalgs.lu}; their signatures are \texttt{lu\textit i(n, A, ldA, blocksize)} with $\texttt{\textit i} \in \{1, 2, 3, 4\}$.

\input{figures/ranking.lu.n}

We use the same method applied to $L \leftarrow L^{-1}$ in the previous section to estimate the performance of these algorithms with the arguments
\begin{center}
    \texttt{%
        lu\textit i(%
        \overparameq{n}{\textit n},
        \overparameq{A}{\textit A},
        \overparameq{ldA}{\textit n},
        \overparameq{blocksize}{96})%
    },
\end{center}
where $\mathtt n \in \{8, 16, \ldots, 1024\}$.
The results of these predictions and measurements of the corresponding algorithm executions are shown in \autoref{fig:ranking.lu.n}, where \metric{efficiency} is given by
$$
    \metric{efficiency} = \frac{\frac13 \mathtt n^3 + \frac12 \mathtt n^2 - \frac56 \mathtt n}{2 \metric{ticks}}.
$$
Since it is almost impossible to distinguish the performance of the five variants in the \metric{ticks} plot, we focus on \metric{efficiency} (Figures \ref{fig:ranking.lu.n:eff} and \ref{fig:ranking.lu.n:effzoom}).
We observe that for the most part our predictions are fairly close to the measurements.
Up to $\mathtt n \approx 900$, they can be used to rank the five algorithms correctly according to their performance.

However, the performance of variant~5~(\ref{fig:ranking.lu.n:var5}) --- the fastest --- is predicted with decreasing accuracy for larger values of \texttt n; the predictions are generally too low and diverge from our measurements.
This leads so far, that the algorithms are incorrectly ranked for problems larger than $n = 900$.
While the other variants are also slightly underestimated, their predictions are more stable and accurate.

        \section{Sylvester Equation: Solving \texorpdfstring{$L X + X U = C$ for $X$}{L X + X U = C for X}}
        \label{sec:ranking.sylv}
        We now study of a more complicated operation: the solution of the Sylvester equation.
This operation is encountered in control theory and is generally of the form $A X + X B = C$, where $A \in \mathbb R^{m \times m}$, $B \in \mathbb R^{n \times n}$, and $C \in \mathbb R^{m \times n}$ are given and $X \in \mathbb R^{m \times n}$ is to be computed.
We consider a special case of this equation, where $A$ and $B$ are lower and upper triangular, respectively: $L X + X U = C$.

With {\sc{Cl\makebox[.58\width][c]{1}ck}} \cite{diego1, diego2}, a tool for the automatic generation of blocked algorithms, we found 16 algorithmic variants for this operation.
Each of them takes three matrices $L$, $U$, and $X$; $X$ initially contains the input matrix $C$ and is overwritten with the solution $X$ by the algorithms.

The update statements for the 16 algorithms are given below.
Within these, $\Omega(L, U, X)$ denotes a recursive invocation of the Sylvester equation solver for a smaller matrix $X$.
The C implementation of the algorithms is given in \appendixref{app:blockedalgs.sylv}; their signature is \texttt{sylv\textit i(m, n, L, ldL, U, ldU, X, ldX, blocksize)} with $\texttt{\textit i} \in \{1, \ldots, 16\}$.
\begin{center}
    \newcommand{\sylv}{\Omega}
    \begin{tabular}{|>{\cellcolor{graybg}}p{.25\textwidth}|}
        \hline
        \multicolumn{1}{|>{\cellcolor{graybg}}c|}{Variant 1} \\
        \hline
         $X_{01} \leftarrow X_{01} - X_{00} U_{01}$ \\
         $X_{10} \leftarrow X_{10} - L_{10} X_{00}$ \\
         $X_{01} \leftarrow \sylv(L_{00}, U_{11}, X_{01})$ \\
         $X_{10} \leftarrow \sylv(L_{11}, U_{00}, X_{10})$ \\
         $X_{11} \leftarrow X_{11} - X_{10} U_{01}$ \\
         $X_{11} \leftarrow X_{11} - L_{10} X_{01}$ \\
         $X_{11} \leftarrow \sylv(L_{11}, U_{11}, X_{11})$ \\
        \hline
    \end{tabular}
    \hspace{.5cm}
    \begin{tabular}{|>{\cellcolor{graybg}}p{.25\textwidth}|}
        \hline
        \multicolumn{1}{|>{\cellcolor{graybg}}c|}{Variant 2} \\
        \hline
         $X_{01} \leftarrow X_{01} - X_{00} U_{01}$ \\
         $X_{10} \leftarrow \sylv(L_{11}, U_{00}, X_{10})$ \\
         $X_{01} \leftarrow \sylv(L_{00}, U_{11}, X_{01})$ \\
         $X_{11} \leftarrow X_{11} - X_{10} U_{01}$ \\
         $X_{20} \leftarrow X_{20} - L_{21} X_{10}$ \\
         $X_{11} \leftarrow X_{11} - L_{10} X_{01}$ \\
         $X_{11} \leftarrow \sylv(L_{11}, U_{11}, X_{11})$ \\
         $X_{21} \leftarrow X_{21} - L_{21} X_{11}$ \\
         $X_{21} \leftarrow X_{21} - L_{20} X_{01}$ \\
        \hline
    \end{tabular}
    \hspace{.5cm}
    \begin{tabular}{|>{\cellcolor{graybg}}p{.25\textwidth}|}
        \hline
        \multicolumn{1}{|>{\cellcolor{graybg}}c|}{Variant 3} \\
        \hline
         $X_{01} \leftarrow X_{01} - X_{00} U_{01}$ \\
         $X_{11} \leftarrow X_{11} - X_{10} U_{01}$ \\
         $X_{21} \leftarrow X_{21} - X_{20} U_{01}$ \\
         $X_{01} \leftarrow \sylv(L_{00}, U_{11}, X_{01})$ \\
         $X_{11} \leftarrow X_{11} - L_{10} X_{01}$ \\
         $X_{11} \leftarrow \sylv(L_{11}, U_{11}, X_{11})$ \\
         $X_{21} \leftarrow X_{21} - L_{21} X_{11}$ \\
         $X_{21} \leftarrow X_{21} - L_{20} X_{01}$ \\
         $X_{21} \leftarrow \sylv(L_{22}, U_{11}, X_{21})$ \\
        \hline
    \end{tabular}

    \vspace{.5cm}

    \begin{tabular}{|>{\cellcolor{graybg}}p{.25\textwidth}|}
        \hline
        \multicolumn{1}{|>{\cellcolor{graybg}}c|}{Variant 4} \\
        \hline
         $X_{01} \leftarrow X_{01} - X_{00} U_{01}$ \\
         $X_{12} \leftarrow X_{12} - X_{10} U_{02}$ \\
         $X_{01} \leftarrow \sylv(L_{00}, U_{11}, X_{01})$ \\
         $X_{11} \leftarrow X_{11} - L_{10} X_{01}$ \\
         $X_{11} \leftarrow \sylv(L_{11}, U_{11}, X_{11})$ \\
         $X_{21} \leftarrow X_{21} - L_{21} X_{11}$ \\
         $X_{21} \leftarrow X_{21} - L_{20} X_{01}$ \\
         $X_{21} \leftarrow \sylv(L_{22}, U_{11}, X_{21})$ \\
         $X_{22} \leftarrow X_{22} - X_{21} U_{12}$ \\
        \hline
    \end{tabular}
    \hspace{.5cm}
    \begin{tabular}{|>{\cellcolor{graybg}}p{.25\textwidth}|}
        \hline
        \multicolumn{1}{|>{\cellcolor{graybg}}c|}{Variant 5} \\
        \hline
         $X_{01} \leftarrow \sylv(L_{00}, U_{11}, X_{01})$ \\
         $X_{10} \leftarrow X_{10} - L_{10} X_{00}$ \\
         $X_{02} \leftarrow X_{02} - X_{01} U_{12}$ \\
         $X_{10} \leftarrow \sylv(L_{11}, U_{00}, X_{10})$ \\
         $X_{11} \leftarrow X_{11} - X_{10} U_{01}$ \\
         $X_{11} \leftarrow X_{11} - L_{10} X_{01}$ \\
         $X_{11} \leftarrow \sylv(L_{11}, U_{11}, X_{11})$ \\
         $X_{12} \leftarrow X_{12} - X_{11} U_{12}$ \\
         $X_{12} \leftarrow X_{12} - X_{10} U_{02}$ \\
        \hline
    \end{tabular}
    \hspace{.5cm}
    \begin{tabular}{|>{\cellcolor{graybg}}p{.25\textwidth}|}
        \hline
        \multicolumn{1}{|>{\cellcolor{graybg}}c|}{Variant 6} \\
        \hline
         $X_{01} \leftarrow \sylv(L_{00}, U_{11}, X_{01})$ \\
         $X_{10} \leftarrow \sylv(L_{11}, U_{00}, X_{10})$ \\
         $X_{02} \leftarrow X_{02} - X_{01} U_{12}$ \\
         $X_{11} \leftarrow X_{11} - X_{10} U_{01}$ \\
         $X_{20} \leftarrow X_{20} - L_{21} X_{10}$ \\
         $X_{11} \leftarrow X_{11} - L_{10} X_{01}$ \\
         $X_{11} \leftarrow \sylv(L_{11}, U_{11}, X_{11})$ \\
         $X_{12} \leftarrow X_{12} - X_{11} U_{12}$ \\
         $X_{21} \leftarrow X_{21} - L_{21} X_{11}$ \\
         $X_{12} \leftarrow X_{12} - X_{10} U_{02}$ \\
         $X_{21} \leftarrow X_{21} - L_{20} X_{01}$ \\
        \hline
    \end{tabular}

    \vspace{.5cm}

    \begin{tabular}{|>{\cellcolor{graybg}}p{.25\textwidth}|}
        \hline
        \multicolumn{1}{|>{\cellcolor{graybg}}c|}{Variant 7} \\
        \hline
         $X_{01} \leftarrow \sylv(L_{00}, U_{11}, X_{01})$ \\
         $X_{11} \leftarrow X_{11} - X_{10} U_{01}$ \\
         $X_{21} \leftarrow X_{21} - X_{20} U_{01}$ \\
         $X_{02} \leftarrow X_{02} - X_{01} U_{12}$ \\
         $X_{11} \leftarrow X_{11} - L_{10} X_{01}$ \\
         $X_{11} \leftarrow \sylv(L_{11}, U_{11}, X_{11})$ \\
         $X_{12} \leftarrow X_{12} - X_{11} U_{12}$ \\
         $X_{21} \leftarrow X_{21} - L_{21} X_{11}$ \\
         $X_{12} \leftarrow X_{12} - X_{10} U_{02}$ \\
         $X_{21} \leftarrow X_{21} - L_{20} X_{01}$ \\
         $X_{21} \leftarrow \sylv(L_{22}, U_{11}, X_{21})$ \\
        \hline
    \end{tabular}
    \hspace{.5cm}
    \begin{tabular}{|>{\cellcolor{graybg}}p{.25\textwidth}|}
        \hline
        \multicolumn{1}{|>{\cellcolor{graybg}}c|}{Variant 8} \\
        \hline
         $X_{01} \leftarrow \sylv(L_{00}, U_{11}, X_{01})$ \\
         $X_{02} \leftarrow X_{02} - X_{01} U_{12}$ \\
         $X_{11} \leftarrow X_{11} - L_{10} X_{01}$ \\
         $X_{11} \leftarrow \sylv(L_{11}, U_{11}, X_{11})$ \\
         $X_{12} \leftarrow X_{12} - X_{11} U_{12}$ \\
         $X_{21} \leftarrow X_{21} - L_{21} X_{11}$ \\
         $X_{21} \leftarrow X_{21} - L_{20} X_{01}$ \\
         $X_{21} \leftarrow \sylv(L_{22}, U_{11}, X_{21})$ \\
         $X_{22} \leftarrow X_{22} - X_{21} U_{12}$ \\
        \hline
    \end{tabular}
    \hspace{.5cm}
    \begin{tabular}{|>{\cellcolor{graybg}}p{.25\textwidth}|}
        \hline
        \multicolumn{1}{|>{\cellcolor{graybg}}c|}{Variant 9} \\
        \hline
         $X_{10} \leftarrow X_{10} - L_{10} X_{00}$ \\
         $X_{10} \leftarrow \sylv(L_{11}, U_{00}, X_{10})$ \\
         $X_{11} \leftarrow X_{11} - X_{10} U_{01}$ \\
         $X_{11} \leftarrow X_{11} - L_{10} X_{01}$ \\
         $X_{11} \leftarrow \sylv(L_{11}, U_{11}, X_{11})$ \\
         $X_{12} \leftarrow X_{12} - X_{11} U_{12}$ \\
         $X_{12} \leftarrow X_{12} - X_{10} U_{02}$ \\
         $X_{12} \leftarrow X_{12} - L_{10} X_{02}$ \\
         $X_{12} \leftarrow \sylv(L_{11}, U_{22}, X_{12})$ \\
        \hline
    \end{tabular}

    \vspace{.5cm}

    \begin{tabular}{|>{\cellcolor{graybg}}p{.25\textwidth}|}
        \hline
        \multicolumn{1}{|>{\cellcolor{graybg}}c|}{Variant 10} \\
        \hline
         $X_{10} \leftarrow X_{10} - L_{10} X_{00}$ \\
         $X_{21} \leftarrow X_{21} - L_{20} X_{01}$ \\
         $X_{10} \leftarrow \sylv(L_{11}, U_{00}, X_{10})$ \\
         $X_{11} \leftarrow X_{11} - X_{10} U_{01}$ \\
         $X_{11} \leftarrow \sylv(L_{11}, U_{11}, X_{11})$ \\
         $X_{12} \leftarrow X_{12} - X_{11} U_{12}$ \\
         $X_{12} \leftarrow X_{12} - X_{10} U_{02}$ \\
         $X_{12} \leftarrow \sylv(L_{11}, U_{22}, X_{12})$ \\
         $X_{22} \leftarrow X_{22} - L_{21} X_{12}$ \\
        \hline
    \end{tabular}
    \hspace{.5cm}
    \begin{tabular}{|>{\cellcolor{graybg}}p{.25\textwidth}|}
        \hline
        \multicolumn{1}{|>{\cellcolor{graybg}}c|}{Variant 11} \\
        \hline
         $X_{10} \leftarrow \sylv(L_{11}, U_{00}, X_{10})$ \\
         $X_{11} \leftarrow X_{11} - X_{10} U_{01}$ \\
         $X_{20} \leftarrow X_{20} - L_{21} X_{10}$ \\
         $X_{11} \leftarrow X_{11} - L_{10} X_{01}$ \\
         $X_{11} \leftarrow \sylv(L_{11}, U_{11}, X_{11})$ \\
         $X_{12} \leftarrow X_{12} - X_{11} U_{12}$ \\
         $X_{21} \leftarrow X_{21} - L_{21} X_{11}$ \\
         $X_{12} \leftarrow X_{12} - X_{10} U_{02}$ \\
         $X_{21} \leftarrow X_{21} - L_{20} X_{01}$ \\
         $X_{12} \leftarrow X_{12} - L_{10} X_{02}$ \\
         $X_{12} \leftarrow \sylv(L_{11}, U_{22}, X_{12})$ \\
        \hline
    \end{tabular}
    \hspace{.5cm}
    \begin{tabular}{|>{\cellcolor{graybg}}p{.25\textwidth}|}
        \hline
        \multicolumn{1}{|>{\cellcolor{graybg}}c|}{Variant 12} \\
        \hline
         $X_{10} \leftarrow \sylv(L_{11}, U_{00}, X_{10})$ \\
         $X_{11} \leftarrow X_{11} - X_{10} U_{01}$ \\
         $X_{20} \leftarrow X_{20} - L_{21} X_{10}$ \\
         $X_{11} \leftarrow \sylv(L_{11}, U_{11}, X_{11})$ \\
         $X_{12} \leftarrow X_{12} - X_{11} U_{12}$ \\
         $X_{21} \leftarrow X_{21} - L_{21} X_{11}$ \\
         $X_{12} \leftarrow X_{12} - X_{10} U_{02}$ \\
         $X_{12} \leftarrow \sylv(L_{11}, U_{22}, X_{12})$ \\
         $X_{22} \leftarrow X_{22} - L_{21} X_{12}$ \\
        \hline
    \end{tabular}

    \vspace{.5cm}

    \begin{tabular}{|>{\cellcolor{graybg}}p{.25\textwidth}|}
        \hline
        \multicolumn{1}{|>{\cellcolor{graybg}}c|}{Variant 13} \\
        \hline
         $X_{11} \leftarrow X_{11} - X_{10} U_{01}$ \\
         $X_{21} \leftarrow X_{21} - X_{20} U_{01}$ \\
         $X_{11} \leftarrow X_{11} - L_{10} X_{01}$ \\
         $X_{11} \leftarrow \sylv(L_{11}, U_{11}, X_{11})$ \\
         $X_{12} \leftarrow X_{12} - X_{11} U_{12}$ \\
         $X_{21} \leftarrow X_{21} - L_{21} X_{11}$ \\
         $X_{12} \leftarrow X_{12} - X_{10} U_{02}$ \\
         $X_{21} \leftarrow X_{21} - L_{20} X_{01}$ \\
         $X_{12} \leftarrow X_{12} - L_{10} X_{02}$ \\
         $X_{21} \leftarrow \sylv(L_{22}, U_{11}, X_{21})$ \\
         $X_{12} \leftarrow \sylv(L_{11}, U_{22}, X_{12})$ \\
        \hline
    \end{tabular}
    \hspace{.5cm}
    \begin{tabular}{|>{\cellcolor{graybg}}p{.25\textwidth}|}
        \hline
        \multicolumn{1}{|>{\cellcolor{graybg}}c|}{Variant 14} \\
        \hline
         $X_{11} \leftarrow X_{11} - X_{10} U_{01}$ \\
         $X_{21} \leftarrow X_{21} - X_{20} U_{01}$ \\
         $X_{11} \leftarrow \sylv(L_{11}, U_{11}, X_{11})$ \\
         $X_{12} \leftarrow X_{12} - X_{11} U_{12}$ \\
         $X_{21} \leftarrow X_{21} - L_{21} X_{11}$ \\
         $X_{12} \leftarrow X_{12} - X_{10} U_{02}$ \\
         $X_{21} \leftarrow \sylv(L_{22}, U_{11}, X_{21})$ \\
         $X_{12} \leftarrow \sylv(L_{11}, U_{22}, X_{12})$ \\
         $X_{22} \leftarrow X_{22} - L_{21} X_{12}$ \\
        \hline
    \end{tabular}
    \hspace{.5cm}
    \begin{tabular}{|>{\cellcolor{graybg}}p{.25\textwidth}|}
        \hline
        \multicolumn{1}{|>{\cellcolor{graybg}}c|}{Variant 15} \\
        \hline
         $X_{11} \leftarrow X_{11} - L_{10} X_{01}$ \\
         $X_{11} \leftarrow \sylv(L_{11}, U_{11}, X_{11})$ \\
         $X_{12} \leftarrow X_{12} - X_{11} U_{12}$ \\
         $X_{21} \leftarrow X_{21} - L_{21} X_{11}$ \\
         $X_{12} \leftarrow X_{12} - L_{10} X_{02}$ \\
         $X_{21} \leftarrow X_{21} - L_{20} X_{01}$ \\
         $X_{12} \leftarrow \sylv(L_{11}, U_{22}, X_{12})$ \\
         $X_{21} \leftarrow \sylv(L_{22}, U_{11}, X_{21})$ \\
         $X_{22} \leftarrow X_{22} - X_{21} U_{12}$ \\
        \hline
    \end{tabular}

    \vspace{.5cm}

    \begin{tabular}{|>{\cellcolor{graybg}}p{.25\textwidth}|}
        \hline
        \multicolumn{1}{|>{\cellcolor{graybg}}c|}{Variant 16} \\
        \hline
         $X_{11} \leftarrow \sylv(L_{11}, U_{11}, X_{11})$ \\
         $X_{12} \leftarrow X_{12} - X_{11} U_{12}$ \\
         $X_{21} \leftarrow X_{21} - L_{21} X_{11}$ \\
         $X_{12} \leftarrow \sylv(L_{11}, U_{22}, X_{12})$ \\
         $X_{21} \leftarrow \sylv(L_{22}, U_{11}, X_{21})$ \\
         $X_{22} \leftarrow X_{22} - X_{21} U_{12}$ \\
         $X_{22} \leftarrow X_{22} - L_{21} X_{12}$ \\
        \hline
    \end{tabular}
\end{center}

These algorithms differ in a few ways from those in the previous sections:
\begin{itemize}
    \item They operate on three matrices, overwriting one of them with the output.
    \item The input matrices are of different sizes: $L \in \mathbb R^{\mathtt m \times \mathtt m}$, $U \in \mathbb R^{\mathtt n \times \mathtt n}$, and $X \in \mathbb R^{\mathtt m \times \mathtt n}$.
        The matrices are traversed along the diagonal as far as possible and then along the remaining dimension (see \autoref{sec:intro.blockedalgs.general}).
    \item There are three recursive calls $\Omega$ in each step of the matrix traversal.
        These operate not only on the $X_{11} \in \mathbb R^{\mathtt{blocksize} \times \mathtt{blocksize}}$ but also on the matrix panels $X_{01}$, $X_{10}$, $X_{12}$, and $X_{21}$.
        For the latter, our C implementation invokes the blocked algorithms recursively; only the small matrices $X_{11}$ trigger their unblocked versions.
\end{itemize}

\input{figures/ranking.sylv.n.tex}

For our performance predictions, we focus on the case
\begin{center}
    \texttt{%
        sylv\textit i(%
        \overparameq{m}{\textit n},
        \overparameq{n}{\textit n},
        \overparameq{L}{\textit L},
        \overparameq{ldL}{\textit n},
        \overparameq{U}{\textit U},
        \overparameq{ldU}{\textit n},
        \overparameq{X}{\textit X},
        \overparameq{ldX}{\textit n},
        \overparameq{blocksize}{96})%
    }.
\end{center}
All matrices are of size $\mathtt n \times \mathtt n$, $\mathtt n \in \{8, 16, \ldots, 1024\}$  and we use the block-size $96$.
\autoref{fig:ranking.sylv.n} compares our predictions for these algorithms with corresponding measurements of their implementations, where
$$
    \metric{efficiency} = \frac{\mathtt n^3 + \mathtt n^2}{2 \metric{ticks}}.
$$
We observe significantly different performances across algorithms: At $\mathtt n = 1024$ variant~1~(\ref{fig:ranking.sylv.n:var1}) is 20 times faster than variant~13~(\ref{fig:ranking.sylv.n:var13}).
In the \metric{ticks} plots, we see that our predictions separate the fast and slow variants very well.
To rank the top candidates, we turn to \metric{efficiency}; here, we see that the prediction quality is not as good as in our previous studies.
However, the algorithms are correctly ranked.

    \chapter{Conclusion and Future Work}
    \label{sec:conclusion}
    In this thesis, we have introduced methods and tools to analyze and model the performance of dense linear algebra routines.
Our goal was to rank sets of blocked algorithms according to their performance without executing them.
Towards this goal, we created a set of automatic performance analysis and modeling tools.
The Sampler measures the performance of routine executions; based on this tool, the Modeler generates performance models for specified routines.
With these models, we were able to accurately predict the performance of blocked algorithms, rank them correctly, and determine the optimal algorithmic block-size.

\paragraph{Sampler.}
As a base for our performance analysis, we designed a performance measurement tool: the Sampler (\autoref{sec:sampling.tool}).
This flexible tool can sample any dense linear algebra routine execution on any system.
Provided with library or object files for a set of routines and according header files, its build system automatically generates an executable that allows to measure the performance of these routines.

The cycle-accurate time stamp counter (\texttt{RDTSC}) serves as the base of its execution time measurements \metric{ticks}.
The Performance Application Programming Interface (PAPI) is optionally\footnote{
    requiring a kernel patch.
} used to access a wide range of other performance counters, such as \metric{flops} or \metric{L1misses}.
These performance counters and further properties of the Sampler, such as memory handling policies, are specified through a simple configuration file (\autoref{sec:sampling.tool.usage.config}).

The Sampler can be interfaced with other programs through its standard input and output streams.
On its input stream, it expects a series of sampling requests, consisting of routine names and argument values; on the output stream, it provides performance measurements for these routine invocations.

Thanks to its flexibility, functionalities, and simple interface, the Sampler serves as an ideal base for performance analysis in dense linear algebra.
Apart form performance modeling, it can be used for tasks such as performance debugging or tuning.

\paragraph{Modeler.}
Based on the Sampler, we constructed an automatic performance modeling tool: the Modeler (\autoref{sec:modeling.tool}).
For dense linear algebra routines, this tool builds models that represent the routines' performance as a function of a set of discrete (e.g., \texttt{side}, \texttt{transA}) and continuous (e.g., \texttt m, \texttt n) arguments.
The generated models yield probabilistic performance estimates in the form of statistical quantities, such as minimum, median, or average.

The Modeler, written in Python, is highly flexible:
It provides a configuration system to specify a wide range of settings.
These include but are not limited to the following:
\begin{itemize}
    \item The choice of the Sampler instance and its configuration;
    \item The routines to be modeled;
    \item The treatment of the routine arguments;
    \item The performance counters to be modeled;
    \item The type of polynomial modeling to be used;
    \item The accuracy and resolution of the polynomial models.
\end{itemize}

The Modeler aims at generating highly accurate performance models, while at the same time using as few measurements as possible.
This poses a trade-off: reducing the number of samples generally decreases the model accuracy and vice versa.
The Modeler configuration allows to balance these two aspects according to the desires of the user (see \autoref{sec:modeling.res.ticks}).

\paragraph{Ranking.}
With the models automatically generated by the Modeler, we can predict the performance of blocked algorithms execution-less (\autoref{sec:ranking}).
Our accurate predictions allow us to
\begin{itemize}
    \item Rank several algorithms according to their performance, and
    \item Determine the optimal algorithmic block-size.
\end{itemize}

        \section{Outlook}
        \label{sec:conclusion.outlook}
        The work presented in this thesis offers a number of possible directions for further research.

\paragraph{Prediction of Other Performance Metrics.}
In our study of blocked algorithms, we focused on predicting the execution time \metric{ticks} --- one of many performance metrics.
Our models, however, can describe a large variety of performance counters, all of which can be included in the prediction process.
Of particular interest might be the cache miss counters \metric{L1misses} through \metric{L3misses}, since these are closely related to the processor's energy consumption.

\paragraph{Other Algorithm Types.}
We limited ourselves to the performance prediction of blocked algorithms.
These, however, are only one type of algorithms used in dense linear algebra.
Other algorithm types include algorithms-by-block and recursive algorithms.
Like blocked algorithms, these are based on BLAS Level-3 routines; our models for these routines can be used in the performance prediction of these algorithms.

\paragraph{Shared Memory Parallelism.}
We were concerned with analyzing the performance of single core algorithms.
It is crucial to understand the phenomena and behaviors that appear in this scenario before we target parallel algorithms.
Since in the field of high performance computing CPUs usually provide several cores, considering shared memory parallelism is a natural next step.

For blocked algorithms, parallelism can be exploited through multi-threaded BLAS implementations.
A possible approach would be to apply our measurement and modeling framework to these parallel BLAS routines and rank the algorithms accordingly.

\paragraph{Extrapolation.}
In this thesis, we build performance models for routines that represent a certain range of size arguments.
We then use these models to predict the performance of blocked algorithms within the same range.
In principle, the polynomials of our models can be extrapolated to estimate the performance of routine executions beyond their scope.
Designed for a limited range of argument values, the estimation quality of our models will deteriorate when we extrapolate.
However, extrapolation is an interesting usage scenario of performance models and performance modeling in general.

    \bibliography{references}
    \bibliographystyle{plain}

    \listoffigures

    \appendix

    \chapter{Introduction to BLAS}
    \label{app:blas}
    In this appendix, we discuss the Basic Linear Algebra Subprograms (BLAS) and their usage.
BLAS as a low level interface specification, which can be accessed directly from within C or Fortran.
However, virtually all dense linear algebra libraries in any other programming language is built on top of BLAS.

BLAS is separated into three levels:
\begin{itemize}
    \item BLAS Level-1 routines implement vector operations (e.g., scaling or addition of vectors); 
    \item BLAS Level-2 routines implement matrix-vector operations (e.g., matrix-vector multiplication or solution of a triangular linear system);
    \item BLAS Level-3 routines implement matrix-matrix operations (e.g., matrix-matrix multiplication or solution of a triangular linear system with multiple right hand sides).
\end{itemize}
Both BLAS Level-1 and BLAS Level-2 operations are memory bound; they cannot possibly reach the theoretical peak performance of a CPU.
Only BLAS Level-3 operations are compute bound; carefully implemented, they can reach up to 99\% of the peak performance.

We focus on discussing the following properties of the interfaces for C and Fortran:
\begin{itemize}
    \item The routine names (\autoref{app:blas.naming});
    \item The routine arguments and matrix storage (\autoref{app:blas.arguments});
    \item Routine invocation and results (\autoref{app:blas.invocation});
    \item Available implementation (\autoref{app:blas.implementations}).
\end{itemize}

Throughout this appendix, we consider \texttt{dtrsm} as an exemplary BLAS routine.
It computes the solution of a triangular linear system with multiple right hand sides, $B \leftarrow \alpha A^{-1} B$.

        \section{Routine names}
        \label{app:blas.naming}
        Due to restrictions in early versions of Fortran, BLAS routine names consist of a maximum of 6 characters.
They are chosen such that they represent both the operation and the data types.

We now study the routine name \texttt{dtrsm} and show why it corresponds to the operation $B \leftarrow \alpha A^{-1} B$.
Character my character, \texttt{dtrsm} has the following meaning:
\begin{itemize}
    \item \texttt d: The first character identifies the data type of all scalar, vector, and matrix arguments.
        Every BLAS routine is available for the following data types:
        \begin{itemize}
            \item \texttt s: single precision real;
            \item \texttt d: double precision real;
            \item \texttt c: single precision complex (real and imaginary part are stored as two consecutive single precision numbers);
            \item \texttt z: double precision complex (two consecutive doubles).
        \end{itemize}
    \item \texttt{tr}: These two characters indicate that a certain matrix is involved.
        There are again four possibilities:
        \begin{itemize}
            \item \texttt{ge}: general matrix;
            \item \texttt{sy}: symmetric matrix;
            \item \texttt{he}: hermetian matrix;
            \item \texttt{tr}: triangular matrix.
        \end{itemize}
    \item \texttt s: This character describes the type of operation:
        \begin{itemize}
            \item \texttt m: multiplication;
            \item \texttt s: solution of a linear system.
        \end{itemize}
    \item \texttt m: The last character indicates the type of the second operand:
        \begin{itemize}
            \item \texttt v: vector;
            \item \texttt m: matrix.
        \end{itemize}
\end{itemize}
\newcommand{\ttunderbrace}[2]{%
\tikz[baseline=(n.base)] {
        \node[inner sep=0pt] (n) {\vphantom{g}#1};
        \draw[decorate, decoration=brace]
            (n.south east) -- (n.south west) node[midway, anchor=north] {\texttt{\vphantom{t}#2}};
    }%
}
Together, \texttt{dtrsm} stands for:
\begin{center}
    \ttunderbrace{double precision}{d}
    \ttunderbrace{triangular}{tr}
    \ttunderbrace{linear system}{s}
    with
    \ttunderbrace{multiple right hand sides}{m}
    ($B \leftarrow \alpha A^{-1} B$).
\end{center}

Most  BLAS Level-2 and Level-3 routines follow this naming scheme; BLAS Level-1 operations use other similarly representative routine names.
Examples of other routine names and the corresponding operations are the following:
\begin{itemize}
    \item \texttt{zgemv}:
        \ttunderbrace{double precision complex}{z}
        \ttunderbrace{general}{ge}
        \ttunderbrace{matrix}{m}
        \ttunderbrace{vector}{v}
        product
        ($\mathbf y \leftarrow \alpha A \mathbf x + \beta \mathbf y$).
    \item \texttt{csyrk}:
        \ttunderbrace{single precision complex}{c}
        \ttunderbrace{symmetric}{sy}
        \ttunderbrace{rank k}{rk}
        update
        ($C \leftarrow \alpha A A^T + \beta C $).
    \item \texttt{saxpy}:
        \ttunderbrace{single precision}{s}
        \ttunderbrace{scalar}{a}
        times
        \ttunderbrace{vector}{x}
        \ttunderbrace{plus}{p}
        \ttunderbrace{vector}{y}
        ($\mathbf y \leftarrow \alpha \mathbf x + \mathbf y$).
\end{itemize}

Since Fortran appends an underscore \texttt\_ to all function names in its object files, this underscore needs to be appended to all BLAS routines, when they are used in C (e.g., \texttt{dtrsm} is available as \texttt{dtrsm\_}).

        \section{Arguments and Matrix Storage}
        \label{app:blas.arguments}
        All arguments of BLAS routines are passed by reference, that is, in the form of pointers.
This again is necessary for compatibility with Fortran, where arguments are always passed by reference.

For each BLAS routine, we distinguish three groups of arguments: discrete, size, and data arguments.
For example, \texttt{dtrsm} has the following signature:
\begin{center}
    \begin{tikzpicture}
        \node (pos) {};
        \def\oldlabel{pos}
        \foreach \label/\text in {
            head/{dtrsm(},
            disc/{side, uplo, transA, diag}, c/{,\ \ },
            size/{m, n}, c/{,\ \ },
            alpha/{alpha}, c/{,\ \ },
            A/{A}, c/{,\ \ }, ldA/{ldA}, c/{,\ \ },
            B/{B}, c/{,\ \ }, ldB/{ldB}, tail/{).}
        } {
            \node[inner sep=0, anchor=base west, base right=0 of \oldlabel.base east]
                (\label) {\texttt{\vphantom{p}\text}};
            \xdef\oldlabel{\label}
        }

        \draw[decorate, decoration=brace] (disc.south east) -- (disc.south west)
            node[midway, anchor=north] {\vphantom{T}discrete};
        \draw[decorate, decoration=brace] (size.south east) -- (size.south west)
            node[midway, anchor=north] {\vphantom{T}size};
        \draw[decorate, decoration=brace] (ldB.south east) -- (alpha.south west)
            node[midway, anchor=north] {\vphantom{T}data};

        \path (alpha.south)   ++(0, -.75)
            node[anchor=north, inner sep=0] (scal) {\vphantom{T}scalar};
        \path (A.south west)   ++(0, -.75)
            node[anchor=north west, inner sep=0] (matrix) {\vphantom{T}matrix};
        \path (ldB.south east) ++(0, -.75)
            node[anchor=north east, inner sep=0, outer sep=0, label=below:dimension] (ld) {\vphantom{T}leading};
        \draw[->, gray] (scal) -- (alpha.south |- 0, -.2);
        \draw[->, gray] (matrix) -- (A.south |- 0, -.2);
        \draw[->, gray] (matrix) -- (B.south west |- 0, -.2);
        \draw[->, gray] (ld) -- (ldA.south |- 0, -.2);
        \draw[->, gray] (ld) -- (ldB.south |- 0, -.2);
        
        \path[gray]
            (scal.west |- 0, .3) -- (scal.east |- 0, .3)
            node[midway, anchor=south] {\vphantom{g}$\alpha$};
        \draw[decorate, decoration=brace, gray]
            (A.west |- 0, .3) -- (ldA.east |- 0, .3)
            node[midway, anchor=south] {\vphantom{g}$A$};
        \draw[decorate, decoration=brace, gray]
            (B.west |- 0, .3) -- (ldB.east |- 0, .3)
            node[midway, anchor=south] {\vphantom{g}$B$};
    \end{tikzpicture}
\end{center}
\begin{itemize}
    \item The \emph{discrete arguments} \texttt{side}, \texttt{uplo}, \texttt{transA}, and \texttt{diag} specify the exact operation:
        \begin{itemize}
            \item $\mathtt{side} \in \{\mathtt{Left}, \mathtt{Right}\}$ identify on which side of $B$ $A$ appears: $B \leftarrow \alpha A^{-1} B$ or $B \leftarrow \alpha B A^{-1}$.
            \item $\mathtt{uplo} \in \{\mathtt{Lower triangular}, \mathtt{Upper triangular}\}$ specifies whether $A$ is lower or upper triangular.
                In operations involving symmetric matrices, \texttt{uplo} indicates, which part of the matrix is stored in memory --- the lower or upper triangular part.
            \item $\mathtt{transA} \in \{\mathtt{No transpose}, \mathtt{Transpose}$ indicates if the system is solved with $A$ or $A^T$.
                For complex valued routines, \texttt{Transpose} is replaced by \texttt{Complex-conjugate}.
            \item $\mathtt{diag} \in \{\mathtt{Non-Unit triangular}, \mathtt{Unit triangular}$ declares whether the entries of the diagonal of $A$ are to be treated as ones (\texttt{Unit triangular}) independent of the data stored in memory.
                Using this argument, it is possible to store both an upper triangular and a lower triangular matrix in the memory of a full matrix, when one of them is unit triangular.
        \end{itemize}
        The values of discrete arguments are identified by their first character (e.g., $\mathtt{side} = \mathtt L$ for \texttt{Left}).
        All 16 value combinations for the four discrete argument in \texttt{dtrsm} are allowed.
    \item The \emph{size arguments} \texttt m and \texttt n specify the sizes of the matrix (and vector) operands.
        For \texttt{dtrsm}, we have $B \in \mathbb R^{\mathtt m \times \mathtt n}$;
        depending on the value of \texttt{side}, $A$ is of size $\mathtt m \times \mathtt m$ (\texttt L) or $\mathtt n \times \mathtt n$ (\texttt R).
    \item The \emph{data arguments} \texttt{alpha}, \texttt A, \texttt{ldA}, \texttt B, and \texttt{ldB} determine the storage locations of the operands:
        \begin{itemize}
            \item The \emph{scalar argument} \texttt{alpha} is (a pointer to) the value of $\alpha$.
            \item The \emph{matrix arguments} \texttt A and \texttt B are (pointers to) these matrices in memory.
                As in Fortran, matrices are expected to be stored in column-major format: the columns of the matrix are continuous in memory; matrices consist of a set of consecutive columns.
                \begin{figure}[t]
    \centering
    \tikzfading[name=fade right, left color=transparent!0, right color=transparent]
    \begin{tikzpicture}

        \path
            ++(4, 2)  node (a bl) {}  +(0, 4)   node (a tl) {}
             +(4.5, 0) node (a br) {} +(4.5, 4) node (a tr) {}
            ++(1, .5)  node (A bl) {}  +(0, 2.5) node (A tl) {}
             +(2, 0)  node (A br) {}  +(2, 2.5) node (A tr) {};

        \path (-.5, 0 |- a tl) -- (a tl) node[midway, anchor=north, text width=3.3cm, align=center] {
            position of $A$ within a larger matrix:
        };

        \fill[graybg] (a bl) rectangle (a tr);

        \node (pos) at (A tl) {};
        \foreach \s in {80, 65, 50, 35, 20} {
            \foreach \c in {plot1, plot2, plot3, plot4}
                \fill[\c!\s] (pos) rectangle +(.5, -.5) ++(.5, 0) node (pos) {};
            \path (pos) ++(-2, -.5) node (pos) {};
        }

        \draw[step=.5cm, help lines] (a bl) grid (a tr) (a bl) rectangle (a tr);
        \draw[step=.5cm] (A bl) grid (A tr) (A bl) rectangle (A tr);
        
        \draw[decorate, decoration=brace] (a bl) ++(-.1, 0) node (i) {} (a bl -| i) -- (a tl -| i) node[midway, anchor=east] {$M$};
        \draw[decorate, decoration=brace] (A bl) ++(-.1, 0) node (i) {} (A bl -| i) -- (A tl -| i) node[midway, anchor=east] {$m$};
        \draw[decorate, decoration=brace] (a tl) ++(0, .1)  node (i) {} (a tl |- i) -- (a tr |- i) node[midway, anchor=south] {$N$};
        \draw[decorate, decoration=brace] (A tl) ++(0, .1)  node (i) {} (A tl |- i) -- (A tr |- i) node[midway, anchor=south] {$n$};
        \path (A bl) ++(1, 1.25) node {$A$};

        \fill[graybg] (-.25, 0) rectangle (12.25, .5);

        \node (pos) at (.5, 0) {};
        \foreach \s in {80, 65, 50, 35, 20}
            \fill[plot1!\s]   (pos) rectangle +(.5, .5) ++(.5, 0) node (pos) {};
        \path (pos) ++(1.5, 0) node (pos) {};
        \foreach \s in {80, 65, 50, 35, 20}
            \fill[plot2!\s] (pos) rectangle +(.5, .5) ++(.5, 0) node (pos) {};
        \path (pos) ++(1.5, 0) node (pos) {};
        \foreach \s in {80, 65, 50, 35, 20}
            \fill[plot3!\s]  (pos) rectangle +(.5, .5) ++(.5, 0) node (pos) {};

        \draw[step=.5cm, help lines] (-.25, 0) grid (12.25, .5);
        \node[anchor=east] at (0, .25) {$\cdots$};
        \node[anchor=west] at (12, .25) {$\cdots$};
        \draw[step=.5cm] (.5, 0)  grid ++(2.5, .5) (.5, 0)  rectangle ++(2.5, .5);
        \draw[step=.5cm] (4.5, 0) grid ++(2.5, .5) (4.5, 0) rectangle ++(2.5, .5);
        \draw[step=.5cm] (8.5, 0) grid ++(2.5, .5) (8.5, 0) rectangle ++(2.5, .5);

        \draw[decorate, decoration=brace]
            (.5, .6) -- ++(2.5, 0) node[midway, anchor=south] {\texttt{n}};
        \draw[decorate, decoration=brace]
            (4.5, -.1) -- (.5, -.1) node[midway, anchor=north] {$\mathtt{ldA} = M$};
        \draw[<-] (.75, .25) -- ++(-.25, .75) node[anchor=south] {\texttt{A}};
        \node at (6, 1) {$A$ in memory:};
    \end{tikzpicture}
    \caption{$A \in \mathbb R^{m \times n}$ as part of an $M \times N$ matrix.}
    \label{fig:base.blas.arguments.storage}
    \tikzset{external/export=false}
\end{figure}
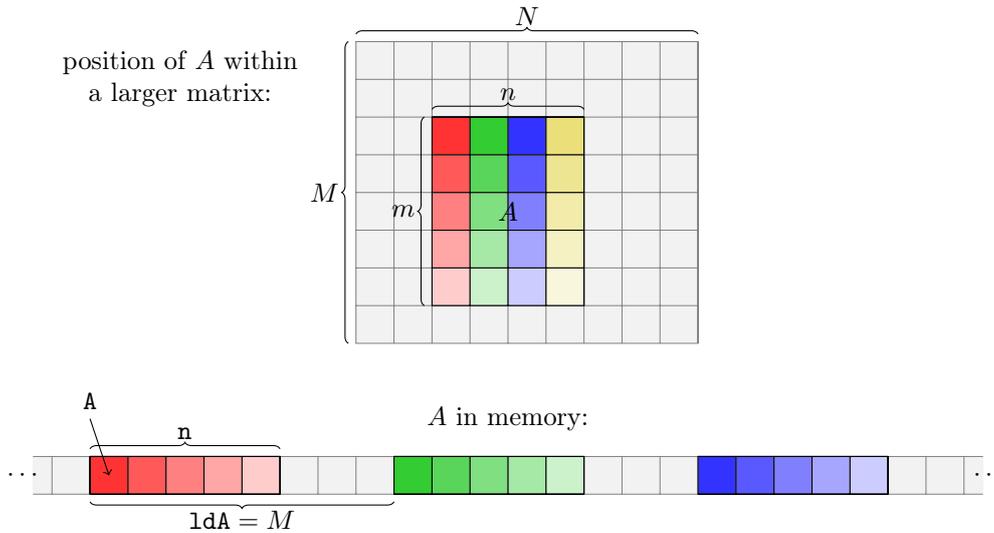
            \item The \emph{leading dimension} arguments \texttt{ldA} and \texttt{ldB}, immediately following the corresponding matrix arguments \texttt A and \texttt B,  give the distance between two adjacent matrix entries in the same row.
                \autoref{fig:base.blas.arguments.storage} shows how this can be used to address a matrix as part of a larger matrix.
        \end{itemize}

        For vector-operations there are two further argument types:
        \begin{itemize}
            \item \emph{Vector arguments} (e.g., \texttt x) are (pointers to) vector in memory.
            \item \emph{Increment arguments} (e.g., \texttt{incx}) immediately follow the vector arguments and specify the distance between two consecutive elements of a vector in memory.
                These arguments allow to access both columns and rows of matrices as vectors.
        \end{itemize}

        The order of data arguments corresponds to the order of the corresponding operands in the basic form of the operation; for \texttt{dtrsm}, \texttt{alpha}, \texttt A, \texttt{ldA}, \texttt B, and \texttt{ldB} are in the order of $B \leftarrow \alpha A^{-1} B$.
\end{itemize}

        \section{Routine Invocation and Results}
        \label{app:blas.invocation}
        Most BLAS routines do not have an explicit return values, like functions do.
Instead, one of the routine arguments is overwritten with the result of the operation --- usually the last one.
Exceptions are BLAS routines that compute scalar quantities; for example, the inner product of two vectors \texttt{ddot} returns $x^T y$.

        \section{Implementations}
        \label{app:blas.implementations}
        Since BLAS was introduced in 1979 \cite{blas}, numerous implementations were developed.
In this section, we will list the most prominent implementations for x86 processors and compare their efficiencies.

\begin{itemize}
    \item The \emph{reference implementation} \cite{refblaspage} is a minimal implementation of BLAS.
        Its purpose is to serve as a reference for both BLAS developers and users.
        The routines are well documented and easy to read and understand.
        Since this implementation was not designed with performance in mind, it is not suitable for scientific codes due to its poor performance.
    \item \emph{\gotoblas} \cite{gotoblas, gotopage} is an open source implementation developed at the Texas Advanced Computing Center (University of Texas at Austin); its name stems from its initiator and former developer Kazushige Goto.
        It is written in C and assembly and contains highly optimized kernels for many CPU architectures.
    \item The \emph{Automatically Tuned Linear Algebra Software} (ATLAS) \cite{atlas1, atlas2, atlaspage} is another open source implementation.
        During its building process, this library optimizes itself for the CPU architecture; for this, it uses empirical techniques to estimate the best configuration of its routines.
    \item Intel's commercial \emph{Math Kernel Library} (MKL) \cite{mklpage} offers (among other functionality) a high performance implementation of BLAS for Intel architectures.
\end{itemize}

\begin{figure}[t]
    \tikzset{external/export=true}
    \centering
    \subfloat[Execution time]{
        \label{fig:base.blas.implementations.perf:tb}
        \begin{tikzpicture}
            \begin{axis}[
                twocolplot,
                xlabel={matrix size $n$},
                ylabel={Time [min]},
                legend to name=fig:base.blas.implementations.perf:legend,
                legend columns=-1,
                xtick={0,2048,...,8192}
            ]
                \addlegendimage{plot1, only marks}
                \label{fig:base.blas.implementations.perf:goto}
                \addlegendentry{\gotoblas}
                \addlegendimage{plot2, only marks}
                \label{fig:base.blas.implementations.perf:mkl}
                \addlegendentry{MKL}
                \addlegendimage{plot3, only marks}
                \label{fig:base.blas.implementations.perf:atlas}
                \addlegendentry{ATLAS}
                \addlegendimage{plot4, only marks}
                \label{fig:base.blas.implementations.perf:ref}
                \addlegendentry{reference implementation}

                \addplot[color=plot1, mark size=.2pt, only marks] file {figures/data/base.blas.implementations.perf/goto.time.dat};
                \addplot[color=plot2, mark size=.2pt, only marks] file {figures/data/base.blas.implementations.perf/mkl.time.dat};
                \addplot[color=plot3, mark size=.2pt, only marks] file {figures/data/base.blas.implementations.perf/atlas.time.dat};
                \addplot[color=plot4, mark size=.2pt, only marks] file {figures/data/base.blas.implementations.perf/ref.time.dat};
            \end{axis}
        \end{tikzpicture}
    }
    \hfill
    \subfloat[Efficiency]{
        \label{fig:base.blas.implementations.perf:eb}
        \begin{tikzpicture}
            \begin{axis}[
                twocolplot,
                xlabel={matrix size $n$},
                ylabel={Efficiency [\%]},
                ymax=100,
                xtick={0,2048,...,8192},
            ]
                \addplot[color=plot1, mark size=.2pt, only marks] file {figures/data/base.blas.implementations.perf/goto.eff.dat};
                \addplot[color=plot2, mark size=.2pt, only marks] file {figures/data/base.blas.implementations.perf/mkl.eff.dat};
                \addplot[color=plot3, mark size=.2pt, only marks] file {figures/data/base.blas.implementations.perf/atlas.eff.dat};
                \addplot[color=plot4, mark size=.2pt, only marks] file {figures/data/base.blas.implementations.perf/ref.eff.dat};
            \end{axis}
        \end{tikzpicture}
    }

    \vspace{.5cm} 

    \tikzset{external/export=false}
    \ref*{fig:base.blas.implementations.perf:legend}
    \caption{Performance of \texttt{dgemm} in different BLAS implementations.}
    \label{fig:base.blas.implementations.perf}
\end{figure}
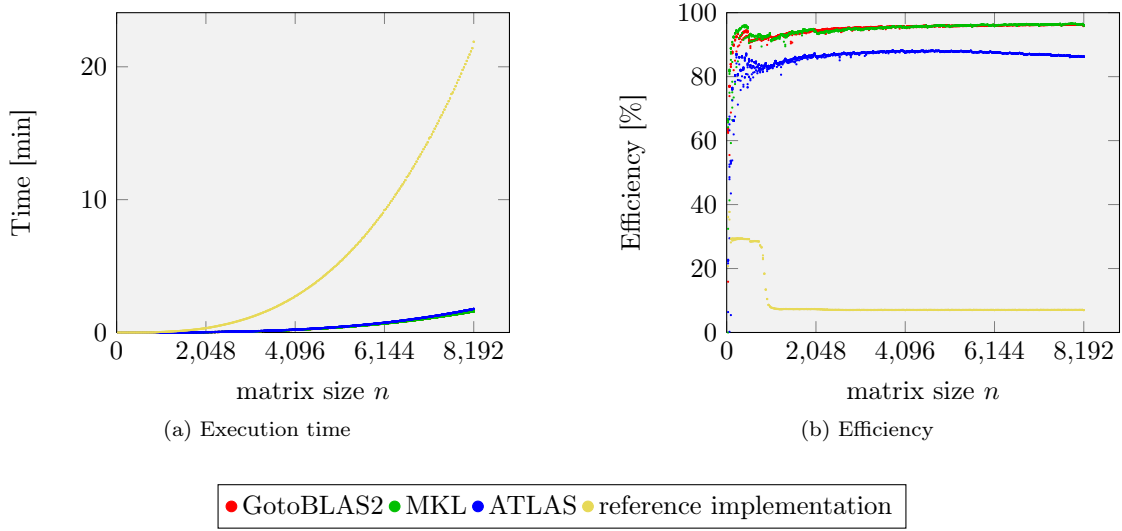

We now compare the performance of these BLAS implementations by considering \texttt{dgemm} ($C \leftarrow \alpha A B + \beta C$).
This routine is usually \emph{the} most optimized routine in any high performance library (In many implementations, other BLAS Level-3 routines are based on \texttt{dgemm} kernels \cite{gemmanatomy}).
We use the Sampler (\autoref{sec:sampling.tool}) on an Intel Harpertown E5450 processor \cite{e5450} running at $2.99 \mathrm{GHz}$ to measure the \metric{ticks} (execution time) of \texttt{dgemm} with the following parameters:
\begin{center}
    \texttt{%
        dgemm(%
        \overparameq{transA}{N}
        \overparameq{transB}{N}
        \overparameq{m}{\textit m},
        \overparameq{n}{\textit n},
        \overparameq{k}{\textit k},
        \overparameq{alpha}{0.5},
        \overparameq{A}{\textit A},
        \overparameq{ldA}{\textit m},
        \overparameq{B}{\textit B},
        \overparameq{ldB}{\textit k},
        \overparameq{beta}{0.5},
        \overparameq{C}{\textit C},
        \overparameq{ldC}{\textit m})%
    },
\end{center}
that is, $C \leftarrow 0.5 A B + 0.5 C$ with $A, B, C \in \mathbb R^{n \times n}$.
In \autoref{fig:base.blas.implementations.perf} we show the \metric{ticks} and \metric{efficiency} of these executions, where
$$
    \metric{efficiency} = \frac{n^3 + 2 n^2}{2 \metric{ticks}}.
$$

As expected, the reference implementation~(\ref{fig:base.blas.implementations.perf:ref}) attains the poorest performance.
For matrices that fit in the $12 \mathrm{MB}$ cache of the processor, it reaches $28\%$ of the peak performance; for larger matrices, the performance drops rapidly to $7\%$.
The high performance implementations \gotoblas~(\ref{fig:base.blas.implementations.perf:goto}), MKL~(\ref{fig:base.blas.implementations.perf:mkl}), and ATLAS~(\ref{fig:base.blas.implementations.perf:atlas}) attain efficiencies that exceed the reference implementation by more than a factor of 10.
With an efficiencies around $96\%$, the hand optimized \gotoblas~(\ref{fig:base.blas.implementations.perf:goto}) and MKL~(\ref{fig:base.blas.implementations.perf:mkl}) outperform the automatically tuned ATLAS~(\ref{fig:base.blas.implementations.perf:atlas}), which reaches $87\%$.

    \chapter{Implementations of Blocked Algorithms}
    \label{app:blockedalgs}
    This appendix lists the implementations of the blocked algorithms used throughout the thesis.
All algorithms are written in C and need to linked with a BLAS library.

We implemented only the blocked version of each algorithm.
The unblocked version is implicitly given by passing $\mathtt{blocksize} = 1$.

        \section{Triangular Inverse \texorpdfstring{$L \leftarrow L^{-1}$}{L <- inv(L)}}
        \label{app:blockedalgs.trinv}
\begin{lstlisting}[caption={\texttt{trinv.c} --- inversion of a triangular matrix.}]
extern void dtrmm_(char*, char*, char*, char*, int*, int*, double*, double*, int*, double*, int*);
extern void dtrsm_(char*, char*, char*, char*, int*, int*, double*, double*, int*, double*, int*);
extern void dgemm_(char*, char*, int*, int*, int*, double*, double*, int*, double*, int*, double*, double*, int*);

int trinv1_(char* diag, int* n, double* A, int* ldA, int* blocksize) {
	if (*n == 1) {
		if (diag[0] == 'N')
			*A = 1 / *A;
		return 0;
	}

	int ione = 1;
   	double one = 1;
	double minusone = -1;

    int p = 0;
    while (p < *n) {
 	    int b = *blocksize;
 	    if (p + b > *n)
		    b = *n - p;
		int r = *n - (p + b);

		// A00 
		// A10 A11 
		// A20 A21 A22
#define A00 (A)
#define A10 (A                  + p)
#define A11 (A + *ldA * p       + p)
#define A20 (A                  + p + b)
#define A21 (A + *ldA * p       + p + b)
#define A22 (A + *ldA * (p + b) + p + b)

	    // A10 = A10 * A00
	    dtrmm_("R", "L", "N", diag, &b, &p, &one, A00, ldA, A10, ldA);
	    // A10 = -A11 \ A10
	    dtrsm_("L", "L", "N", diag, &b, &p, &minusone, A11, ldA, A10, ldA);
	    // A11 = 1 / A11
		trinv1_(diag, &b, A11, ldA, &ione);

	    p += b;
    }
	return 0;
}

int trinv2_(char* diag, int* n, double* A, int* ldA, int* blocksize) {
	if (*n == 1) {
		if (diag[0] == 'N')
			*A = 1 / *A;
		return 0;
	}

	int ione = 1;
   	double one = 1;
	double minusone = -1;

    int p = 0;
    while (p < *n) {
	    int b = *blocksize;
	    if (p + b > *n)
	        b = *n - p;
		int r = *n - (p + b);

		// A00 
		// A10 A11 
		// A20 A21 A22
#define A00 (A)
#define A10 (A                  + p)
#define A11 (A + *ldA * p       + p)
#define A20 (A                  + p + b)
#define A21 (A + *ldA * p       + p + b)
#define A22 (A + *ldA * (p + b) + p + b)

	    // A21 = A22 \ A21
	    dtrsm_("L", "L", "N", diag, &r, &b, &one, A22, ldA, A21, ldA);
		// A21 = -A21 / A11
	    dtrsm_("R", "L", "N", diag, &r, &b, &minusone, A11, ldA, A21, ldA);
		// A11 = 1 / A11
		trinv2_(diag, &b, A11, ldA, &ione);

	    p += b;
    }
	return 0;
}

int trinv3_(char* diag, int* n, double* A, int* ldA, int* blocksize) {
	if (*n == 1) {
		if (diag[0] == 'N')
			*A = 1 / *A;
		return 0;
	}
	
	int ione = 1;
   	double one = 1;
	double minusone = -1;

    int p = 0;
    while (p < *n) {
	    int b = *blocksize;
	    if (p + b > *n)
			b = *n - p;
		int r = *n - (p + b);

		// A00 
		// A10 A11 
		// A20 A21 A22
#define A00 (A)
#define A10 (A                  + p)
#define A11 (A + *ldA * p       + p)
#define A20 (A                  + p + b)
#define A21 (A + *ldA * p       + p + b)
#define A22 (A + *ldA * (p + b) + p + b)

		// A21 = -A21 / A11
	    dtrsm_("R", "L", "N", diag, &r, &b, &minusone, A11, ldA, A21, ldA);
		// A20 = A21 * A10 + A20
		dgemm_("N", "N", &r, &p, &b, &one, A21, ldA, A10, ldA, &one, A20, ldA);
		// A10 = A11 \ A10
	    dtrsm_("L", "L", "N", diag, &b, &p, &one, A11, ldA, A10, ldA);
		// A11 = 1 / A11
		trinv3_(diag, &b, A11, ldA, &ione);

	    p += b;
    }
	return 0;
}

int trinv4_(char* diag, int* n, double* A, int* ldA, int* blocksize) {
	if (*n == 1) {
		if (diag[0] == 'N')
			*A = 1 / *A;
		return 0;
	}
	
	int ione = 1;
   	double one = 1;
	double minusone = -1;

    int p = 0;
    while (p < *n) {
	    int b = *blocksize;
	    if (p + b > *n)
	        b = *n - p;
		int r = *n - (p + b);

		// A00 
		// A10 A11 
		// A20 A21 A22
#define A00 (A)
#define A10 (A                  + p)
#define A11 (A + *ldA * p       + p)
#define A20 (A                  + p + b)
#define A21 (A + *ldA * p       + p + b)
#define A22 (A + *ldA * (p + b) + p + b)

		// A21 = -A22 \ A21
	    dtrsm_("L", "L", "N", diag, &r, &b, &minusone, A22, ldA, A21, ldA);
		// A20 = -A21 * A10 + A20
		dgemm_("N", "N", &r, &p, &b, &minusone, A21, ldA, A10, ldA, &one, A20, ldA);
		// A10 = A10 * A00
	    dtrmm_("R", "L", "N", diag, &b, &p, &one, A00, ldA, A10, ldA);
		// A11 = 1 / A11
	    trinv4_(diag, &b, A11, ldA, &ione);

	    p += b;
    }
	return 0;
}
\end{lstlisting}

        \section{LU Decomposition \texorpdfstring{$L U \leftarrow A$}{L U <- A}}
        \label{app:blockedalgs.lu}
\begin{lstlisting}[caption={\texttt{lu.c} --- LU decomposition.}]
extern void dtrmm_(char*, char*, char*, char*, int*, int*, double*, double*, int*, double*, int*);
extern void dtrsm_(char*, char*, char*, char*, int*, int*, double*, double*, int*, double*, int*);
extern void dgemm_(char*, char*, int*, int*, int*, double*, double*, int*, double*, int*, double*, double*, int*);

int lu1_(int* n, double* A, int* ldA, int* blocksize) {
	if (*n == 1)
		return 0;

	int ione = 1;
   	double one = 1;
	double minusone = -1;

    int p = 0;
    while (p < *n) {
 	    int b = *blocksize;
 	    if (p + b > *n)
		    b = *n - p;
		int r = *n - (p + b);

		// A00 A01 A02
		// A10 A11 A12
		// A20 A21 A22
#define A00 (A)
#define A01 (A + *ldA * p)
#define A02 (A + *ldA * (p + b))
#define A10 (A                  + p)
#define A11 (A + *ldA * p       + p)
#define A12 (A + *ldA * (p + b) + p)
#define A20 (A                  + p + b)
#define A21 (A + *ldA * p       + p + b)
#define A22 (A + *ldA * (p + b) + p + b)

    	// A01 = trilu( A00 ) \ A01
		dtrsm_("L", "L", "N", "U", &p, &b, &one, A00, ldA, A01, ldA);
		// A10 = A10 / triu( A00 )
	    dtrsm_("R", "U", "N", "N", &b, &p, &one, A00, ldA, A10, ldA);
		// A11 = A11 - A10 * A01
		dgemm_("N", "N", &b, &b, &p, &minusone, A10, ldA, A01, ldA, &one, A11, ldA);
		// A11 = LU(A11)
		lu1_(&b, A11, ldA, &ione);

	    p += b;
    }
	return 0;
}

int lu2_(int* n, double* A, int* ldA, int* blocksize) {
	if (*n == 1)
		return 0;

	int ione = 1;
   	double one = 1;
	double minusone = -1;

    int p = 0;
    while (p < *n) {
 	    int b = *blocksize;
 	    if (p + b > *n)
		    b = *n - p;
		int r = *n - (p + b);

		// A00 A01 A02
		// A10 A11 A12
		// A20 A21 A22
#define A00 (A)
#define A01 (A + *ldA * p)
#define A02 (A + *ldA * (p + b))
#define A10 (A                  + p)
#define A11 (A + *ldA * p       + p)
#define A12 (A + *ldA * (p + b) + p)
#define A20 (A                  + p + b)
#define A21 (A + *ldA * p       + p + b)
#define A22 (A + *ldA * (p + b) + p + b)

		// A10 = A10 / triu( A00 )
	    dtrsm_("R", "U", "N", "N", &b, &p, &one, A00, ldA, A10, ldA);
		// A11 = A11 - A10 * A01
		dgemm_("N", "N", &b, &b, &p, &minusone, A10, ldA, A01, ldA, &one, A11, ldA);
		// A11 = LU(A11)
		lu2_(&b, A11, ldA, &ione);
		// A12 = A12 - A10 * A02
		dgemm_("N", "N", &b, &r, &p, &minusone, A10, ldA, A02, ldA, &one, A12, ldA);
		// A12 = trilu( A11 ) \ A12
		dtrsm_("L", "L", "N", "U", &b, &r, &one, A11, ldA, A12, ldA);
	
	    p += b;
    }
	return 0;
}

int lu3_(int* n, double* A, int* ldA, int* blocksize) {
	if (*n == 1)
		return 0;

	int ione = 1;
   	double one = 1;
	double minusone = -1;

    int p = 0;
    while (p < *n) {
 	    int b = *blocksize;
 	    if (p + b > *n)
		    b = *n - p;
		int r = *n - (p + b);

		// A00 A01 A02
		// A10 A11 A12
		// A20 A21 A22
#define A00 (A)
#define A01 (A + *ldA * p)
#define A02 (A + *ldA * (p + b))
#define A10 (A                  + p)
#define A11 (A + *ldA * p       + p)
#define A12 (A + *ldA * (p + b) + p)
#define A20 (A                  + p + b)
#define A21 (A + *ldA * p       + p + b)
#define A22 (A + *ldA * (p + b) + p + b)


		// A01 = trilu( A00 ) \ A01
		dtrsm_("L", "L", "N", "U", &p, &b, &one, A00, ldA, A01, ldA);
		// A11 = A11 - A10 * A01
		dgemm_("N", "N", &b, &b, &p, &minusone, A10, ldA, A01, ldA, &one, A11, ldA);
		// A11 = LU(A11)
		lu3_(&b, A11, ldA, &ione);
		// A21 = A21 - A20 * A01
		dgemm_("N", "N", &r, &b, &p, &minusone, A20, ldA, A01, ldA, &one, A21, ldA);
		// A21 = A21 / triu( A11 )
	    dtrsm_("R", "U", "N", "N", &r, &b, &one, A11, ldA, A21, ldA);

	    p += b;
    }
	return 0;
}

int lu4_(int* n, double* A, int* ldA, int* blocksize) {
	if (*n == 1)
		return 0;

	int ione = 1;
   	double one = 1;
	double minusone = -1;

    int p = 0;
    while (p < *n) {
 	    int b = *blocksize;
 	    if (p + b > *n)
		    b = *n - p;
		int r = *n - (p + b);

		// A00 A01 A02
		// A10 A11 A12
		// A20 A21 A22
#define A00 (A)
#define A01 (A + *ldA * p)
#define A02 (A + *ldA * (p + b))
#define A10 (A                  + p)
#define A11 (A + *ldA * p       + p)
#define A12 (A + *ldA * (p + b) + p)
#define A20 (A                  + p + b)
#define A21 (A + *ldA * p       + p + b)
#define A22 (A + *ldA * (p + b) + p + b)

		// A11 = A11 - A10 * A01
		dgemm_("N", "N", &b, &b, &p, &minusone, A10, ldA, A01, ldA, &one, A11, ldA);
		// A11 = LU(A11)
		lu4_(&b, A11, ldA, &ione);
		// A12 = A12 - A10 * A02
		dgemm_("N", "N", &b, &r, &p, &minusone, A10, ldA, A02, ldA, &one, A12, ldA);
		// A12 = trilu( A11 ) \ A12
		dtrsm_("L", "L", "N", "U", &b, &r, &one, A11, ldA, A12, ldA);
		// A21 = A21 - A20 * A01
		dgemm_("N", "N", &r, &b, &p, &minusone, A20, ldA, A01, ldA, &one, A21, ldA);
		// A21 = A21 / triu( A11 )
	    dtrsm_("R", "U", "N", "N", &r, &b, &one, A11, ldA, A21, ldA);

	    p += b;
    }
	return 0;
}

int lu5_(int* n, double* A, int* ldA, int* blocksize) {
	if (*n == 1)
		return 0;

	int ione = 1;
   	double one = 1;
	double minusone = -1;

    int p = 0;
    while (p < *n) {
 	    int b = *blocksize;
 	    if (p + b > *n)
		    b = *n - p;
		int r = *n - (p + b);

		// A00 A01 A02
		// A10 A11 A12
		// A20 A21 A22
#define A00 (A)
#define A01 (A + *ldA * p)
#define A02 (A + *ldA * (p + b))
#define A10 (A                  + p)
#define A11 (A + *ldA * p       + p)
#define A12 (A + *ldA * (p + b) + p)
#define A20 (A                  + p + b)
#define A21 (A + *ldA * p       + p + b)
#define A22 (A + *ldA * (p + b) + p + b)

		// A11 = LU(A11)
		lu5_(&b, A11, ldA, &ione);
		// A12 = trilu( A11 ) \ A12
		dtrsm_("L", "L", "N", "U", &b, &r, &one, A11, ldA, A12, ldA);
		// A21 = A21 / triu( A11 )
	    dtrsm_("R", "U", "N", "N", &r, &b, &one, A11, ldA, A21, ldA);
		// A22 = A22 - A21 * A12
		dgemm_("N", "N", &r, &r, &b, &minusone, A21, ldA, A12, ldA, &one, A22, ldA);

	    p += b;
    }
	return 0;
}
\end{lstlisting}

        \section{Sylvester Equation: Solving \texorpdfstring{$L X + X U = C$ for $X$}{L X + X U = C for X}}
        \label{app:blockedalgs.sylv}
        Since the file containing all 16 variants of the Sylvester Equation solver spans 1,127 lines, we only give the full code up to and including the first variant.
For variants 2 through 16, we list only the update statements; the surrounding part of the algorithms is, up to the name, identical to the first variant.

\begin{lstlisting}[caption={\texttt{sylv.c} (extract up to first algorithm) --- Solution of Sylvester Equation.}]
extern void dgemm_(char*, char*, int*, int*, int*, double*, double*, int*, double*, int*, double*, double*, int*);

		// L00 L01 L02
		// L10 L11 L12
		// L20 L21 L22
#define L00m Lp 
#define L01m Lp 
#define L02m Lp 
#define L10m Lb 
#define L11m Lb 
#define L12m Lb 
#define L20m Lr 
#define L21m Lr 
#define L22m Lr 
#define L00n Lp 
#define L01n Lb 
#define L02n Lr 
#define L10n Lp 
#define L11n Lb 
#define L12n Lr 
#define L20n Lp 
#define L21n Lb 
#define L22n Lr 
#define L00 (L)
#define L01 (L + *ldL * Lp)
#define L02 (L + *ldL * (Lp + Lb))
#define L10 (L                    + Lp)
#define L11 (L + *ldL * Lp        + Lp)
#define L12 (L + *ldL * (Lp + Lb) + Lp)
#define L20 (L                    + Lp + Lb)
#define L21 (L + *ldL * Lp        + Lp + Lb)
#define L22 (L + *ldL * (Lp + Lb) + Lp + Lb)

		// U00 U01 U02
		// U10 U11 U12
		// U20 U21 U22
#define U00m Up 
#define U01m Up 
#define U02m Up 
#define U10m Ub 
#define U11m Ub 
#define U12m Ub 
#define U20m Ur 
#define U21m Ur 
#define U22m Ur 
#define U00n Up 
#define U01n Ub 
#define U02n Ur 
#define U10n Up 
#define U11n Ub 
#define U12n Ur 
#define U20n Up 
#define U21n Ub 
#define U22n Ur 
#define U00 (U)
#define U01 (U + *ldU * Up)
#define U02 (U + *ldU * (Up + Ub))
#define U10 (U                    + Up)
#define U11 (U + *ldU * Up        + Up)
#define U12 (U + *ldU * (Up + Ub) + Up)
#define U20 (U                    + Up + Ub)
#define U21 (U + *ldU * Up        + Up + Ub)
#define U22 (U + *ldU * (Up + Ub) + Up + Ub)

		// X00 X01 X02
		// X10 X11 X12
		// X20 X21 X22
#define X00m Lp 
#define X01m Lp 
#define X02m Lp 
#define X10m Lb 
#define X11m Lb 
#define X12m Lb 
#define X20m Lr 
#define X21m Lr 
#define X22m Lr 
#define X00n Up 
#define X01n Ub 
#define X02n Ur 
#define X10n Up 
#define X11n Ub 
#define X12n Ur 
#define X20n Up 
#define X21n Ub 
#define X22n Ur 
#define X00 (X)
#define X01 (X + *ldX * Up)
#define X02 (X + *ldX * (Up + Ub))
#define X10 (X                    + Lp)
#define X11 (X + *ldX * Up        + Lp)
#define X12 (X + *ldX * (Up + Ub) + Lp)
#define X20 (X                    + Lp + Lb)
#define X21 (X + *ldX * Up        + Lp + Lb)
#define X22 (X + *ldX * (Up + Ub) + Lp + Lb)


int sylv1_(int* m, int* n, double* L, int* ldL, double* U, int* ldU, double* X, int* ldX, int* blocksize) {
#undef sylv_
#define sylv_ sylv1_
	if (*m * *n == 0)
		return 0;
	if (*m * *n == 1) {
		*X /= *L + *U;
		return 0;
	}

	int ione = 1;
   	double one = 1;
	double mone = -1;

    int p = 0;
 	int b = *blocksize;
	if (b >= *m && b >= *n)
		b = 1;
    while (p < *n || p < *m) {

		int Lp = p;
		int Lb = b;
		if (Lp >= *m) {
			Lp = *m;
			Lb = 0;
		} else
			if (Lp + Lb > *m)
				Lb = *m - Lp;
		int Lr = *m - (Lp + Lb);

		int Up = p;
		int Ub = b;
		if (Up >= *n) {
			Up = *n;
			Ub = 0;
		} else
			if (Up + Ub > *n)
				Ub = *n - Up;
		int Ur = *n - (Up + Ub);

		// X01 = -X00 U01 + X01
		dgemm_("N", "N", &X01m, &X01n, &X00n, &mone, X00, ldX, U01, ldU, &one, X01, ldX);
		// X10 = -L10 X00 + X10
		dgemm_("N", "N", &X10m, &X10n, &L10n, &mone, L10, ldL, X00, ldX, &one, X10, ldX);
		// X01 = sylv(L00, U11, X01)
		sylv_(&X01m, &X01n, L00, ldL, U11, ldU, X01, ldX, blocksize);
		// X10 = sylv(L11, U00, X10)
		sylv_(&X10m, &X10n, L11, ldL, U00, ldU, X10, ldX, blocksize);
		// X11 = -X10 U01 + X11
		dgemm_("N", "N", &X11m, &X11n, &X10n, &mone, X10, ldX, U01, ldU, &one, X11, ldX);
		// X11 = -L10 X01 + X11
		dgemm_("N", "N", &X11m, &X11n, &L10n, &mone, L10, ldL, X01, ldX, &one, X11, ldX);
		// X11 = sylv(L11, U11, X11)
		sylv_(&X11m, &X11n, L11, ldL, U11, ldU, X11, ldX, blocksize);

	    p += b;
    }
	return 0;
}
\end{lstlisting}

\begin{lstlisting}[caption={\texttt{sylv.c} (2\textsuperscript{nd} algorithm updates) --- Solution of Sylvester Equation.},firstnumber=197]
// X01 = -X00 U01 + X01
dgemm_("N", "N", &X01m, &X01n, &X00n, &mone, X00, ldX, U01, ldU, &one, X01, ldX);
// X10 = sylv(L11, U00, X10)
sylv_(&X10m, &X10n, L11, ldL, U00, ldU, X10, ldX, blocksize);
// X01 = sylv(L00, U11, X01)
sylv_(&X01m, &X01n, L00, ldL, U11, ldU, X01, ldX, blocksize);
// X11 = -X10 U01 + X11
dgemm_("N", "N", &X11m, &X11n, &X10n, &mone, X10, ldX, U01, ldU, &one, X11, ldX);
// X20 = -L21 X10 + X20
dgemm_("N", "N", &X20m, &X20n, &L21n, &mone, L21, ldL, X10, ldX, &one, X20, ldX);
// X11 = -L10 X01 + X11
dgemm_("N", "N", &X11m, &X11n, &L10n, &mone, L10, ldL, X01, ldX, &one, X11, ldX);
// X11 = sylv(L11, U11, X11)
sylv_(&X11m, &X11n, L11, ldL, U11, ldU, X11, ldX, blocksize);
// X21 = -L21 X11 + X21
dgemm_("N", "N", &X21m, &X21n, &L21n, &mone, L21, ldL, X11, ldX, &one, X21, ldX);
// X21 = -L20 X01 + X21
dgemm_("N", "N", &X21m, &X21n, &L20n, &mone, L20, ldL, X01, ldX, &one, X21, ldX);
\end{lstlisting}
\begin{lstlisting}[caption={\texttt{sylv.c} (3\textsuperscript{rd} algorithm updates) --- Solution of Sylvester Equation.},firstnumber=261]
// X01 = -X00 U01 + X01
dgemm_("N", "N", &X01m, &X01n, &X00n, &mone, X00, ldX, U01, ldU, &one, X01, ldX);
// X11 = -X10 U01 + X11
dgemm_("N", "N", &X11m, &X11n, &X10n, &mone, X10, ldX, U01, ldU, &one, X11, ldX);
// X21 = -X20 U01 + X21
dgemm_("N", "N", &X21m, &X21n, &X20n, &mone, X20, ldX, U01, ldU, &one, X21, ldX);
// X01 = sylv(L00, U11, X01)
sylv_(&X01m, &X01n, L00, ldL, U11, ldU, X01, ldX, blocksize);
// X11 = -L10 X01 + X11
dgemm_("N", "N", &X11m, &X11n, &L10n, &mone, L10, ldL, X01, ldX, &one, X11, ldX);
// X11 = sylv(L11, U11, X11)
sylv_(&X11m, &X11n, L11, ldL, U11, ldU, X11, ldX, blocksize);
// X21 = -L21 X11 + X21
dgemm_("N", "N", &X21m, &X21n, &L21n, &mone, L21, ldL, X11, ldX, &one, X21, ldX);
// X21 = -L20 X01 + X21
dgemm_("N", "N", &X21m, &X21n, &L20n, &mone, L20, ldL, X01, ldX, &one, X21, ldX);
// X21 = sylv(L22, U11, X21)
sylv_(&X21m, &X21n, L22, ldL, U11, ldU, X21, ldX, blocksize);
\end{lstlisting}
\begin{lstlisting}[caption={\texttt{sylv.c} (4\textsuperscript{th} algorithm updates) --- Solution of Sylvester Equation.},firstnumber=325]
// X01 = -X00 U01 + X01
dgemm_("N", "N", &X01m, &X01n, &X00n, &mone, X00, ldX, U01, ldU, &one, X01, ldX);
// X12 = -X10 U02 + X12
dgemm_("N", "N", &X12m, &X12n, &X10n, &mone, X10, ldX, U02, ldU, &one, X12, ldX);
// X01 = sylv(L00, U11, X01)
sylv_(&X01m, &X01n, L00, ldL, U11, ldU, X01, ldX, blocksize);
// X11 = -L10 X01 + X11
dgemm_("N", "N", &X11m, &X11n, &L10n, &mone, L10, ldL, X01, ldX, &one, X11, ldX);
// X11 = sylv(L11, U11, X11)
sylv_(&X11m, &X11n, L11, ldL, U11, ldU, X11, ldX, blocksize);
// X21 = -L21 X11 + X21
dgemm_("N", "N", &X21m, &X21n, &L21n, &mone, L21, ldL, X11, ldX, &one, X21, ldX);
// X21 = -L20 X01 + X21
dgemm_("N", "N", &X21m, &X21n, &L20n, &mone, L20, ldL, X01, ldX, &one, X21, ldX);
// X21 = sylv(L22, U11, X21)
sylv_(&X21m, &X21n, L22, ldL, U11, ldU, X21, ldX, blocksize);
// X22 = -X21 U12 + X22
dgemm_("N", "N", &X22m, &X22n, &X21n, &mone, X21, ldX, U12, ldU, &one, X22, ldX);
\end{lstlisting}
\begin{lstlisting}[caption={\texttt{sylv.c} (5\textsuperscript{th} algorithm updates) --- Solution of Sylvester Equation.},firstnumber=389]
// X01 = sylv(L00, U11, X01)
sylv_(&X01m, &X01n, L00, ldL, U11, ldU, X01, ldX, blocksize);
// X10 = -L10 X00 + X10
dgemm_("N", "N", &X10m, &X10n, &L10n, &mone, L10, ldL, X00, ldX, &one, X10, ldX);
// X02 = -X01 U12 + X02
dgemm_("N", "N", &X02m, &X02n, &X01n, &mone, X01, ldX, U12, ldU, &one, X02, ldX);
// X10 = sylv(L11, U00, X10)
sylv_(&X10m, &X10n, L11, ldL, U00, ldU, X10, ldX, blocksize);
// X11 = -X10 U01 + X11
dgemm_("N", "N", &X11m, &X11n, &X10n, &mone, X10, ldX, U01, ldU, &one, X11, ldX);
// X11 = -L10 X01 + X11
dgemm_("N", "N", &X11m, &X11n, &L10n, &mone, L10, ldL, X01, ldX, &one, X11, ldX);
// X11 = sylv(L11, U11, X11)
sylv_(&X11m, &X11n, L11, ldL, U11, ldU, X11, ldX, blocksize);
// X12 = -X11 U12 + X12
dgemm_("N", "N", &X12m, &X12n, &X11n, &mone, X11, ldX, U12, ldU, &one, X12, ldX);
// X12 = -X10 U02 + X12
dgemm_("N", "N", &X12m, &X12n, &X10n, &mone, X10, ldX, U02, ldU, &one, X12, ldX);
\end{lstlisting}
\begin{lstlisting}[caption={\texttt{sylv.c} (6\textsuperscript{th} algorithm updates) --- Solution of Sylvester Equation.},firstnumber=453]
// X01 = sylv(L00, U11, X01)
sylv_(&X01m, &X01n, L00, ldL, U11, ldU, X01, ldX, blocksize);
// X10 = sylv(L11, U00, X10)
sylv_(&X10m, &X10n, L11, ldL, U00, ldU, X10, ldX, blocksize);
// X02 = -X01 U12 + X02
dgemm_("N", "N", &X02m, &X02n, &X01n, &mone, X01, ldX, U12, ldU, &one, X02, ldX);
// X11 = -X10 U01 + X11
dgemm_("N", "N", &X11m, &X11n, &X10n, &mone, X10, ldX, U01, ldU, &one, X11, ldX);
// X20 = -L21 X10 + X20
dgemm_("N", "N", &X20m, &X20n, &L21n, &mone, L21, ldL, X10, ldX, &one, X20, ldX);
// X11 = -L10 X01 + X11
dgemm_("N", "N", &X11m, &X11n, &L10n, &mone, L10, ldL, X01, ldX, &one, X11, ldX);
// X11 = sylv(L11, U11, X11)
sylv_(&X11m, &X11n, L11, ldL, U11, ldU, X11, ldX, blocksize);
// X12 = -X11 U12 + X12
dgemm_("N", "N", &X12m, &X12n, &X11n, &mone, X11, ldX, U12, ldU, &one, X12, ldX);
// X21 = -L21 X11 + X21
dgemm_("N", "N", &X21m, &X21n, &L21n, &mone, L21, ldL, X11, ldX, &one, X21, ldX);
// X12 = -X10 U02 + X12
dgemm_("N", "N", &X12m, &X12n, &X10n, &mone, X10, ldX, U02, ldU, &one, X12, ldX);
// X21 = -L20 X01 + X21
dgemm_("N", "N", &X21m, &X21n, &L20n, &mone, L20, ldL, X01, ldX, &one, X21, ldX);
\end{lstlisting}
\begin{lstlisting}[caption={\texttt{sylv.c} (7\textsuperscript{th} algorithm updates) --- Solution of Sylvester Equation.},firstnumber=521]
// X01 = sylv(L00, U11, X01)
sylv_(&X01m, &X01n, L00, ldL, U11, ldU, X01, ldX, blocksize);
// X11 = -X10 U01 + X11
dgemm_("N", "N", &X11m, &X11n, &X10n, &mone, X10, ldX, U01, ldU, &one, X11, ldX);
// X21 = -X20 U01 + X21
dgemm_("N", "N", &X21m, &X21n, &X20n, &mone, X20, ldX, U01, ldU, &one, X21, ldX);
// X02 = -X01 U12 + X02
dgemm_("N", "N", &X02m, &X02n, &X01n, &mone, X01, ldX, U12, ldU, &one, X02, ldX);
// X11 = -L10 X01 + X11
dgemm_("N", "N", &X11m, &X11n, &L10n, &mone, L10, ldL, X01, ldX, &one, X11, ldX);
// X11 = sylv(L11, U11, X11)
sylv_(&X11m, &X11n, L11, ldL, U11, ldU, X11, ldX, blocksize);
// X12 = -X11 U12 + X12
dgemm_("N", "N", &X12m, &X12n, &X11n, &mone, X11, ldX, U12, ldU, &one, X12, ldX);
// X21 = -L21 X11 + X21
dgemm_("N", "N", &X21m, &X21n, &L21n, &mone, L21, ldL, X11, ldX, &one, X21, ldX);
// X12 = -X10 U02 + X12
dgemm_("N", "N", &X12m, &X12n, &X10n, &mone, X10, ldX, U02, ldU, &one, X12, ldX);
// X21 = -L20 X01 + X21
dgemm_("N", "N", &X21m, &X21n, &L20n, &mone, L20, ldL, X01, ldX, &one, X21, ldX);
// X21 = sylv(L22, U11, X21)
sylv_(&X21m, &X21n, L22, ldL, U11, ldU, X21, ldX, blocksize);
\end{lstlisting}
\begin{lstlisting}[caption={\texttt{sylv.c} (8\textsuperscript{th} algorithm updates) --- Solution of Sylvester Equation.},firstnumber=589]
// X01 = sylv(L00, U11, X01)
sylv_(&X01m, &X01n, L00, ldL, U11, ldU, X01, ldX, blocksize);
// X02 = -X01 U12 + X02
dgemm_("N", "N", &X02m, &X02n, &X01n, &mone, X01, ldX, U12, ldU, &one, X02, ldX);
// X11 = -L10 X01 + X11
dgemm_("N", "N", &X11m, &X11n, &L10n, &mone, L10, ldL, X01, ldX, &one, X11, ldX);
// X11 = sylv(L11, U11, X11)
sylv_(&X11m, &X11n, L11, ldL, U11, ldU, X11, ldX, blocksize);
// X12 = -X11 U12 + X12
dgemm_("N", "N", &X12m, &X12n, &X11n, &mone, X11, ldX, U12, ldU, &one, X12, ldX);
// X21 = -L21 X11 + X21
dgemm_("N", "N", &X21m, &X21n, &L21n, &mone, L21, ldL, X11, ldX, &one, X21, ldX);
// X21 = -L20 X01 + X21
dgemm_("N", "N", &X21m, &X21n, &L20n, &mone, L20, ldL, X01, ldX, &one, X21, ldX);
// X21 = sylv(L22, U11, X21)
sylv_(&X21m, &X21n, L22, ldL, U11, ldU, X21, ldX, blocksize);
// X22 = -X21 U12 + X22
dgemm_("N", "N", &X22m, &X22n, &X21n, &mone, X21, ldX, U12, ldU, &one, X22, ldX);
\end{lstlisting}
\begin{lstlisting}[caption={\texttt{sylv.c} (9\textsuperscript{th} algorithm updates) --- Solution of Sylvester Equation.},firstnumber=653]
// X01 = sylv(L00, U11, X01)
sylv_(&X01m, &X01n, L00, ldL, U11, ldU, X01, ldX, blocksize);
// X02 = -X01 U12 + X02
dgemm_("N", "N", &X02m, &X02n, &X01n, &mone, X01, ldX, U12, ldU, &one, X02, ldX);
// X11 = -L10 X01 + X11
dgemm_("N", "N", &X11m, &X11n, &L10n, &mone, L10, ldL, X01, ldX, &one, X11, ldX);
// X11 = sylv(L11, U11, X11)
sylv_(&X11m, &X11n, L11, ldL, U11, ldU, X11, ldX, blocksize);
// X12 = -X11 U12 + X12
dgemm_("N", "N", &X12m, &X12n, &X11n, &mone, X11, ldX, U12, ldU, &one, X12, ldX);
// X21 = -L21 X11 + X21
dgemm_("N", "N", &X21m, &X21n, &L21n, &mone, L21, ldL, X11, ldX, &one, X21, ldX);
// X21 = -L20 X01 + X21
dgemm_("N", "N", &X21m, &X21n, &L20n, &mone, L20, ldL, X01, ldX, &one, X21, ldX);
// X21 = sylv(L22, U11, X21)
sylv_(&X21m, &X21n, L22, ldL, U11, ldU, X21, ldX, blocksize);
// X22 = -X21 U12 + X22
dgemm_("N", "N", &X22m, &X22n, &X21n, &mone, X21, ldX, U12, ldU, &one, X22, ldX);
\end{lstlisting}
\begin{lstlisting}[caption={\texttt{sylv.c} (10\textsuperscript{th} algorithm updates) --- Solution of Sylvester Equation.},firstnumber=717]
// X10 = -L10 X00 + X10
dgemm_("N", "N", &X10m, &X10n, &L10n, &mone, L10, ldL, X00, ldX, &one, X10, ldX);
// X21 = -L20 X01 + X21
dgemm_("N", "N", &X21m, &X21n, &L20n, &mone, L20, ldL, X01, ldX, &one, X21, ldX);
// X10 = sylv(L11, U00, X10)
sylv_(&X10m, &X10n, L11, ldL, U00, ldU, X10, ldX, blocksize);
// X11 = -X10 U01 + X11
dgemm_("N", "N", &X11m, &X11n, &X10n, &mone, X10, ldX, U01, ldU, &one, X11, ldX);
// X11 = sylv(L11, U11, X11)
sylv_(&X11m, &X11n, L11, ldL, U11, ldU, X11, ldX, blocksize);
// X12 = -X11 U12 + X12
dgemm_("N", "N", &X12m, &X12n, &X11n, &mone, X11, ldX, U12, ldU, &one, X12, ldX);
// X12 = -X10 U02 + X12
dgemm_("N", "N", &X12m, &X12n, &X10n, &mone, X10, ldX, U02, ldU, &one, X12, ldX);
// X12 = sylv(L11, U22, X12)
sylv_(&X12m, &X12n, L11, ldL, U22, ldU, X12, ldX, blocksize);
// X22 = -L21 X12 + X22
dgemm_("N", "N", &X22m, &X22n, &L21n, &mone, L21, ldL, X12, ldX, &one, X22, ldX);
\end{lstlisting}
\begin{lstlisting}[caption={\texttt{sylv.c} (11\textsuperscript{th} algorithm updates) --- Solution of Sylvester Equation.},firstnumber=781]
// X10 = -L10 X00 + X10
dgemm_("N", "N", &X10m, &X10n, &L10n, &mone, L10, ldL, X00, ldX, &one, X10, ldX);
// X21 = -L20 X01 + X21
dgemm_("N", "N", &X21m, &X21n, &L20n, &mone, L20, ldL, X01, ldX, &one, X21, ldX);
// X10 = sylv(L11, U00, X10)
sylv_(&X10m, &X10n, L11, ldL, U00, ldU, X10, ldX, blocksize);
// X11 = -X10 U01 + X11
dgemm_("N", "N", &X11m, &X11n, &X10n, &mone, X10, ldX, U01, ldU, &one, X11, ldX);
// X11 = sylv(L11, U11, X11)
sylv_(&X11m, &X11n, L11, ldL, U11, ldU, X11, ldX, blocksize);
// X12 = -X11 U12 + X12
dgemm_("N", "N", &X12m, &X12n, &X11n, &mone, X11, ldX, U12, ldU, &one, X12, ldX);
// X12 = -X10 U02 + X12
dgemm_("N", "N", &X12m, &X12n, &X10n, &mone, X10, ldX, U02, ldU, &one, X12, ldX);
// X12 = sylv(L11, U22, X12)
sylv_(&X12m, &X12n, L11, ldL, U22, ldU, X12, ldX, blocksize);
// X22 = -L21 X12 + X22
dgemm_("N", "N", &X22m, &X22n, &L21n, &mone, L21, ldL, X12, ldX, &one, X22, ldX);
\end{lstlisting}
\begin{lstlisting}[caption={\texttt{sylv.c} (12\textsuperscript{th} algorithm updates) --- Solution of Sylvester Equation.},firstnumber=849]
// X10 = -L10 X00 + X10
dgemm_("N", "N", &X10m, &X10n, &L10n, &mone, L10, ldL, X00, ldX, &one, X10, ldX);
// X21 = -L20 X01 + X21
dgemm_("N", "N", &X21m, &X21n, &L20n, &mone, L20, ldL, X01, ldX, &one, X21, ldX);
// X10 = sylv(L11, U00, X10)
sylv_(&X10m, &X10n, L11, ldL, U00, ldU, X10, ldX, blocksize);
// X11 = -X10 U01 + X11
dgemm_("N", "N", &X11m, &X11n, &X10n, &mone, X10, ldX, U01, ldU, &one, X11, ldX);
// X11 = sylv(L11, U11, X11)
sylv_(&X11m, &X11n, L11, ldL, U11, ldU, X11, ldX, blocksize);
// X12 = -X11 U12 + X12
dgemm_("N", "N", &X12m, &X12n, &X11n, &mone, X11, ldX, U12, ldU, &one, X12, ldX);
// X12 = -X10 U02 + X12
dgemm_("N", "N", &X12m, &X12n, &X10n, &mone, X10, ldX, U02, ldU, &one, X12, ldX);
// X12 = sylv(L11, U22, X12)
sylv_(&X12m, &X12n, L11, ldL, U22, ldU, X12, ldX, blocksize);
// X22 = -L21 X12 + X22
dgemm_("N", "N", &X22m, &X22n, &L21n, &mone, L21, ldL, X12, ldX, &one, X22, ldX);
\end{lstlisting}
\begin{lstlisting}[caption={\texttt{sylv.c} (13\textsuperscript{th} algorithm updates) --- Solution of Sylvester Equation.},firstnumber=913]
// X11 = -X10 U01 + X11
dgemm_("N", "N", &X11m, &X11n, &X10n, &mone, X10, ldX, U01, ldU, &one, X11, ldX);
// X21 = -X20 U01 + X21
dgemm_("N", "N", &X21m, &X21n, &X20n, &mone, X20, ldX, U01, ldU, &one, X21, ldX);
// X11 = -L10 X01 + X11
dgemm_("N", "N", &X11m, &X11n, &L10n, &mone, L10, ldL, X01, ldX, &one, X11, ldX);
// X11 = sylv(L11, U11, X11)
sylv_(&X11m, &X11n, L11, ldL, U11, ldU, X11, ldX, blocksize);
// X12 = -X11 U12 + X12
dgemm_("N", "N", &X12m, &X12n, &X11n, &mone, X11, ldX, U12, ldU, &one, X12, ldX);
// X21 = -L21 X11 + X21
dgemm_("N", "N", &X21m, &X21n, &L21n, &mone, L21, ldL, X11, ldX, &one, X21, ldX);
// X12 = -X10 U02 + X12
dgemm_("N", "N", &X12m, &X12n, &X10n, &mone, X10, ldX, U02, ldU, &one, X12, ldX);
// X21 = -L20 X01 + X21
dgemm_("N", "N", &X21m, &X21n, &L20n, &mone, L20, ldL, X01, ldX, &one, X21, ldX);
// X12 = -L10 X02 + X12
dgemm_("N", "N", &X12m, &X12n, &L10n, &mone, L10, ldL, X02, ldX, &one, X12, ldX);
// X21 = sylv(L22, U11, X21)
sylv_(&X21m, &X21n, L22, ldL, U11, ldU, X21, ldX, blocksize);
// X12 = sylv(L11, U22, X12)
sylv_(&X12m, &X12n, L11, ldL, U22, ldU, X12, ldX, blocksize);
\end{lstlisting}
\begin{lstlisting}[caption={\texttt{sylv.c} (14\textsuperscript{th} algorithm updates) --- Solution of Sylvester Equation.},firstnumber=981]
// X11 = -X10 U01 + X11
dgemm_("N", "N", &X11m, &X11n, &X10n, &mone, X10, ldX, U01, ldU, &one, X11, ldX);
// X21 = -X20 U01 + X21
dgemm_("N", "N", &X21m, &X21n, &X20n, &mone, X20, ldX, U01, ldU, &one, X21, ldX);
// X11 = sylv(L11, U11, X11)
sylv_(&X11m, &X11n, L11, ldL, U11, ldU, X11, ldX, blocksize);
// X12 = -X11 U12 + X12
dgemm_("N", "N", &X12m, &X12n, &X11n, &mone, X11, ldX, U12, ldU, &one, X12, ldX);
// X21 = -L21 X11 + X21
dgemm_("N", "N", &X21m, &X21n, &L21n, &mone, L21, ldL, X11, ldX, &one, X21, ldX);
// X12 = -X10 U02 + X12
dgemm_("N", "N", &X12m, &X12n, &X10n, &mone, X10, ldX, U02, ldU, &one, X12, ldX);
// X21 = sylv(L22, U11, X21)
sylv_(&X21m, &X21n, L22, ldL, U11, ldU, X21, ldX, blocksize);
// X12 = sylv(L11, U22, X12)
sylv_(&X12m, &X12n, L11, ldL, U22, ldU, X12, ldX, blocksize);
// X22 = -L21 X12 + X22
dgemm_("N", "N", &X22m, &X22n, &L21n, &mone, L21, ldL, X12, ldX, &one, X22, ldX);
\end{lstlisting}
\begin{lstlisting}[caption={\texttt{sylv.c} (15\textsuperscript{th} algorithm updates) --- Solution of Sylvester Equation.},firstnumber=1045]
// X11 = -L10 X01 + X11
dgemm_("N", "N", &X11m, &X11n, &L10n, &mone, L10, ldL, X01, ldX, &one, X11, ldX);
// X11 = sylv(L11, U11, X11)
sylv_(&X11m, &X11n, L11, ldL, U11, ldU, X11, ldX, blocksize);
// X12 = -X11 U12 + X12
dgemm_("N", "N", &X12m, &X12n, &X11n, &mone, X11, ldX, U12, ldU, &one, X12, ldX);
// X21 = -L21 X11 + X21
dgemm_("N", "N", &X21m, &X21n, &L21n, &mone, L21, ldL, X11, ldX, &one, X21, ldX);
// X12 = -L10 X02 + X12
dgemm_("N", "N", &X12m, &X12n, &L10n, &mone, L10, ldL, X02, ldX, &one, X12, ldX);
// X21 = -L20 X01 + X21
dgemm_("N", "N", &X21m, &X21n, &L20n, &mone, L20, ldL, X01, ldX, &one, X21, ldX);
// X12 = sylv(L11, U22, X12)
sylv_(&X12m, &X12n, L11, ldL, U22, ldU, X12, ldX, blocksize);
// X21 = sylv(L22, U11, X21)
sylv_(&X21m, &X21n, L22, ldL, U11, ldU, X21, ldX, blocksize);
// X22 = -X21 U12 + X22
dgemm_("N", "N", &X22m, &X22n, &X21n, &mone, X21, ldX, U12, ldU, &one, X22, ldX);
\end{lstlisting}
\begin{lstlisting}[caption={\texttt{sylv.c} (16\textsuperscript{th} algorithm updates) --- Solution of Sylvester Equation.},firstnumber=1109]
// X11 = sylv(L11, U11, X11)
sylv_(&X11m, &X11n, L11, ldL, U11, ldU, X11, ldX, blocksize);
// X12 = -X11 U12 + X12
dgemm_("N", "N", &X12m, &X12n, &X11n, &mone, X11, ldX, U12, ldU, &one, X12, ldX);
// X21 = -L21 X11 + X21
dgemm_("N", "N", &X21m, &X21n, &L21n, &mone, L21, ldL, X11, ldX, &one, X21, ldX);
// X12 = sylv(L11, U22, X12)
sylv_(&X12m, &X12n, L11, ldL, U22, ldU, X12, ldX, blocksize);
// X21 = sylv(L22, U11, X21)
sylv_(&X21m, &X21n, L22, ldL, U11, ldU, X21, ldX, blocksize);
// X22 = -X21 U12 + X22
dgemm_("N", "N", &X22m, &X22n, &X21n, &mone, X21, ldX, U12, ldU, &one, X22, ldX);
// X22 = -L21 X12 + X22
dgemm_("N", "N", &X22m, &X22n, &L21n, &mone, L21, ldL, X12, ldX, &one, X22, ldX);
\end{lstlisting}

\end{document}